\documentclass[
twoside,openright,
12pt,
openbib,
a4paper,english,catalan,boldpropi]{thesis}
\usepackage{epsf,epsfig,amsfonts}
\usepackage[catalan]{babel}
\usepackage{amsmath,amssymb}
\usepackage[latin1]{inputenc}

\renewcommand{\bigskip}{\vspace{.8cm}}

\newcommand{\be}{\begin{equation}}
\newcommand{\ee}{\end{equation}}
\newcommand{\bs}{\begin{subequations}}
\newcommand{\es}{\end{subequations}}
\newcommand{\bea}{\begin{eqnarray}}
\newcommand{\eea}{\end{eqnarray}}

\newcommand{\ket}[1]{{| #1 \rangle}}

\newcommand{\sac}{\, , \qquad}
\newcommand{\eg}{{\it e.g. }}
\newcommand{\ie}{{\it i.e. }}
\newcommand{\ads}[1]{\mbox{${AdS}_{#1}$}}
\newcommand{\adss}[2]{\mbox{$AdS_{#1}\times {S}^{#2}$}}
\def\CR{\mathbb{R}}
\def\CM{\mathcal{M}}
\def\CC{\mathcal{C}}

\def\Poin{Poincar\'e }
\newcommand{\bref}[1]{(\ref{#1})}
\def\bitem{\begin{itemize}}
\def\eitem{\end{itemize}}
\def\rvec{\vec{r}}
\def\Avec{\vec{A}}
\def\vvec{\vec{v}}
\def\pivec{\vec{\pi}}
\def\pvec{\vec{p}}
\def\undos{\frac{1}{2}}
\def\l {\lambda}
\def\a {\alpha}
\def\ap {\alpha'}
\def\b {\beta}
\def\h {\eta}
\def\g {\gamma}
\def\G {\Gamma}
\def\d {\delta}
\def\r {\rho}
\def\s {\sigma}
\def\e {\epsilon}

\def\vp {\varphi}
\def\T {\Theta}
\def\w{\omega}
\def\O{\Omega}
\def\L{\Lambda}
\def\t{\theta}
\def\espai{\;\;\;\;\;\;\;}
\def\zespai{\;\;\;\;}
\def\hO{\hat{O}}
\def\hF{\hat{F}}
\def\hA{\hat{A}}

\def\hx{\hat{x}}
\def\hp{\hat{p}}
\def\hl{\hat{\lambda}}
\def\tE{\tilde{E}}
\def\tL{\tilde{L}}
\def\pa{\partial}
\def\cpa{\bar{\pa}}
\def\cz{\bar{z}}
\def\punt{\, \cdot}
\def\pop{p \, \circ p}
\def\zavall{\vspace{0.2cm}}
\def\avall{\vspace{1cm}}
\def\enub{\begin{enumerate}}
\def\enue{\end{enumerate}}

\def\4n{$\caln=4$}
\def\gym{g_{YM}}
\def\Tr{\mbox{Tr}}
\def\0a{\ap \rightarrow 0}
\def\5ads{$AdS_5 \times S^5$}
\def\we{\wedge}
\def\str{\mbox{str}}
\def\go0{\rightarrow 0}
\def\sothat{\,\,\, \Rightarrow \,\,\,}
\def\dreta{\rightarrow}
\def\S{\Sigma}
\def\kah{K\"{a}hler }
\def\su2{SU(2)_L \times SU(2)_R}
\def\nun{N\'u\~{n}ez }

\newcommand{\caixa}[1]{\noindent\fbox{\noindent\parbox{5.2in}{#1}}}
\newcommand{\cent}[1]{\begin{center}{#1}\end{center}}
\newcommand{\petit}[1]{\small{#1}\normalsize}
\newcommand{\num}[1]{\enub {#1} \enue}
\newcommand{\tem}[1]{\bitem {#1} \eitem}
\newcommand{\figu}[3]{
    \begin{figure}[here] \begin{center}
    \includegraphics[width=#1 cm,height=#2 cm]{#3}
    \end{center} \end{figure}}
\newcommand{\cfigu}[4]{
    \begin{figure}[here] \begin{center}
    \includegraphics[width=#1 cm,height=#2 cm]{#3}
    \caption{#4}
    \end{center} \end{figure}}
\newcommand{\clfigu}[5]{
    \begin{figure}[here] \begin{center}
    \includegraphics[width=#1 cm,height=#2 cm]{#3}
    \caption{#4 \label{#5}}
    \end{center} \end{figure}}
\newcommand{\cltfigu}[5]{
    \begin{figure}[t] \begin{center}
    \includegraphics[width=#1 cm,height=#2 cm]{#3}
    \caption{#4}
    \label{#5}
    \end{center} \end{figure}}

\newcommand{\para}[1]{
    \medskip
    \medskip

    \noindent

    {\bf #1}
    \noindent }

\newcommand{\ba}[2]{\begin{array}{#1} #2 \end{array}}

\newcommand{\nn}{\nonumber \\}

\newcommand{\cala}{\mbox{${\cal A}$}}
\newcommand{\calb}{\mbox{${\cal B}$}}
\newcommand{\calc}{\mbox{${\cal C}$}}
\newcommand{\cald}{\mbox{${\cal D}$}}
\newcommand{\cale}{\mbox{${\cal E}$}}
\newcommand{\calf}{\mbox{${\cal F}$}}

\newcommand{\call}{\mbox{${\cal L}$}}
\newcommand{\calm}{\mbox{${\cal M}$}}
\newcommand{\caln}{\mbox{${\cal N}$}}
\newcommand{\calo}{\mbox{${\cal O}$}}
\newcommand{\calp}{\mbox{${\cal P}$}}
\newcommand{\calq}{\mbox{${\cal Q}$}}
\newcommand{\calr}{\mbox{${\cal R}$}}

\newcommand{\calt}{\mbox{${\cal T}$}}

\newcommand{\calz}{\mbox{${\cal Z}$}}

\def\dq{\dot{q}}
\def\lag{Lagrangian }

\def\ham{Hamiltonian }
\def\cf{{\cal K}}
\newcommand{\Lam}{\Lambda}
\def\dpo{d {\hskip -0.02cm{+}} {\hskip -0.05cm{1}}}
\def\6{\partial}
\def\7{\tilde}
\def\8{\widehat}
\def\bx{{\bf x}}
\newcommand{\vep}{\varepsilon}
\newcommand{\iPi}{{\it\Pi}}

\def\bl{\biggl(}
\def\br{\biggr)}
\def\dphi{\phi^{\dagger}}
\def\kvec{\vec{k}}
\def\pvec{\vec{p}}
\def\opvec{\vec{P}}
\def\del3{\delta^{(3)}}
\def\tp{\tilde{P}}
\def\otp{\tilde{p}}

\def\undos{{1 \over 2}}
\def\pa{\partial}

\newcommand{\fc}{\frac}
\def\rl {\sqrt{\lambda}}
\newcommand{\bbi}[1]{\mbox{${\mathbb I}_{#1}$}}
\newcommand{\ra}{\rightarrow}
\def\xsac{,\,\,}

\def\eM{$\CM_8$}
\def\eeM{$\CM_8$ }
\def\otaula{\begin{tabular}}
\def\ctaula{\end{tabular}}
\def\k{\kappa}
\def\te{\tilde{e}}
\def\tte{\tilde{\e}}
\def\ts{\tilde{*}}

\def\euc{\mathbb{E}}

\newcommand{\U}{\mathop{\rm {}U}}
\def\rme{{\rm e}}

\def\rmd{{\rm d}}
\def\1u{\underline{1}}
\def\2u{\underline{2}}

\def\target{$\CR^{1,1}\times \mathcal{M}_8$ }
\def\target2{$\CR^{1,1}\times \mathcal{M}_8$,}
\def\9G{\G_{\underline{9}}}
\def\sign{\mathop{\rm sign}\nolimits}
\def\op{\oplus}

\newsavebox{\uuunit}
\sbox{\uuunit}
    {\setlength{\unitlength}{0.825em}
     \begin{picture}(0.6,0.7)
        \thinlines
        \put(0,0){\line(1,0){0.5}}
        \put(0.15,0){\line(0,1){0.7}}
        \put(0.35,0){\line(0,1){0.8}}
       \multiput(0.3,0.8)(-0.04,-0.02){12}{\rule{0.5pt}{0.5pt}}
     \end {picture}}
\newcommand {\unity}{\mathord{\!\usebox{\uuunit}}}
\def\trace{\mathop{\rm Tr}\nolimits}
\newcommand{\QED}{{\hspace*{\fill}\rule{2mm}{2mm}}}
\def\yvec{\vec{y}}
\newcommand{\ft}[2]{{\textstyle\frac{#1}{#2}}}

\def\met{g_{\mu\nu}}

\def\pamu{\partial_{\mu}}

\def\amu{A_{\mu}}

\def\ut{\underline{3}}
\def\uu{\underline{1}}
\def\ud{\underline{2}}
\def\uz{\underline{0}}
\def\ur{\underline{r}}

\def\tth{\tilde{\theta}}
\def\tphi{\tilde{\phi}}
\def\tpsi{\tilde{\psi}}

\def\et{\tilde{e}}
\def\tH{\tilde{H}}
\def\th{\tilde{h}}
\def\oz{\overline{z}}
\def\2ka{K\"ahler}

\def\tPhi{\hat{\Phi}}
\def\hg{\hat{g}}
\def\cO{\mathcal{O}}
\def\cC{\mathcal{C}}

\def\dtau{\dot{\tau}}
\def\dsigma{\dot{\sigma}}

\def\tte{\tilde{\tilde{\epsilon}}}
\def\tsp{\tilde{\e}}

\begin{document}
\setlength{\textwidth}{13.5cm} 

\thispagestyle{empty}

\vspace*{2 cm}

\begin{center}
\large Ph. D. Thesis on \\
\vspace{0.4cm}
\Large
\textbf{
D-branes, gauge/string duality \\ and noncommutative theories}\large

\vspace{2.5cm}
Toni Mateos \\\large
\normalsize
\vspace{2.5cm}
Advisor: Joaquim Gomis Torn\'e \\
Departament d'Estructura i Constituents de la Matèria\\
Universitat de Barcelona \\
\vspace{1cm}
Barcelona, April 2004 \\
\vspace{1cm}
Thesis defended on June 19th 2004 \\
\end{center}\normalsize

\newpage\thispagestyle{empty}\phantom{a}
\newpage\thispagestyle{empty}

\frontmatter

\thispagestyle{empty}\phantom{a}

This thesis is mainly based on the following published articles:

\begin{enumerate}

\item

J.~Gomis, K.~Kamimura and T.~Mateos,
``Gauge and BRST generators for space-time non-commutative U(1) theory,''
JHEP {\bf 0103} (2001) 010
[arXiv:hep-th/0009158].

\item

T.~Mateos and A.~Moreno,
``A note on unitarity of non-relativistic non-commutative theories,''
Phys.\ Rev.\ D {\bf 64} (2001) 047703
[arXiv:hep-th/0104167].

\item

J.~Gomis and T.~Mateos,
``D6 branes wrapping Kaehler four-cycles,''
Phys.\ Lett.\ B {\bf 524} (2002) 170
[arXiv:hep-th/0108080].

\item

J.~Brugues, J.~Gomis, T.~Mateos and T.~Ramirez,
``Supergravity duals of noncommutative wrapped D6 branes and supersymmetry without supersymmetry,''
JHEP {\bf 0210} (2002) 016
[arXiv:hep-th/0207091].

\item

T.~Mateos, J.~M.~Pons and P.~Talavera,
``Supergravity dual of noncommutative N = 1 SYM,''
Nucl.\ Phys.\ B {\bf 651} (2003) 291
[arXiv:hep-th/0209150].

\item

J.~Brugues, J.~Gomis, T.~Mateos and T.~Ramirez,
``Commutative and noncommutative N = 2 SYM in 2+1 from wrapped D6-branes,''
Class.\ Quant.\ Grav.\  {\bf 20} (2003) S441
[arXiv:hep-th/0212179].

\item

J.~Gomis, T.~Mateos, P.~J.~Silva and A.~Van Proeyen,
``Supertubes in reduced holonomy manifolds,''
Class.\ Quant.\ Grav.\  {\bf 20} (2003) 3113
[arXiv:hep-th/0304210].

\item

D.~Mateos, T.~Mateos and P.~K.~Townsend,
``Supersymmetry of tensionless rotating strings in $AdS_5 \times S^5$, and nearly-BPS operators,''
JHEP {\bf 0312} (2003) 017
[arXiv:hep-th/0309114].

\item

D.~Mateos, T.~Mateos and P.~K.~Townsend,
``More on supersymmetric tensionless rotating strings in $AdS_5 \times S^5$,''
arXiv:hep-th/0401058.

\end{enumerate}

\newpage\thispagestyle{empty}\phantom{a}
\newpage\thispagestyle{empty}


\thispagestyle{empty}\phantom{a}

\vspace*{40pt}

\begin{center}
\large
\textbf{ ACKNOWLEDGEMENTS}\normalsize
\end{center}

\medskip
\medskip

\noindent

Here comes the most pleasant part of the writing of this thesis, a part
for which I have written down mentally so many little notes throughout
all these years, trying not to forget anyone.

The first person I would like to thank is my advisor Joaquim Gomis.
I still remember that it was him who wrote for me the first field theory
action I had seen in my life. He said that thanks to the fact that
it was two-dimensional, it enjoyed a symmetry called 'conformal' which
happened to be infinite-dimensional, and that the absence of a certain
anomaly called 'Weyl' implied that the world had to have 26 dimensions.
At that moment I just wondered how long would it take for me to start distinguishing
String Theory from Chinese. Thanks Quim for having helped me so much
with this enterprize, putting pressure on me in the right moments.
Thanks as well for having been a friend and for creating such a good
atmosphere in the department.

The second person I would like to thank is Paul Townsend, with whom
I am also indebted. Thank you so much for hosting me in Cambridge
and for all those 'sobremesas' with dissertations about life, the huge damage
caused by the prehistorical agriculture or the role of Kings in modern democracies.
It has been really fascinating to get to know your human side.
From an academical point of view, I had the feeling that my learning
of string theory speeded up every time we discussed in the blackboard,
be sure that your way of viewing physics has left a deep fingerprint on me.
I really hope to have the chance to keep learning from you in the future.

Next I would like to thank some other persons with whom I had the opportunity
to collaborate. I would like to thank Antoine Van Proeyen for those two
concentrated weeks in which, together with Quim and Pedro,
we ran against time to finish a project. Thank you too for helping me
every time I needed it, and for all those suggestions and improvements on
the manuscript of this thesis.
Thanks Pedro for always being full of projects and for always listening to
my crazy ideas. Thank you too for your friendship,
I hope we manage to coincide more than two weeks together
in the future! A special mention goes to the meson formed by
Josep M. Pons and Pere Talavera. Sharing our first steps in string theory
was a wonderful experience. I hope that the next time we collaborate we will
know a little bit of what these guys are talking about! It is also a pleasure
to thank Alfonso Ramallo for sharing his mythical notebooks with me
and for so many 'tertulias' at lunches and dinners.
Thanks as well for sharing those early days to Alex Moreno, to whom I seem to have scared to
the point of quitting physics! Thanks too to the Jedi knights Jan and Tonir,
who have just started to feel the Force.

Thanks to Jos\'e Edelstein, Roberto Emparan, Javier Mas,
Carlos N\'u\~{n}ez, Prem Kumar and Jorge Russo for many
discussions and valuable comments, and to the professors of my
department Domènec Espriu, Josep I. Latorre, Josep Taron,
Joan Soto, Rolf Tarrach and Enric Verdaguer for always
being available to solve my doubts in four dimensions.

I am also in debt with the Persian Gang (Saman, Ahmed, Amir, Ali, Hussein, Nazdereh...)
who made my stay in Iran unforgettable. Since I came I have been planning to travel back
there again every year, still without success. I wish you good luck with your lifes and
with your country.

A huge hug to the Parisian Gang (Antonio, Nicco, Aldino and Aldina, Fabio,
Martina, crazy Paskal and even more crazy Tasos, Steffi,
my fairy godmother Liattina...).
You know that I do not exaggerate when I say that those were
possibly the best three months of my life. It is great that we all
remain in Europe and that we keep meeting every now and then.

Thanks as well to the Cambridge Gang (Rub\'en, Sean, Guishermito,
Marta, Christophe, Maruxa) for making my stay there so good too.
Thanks to all the students I have met in my department: Aleix, \`Alex,
Luca, Juli\'an, Enrique, Diego, Ernest, Miriam, Roman,
Xavi, Jaume, David, Dani, Joan, Majo and very specially
to Lluis, Toninho Ac\'{\i}n, Enric Jan\'e and Adam Love. I am sure
we will manage to keep our friendship in the future. Indeed, I have
gone with Lluis through so many extreme circumstances that it seems
unbelievable that we both will finish the PhD alive. I hope our life
will be easier in England!

Last and most important, thanks to my family, specially to my brother
David, to whom I love and admire
and with whom I have had the pleasure to share both life and physics.

\thispagestyle{empty}\phantom{a}

\newpage\thispagestyle{empty}

\selectlanguage{english} \tableofcontents
\mainmatter
\setcounter{chapter}{0}
\selectlanguage{english}

\chapter{Introduction}\label{ch:introduction}

It has been almost ten years since the discovery of D-branes~\cite{Polchinski:1995mt},
and it is fair to say that nothing has been the same anymore. Polchinski
used to start his talks with a transparency like this,
\medskip \bigskip

\be
    \includegraphics[width=6.1 cm,height=5.3 cm]{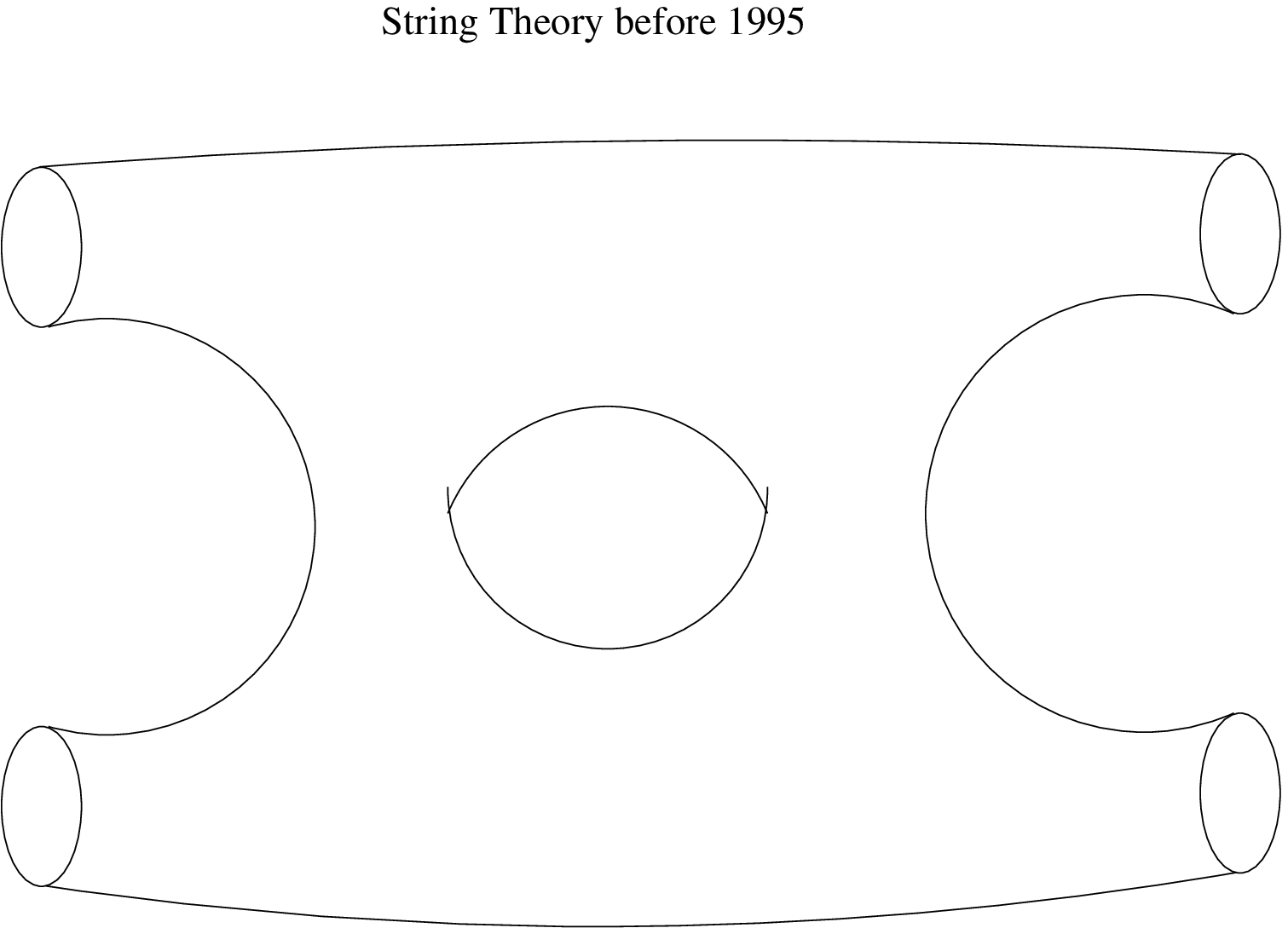}
    ~~~  ~~~ ~~~
    \includegraphics[width=6.1 cm,height=5.3 cm]{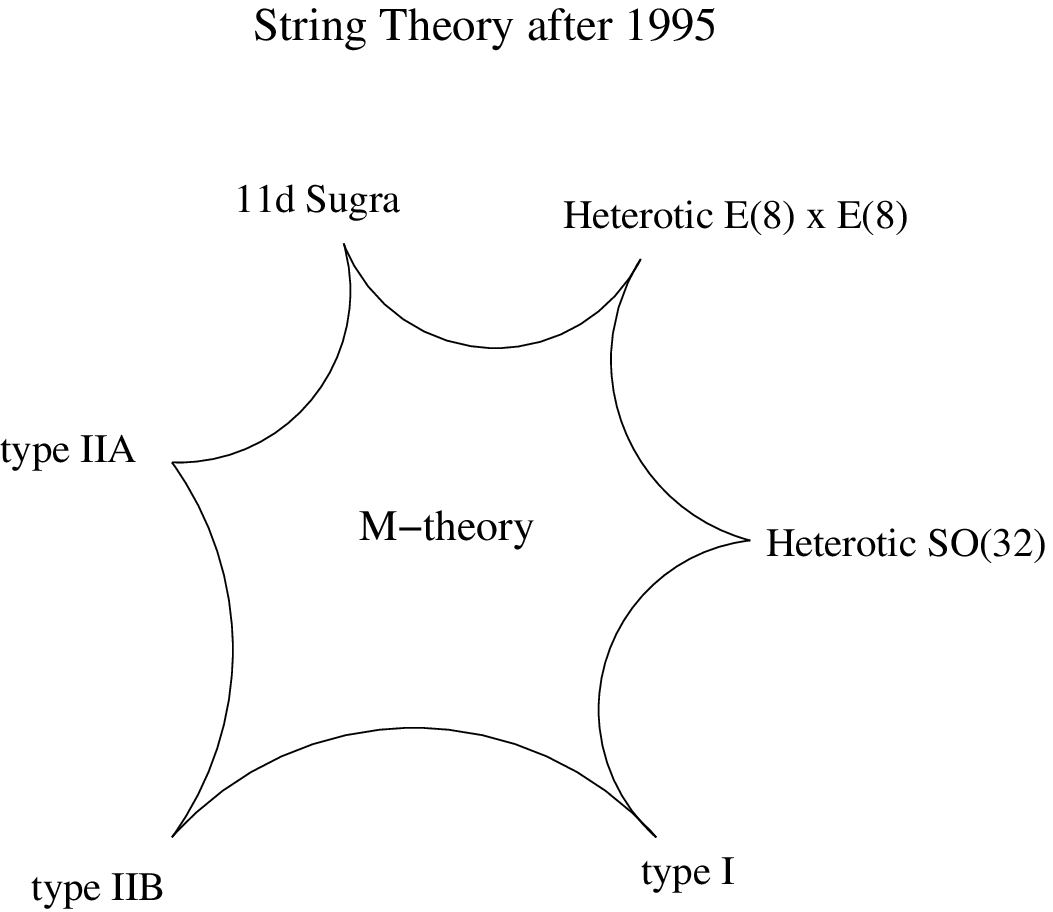}
\nonumber
\ee
\medskip

\noindent
which encodes the deep transformation of our view of String Theory that
took place during 1995-1997. This thesis deals about D-branes and
some of the main new lines of research that they opened.

Polchinski's diagram refers to the development of
a series of dualities that allowed to relate the five 10d superstring
theories which are known to be free from anomalies. Some of these
dualities mapped the strong coupling regime of one of the string
theories to the weak coupling of another one. This is the example
of the type IIB S-duality, a case in which the original and final
theories are the same. For cases like these, D-branes turned out to be
fundamental, as they provided the non-perturbative states needed to
complete the net of connections between the various Hilbert
spaces that we observe at the vertices of the M-theory diagram.

Dualities were one of the first developments in which D-branes played
a crucial role, and they are probably among the most important
achievements towards the understanding of what String Theory really is.
There was however another key property of D-branes that was just
waiting for its exploitation, a property that turned String Theory
into one of the most multidisciplinary fields of physics: the fact
that at low energies they can be described by ordinary gauge field
theories\footnote{To be strict, this is not always the case as sometimes there
are massive excitations that cannot be decoupled.}.
If the spatial dimension of the D-brane is $p$, in which case we talk
about a D$p$-brane, one is led to consider gauge theories in $p+1$ dimensions.
Progressively, most of the field theory phenomena that we were familiar with
acquired a geometrical interpretation in terms of how a particular setup
of D-branes is embedded in a certain 10d manifold. The list of
examples of this reinterpretation is uncountable.\footnote{A basic review of
some of these is given in chapter \ref{ch:dbranes}.}
 Let us just mention
some of the most intuitive ones:

\tem{
\item scalars in the field theory are reinterpreted as giving the embedding
of the D-brane in its transverse space; they are actually the Goldstone
bosons corresponding to the background symmetries broken by the presence
of the D-brane,
\item the $R$-symmetry group is reinterpreted as the group of isometries
in the D-brane transverse space,
\item (the breaking of) the Poincar\'e group of the field theory corresponds
to (the breaking of) the Poincar\'e isometry along the D-brane directions,
\item many field theory instantons and monopoles can be interpreted as
various strings/branes ending or intersecting D-branes.
}

But in order for a reinterpretation to be useful and not just philosophy,
it must be able to provide new results. It turns out that the {\it gauge theory phenomena
$\leftrightarrow$ D-brane} relation has been extremely fruitful; it has allowed
to find new results in gauge theories based on stringy intuition and,
conversely, to find new results in D-brane physics based on a purely
field-theoretical approach. This thesis contains examples of results in both
directions, the most sophisticated one possibly being the way that D-branes implement
the field-theory phenomenon known as {\it twisting}, to be discussed later.

\section{AdS/CFT}

Apart from carrying gauge theories on their worldvolumes,
there is another crucial property of D-branes: they are charged
under the Ramond-Ramond (RR) field potentials.
This allowed for an identification of D-branes with
the supergravity solitons with nonzero RR-potentials
that had been known for some years~\cite{Horowitz:1991cd}.
Despite being solutions of the
low energy effective action of the various string theories, they had
been waiting for an interpretation; they could not describe the backreaction
of any state in the perturbative Hilbert space of string theory, as none
of them couples minimally to the RR fields. D-branes came to fill this
gap bringing supergravity back to the game. Somehow, there was a transition
between the description of D-branes as 2d conformal field theories
with boundaries and their description in terms of supergravity solutions.
For example, by putting more and more branes on top of each other, the
gravitational scale of the system starts growing, the D-branes become 'fat'
and the backreaction cannot be neglected.
People started to realize that these two points of view could be
made functional. Some observables corresponding to the low energy
gauge theory on the D-branes started to be computable from
the supergravity side.

It was Maldacena \cite{Maldacena:1998re}
who finally made the conjecture\footnote{This conjecture
is motivated and explained in detail in chapter \ref{ch:ads-cft}.}
that, at least in
the case of $N$ D3-branes, its low energy $\caln=4$ $SU(N)$ superconformal
Yang-Mills was dual to type IIB string theory in the near horizon region
of its supergravity solution, \5ads. This provided
the first concrete example of the conjecture made decades earlier
by 't Hooft  that non-abelian gauge theories could be described
in terms of string theories at least at large $N$. Maldacena's
duality was even more surprising as the particular string dual of the 4d
gauge theory was actually a string theory in {\it ten dimensions.}
Somehow, the degrees of freedom of the type IIB strings had to be 'holographically
projected' to its boundary, which is conformal to 4d Minkowski space.
This made the ideas of holography, a discipline that had been proposed
independently of string theory, enter the game as well.

Maldacena's conjecture, which is also referred to as AdS/CFT duality,
had better be of a weak/strong nature, as we do not see anything like closed strings
or branes in a perturbative analysis of SYM. That this is so made the conjecture
both powerful and difficult test, not to mention 'prove'. In principle, it allowed to compute
the same observable in both sides (although using different languages) and say
that the results do not necessarily match, as they correspond to opposite
regimes of either the field or the string theory. However, this encounters
a huge obstacle from its very beginning: how do we translate the degrees of
freedom from one theory to the other and  confirm that we
are computing the same observable? The answer to this question is not straightforward
at all and it is fair to say that the states of both Hilbert spaces that have
been able to be mapped are of zero measure compared to the number of total states.
To construct this dictionary it is good to have some symmetries at hand, like
the bosonic group of global symmetries of the SYM theory, $SO(2,4)\times SO(6)$
or its full supersymmetry group $PSU(2,2|4)$.
These symmetries should be present for all values of the coupling, thus
they should be visible in the closed string side; they indeed correspond
to the isometries of the \5ads background. Classifying states in irreps of
these groups helps building the dictionary as we will repeatedly see
during this thesis.

But even if we had the complete dictionary, the aim of testing the duality
would remain almost unreachable if all that we can use is perturbative SYM
and supergravity. The latter is the only approximation that we can deal
with in the string side as the IIB sigma model in \5ads is, to date,
not possible to quantize. There are however a series of observables
that do allow for a comparison, those whose values are known not to
depend on the coupling. These observables are typically related to
BPS states in supergravity and BPS operators in the SCFT. The energy
of the former and the conformal dimension of the latter are completely
fixed by the underlying superconformal algebra, which relates them
to the other charges that they may have. Consider the example of the operators
\be \setcounter{equation}{1}\label{opes}
\calo = \Tr \left(X^J\right) \sac X \equiv \phi_1 + i \phi_2 \,,
\ee
where $\phi_{1,2}$ are two of the scalars of the $\caln=4$ supermultiplet.
These operators are invariant under half of the Poincar\'e supersymmetries
and a straightforward argument based on the $PSU(2,2|4)$
superalgebra\footnote{See section \ref{section-bound} for a prove of this and similar but
more general relations, and the appendix \ref{ch2:sec:algebra} for a detailed discussion
on BPS operators.}
shows that their conformal dimension $\Delta$ must be $\Delta=J$. These
operators are dual to supergravity excitations with angular momenta $J$
along the $S^5$, which are also 1/2-BPS and their energy is $E=J$.

The impossibility of testing the AdS/CFT duality beyond BPS-protected
quantities was enormously improved in the work of Berenstein,
Maldacena and Nastase (BMN)~\cite{Berenstein:2002jq}. They showed that
the Penrose limit of \5ads along a null geodesic in the $S^5$ was
dual to a subset of the $\caln=4$ operators with large charge $J$
under a $U(1)$ subgroup of the $SO(6)$ $R$-symmetry. This Penrose
limit leads to the only maximally supersymmetric IIB background~\cite{Figueroa-O'Farrill:2002ft}
that remained to be exploited: a certain class of pp-wave. The simplicity
of this background allows for a quantization of the string $\s$-model,
and this opens a huge region in the parameter space of the two theories
where both are simultaneously accessible with our present techniques.
Let us remark that the operators that survive the BMN limit are,
despite being non-BPS, very close to those in \bref{opes};
they have a number of insertions of other fields which is small compared to $J$.

Soon after the work of BMN, a shortcut was provided by Gubser, Klebanov
and Polyakov  in a paper \cite{Gubser:2002tv} where they proposed
that some $\s$-model solitons in \5ads were able to provide
similar answers within a classical approximation, bypassing the need
to quantize in RR backgrounds. Their ideas were immediately applied to
solitons in many other complicated backgrounds which are believed to have a
gauge theory dual, yielding a number of predictions for the strong
coupling behavior of some of their observables.

A qualitatively new set of results started with the papers of
Frolov and Tseytlin \cite{Frolov:2003qc,Frolov:2003tu} where they
considered $\s$-model solitons that carried three angular momenta
$(J_1,J_2,J_3)$ along the $S^5$. The novelty was that they were able to match
exactly~\cite{Beisert:2003xu,Frolov:2003xy,Arutyunov:2003uj,Beisert:2003ea,Arutyunov:2003rg,Engquist:2003rn,Kristjansen:2004ei}
the classical energy $E(J_i)$ of the solitons to a one-loop computation
of the conformal dimension of the operators
\be
\calo = \Tr \left(X^{J_1} Y^{J_2} Z^{J_3} \right) \, + \, \mbox{permutations,}
\ee
with $Y=\phi_3+i\phi_4$ and  $Y=\phi_5+i\phi_6$, by interpreting the
one-loop anomalous dimension matrix as an integrable spin-chain
Hamiltonian~\cite{Minahan:2002ve,Beisert:2003yb}.
These operators, and
their corresponding string theory states, are very far from the
1/2-BPS BMN operators, so it looks like supersymmetry has nothing to do with
these tests. If this was so, then there are a number of immediate difficulties
and still open questions that progress along these lines is having to
face. The first problem is that many of the solitons that provide successful comparisons
were shown to be unstable. The second problem is that one has to justify
why are quantum $\s$-model corrections negligible against the classical result.
We will see that this is a very involved problem that has only been answered
to one loop and for a particular class of solutions\footnote{Note that some of these
quantum corrections were performed about an unstable vacuum!}.
We will investigate this subject in great detail
in section \ref{ch2:sec:ads-beyond} and we will give arguments
why this particular correspondence is being so successful.
Some recent results~\cite{Bigazzi:2004yt} seem to support our proposal
as we discuss in the conclusions section.
A deeper understanding is, however, still required.

\medskip

The AdS/CFT duality and more generally the relation between
the two open/closed descriptions of D-branes
has also provided new ways to look at a problem.
Consider a set of $N$ Dp-branes and $M$ Dq-branes in flat space.
The probe picture is consistent at weak coupling as long as
both $N$ and $M$ are small, so that the backreaction can be neglected.
One can compute the interaction among them by the standard techniques
and then conclude whether they attract, repel or do not feel
the others' presence at all. The leading order of this interaction
typically involve a one-loop diagram of open strings or, equivalently,
a tree level diagram of closed strings with sources (boundary states).

The open/closed string description comes into the game when we let $N$
grow. At some point, the Dp-branes description is more adequate in terms
of their supergravity solution. If in this process we kept $M$ fixed,
we end up with $M$ Dq-probes in the background of $N$ Dp-branes. We will
explore a wide set configurations in which the original setup is such
that the final $M$ Dq-probes are embedded as an $AdS\times \S$
submanifold in the near horizon region of the $N$ Dp-branes background,
with $\S$ a compact submanifold. The standard case is a setup
of D3/D5 branes as the following array indicates
\begin{center}
\[
\begin{array}{c | c c c c c c c c c c}
{\rm IIB} &x^0 & x^1 & x^2 & x^3 & x^4 & x^5 & x^6 & x^7 &x^8 & x^9 \\ \hline
{\rm D3}& -&-&-&-&&&&&& \\
{\rm D5}&-&- & -&&-&-&-&&&\\
\end{array}\]
\end{center}
We will see in section~\ref{ch2:sec:stable-branes}
that substituting the D3's by their \5ads background can lead
to an embedding of the D5's as an $AdS_4\times S^2$ submanifold in
which the $S^2$ has maximal volume within the $S^5$. Note that there
is no topological obstruction for the $S^2$ to collapse to a point
in the $S^5$. However, the embedding must be stable as the original
set of branes is known to be 1/4- supersymmetric. The
apparent paradox can be resolved by noting that the tachyonic
instabilities of the $S^2$ embedding have masses above the
Breitenlohner-Friedman bound~\cite{Breitenlohner:1982bm,Breitenlohner:1982jf,Mezincescu:1985ev}
from the point of view of the field theory in the $AdS_4$ factor.

We will study many other examples and show that the type of
interaction between D- or M-branes can be understood
in terms of tachyonic masses being above or below the corresponding
BF bound. This analysis will also lead us to the possibility
of introducing non-supersymmetric but stable D-branes in $AdS\times S$
backgrounds. According to the AdS/dCFT duality (where $d$ stands for
{\it defect}), the probes correspond to the addition of matter multiplets
in the dual CFT; these are confined to live in the submanifold where
the probe intersects de $AdS$ boundary and hence the name of {\it defect}.
Despite being still work in progress, we will present candidates for such stable
but non-supersymmetric embeddings in which the D-/M-brane probes are $AdS$-filling.
This should correspond to the addition of non-supersymmetric matter in the
dual theory without any confining restriction, \ie ordinary matter.
We cannot be conclusive at this
stage yet about the stability of these embeddings, but we hope to report
on it in the near future.

\section{Beyond AdS/CFT: the gauge/string duality}

The possibility of having a strong coupling dual of a theory like QCD
motivated a lot of effort in trying to extend the AdS/CFT duality
to field theories other than the $\caln=4$. Any such extension
has finally earned the name of {\it gauge/string duality}, reserving
'AdS/CFT' for those cases in which the field theory involved
is conformal.
It was clear that
the original picture of $N$ D3-branes in flat 10d space had
to be made more sophisticated if one wanted to end with less
than 16 supercharges. Some attempts were initially based on replacing
the $S^5$ background by cosets $S^5/\G$ or by cones over other Einstein
5d manifolds. Some other attempts introduced small perturbations to the $\caln=4$
Lagrangian, which are dual to deformations of the $AdS_5$ that
do not change its asymptotics. We comment on these in section \ref{ch5:sec:away-from-flatness}.

We will mostly consider a different approach in which
the flatness of  both the ambient space and of the
brane's worldvolume is completely abandoned. The preservation
of some fraction of supersymmetry by the background will lead
us to the concept of special holonomy manifolds,
whereas the preservation by the embedded worldvolume will
lead us to the concept of calibrated cycles. We will see that
the particular way in which the worldvolume gauge theory
of the brane manages to preserve supersymmetry
 had actually been discovered 15 years ago
by Witten~\cite{Witten:1988ze}. By means of a mechanism called
{\it twisting}, one is able to put supersymmetric field
theories in some curved backgrounds. The number of preserved supersymmetries
turns out to be less than the corresponding theory in flat space,
which is precisely what we were looking for.

This field theory intuition is crucial in order to build the
closed string duals of these less than maximally supersymmetric
theories, and it constitutes one of the most sophisticated
examples of the interplay between gauge theories and D-branes
that we mentioned above. This will help us to construct
the closed string dual of an $\caln=2$ $SU(N)$ SYM theory in 2+1 dimensions
without any matter other than the vector multiplet. We will analyze
what string theory can tell us about its moduli space and
discuss that it is tentative to interpret it as an all-loops
resummation.

Indeed, because the closed string dual is constructed with D6-branes,
the uplift of this solution to 11d supergravity will produce
an explicit metric for an eight-dimensional Calabi-Yau space~\cite{Gomis:2001vk}.
Indeed, it is through this D-brane intuition that so many
metrics with special holonomy have been built in the recent
years. Whereas for Calabi-Yau spaces we have Yau's theorem
guaranteing the existence of a unique Ricci flat
metric with $SU(n)$ holonomy in each \kah class, there is no such theorem for
$Spin(7)$ and $G_2$ manifolds. By wrapping D-branes it has
been able to prove the existence of some of such metrics
by simply constructing them. Thus starting from the twist
of field theories we have ended with a purely mathematical
progress!
\medskip
\medskip
\medskip

\scriptsize{\cent{\begin{tabular}{ccccc}
Field theory intuition & $\dreta$ & String Th. intuition & $\dreta$& Maths result \\
&&&&\\
Susy field theories & & D6-branes can be  & & Explicit construction \\
can be put in curved spaces &$\dreta$ & wrapped in special  &$\dreta$ & of metrics with \\
by twisting them & & holonomy manifolds & & special holonomy
\end{tabular}}}
\normalsize

\medskip
\medskip
\medskip
\noindent
There are a couple of features of the duals that extend the AdS/CFT correspondence
that must be stressed. The first one is technical and refers to the fact
that what ultimately simplified the construction of the supergravity
solutions was the use of gauged supergravities, as proposed by Maldacena
and \nun~\cite{Maldacena:2000mw}. These are much simpler than the
IIA/IIB/11d supergravities as they arise after a truncation of an infinite
number of modes. Furthermore, D-branes arise as domain-wall solutions of
them, a fact that dramatically simplifies the ansatz.
We will see during this thesis that, unfortunately, gauged supergravities cannot
always be used.
The second point is
actually a drawback common to most of the AdS/CFT extensions achieved until
now. It turns out that in the limit in which supergravity is valid, and
we recall that this is the only possibility due to the incapability of
quantizing the corresponding $\s$-models, the dual field theory is not just
what one was looking for at the beginning but it contains an infinite
number of undecoupled degrees of freedom.
For example, when the field theory involved is a deformation of the $\caln=4$
SYM with operators of typical mass $M$, the supergravity approximation is valid
only when $M$ is of the same order as the dynamically generated scale $\L_{QCD}$;
thus the confining or strong coupling phase does not correspond to the QCD-like
theory alone.

In the examples of wrapped branes, one expects to recover an ordinary SYM
theory in the non-compact part of the D-brane when the volume of the cycle
that they wrap tends to zero, \ie in the IR. However, such small cycles typically
imply that the background curvature is larger than the string scale, which
renders the supergravity approximation invalid. Insisting on the use of supergravity
means that the physics on the non-compact part of the brane contains an infinite
set of undecoupled Kaluza-Klein modes.

Everything we have mentioned in this section is expanded and discussed in
detail in chapter \ref{ch:wrapped}.

\section{Noncommutative theories in string theory}

Having exploited D-branes to obtain AdS/CFT-like dualities,
let us change subject and analyze another branch of String
Theory that D-branes allowed to open.
Whereas the quantization of the string $\s$-model in flat
space is rather straightforward as it is essentially gaussian,
it becomes a difficult problem as soon the background becomes
more involved. Finding quantizable backgrounds is an important
task, as some of them can lead to a better understanding
of string theory in different regions of its moduli; we already
saw above the great relevance of the quantization in
the IIB pp-wave background.

In \cite{Seiberg:1999vs}, quantization of the open string $\s$-model with
D-brane boundary conditions in a
background with a constant NS-NS $B_2$-field was achieved,
and this led to some new surprises: the low energy limit
turned out to be described by noncommutative (NC) gauge theories.
To be more specific, only those with magnetic or light-like $B_2$-fields
admit a consistent field theory limit, whereas those with electric ones
do not admit a decoupling of all the string massive modes\footnote{We
will see however in section \ref{ch3:sec:e-bckg}
that in electric backgrounds there is a different
limit that leads to a theory in which open strings decouple from closed
string.}.
The NC actions are obtained from the usual commutative ones by replacing
the standard product of functions by a $*$-product defined as
\be \label{0-Moyal}
(F * G)(x) = F(x) \exp \left[
{{i\over 2} \theta^{\mu\nu} \left( \overleftarrow{\partial_{x^{\mu}}}
\overrightarrow{\partial_{x^{\nu}}} -\overleftarrow{\partial_{x^{\nu}}} \overrightarrow{\partial_{x^{\mu}}}
\right)} \right]
G(x) \,,
\ee
where $\t^{\mu\nu}$ measures the intensity of the noncommutativity
between spacetime coordinates,
\be
x^{\mu} * x^{\nu} - x^{\nu} * x^{\mu} = i\t^{\mu\nu} \,.
\ee
The uncertainty principle states then that an attempt
to localize a wave-function in one direction makes it increasingly
delocalized in another one. Maybe a better way to understand it
is that given two functions $f$ and $g$ with support in a region
of small size $\Delta$, the $*$-product $h=f*g$ is supported
in a region of size $\t/\Delta$. The extreme example is the $*$-product
of two delta functions, which gives a constant function with infinite
support.

This property is behind one of the most intriguing aspects of NC
theories, an aspect which arises at the quantum level when trying
to compute loop corrections to observables. It turns out that
the IR and the UV physics of the theory are completely undecoupled,
a property that frontally jeopardizes the Wilsonian approach to
renormalization\footnote{See \cite{Nicholson:2003wp} for a recent
PhD thesis on this subject.}. Perhaps the simplest example is a particular diagram
that contributes to the 1-loop self energy of a NC $\phi^4$ theory as
\be
\Gamma^{2}(p)
 =\frac{\lambda}{96\pi^{2}}\left[\Lambda_{\text{eff}}%
^{2}-m^{2}\ln\left(\frac{\Lambda_{\text{eff}}^{2}}{m^{2}}\right)\right]
\,,\ee
where
\begin{equation} 
\Lambda_{\text{eff}}^{2}=\frac{1}{1/\Lambda^{2}+\pop} \,,
\end{equation}
$\L$ is a UV hard cutoff, and $\pop=-p^{\mu} \t^2_{\mu\nu} p^{\nu}$.
This result seems to be finite if we just send the cutoff to infinity,
\ie if we include arbitrarily high energy modes;
however, this leaves us with an IR divergence as $p\go0$. Similarly,
the contribution is then divergent as $\t \go0$, which means that the
commutative limit of the quantum NC theory is not the commutative
quantum theory. These IR divergences would not be present if we kept $\L$
finite, which suggests that they are actually caused by modes in the UV.

This phenomenon, known as UV/IR mixing, motivated  a critic examination
of these theories. Were they actually sensible at all? It was soon found
that those NC theories that did arise as consistent field theory limits
of string theory inherited its unitarity~\cite{Gomis:2000zz} and
causality~\cite{Seiberg:2000gc}, providing stronger
evidence that they are solid quantum field theories on their own, and
that the UV/IR mixing required more study.

Chapter \ref{ch:nc-theories} deals with NC theories at the classical
and the quantum level. We will study the unitarity at one-loop
of some NC scalar field theories and confirm that it is not violated
unless electric components of the $B$-field are turned on. We will
also examine a possible way in which the non-decoupling
of the stringy modes in these cases can be traced into the lack
of unitarity. In particular, we will try to restore it by enlarging
the asymptotic Hilbert space of the field theory (adding the so-called
$\chi$-particles). We will see that this is specially difficult in
a toy model of a non-relativistic NC $\phi^4$ theory in 2+1 dimensions.

The enormous qualitative difference between magnetic and electric
backgrounds can be understood by the appearance of non-locality
involving time in the latter. The quantization of theories
which are non-local in time is not straightforward at all and
most approaches consider their Lagrangians as a function
of a field and (at best) all its time derivatives. In chapter \ref{ch:hamiltonian}
we will
thoroughly discuss a more solid method for constructing a Hamiltonian
formalism for time non-local theories which is based on the original
idea of Llosa and Vives \cite{Llosa:1993sj}, further developed in~\cite{Gomis:2000gy}.
We will then apply it to
settle a consistent Hamiltonian and BRST formalism for
a NC $U(1)$ gauge theory in four dimensions in which the notions
of conserved charges and symmetry generators appear naturally.
We remark that our analysis does not apply only to NC theories,
but to any theory which is non-local in time. In particular, it
has been recently applied to the study of tachyon condensation
within the $p$-adic string and String Field Theory~\cite{Gomis:2003xv}.

\section{Linking NC theories, AdS/CFT and gauge/string duality}

The suspicion that the UV/IR mixing of NC theories may be an
artifact of the Feynman diagrammatic expansion is just one
of the motivations to study them by alternative methods\footnote{We have
in mind now only the magnetic cases which do arise in string theory.}.
If dual closed string backgrounds could be found, they could
shed some new light to this phenomenon and provide new
non-perturbative information.

The first duals were constructed
by Maldacena and Russo~\cite{Maldacena:1999mh}, and
Hashimoto and Itzhaki~\cite{Hashimoto:1999ut}. In particular
they were able to study the magnetic NC deformation of
the superconformal $\caln=4$ in four dimensions. Some expected
features of noncommutativity were visible in this background;
in particular, UV/IR mixing seems to slightly modify the geometry
in the IR but it completely disappears in the deep infrared.
The disadvantage of the $\caln=4$ in this respect is that it is absent
of UV divergences in its perturbative diagrams. This implies that the
UV/IR phenomenon does not seem to be present, at least in perturbation
theory, which would perfectly fit with the prediction from the
supergravity dual\footnote{The issue of whether the IR or the commutative
limit of the $\caln=4$ is actually smooth requires a more careful
study. Similarly, it is not always the case that the planar limit
of a NC theory corresponds to the commutative one. See~\cite{Armoni:2000xr}
for a good discussion on this issues.}.

Thus we are led to the enterprize of finding supergravity solutions
of NC theories with less than maximal supersymmetry. Fortunately,
we found that the same ideas that allowed for a reduction of
supersymmetry by wrapping
D-branes in special holonomy manifolds extend to backgrounds in
which we turn on magnetic $B$-fields. The incompatibility is avoided
as long as the $B$-field flux along the special holonomy manifold
vanishes. This is not an impediment for our purposes, as in the IR
we want to end up with a NC field theory on the flat noncompact
part of the D-brane. The schematic picture is as follows,
\medskip \medskip
\be
\includegraphics[width=10 cm,height=5 cm]{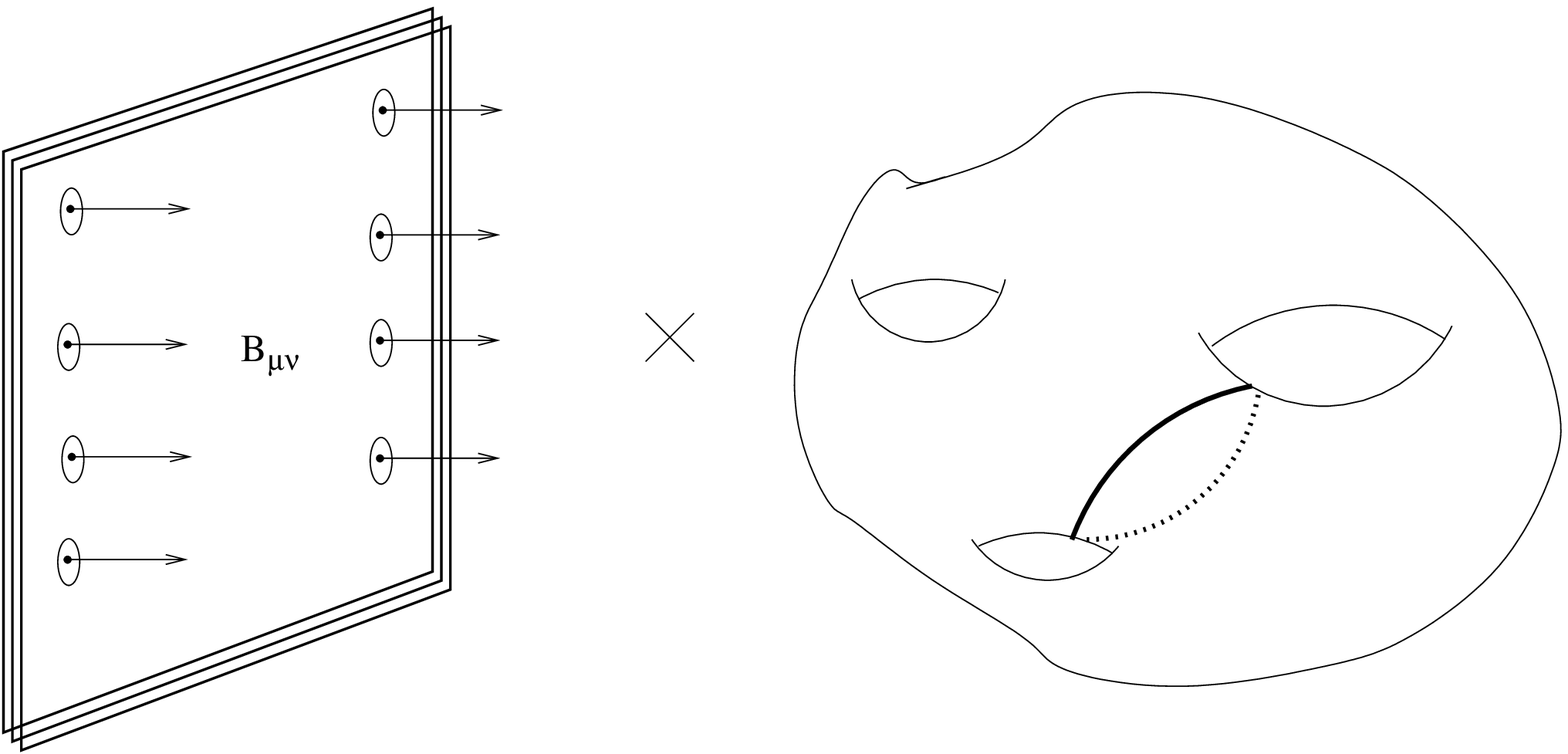}
\nonumber
\ee
\medskip

\noindent
We will be able to construct the supergravity duals of two NC
theories\footnote{Note that we always consider $U(N)$ instead
of $SU(N)$ gauge groups when dealing with NC theories. This is
because, unlike in commutative theories, the $U(1)$ photon couples
to rest of fields in the gauge multiple, see page \pageref{nc-photon}.}:
\tem{
\item a $U(N)$ NC $\caln=1$ SYM in 3+1 (section~\ref{ch6:sec:nc-mn}),
\item a $U(N)$ NC $\caln=2$ SYM in 2+1 (section~\ref{ch6:sec:nc-gm}).
}
For the first theory we discuss a good amount of nonperturbative
properties derived from the closed string dual: the presence of
UV/IR mixing, confinement, the $\b$-function and chiral-symmetry
breaking. We will see an interesting property which is absent
from the commutative counterpart: the new scale introduced by
the noncommutativity can be fine-tuned so that it allows for
a decoupling of the KK modes. We advance here that such a decoupling
can only be achieved by setting the NC scale to be the largest one
in the problem, thus it does not allow to end up with a 'realistic' theory
anyway.
In constructing the dual of the second theory,
we will encounter the unexpected problem that
the phenomenon known as 'supersymmetry without
supersymmetry' affects NC theories in a completely
different manner to their commutative counterparts.
We will prove that gauged supergravities are useless in
constructing NC backgrounds and provide an alternative method
which involves a series of T-dualities.

\section{Map of the thesis}

We have introduced the three main subjects in which
the whole work of this thesis is embedded:
\num{
\item D-branes,
\item AdS/CFT duality and its extension to less supersymmetric theories,
\item NC field theories.}
Let us sketch what how the original work is distributed along the thesis.

\tem{
\item Chapter \ref{ch:dbranes} includes the part of the
work that deals purely with D-branes, in particular with the possibility
of constructing supertubes~\cite{Mateos:2001qs} in a large class of curved manifolds~\cite{Gomis:2003zw}.
The chapter includes the introductory material to D-branes that will be needed
in the rest of the thesis.

\item Chapter \ref{ch:ads-cft} contains the part of the work
that deals purely with the AdS/CFT. We study the possibility
of testing the duality beyond supergravity and supersymmetry,
as reported in~\cite{Mateos:2003de,Mateos:2004rn}. We include
unpublished work in collaboration with D. Mateos and P.K. Townsend
about the various possibilities of embedding D-brane probes in
$AdS\times \S$ submanifolds of 10d and 11d $AdS\times S$ backgrounds.
We will see that the Breitenlohner-Freedman bound of the field theory
that lives in the $AdS$ factor of the D-brane is able to tell us whether
the various involved D-branes attract, repel or do not feel any force at all.

\item Chapter \ref{ch:wrapped} includes the part of the work
that deals purely with the extension of the AdS/CFT to less supersymmetric
field theories. After introducing all the necessary concepts with
some detail, we describe the construction of the supergravity
dual of an $SU(N)$ $\caln=2$ SYM theory in
three dimensions.
Its 11d description provides a metric for an 8d noncompact Calabi-Yau
manifold. We analyze the its moduli space from the supergravity side,
based on the results reported in~\cite{Gomis:2001vg,Brugues:2002pm}.

\item Chapter \ref{ch:nc-theories} contains the part of the work
that deals  noncommutative {\it field} theories. After a brief
review of how they where introduced in string theory, we analyze some
of its main quantum properties. In particular, as reported in
\cite{Mateos:2001tj}, we study the unitarity
of a non-relativistic NC $\phi^4$ theory and the possibility of
adding $\chi$-particles to restore unitarity in the electric case.

\item Chapter \ref{ch:nc-sugra} includes the part of the work
devoted to link the extensions of AdS/CFT via wrapped branes
with the NC theories. It is based on the results reported
in~\cite{Brugues:2002ff,Mateos:2002rx,Brugues:2002pm}.

\item Finally, chapter \ref{ch:hamiltonian} contains the
work devoted to settle a Hamiltonian and BRST formalism for
any non-local field theory in time,
such as the electric NC theories described above. It is based
on the results of~\cite{Gomis:2000sp}.
}

\bigskip

\cfigu{13}{10}{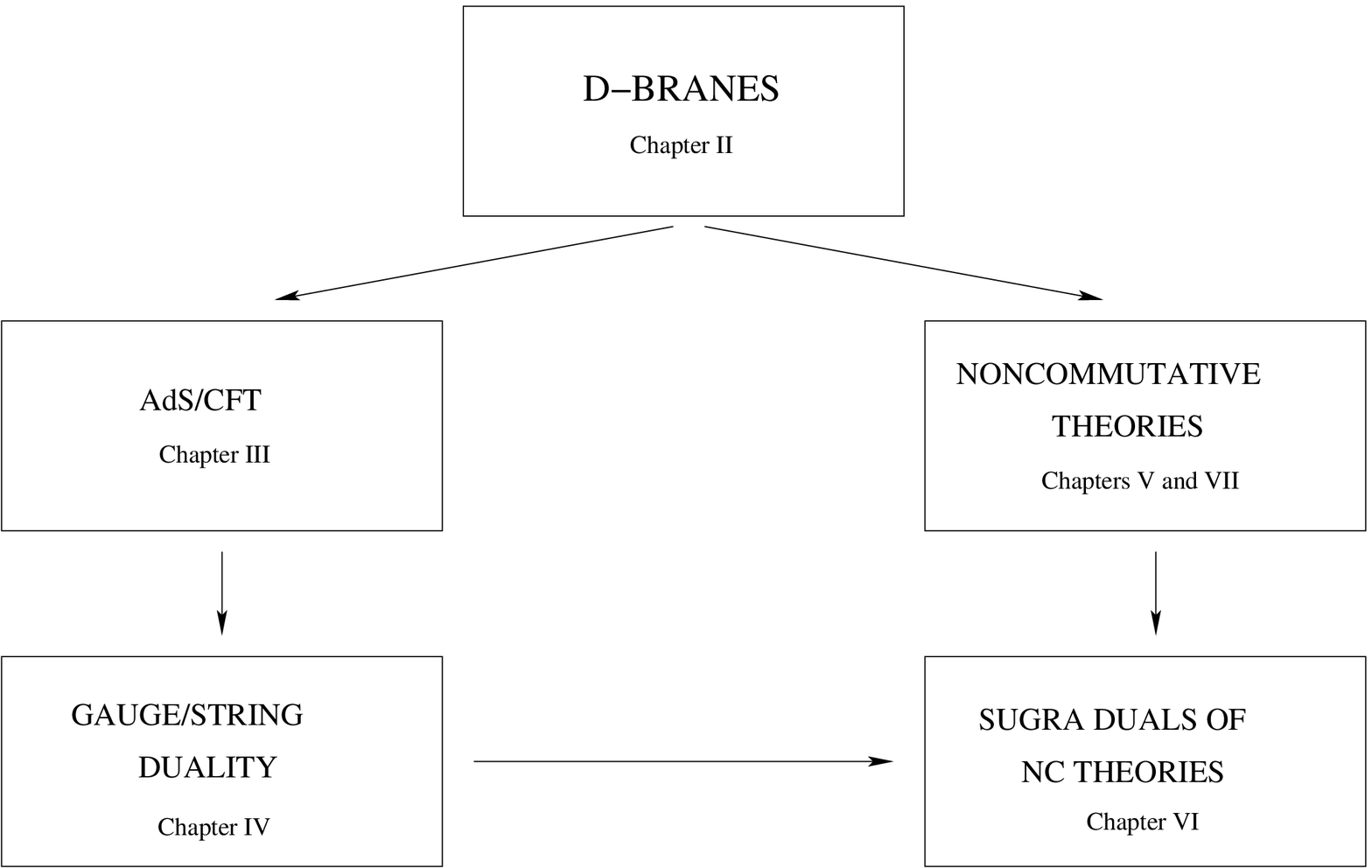}{The thesis at a glance.}

\chapter{Physics of D-branes}\label{ch:dbranes}

This chapter covers some basics of the physics of D-branes,
giving a special emphasis to those topics that will be needed in
the thesis.
We will mainly concentrate on the conceptual issues,
trying to build a self-consistent base for the three
topics of the next chapters: NC theories, AdS/CFT correspondence
and gauge/string duality. There are excellent reviews in the literature
(\eg \cite{Polchinski:1996na,Johnson:2000ch}) and
we refer the reader to them for
technical details and extra material.

After this short review,
and as part of the work during this thesis concerning only D-brane physics, we
introduce the supertubes in section~\ref{ch1:sec:supertubes}.
As will be shown, supertubes intensively exploit the sophisticated
dynamics of D-branes. We extend the original construction of
\cite{Mateos:2001qs,Mateos:2001pi} and show that supertubes can be constructed supersymmetrically
in a huge variety of curved spaces enabling, among other things,
the construction of their closed strings description in terms
of IIB supergravity backgrounds preserving from $1/4$ to $1/32$ supersymmetries.

\section{Perturbative definition and spectrum of a single D-brane} \label{ch1:sec:perturbative-D}

In string perturbation theory, Dp-branes are defined as (p+1)-dimensional
hypersurfaces (let us call them $\Sigma_{p+1}$) where open strings are allowed to end. Their dynamics are
therefore described by the excitations of open strings with a mixed
set of boundary conditions
\bea
\mbox{Neumann BC's along $\Sigma_{p+1}$: }& &\partial_{n} X^{\mu} = 0 \sac \mu=0,...,p \,, \nonumber \\
\mbox{Dirichlet BC's normal to $\Sigma_{p+1}$: }& & \partial_{t} X^{i} = 0 \sac i=p+1,...,(D-1) \nonumber \,,
\eea
where $\partial_{n}$ and $\partial_{t}$ stand for normal and tangent derivatives to the
surface swept by the string worldsheet in a $D$-dimensional spacetime.

\clfigu{5}{7}{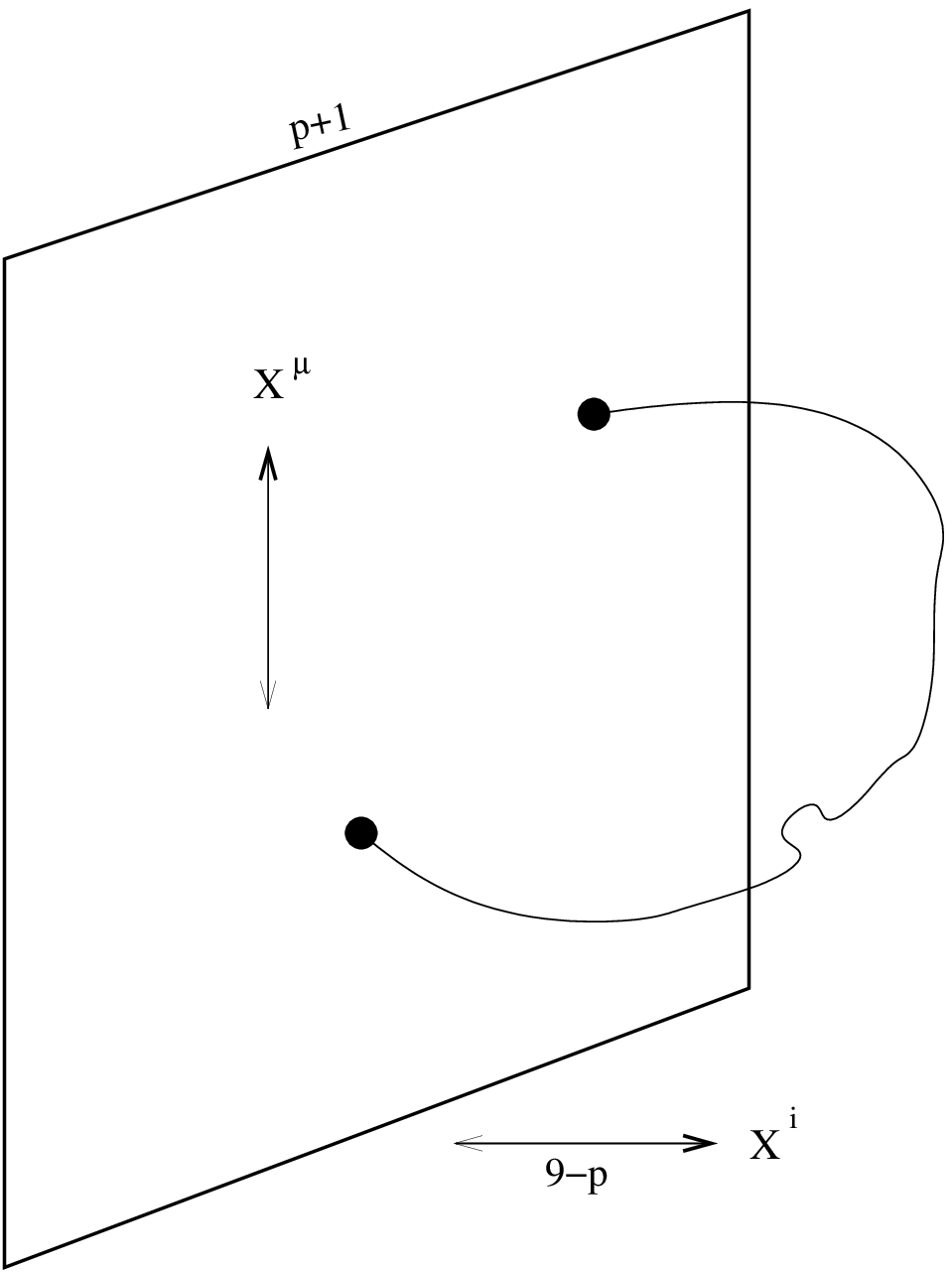}{Definition of D-branes in perturbative open string theory}{typical}

Having such a perturbative definition, let us ask what the spectrum of the open strings
subject to D-brane boundary conditions is.
Whereas D-branes exist in all string theories containing open strings, the detailed
spectrum of their fluctuations varies. All along this thesis, we will be mainly
concerned with branes in IIA and IIB superstring theories. To describe their spectra,
it is worth noticing first that their branes can preserve, at most, 16 supersymmetries.
This is easy to understand since the left/right worldsheet supercharges of type II theories
are separately conserved in closed topologies, but they must be identified in open topologies
due to the boundary conditions. As a result only one half of the background supercharges can
be preserved (at most), and this is the case of \eg flat D-branes in Minkowsky space.
This is one of the simplest and best understood brane configurations in string theory,
so let us take a pause to study them a bit further.

\subsection{Low energy effective action for a single D-brane} \label{ch1:sec:leff-action}

We start reviewing the spectrum of open superstrings in 10d Minkowsky space $\calm_{10}$ with
Neumann BC's for all of the scalar fields, which is identified with a D9-brane. We still have
the option to choose NS or R BC's for the fermion fields on
the worldsheet.
The vacuum in the NS sector is a tachyon, which is GSO-projected out;
the next physical states form a
vector representation of $SO(8)$, and therefore provide us with the on-shell degrees of
freedom of a massless spacetime abelian vector field $A_\mu$.
On the other hand the vacuum in the R sector is, after the GSO projection, just a
Majorana-Weyl spinor representation of $SO(8)$. As expected by the argument above, this is precisely
the content of the unique \caln=1 vector multiplet in ten dimensions. All other massive
modes in each sector have masses of the order $1/l_{s}$, and they form independent
supersymmetric representations at each mass level. Finally, the spectrum of any
other flat Dp-brane in $\calm_{10}$ can be found by dimensionally reducing the
just mentioned 10d spectrum to p+1 dimensions. It is very important to realize that various
polarizations of the vector $A_{\mu}$ are transformed into scalar degrees of
freedom for lower-dimensional branes. The expectation
values of these scalars can be interpreted as parameterizing the position of the brane
in its transverse space; indeed, they are precisely the massless Goldstone bosons
associated to the breaking of the global background \Poin symmetry by the presence
of the hyperplane. This interpretation can be supported in a number of different ways,
as we will keep encountering in the rest of this thesis.

Summarizing, the massless spectrum of open strings ending on a single D-brane consists of a $U(1)$
gauge multiplet with 16 supercharges in (p+1) dimensions and
the rest of excitations have spacetime masses
of order $1/l_{s}$. At low energies ($E \ll 1/l_s$) only the massless excitations
remain relevant, and the originally stringy theory reduces to a field theory
governed by the usual (p+1)-dimensional Super Yang-Mills (SYM) actions for the mentioned gauge
 multiplet.
This is typically proven either by analyzing the low energy limit of the various S-matrix processes
or by imposing the vanishing of the Weyl anomaly at first order in $l_s^2$.

Indeed one can do better than just writing the SYM action for these massless modes. Was one to
take into account the interactions of the massless modes with the rest of massive string modes,
the SYM action would then be just the first term of the complete action, thought of as an
expansion in $l_s^2$. It has been possible to resum all such contributions for the case of
constant gauge fields, and the complete bosonic action is of a Dirac-Born-Infield (DBI) type
\be \label{DBI}
S_{DBI} =  -\mu_{p} \int_{\Sigma_{p+1}} d^{p+1}\sigma \,
 e^{-\Phi} \left( \sqrt{ -\det \left( P\left[ G+B \right] +2\pi l_s^2 F
\right) } \right) \,,
\ee
with
\be
\mu_p  = \frac{1}{( 2 \pi )^p l_s^{p+1}} \,,
\ee
and the operator $P$ denoting the pullback of spacetime fields to the worldvolume.

Before completing the discussion of the single-brane effective actions, we need to
take into account that branes must act as sources of closed strings,
as can easily be seen just by worldsheet duality. In particular, branes {\it can} gravitate and
emit dilaton and RR fields quanta.
When the reaction of the background can be neglected, we say that we
are in the probe approximation; its validity depends only on the scale of energies we
want to study. We will later see that one can take limits where this approximation
is never valid (\eg by considering an infinite number of branes on top of each other),
and the backreaction must then be taken into account. Indeed the action \bref{DBI}
already includes the coupling to the background metric and dilaton; it was not
until \cite{Polchinski:1995mt} that the coupling to the RR fields was discovered. The DBI action
must then be supplemented with these new couplings, which turn out to be
of a Wess-Zumino type
\be \label{WZ}
S_{WZ} =\mu_p \int_{\Sigma_{p+1}} P\left[\bigoplus_n C_{n} e^{B} \right] e^{2 \pi l_s^2 F}\,,
\ee
where $C_{n}$ are the various  RR $n$-form fields, and we
have written the action as  a formal sum of forms of different degree; the integration
only picks up those with the correct degree to be integrated over $\Sigma_{p+1}$.

\subsection{Multiple D-branes} \label{ch1:sec:multiple-D}

An interesting phenomenon occurs when we consider $N$ D-branes in flat space
which, by definition, is equivalent to adding Chan-Paton factors to the endpoints
of the open strings.
To begin with, one may worry about whether considering this situation is worth at all.
It could well happen that no such a {\it static} configuration is achievable because
both branes attract or repel; this is what one would expect for
objects that gravitate and couple to gauge fields. Indeed, most part of this thesis
is build over exceptions to this naive expectation.

Let us leave for chapter~\ref{ch:wrapped} the general discussion and concentrate again
on branes in flat space. It is not hard to see that if all the branes are parallel,
then the set of boundary
conditions still preserve the same supersymmetries as a single brane.
Being a supersymmetric configuration, it minimizes the energy and it is therefore stable.
The physical explanation for this is that the interactions among the D-branes due to
interchange of closed string modes is exactly
zero at each mass level. For example, at the massless level, the gravitational
and dilatonic attraction is cancelled by the repulsion due to RR-fields exchange.

What about the spectrum?
Quantisation leads now to a massless spectrum that consists of a $U(1)^N$ gauge
supermultiplet. These massless modes correspond to the low energy excitations of the open
strings with both endpoints in only one D-brane. Quite remarkably, the next massive
states have masses which are now not just proportional to $1/l_s$ but to
$\Delta X/l_s$ times $1/l_s$. These are the lowest energy excitations
of strings with endpoints on different branes, $\Delta X$ being
the distance between them. The states with these masses have precisely the
right quantum numbers to be interpreted as the $W$-bosons for spontaneous symmetry
breaking of $U(N)$ to $U(1)^N$. It is then understandable that when any two branes
are placed on top of each other the $W$-bosons become massless, and
the gauge symmetry is enhanced from $U(1)\times U(1) \rightarrow U(2)$.
\figu{10}{7}{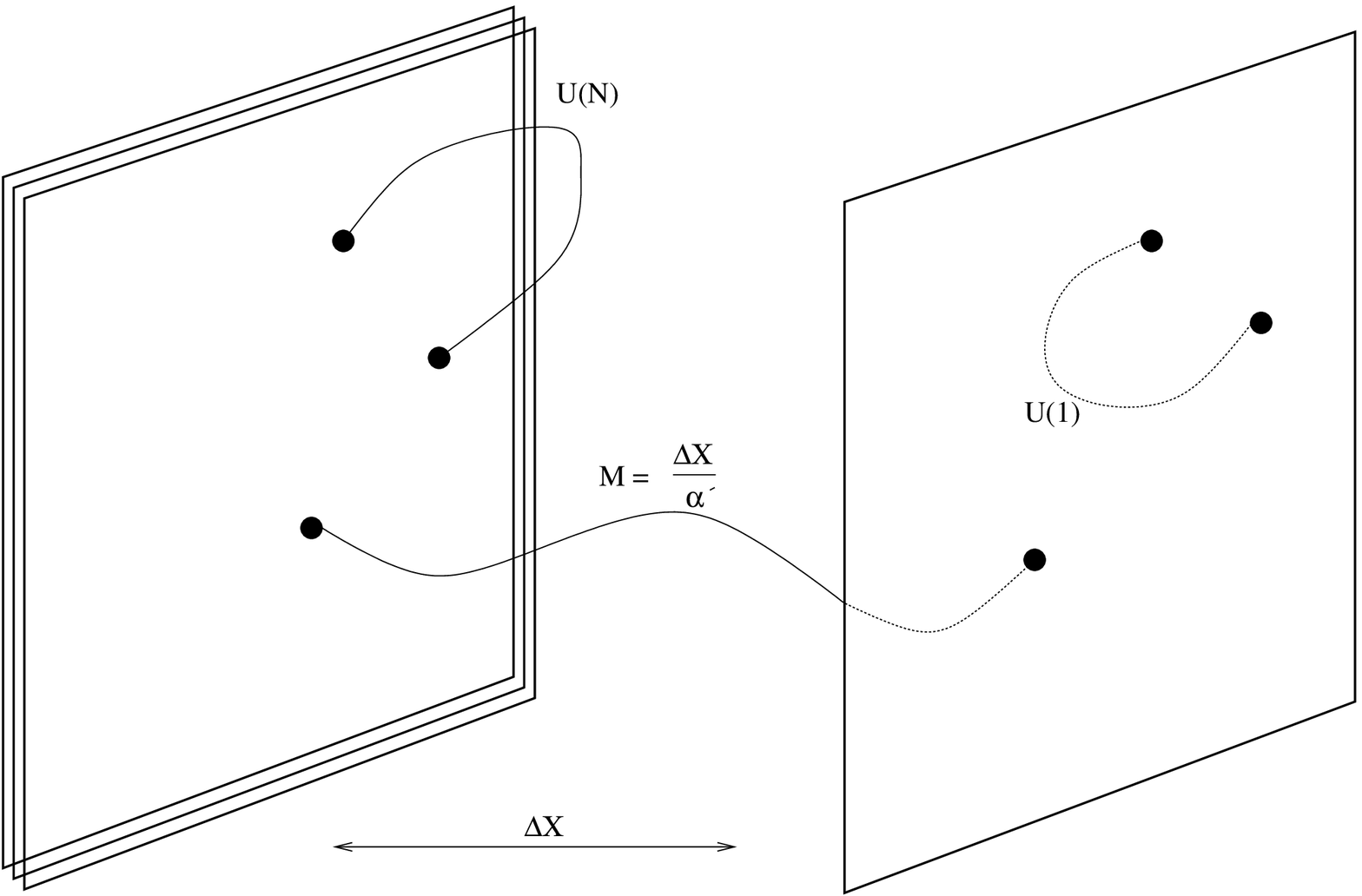}
Putting all of the D-branes together just provides us with a $U(N)$ supermultiplet
in $(p+1)$ dimensions with 16 supercharges, obtainable again from the
ten dimensional one by dimensional reduction. Note that this includes a set of
transverse scalar fields which, being in the same multiplet as the gauge fields,
transform in the adjoint of the gauge group.

What about the dynamics now? Repeating the same arguments above, one finds
that the low energy effective action for the massless fields is governed
by the non-abelian $U(N)$ SYM action in $(p+1)$ with 16 supersymmetries obtainable by
dimensional reduction from the ten dimensional one.
An important feature to note for later reference is that the resulting
YM coupling in terms of the string parameters is
\be
\gym^2 = (2\pi)^{p-2} g_s l_s^{p-3} \,.
\ee
Another important result is that the action
contains a positive-definite potential for the scalar fields $\phi^i$ of the type
\be \label{potentialz}
V \sim \sum_{i,j} [\phi^i,\phi^j]^2 \,.
\ee

Here we meet again a geometrical interpretation of a usual field theory phenomena.
There exists a moduli space of vacua which minimize \bref{potentialz} parametrized
by all the vev of the scalars lying in the Cartan subalgebra of the gauge group.
They are therefore simultaneously diagonalizable and their eigenvalues can be interpreted
again as parametrizing the positions of the $N$ branes in their transverse space.
Higgsing the gauge group just corresponds to giving a vev to one of this scalars,
and therefore to moving one of the branes apart. As mentioned above, the open strings
with one endpoint in the stack and the other in the fugitive brane become the massive
$W$-bosons.

We finish this review of the physics of multiple D-branes by mentioning that
the non-abelian generalization of the DBI action is still not known completely, and it
is not even clear that such an enterprize makes sense at all. However, some extra terms
are well-known and specially the ones involving the RR fields have recently attracted a lot of
attention, since they are capable to provide couplings of low-dimensional branes to RR
fields sourced by higher dimensional ones. We refer the reader to \cite{Myers:2003bw} for
a recent review on this subject.

\subsection{\4n SYM} \label{ch1:sec:4n-sym}

It is worth illustrating the previous general discussion in a particular example.
As this will be one of the most important cases, let us consider in more detail
the configuration of $N$ D3 branes in a flat IIB background. The low energy
theory is a $U(N)$ \4n SYM in 3+1 dimensions.  It turns out that a field theory action
with such properties is uniquely determined by the coupling constant
$\gym$ and the rank of the gauge group $N$.
The field content is: one gluon, 6 scalars and 4 Majorana gluinos. We will not
need to consider the fermions for most part of this thesis, but for the purposes
of writing the action in a simple way, we will group the 4 gluinos into a 10d
16-component Majorana-Weyl spinor. The notation for the fields is then
\be
A_\mu(x), \qquad
\phi_i(x),\quad i=1,\ldots, 6 \qquad
\chi_\alpha(x), \quad \alpha=1,\ldots 16 \,,
\label{fields}
\ee
where all fields are valued in the adjoint of the gauge group.
The action is
\bea
S & = \frac{2}{\gym^2}\, \int d^4x\, \Tr \, \Bigl\{ & \frac{1}{4}\, (F_{\mu\nu})^2
+\frac{1}{2}\, (D_\mu\phi_i)^2 -\frac{1}{4}\, [\phi_i,\phi_j]\,
[\phi_i,\phi_j] \nn &&+ \frac{1}{2}\, \bar\chi D\!\!\!\!/\,\, \chi - \frac{i}{2}
\, \bar \chi\, \Gamma_i\, [\phi_i,\chi]\, \Bigr \} \,,
\label{N4SYM}
\eea
where the $\G$-matrices are the 10d ones.

Although this action can be built just by imposing the mentioned properties,
it turns our that it enjoys a good extra amount of symmetry: superconformal
invariance. The super-Poincar\'e generators $\{P^{\mu}, M_{\mu\nu}, Q_{\a}\}$
and the R-symmetry ones $T^A$
are accompanied by the generators of special conformal transformations $K^{\mu}$,
dilations $D$ and conformal supersymmetries $S_\a$
The whole symmetry group is $PSU(2,2|4)$, whose bosonic
subgroup is $SO(2,4)\times SU(4)_R$. Under the $SU(4)_R$
$R$-symmetry $A_{\mu}$ is a singlet, the 4 fermions are in the fundamental,
and the scalars transform as a vector of the homeomorphic group $SO(6)$.
We will discuss in detail this superalgebra and its representations
in chapter 3 and in the appendix~\ref{ch2:sec:algebra}.

The everyday case is that the classical scale invariance of the action is immediately
broken at the quantum level, which is actually a virtue rather than a problem.
For example, this allows massless QCD to give an approximate
description of the real world, where we observe everything but scale invariance.
There is typically no way to make sense of the UV divergences of a QFT without
introducing a scale. The present \4n SYM theory is an exception since, at least
in perturbation theory, no single correlation function exhibits UV divergences.
Even instanton contributions are finite, and the theory is believed to be
UV finite. As a consequence, the $\b$-function is exactly zero and the
superconformal group remains as a symmetry of the quantum theory.

Although $N$ and $\gym$ completely determine the action, they do not
uniquely determine the theory; one still needs to specify the vacuum
in which he wants to live. Unlike most non-supersymmetric theories,
supersymmetric ones are often unable to dynamically determine the
lowest energy state. Indeed, they typically have a continuous
of such vacua (called moduli space) parametrized by the vacuum
expectation value (vev) of some fields. For our case, if we assume that
the vevs of the fermions and the gauge field are zero, we can give a
set of different vevs to the scalars such that they all minimize
the potential
\be \label{pote}
V \sim  [\phi_i,\phi_j]\, [\phi_i,\phi_j]\,.
\ee
We then speak of different phases or branches of the moduli
space. Being positive definite, minimization of \bref{pote}
is just the equation $V=0$, and this has two types of solutions:
\bitem
\item {\it The superconformal phase} is characterized by
all vevs of the scalars being zero. This clearly preserves
the whole superconformal group.
\item In {\it the Coulomb phase}, one has a nonzero vev
for at least one of the scalars. Note that if there are
more than one such nonzero vevs, we must require them
to be in the Cartan subalgebra of $SU(N)$, so that
these fields commute and we still have $V=0$. Having
$r$ such nonzero vevs will spontaneously break
$SU(N) \rightarrow U(1)^r \times SU(N-r)$. As soon as
this happens, we will observe at large distances the
appearance of massless photons with their usual
Coulomb interactions, and hence the name for this branch.
Needless to say, conformal invariance is also broken
due to the scales introduced by the scalars.
\eitem

We saw in section~\ref{ch1:sec:multiple-D} that the scalars have the
interpretation of giving the position of the D-branes
in their transverse space. Therefore, the Coulomb phase
is associated to having one (or more) branes separated
from the others. The degrees of freedom of the strings
connecting separated branes became the $W$-bosons,
with masses of the order
\be
M_W \sim {\Delta X \over l_s^2} \,.
\ee

\section{D-branes as solutions of closed string theory} \label{ch1:sec:closed-picture}

The picture we wish to present here is similar to
the more familiar picture of electrons and electric fields.
Consider putting an electron in an almost empty space (with weak
background fields). We would start describing the
electron by its worldline relativistic action plus a minimal
coupling
\be
S=\int_{worldine} \left[ ds \, + \, A_{1} \right] \,.
\ee
If the space had been truly empty before putting the electron,
we would have not considered the second term. This is the analog
of the probe approximation we were using in the previous section.
On the other hand, because of its coupling to the photon fields,
we are also used to describing the electron
by the electric field that it produces, which at long distances
behaves as $V \sim -e/r$. It this picture, the electron
is a delta-function source for the potential
\be \label{electron}
\vec{\nabla}^2 V \sim e \cdot \delta^{(3)}(\vec{x}-\vec{x}_0) \,.
\ee
This second point of view is the one we want to adopt here, \ie
the description of D-branes as  closed string backgrounds
with (typically) $\delta$ sources. As the low energy dynamics of
closed strings is supergravity the problem of singularities
is a little more sophisticated than in \bref{electron},
and one has to deal with horizons, proper assymptotics, causal structures,
naked singularities...

So, let us remain in type IIA/IIB string theory
and look for solutions of their corresponding supergravities
describing our D-branes. Our aim here is the most modest one:
we want solutions that describe {\it flat D-branes in flat
Minkowsky space}. We would like to stress that
\bitem
\item
{\it solutions corresponding to flat D-branes in flat space are not flat},
just like the electric field derived from \bref{electron}
is not zero, and it describes an electron in an empty space.
\item
{\it the solutions may even not contain any D-brane!}
The criterium of whether {\it there is a brane or not}
in a supergravity solution is normally answered by
whether it is a solution of the supergravity equations
of motion {\it with or without} source\footnote{See however
\cite{Marolf:2000cb} for a clear discussion of the different
concepts of charge used in the literature.}. According to this,
the background \bref{electron} contains an electron since
it does not solve the Maxwell equations alone but
\be
S  \sim \int d^4x \left( F_{\mu\nu}F^{\mu\nu} + j^{\mu} A_{\mu} \right)
\sac j^{\mu} \sim \delta^{(3)} (\vec{x}-\vec{x}_0) \d^{\mu}_0 \,.
\ee
A typical stringy counter example to this is the
geometry of the D3-brane solution. It is absolutely regular
everywhere and it solves the vacuum (without extra sources)
IIB supergravity equations of motion.
\eitem
Especially in cases where the second point applies, it
is standard to use the expression {\it geometric transition};
the object we started with disappears and only a curved geometry
with fluxes remains.

The solutions we are looking for were found in 1991~\cite{Horowitz:1991cd},
much earlier
than the understanding of Dp-branes in open string theory; they
had been called simply p-branes, and the name is still used
to emphasize the supergravity picture we are describing.
They can be found by requiring that they have the
properties expected for a D-brane:
\enub
\item The background must involve only those massless fields
that couple to the D-brane, \ie the graviton, the dilaton
and a $p+1$ RR potential.
\item It must have the isometries $ISO(1,p)\times SO(9-p)$
corresponding Poincar\'e invariance in the worldvolume
and rotational invariance in the transverse space.
\item It must break preserve 16 supersymmetries.
\enue
The solutions to these requirements are
\bea \label{sugrasol1}
ds^2_{10}&=&H^{-\undos} dx_{0,p}^2+H^{\undos} dx_{p+1,9}^2, \\
\label{sugrasol2}
e^{\phi}&=&g_s H^{3-p \over 4}, \\ A_{p+1}&=&-\undos(H^{-1}-1) dx^0 \wedge ... \wedge dx^{p} \,,
\label{sugrasol3}
\eea
where
\bea
H&=&1+ {R^{7-p}  \over r^{7-p}} \sac r^2=(x^{p+1})^2+...+(x^9)^2 \,,\\
R^{7-p}&=& {d_p \, \gym^2 N \ap^{5-p}} \sac
d_p=2^{7-2p} \pi^{9-3p \over 2} \Gamma\left({7-p\over 2}\right)\,.
\eea
We are using an optimized notation in which
\be
dx_{0,p}^2 ~\equiv~ -(dx^0)^2+(dx^1)^2+...+(dx^p)^2 \,.
\ee
Let us remark that these solutions are adapted to describing a single stack of $N$
p-branes. The function $H$ can be any harmonic function on the transverse space,
and the configuration still solves the e.o.m. and preserves the same number of supersymmetries.
So the obvious way to describe multiple stacks of parallel p-branes is to choose
\be
H= 1+ \sum_i {R_i^{7-p} \over |\vec{r}-\vec{r}_i|^{7-p}}
\sac R_i^{7-p} = {d_p \, \gym^2 N_i \ap^{5-p}} \,.
\ee

In general these metrics present a null curvature singularity at $r=0$.
This is the case of all $p$-branes with $p\neq 3,6$. For $p=6$ the
singularity is time-like and for $p=3$ there is no singularity
at all (it is a coordinate singularity) and one can analytically continue
the solution inside the horizon~\cite{Gibbons:1995vm}.

\medskip

\caixa{{\bf Validity of the solutions:}

It is extremely important to take care of the regimes of
validity of the description just given. A careful case-by-case
analysis was given in~\cite{Itzhaki:1998dd}. In this thesis we will be
more interested in the near-horizon limits rather than
the full solutions themselves. Therefore we
postpone this discussion until chapter~\ref{ch:ads-cft},
after the introduction of the AdS/CFT ideas.}

\newpage

\section{An example of brane dynamics: supertubes} \label{ch1:sec:supertubes}

Having discussed how D-branes appear in string theory,
we are ready to start exploiting the two descriptions
that they admit.

\subsection{Generalities of D-brane stabilization} \label{ch1:sec:stabilization}

All the examples we have seen so far
described flat D-brane configurations in flat space.
We saw that these are completely stable configurations
that preserve a high amount of supersymmetry. It is enough
to deal with such configurations for many purposes, \eg
to motivate the AdS/CFT correspondence.
Many other purposes, however, require the consideration
of more complicated configurations in less trivial
backgrounds. The gauge/string correspondence and
the appearance of NC gauge theories are examples of this.
One can think of different ways of complicating the picture,
like
\num{
\item considering non-flat D-branes,
\item putting them in non-flat backgrounds,
\item intersecting D-branes of (possibly) different dimensions.
}
All three issues have been intensively studied in the literature
and they have led to many interesting insights in different
areas of physics. A general problem which is common to the 3
generalizations is how to stabilize a given D-brane configuration.
Being extended massive and charged objects, different points
interact among each other and with the background, and stability is an exceptional
situation rather than a standard one.

There are cases in which supersymmetry guarantees stability
because supersymmetric states typically have the
minimum possible energy for their given charges. This statement
is powerful because it is normally proven by algebraic methods,
thus they do not depend on the perturbative approximations that
are normally needed to be made. We will see in detail how these
arguments work in chapter~\ref{ch:wrapped}. Note however that
supersymmetry is not always necessary, as there exist examples of
stable but non-supersymmetric brane configurations
(see \eg ~\cite{Sen:1998rg,Sen:1998ii}).

Let us guess which are the best candidates for
being stable but non-trivial D-brane configurations.
We start by keeping the background space flat and trying
to curve the D-branes. As soon as we move away from the
flat hyper-plane case, the D-brane tension will create
a tendency to modify such an embedding; indeed, if a part of the
D-brane is compact, such a tendency will be towards its collapse.
Maybe the simplest option is to change
the background for a topologically non-trivial one,
so that the collapse is prevented because the brane
wraps a non-contractible cycle. This will be
the subject of section~\ref{ch5:sec:wrapping-branes}.

In this chapter we will concentrate on supertubes,
which have the distinctive property that they provide
D-brane stabilization in {\it flat space}. Despite
the difficulties mentioned above, such configurations
are possible precisely because the dynamics of D-branes
are much richer than those of a simple relativistic extended
object.

\subsection{Preliminaries for the construction of the supertube in the open string picture} \label{ch1:sec:prelimaries}

The purpose of this section is to provide the background
and the intuition needed to understand why supertubes
were possible to be constructed. Their generalizations
to curved backgrounds is  also heuristically motivated.
We postpone a more formal treatment to the following sections.

In order to achieve the construction of a curved
brane in flat space we will exploit various couplings
that appear in the low energy dynamics of the open strings
attached to the brane. Recall that the action consists of
two pieces~\ref{DBI}-\ref{WZ}
\bea
S & =&  S_{DBI} + S_{WZ} \,,\nn
\label{DBI2}
S_{DBI} &=&  -\mu_{p} \int_{\Sigma_{p+1}} d^{p+1}\sigma \,
 e^{-\Phi}  \sqrt{ -\det \left( P\left[ G+B \right] +2\pi l_s^2 F
\right) }  \,, \\
\label{WZ2}
S_{WZ} &=&\mu_p \int_{\Sigma_{p+1}} P\left[\bigoplus_n C_{n} e^{B} \right] e^{2 \pi l_s^2 F}\,,
\eea
and that it is exact for constant worldvolume gauge fields $F_2$.

The first point we want to make is that the electric and magnetic
fields in $F_2$ are sources for background D(p-2)-branes and fundamental
strings F1, respectively. The reason for the former is that the presence of a magnetic $F_2$
flux in the worldvolume of the Dp-brane couples to the RR-potential of a D(p-2) through
one of the terms in the WZ action \bref{WZ2} as
\be \label{dp-2}
S_{WZ}|_{C_{p-1}} ~ \sim  ~\int_{\Sigma_{p+1}} F_2 \wedge C_{p-1} ~\sim ~q_{p-2} \int_{\Sigma_{p-1}} C_{p-1} \,,
\ee
where $q_s$ is the flux of the magnetic field on a spatial 2d submanifold of the brane's worldvolume.
The latter is due to the coupling of the electric components of $F_2$ to the electric
components of the background $B_2$-field, which is the source for F1. This coupling
appears when expanding the DBI action \bref{DBI2}
\be \label{f1}
S_{DBI} ~\ni~\int_{\Sigma_{p+1}} F_{\mu\nu}B^{\mu\nu} ~ \sim ~
\int_{\Sigma_{p+1}}  * F_2 \wedge B_2 ~ \sim ~ q_{F1} \int_{\Sigma_{2}} B_2 \,,
\ee
where $q_{F1}$ is the flux of the electric field on a spatial (p-1)-submanifold
of $\Sigma_{p+1}$ and $*$ is the worldvolume Hodge dual.

Indeed, saying that the $F_2$ is a source for such closed string fields requires
a point of view in which the two pieces of the actions \bref{dp-2}-\bref{f1}
are added to the (supergravity) closed strings action. If one takes the opposite
point of view, then we can say that the D(p-2)/F1 background supergravity fields
act as sources of magnetic/electric components of worldvolume gauge field.

These ideas are key to construct a supertube. Imagine taking a
large set of F1's in flat space and trying to form a tube $\CR
\times S^1$ with them. In the continuous limit in which we think
of a constant density of F1's along the cross section $S^1$, this
will look like a tubular D2 brane with dissolved $q_{F1}$ charge.
The D2 tension will try to collapse the tube, so we
could think of trying to stabilize it by making it rotate.
However, momentum tangent to the brane directions is unphysical,
 so we must abandon the idea. However, it
is known that we can link any number of D0-branes in an F1 at the
cost of breaking a half of the original 1/2 supersymmetries
preserved by the string. From the point of view of the D2,
the F1's will be described by an electric $F_2$ and the D0's by a
magnetic one. These fields generate a Poynting vector which, after
a careful choice of the D0/F1 charge densities, emulates the
effect of the necessary angular momentum that prevents
the D2 from collapse. Indeed, it was realized in~\cite{Mateos:2001pi} that one
can achieve any arbitrary cross section $S^1 \rightarrow \calc$
and still have a stable supersymmetric object. It is just
a matter of choosing the right local charge densities that
generate the appropriate Poynting vector; the latter
provides a centrifugal force which compensates the effect of the
tension {\it at every point of $\calc$}.

\medskip

Having heuristically explained how to stabilize supersymmetrically
a supertube in flat space, we can think of whether there is any
chance to do the same in a curved background. After all,
the effect of the background curvature is to modify the
precise value of the force felt by each point of $\calc$ by adding
a gravitational effect.
Qualitatively it does not look too different from the stabilization
of an arbitrary curve in flat space. We will confirm that this
is indeed possible. What it looks much more difficult is to be
able to do it preserving any supersymmetry at all. Two conditions
must be satisfied:
\num{
\item the background itself must leave some supersymmetries unbroken,
\item the supercharges preserved by the supertube must be compatible with
those preserved by the background.
}

The first step is then to choose a supersymmetric background of type IIA
supergravity. We will only consider cases in which the backgrounds
are purely gravitational, so that all fluxes are turned
off.\footnote{Note that this excludes the obvious possibility of putting
a supertube in the background created by a large number of supertubes,
which has been used recently in the study of closed timelike curves
in string theory.} Furthermore, we will only consider backgrounds of the form
$\CR^{1,1} \times \CM_8$, with $\CM_{8}$ a curved manifold. This is
because we want to put the longitudinal (and the timelike) directions of
the tube along the $\CR^{1,1}$ factor. The cross section $\calc$, however,
will completely lie inside $\CM_8$ as shown in the figure.
\begin{figure}[here]
\begin{center}
\includegraphics[width=8cm,height=4.5cm]{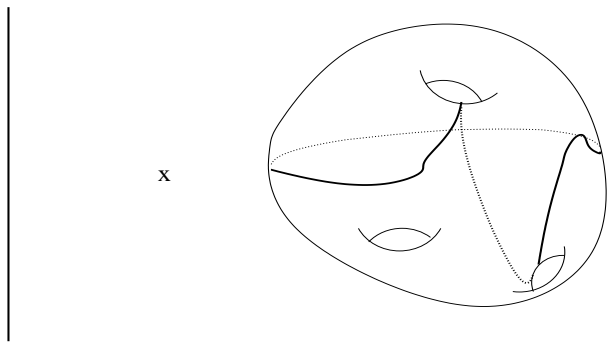}
\caption{The embedding of a supertube in $\CR \times \CM_8$. The
cross section is an arbitrary curve in $\CM_8$.}
\end{center}
\end{figure}

\medskip

In section~\ref{ch5:sec:special-holonomy} we will discuss in detail the
classification of supersymmetric backgrounds of this
type. Let us just cite the results here: it turns out
that \eeM must be one of the usual manifolds with reduced holonomy~\cite{Berger},
\begin{center}
\label{m8-taula}
\otaula {||c||c||} \hline \hline
 \eM  & Fraction of supersymmetry preserved     \\ \hline \hline
$\CR^4 \times CY_2$  &          1/2           \\ \hline
$CY_2 \times CY_2$  &          1/4           \\ \hline
$\CR^2 \times CY_3$  &          1/4           \\ \hline
$CY_4$  &        1/8           \\ \hline
$\CR \times G_2$  &     1/8           \\ \hline
Spin(7)  &      1/8           \\ \hline
Sp(2)  &        3/8           \\ \hline\hline
\ctaula
\end{center}
where we have indicated the fraction of supersymmetry
preserved by \eM; the maximum is 16, in which case \eM$=\CR^8$.

\subsection{Plan and summary of the results} \label{ch1:sec:plan}

We will extend the analysis of~\cite{Mateos:2001qs,Mateos:2001pi} and
show that it is possible to supersymmetrically embed the
supertube in these backgrounds in such a way that its time and
longitudinal directions fill the $\CR^{1,1}$ factor, while its compact
direction can describe an arbitrary curve $\CC$ in \eM.
The problem will be analyzed in two different descriptions.
\tem{
\item
In the first
one, we will perform a worldvolume approach by considering a D2 probe in
these backgrounds with the mentioned embedding and with an
electromagnetic worldvolume gauge field corresponding to the threshold
bound state of D0/F1. With the knowledge of some general properties of
the Killing spinors of the \eeM manifolds, it will be shown, using its
$\k$-symmetry, that the probe bosonic effective action is supersymmetric.
As in flat space supertubes, the only charges and projections involved
correspond to the D0-branes and the fundamental strings, while the D2
ones do not appear anywhere. This is why, in all cases, the preserved
amount of supersymmetry will be 1/4 of the fraction already preserved by
the choice of background.

Note that, in particular, this allows for configurations preserving a
single supercharge, as is shown in one of the examples that we present.%
\footnote{This is not in contradiction with the fact that the minimal
spinors in 2+1 dimensions have 2 independent components since, because of
the non-vanishing electromagnetic field, the theory on the worldvolume of
the D2 is not Lorentz invariant.} In the other example,
we exploit the fact that the curve $\CC$ can now wind around the
non-trivial cycles that the \eeM manifolds have, and construct a supertube
with cylindrical shape $\CR \times S^1$, with the $S^1$ wrapping one of
the non-trivial $S^2$ cycles of an ALE space. In the absence of D0 and F1
charges, $q_0$ and $q_s$ respectively, the $S^1$ is a collapsed point in
one of the poles of the $S^2$. As $|q_0 q_s|$ is increased, the $S^1$
slides down towards the equator. Unlike in flat space, here $|q_0 q_s|$
is bounded from above and it acquires its maximum value precisely when
the $S^1$ is a maximal circle inside the $S^2$.

\item
The second approach will be a spacetime description, where the
back-reaction of the system will be taken into account, and we will be
able to describe the configuration by means of a supersymmetric solution
of type IIA supergravity, the low-energy effective theory of the closed
string sector. Such solutions can be obtained from the original ones,
found in~\cite{Emparan:2001ux}, by simply replacing the 8-dimensional
Euclidean space that appears in the metric by \eM. We will show that this
change is consistent with the supergravity equations of motion as long as
the various functions and one-forms that were harmonic in $\euc^8$ are
now harmonic in \eM. It will also be shown that the supergravity solution
preserves the same amount of supersymmetry that was found by the probe
analysis.
}

The organization of the analysis is as follows: in section~\ref{ch1:sec:WVanal} we analyze
the system where the D2-supertube probes the $\CR^{1,1} \times \CM_8$,
and prove that the effective worldvolume action for the D2 is
supersymmetric using the $\k$-symmetry. In section~\ref{ch1:sec:Hamilanal} we
perform the Hamiltonian analysis of the system. We show that the
supersymmetric embeddings minimize the energy for given D0 and F1
charges, showing that the tension and the gravitational force induced
by the background are locally compensated by the Poynting
vector. In section~\ref{ch1:sec:examples} we give to examples in order to
clarify and illustrate these constructions. Section~\ref{ch1:sec:SGanal} is
devoted to the supergravity analysis of the generalised supertubes. We
prove there the supersymmetry from a spacetime point of view. Conclusions
are given in section~\ref{ch1:sec:concl}.

\subsection{Probe worldvolume analysis}  \label{ch1:sec:WVanal}

In this section we will prove that the curved direction of a supertube
can live in any of the usual manifolds with reduced holonomy, while still
preserving some amount of supersymmetry. The analysis will be based on
the $\k$-symmetry properties of the bosonic worldvolume action, and its
relation with the supersymmetry transformation of the background fields.

\subsubsection{The setup} \label{ch1:sec:setup}

Let us write the target space metric on $\CR^{1,1}\times \CM_8$ as
\begin{equation}
\rmd s^2_{IIA}=-(\rmd x^0)^2+(\rmd x^1)^2+ {e^{\underline{j}}} {e^{\underline{j}}}
\delta_{\underline{ij}}\,, \espai {e^{\underline{j}}}=\rmd y^j e_j{}^{\underline{i}} \,,
\espai i,j = 2,3,...,9\,,
\end{equation}
where ${e^{\underline{j}}}$ is the vielbein of a Ricci-flat metric on \eM.
Underlined indices refer to tangent space objects.
We will embed the supertube in such a way that its time and
longitudinal directions live in $\CR^{1,1}$ while its
curved direction describes an arbitrary curve $\CC$ in $\CM_8$.
By naming the D2 worldvolume coordinates $\{\s^0,\s^1,\s^2\}$,
such an embedding is determined by
\begin{equation} \label{embed}
x^0=\s^0, \espai\espai x^1=\s^1, \espai \espai y^i=y^i(\s^2)\,,
\end{equation}
where $y^i$ are arbitrary functions of $\s^2$.
Let us remark that, in general, the curve $\CC$ will be contractible in
\eM. As a consequence, due to its tension, the compact
direction of the D2 will naturally tend to collapse to a point.

Following~\cite{Mateos:2001qs}, we will stabilize the D2 by turning on an
electromagnetic flux in its worldvolume
\begin{equation} \label{fluxes}
 F_{\it 2}=E\, \rmd \sigma^0 \wedge \rmd \sigma^1
 + B\, \rmd \sigma^1 \wedge \rmd \sigma^2\,,
\end{equation}
which will provide the necessary centrifugal force to compensate the D2
tension and the gravitational effect due to the curvature of $\CM_8$.
We will restrict to static configurations.

The effective action of the D2 is the DBI action (the Wess-Zumino term
vanishes in our purely geometrical backgrounds),
\begin{equation} \label{delta}
S= \int_{\CR^{1,1} \times C} \rmd \sigma^0 \rmd \sigma^1 \rmd \sigma^2
{\cal L}_{DBI}\,, \espai \espai {\cal L}_{DBI}=-\Delta\equiv
-\sqrt{-\det[g+F]}\,,
\end{equation}
where $g$ is the induced metric determined by the embedding $x^M(\sigma
^\mu )$,
and $F_{\mu \nu }$ is the electromagnetic field strength. $M$ denotes the
spacetime components $0,1,\ldots ,9$, and $\mu $ labels the worldvolume
coordinates $\mu =0,1,2$. The $\k$-symmetry imposes restrictions on the
background supersymmetry transformation when only worldvolume bosonic
configurations are considered. Basically we get $\Gamma _\kappa \epsilon
=\epsilon$ (see e.g.~\cite{Bergshoeff:1997kr}), where $\epsilon$ is the
background Killing spinor and $\Gamma_\kappa$ (see
e.g.~\cite{Bergshoeff:1997tu}) is a matrix that squares to~1:
\begin{equation}
\rmd^3\sigma \;  \Gamma _{\kappa }=\Delta ^{-1}\left[ \gamma _{\it
3}+\gamma _{\it 1} \Gamma
  _*\wedge F_{\it 2}\right].
 \label{Gammakappa}
\end{equation}
Here $\G_*$ is the chirality matrix in ten dimensions (in our conventions
it squares to one), and the other definitions are
\begin{eqnarray}
 \gamma _{\it 3} & = & \rmd\sigma ^0\wedge \rmd\sigma ^1\wedge \rmd\sigma ^2
 \,\partial _0x^M \partial _1x^N\partial _2x^P
 e_M{}^{\underline{M}}e_N{}^{\underline{N}}
 e_P{}^{\underline{P}}\Gamma_{\underline{MNP}}\,,
\nonumber\\
 \gamma _{\it 1} & = & \rmd \sigma ^\mu \partial _\mu
 x^Me_M{}^{\underline{M}}\Gamma_{\underline{M}}  \,.
 \label{hulp}
\end{eqnarray}
where $e_M{}^{\underline{M}}$ are the vielbeins of the target space and
$\Gamma_{\underline{M}}$ are the flat gamma matrices. We are using Greek
letters for worldvolume indices and Latin characters for the target
space.

We are now ready to see under which circumstances can the
configuration~(\ref{embed}),~(\ref{fluxes}) be supersymmetric. This is
determined by the condition for $\k$-symmetry, which becomes
\begin{equation} \label{ksym}
[\G_{\uz\1u}\g_{2} + E \g_2 \G_* +B \G_{\underline{0}}\Gamma_*
- \Delta] \e = 0\,,
\end{equation}
where
\begin{equation}
\Delta^2=B^2+y'^{\underline{i}}y'^{\underline{i}}(1-E^2)\,, \qquad
y'^{\underline{i}}=y'^i e_i{}^{\underline{i}}\,,\qquad \gamma
_2=y'^{\underline{i}}\Gamma _{\underline{i}}\sac  y'^i
:=\partial_2y^i\,. \label{gamma2}
\end{equation}
The solutions of~(\ref{ksym}) for $\e$ are the Killing spinors of the
background, determining the remaining supersymmetry.

\subsubsection{Proof of worldvolume supersymmetry} \label{ch1:sec:proof}

In this section we shall prove that the previous configurations always
preserve $1/4$ of the remaining background supersymmetries preserved by
the choice of \eM. We will show that the usual supertube projections are
necessary and sufficient in all cases except when we do not require that
the curve $\CC$ is arbitrary and it lies completely within the flat
directions that \eeM may have. Therefore we first discuss the arbitrary
case, and after that, we deal with the special situation.

\avall

{\bf Arbitrary Curve:} If we demand that the configuration is
supersymmetric for any arbitrary curve in \eM, then all the terms
in~(\ref{ksym}) that contain the derivatives $y'^i(\s^2)$ must vanish
independently of those that do not contain them. The vanishing of the
first ones (those containing $\gamma _2$) give
\begin{equation} \label{F1}
\G_{\uz \1u }\Gamma _*\e=-E \e \espai \Longrightarrow \espai E^2=1\,,
\espai \textrm{and} \espai \G_{\uz \1u}\G_{*}\e=-\sign(E) \epsilon\,,
\end{equation}
which signals the presence of fundamental strings in the
longitudinal direction of the tube. Now, when $E^2=1$,
then $\Delta=|B|$, and the vanishing of the terms independent
of $y'^i(\s^2)$ in~(\ref{ksym}) give
\begin{equation} \label{D0}
\G_{\uz}\Gamma _*\e=\sign(B) \epsilon\,,
\end{equation}
which signals the presence of D0 branes dissolved in the worldvolume of
the supertube. Since both projections, (\ref{F1}) and (\ref{D0}),
commute, the configuration will preserve $1/4$ of the background
supersymmetries {\it as  long as they also commute with all the
projections imposed by the background itself.}

It is easy to prove that this will always be the case. Since the
target space is of the form \target2 the only nontrivial conditions that
its Killing spinors have to fulfil are
\begin{equation} \label{constant}
\nabla_i \e=\left(\partial_i + {1\over 4} w_i{}^{\underline{jk}}
\G_{\underline{jk}} \right)\e=0\,,
\end{equation}
with all indices only on \eeM (which in our ordering, means $2 \leq i
\leq 9$). If one prefers, the integrability
condition can be written as
\begin{equation} \label{integrability-con}
[\nabla_i,\nabla_j]\e = {1 \over 4} R_{ij}{}^{\underline{kl}}
\G_{\underline{kl}}\e =0\,.
\end{equation}
In either form, all the conditions on the background spinors involve
only a sum of terms with two (or none) gamma matrices of \eM. It is then clear
that such projections will always commute with the F1 and the D0
ones, since they do not involve any gamma matrix of \eM.

To complete the proof, one must take into account further possible
problems that could be caused by the fact that the projections considered
so far are applied to background spinors which are not necessarily
constant. To see that this does not change the results, note
that~(\ref{constant}) implies that all the dependence of $\e$ on the \eeM
coordinates $y^i$ must be of the form
\begin{equation}
\e=M(y)\e_0\,,
\end{equation}
with $\e_0$ a constant spinor, and $M(y^i)$ a matrix that involves only products of even number
of gamma matrices on \eeM (it may well happen that $M(y)= \unity $).
Now, any projection on $\e$ can be translated to a projection on
$\e_0$ since
\[
P\e=\e\,, \espai \mbox{with} \espai P^2=\unity \,,\qquad  \trace P=0\,,
\espai \Longrightarrow
\]
\begin{equation}
\tilde{P}\e_0 =\e_0\,, \espai \mbox{with} \espai \tilde{P}\equiv M^{-1}(y) P
M(y)\,, \quad
\tilde{P}^2=\unity \,,\quad  \trace \tilde{P}=0\,.
\end{equation}
The only subtle point here is that, if some of the $\e_0$ have to
survive, the product of $ M^{-1}(y) P M(y)$ must be a constant
matrix\footnote{Note that it is not necessary that $P$ commutes with
$M(y)$.}. But this is always the case for all the projections
related to the presence of \eM, since we
know that such spaces preserve some Killing spinors.
Finally, it is also the case for the F1 and D0 projections, since
they commute with any even number of gamma matrices on \eM.

The conclusion is that, for an arbitrary curve in \eeM to preserve
supersymmetry, it is necessary and sufficient to impose the F1 and D0
projections. In all cases, it will preserve $1/4$ of the background
supersymmetry. We will illustrate this with particular examples in
section~\ref{ch1:sec:examples}.

\avall

{\bf Non-Arbitrary Curve:} If we now give up the restriction that the
curve must be arbitrary, we can still show that the F1 and D0 projection
are necessary and sufficient, except for those cases in which the curve
lies entirely in the flat directions that \eeM may have. Of course, the
former discussion shows that such projections are always sufficient, so
we will now study in which cases they are necessary as well.

In order to proceed, we need to prove an intermediate result.

\noindent {\it Lemma: If the velocity of the curve does not point in a
flat direction of \eM, then the background spinor always satisfies at
least one projection like
\begin{equation} \label{esquematic}
P\e=Q\e   \,, \qquad \mbox{such that}\qquad [P,\gamma_2]=0 \,,\qquad
\{Q,\gamma_2\}=0\,,
\end{equation}
with $P$ and $Q$ a non-vanishing sum of terms involving only an even
number of gamma matrices, and $Q$ invertible}.

To prove this, we move to a point of the curve that lies in a curved
direction of \eM, i.e. a point where not all components of
$R_{ij}{}^{\underline{kl}}$ are zero. We perform a rotation in the
tangent space such that the velocity of the curve points only in one of
the curved directions, {\it e.g.}
\begin{equation}
y'^{\underline{9}}\neq 0 \,, \espai \espai y'^{\underline{a}}=0 \,,
\espai \espai a=2,...,8 \,,\qquad R_{ij}{}^{a9}\neq 0\,,
\label{9direction}
\end{equation}
for at least one choice of $i$, $j$ and $a$, and where we use the definitions
of~(\ref{gamma2}). With this choice, $\g_2$ becomes simply
$\gamma_2=y'^{\underline{9}} \G_{\underline{9}}$. Therefore, at least one
of the equations in~(\ref{integrability-con}) can be split in
\begin{equation}
\left( R_{ij}{}^{\underline{ab}}\G_{\underline{ab}} +
R_{ij}{}^{\underline{a9}}\G_{\underline{a9}} \right) \e=0 \,,
\end{equation}
with the definitions
\begin{equation} \label{define}
P= R_{ij}{}^{\underline{ab}}\G_{\underline{ab}} \,,  \espai\espai Q=
-R_{ij}{}^{\underline{a9}}\G_{\underline{a9}} \,.
\end{equation}
The assumption~(\ref{9direction}) implies that $Q$ is nonzero and
invertible, as the square of $Q$ is a negative definite multiple of the
unit matrix. This implies that also $P$ is non-zero since, otherwise,
$\e$ would have to be zero and this is against the fact that all the
listed \eeM manifolds admit covariantly constant spinors. It is now
immediate to check that $\gamma_2$ commutes with $P$ while it
anticommutes with $Q$, which completes the proof. \QED

\avall

We can now apply this lemma and rewrite one of the conditions
in~(\ref{integrability-con}) as an equation of the kind~(\ref{esquematic}).
We then multiply the $\k$-symmetry condition~(\ref{ksym}) by $P-Q$.
Clearly only the first two terms survive, and we can write
\begin{equation}
0=\left[  \Gamma _{\underline{01}}-E \Gamma _*\right](P-Q)\gamma _2
\epsilon =- 2 \left[  \Gamma _{\underline{01}}-E \Gamma _*\right]\gamma
_2 Q\epsilon=-2\gamma _2 Q\left[  \Gamma _{\underline{01}}+E \Gamma
_*\right]\epsilon\,.
 \label{Pkappa}
\end{equation}
Since $(\gamma_2)^2=y'_iy'^i$ cannot be zero if the curve is not
degenerate, we just have to multiply with $Q^{-1}\gamma _2$ to find
again~(\ref{F1}). Plugging this back into~(\ref{ksym}) gives the
remaining D0 condition~(\ref{D0}).

\avall

Summarizing, the usual supertube conditions are always necessary and
sufficient except for those cases where the curve is not required to be
arbitrary and lives entirely in flat space; then, they are just
sufficient. For example, one could choose $\CC$ to be a straight line in
one of the $\CR$ factors that some of the \eM have, and take a constant
$B$, which would correspond to a planar D2-brane preserving $1/2$ of the
background supersymmetry.

\subsection{Hamiltonian analysis} \label{ch1:sec:Hamilanal}

We showed that in order for the supertube
configurations~(\ref{embed}),~(\ref{fluxes}) to be supersymmetric we
needed $E^2=1$, but we found no restriction on the magnetic field
$B(\s^1,\s^2)$. We shall now check that some conditions must hold in
order to solve the equations of motion of the Maxwell fields. We will go
through the Hamiltonian analysis which will enable us to show that these
supertubes saturate a BPS bound which, in turn, implies the second-order
Lagrange equations on the submanifold determined by the constraints. We
will restrict to time-independent configurations, which we have checked
to be compatible with the full equations of motion. The Lagrangian is
then given by~(\ref{delta})
\begin{equation}
  {\cal L}=-\Delta =-\sqrt{B^2+R^2(1-E^2)}\,,
 \label{Lgeneral}
\end{equation}
where we have defined $R^2=y'^{\underline{i}}y'_{\underline{i}}$, and
$R>0$. To obtain the Hamiltonian we first need the displacement field,
\begin{equation}
  \Pi= \frac{\partial {\cal L}}{\partial
  E}\,=\frac{E R^2}{\sqrt{B^2+(1-E^2) R^2}}\,,
 \label{valuePi}
\end{equation}
which can be inverted to give
\begin{equation}
E=\frac{\Pi}{R} \sqrt{\frac{B^2+R^2}{R^2+\Pi ^2}}\,,\qquad \Delta=R
\sqrt{\frac{B^2+R^2}{ R^2+\Pi ^2}}\,.
 \label{E2}
\end{equation}

The Lagrange equations for $A_0$ and $A_2$ give two constraints
\begin{equation}
\partial_1 \Pi=0 \,, \espai\espai
  \partial _1\left(\frac{B}{ R}\sqrt{\frac{ R^2+\Pi ^2}{B^2+
  R^2}}\right)=0\,,
\label{gausslaw}
\end{equation}
the first one being the usual Gauss law. Together, they imply that
$\partial _1B=0$, i.e., the magnetic field can only depend on $\sigma^2$.
Finally, the equations for $A_1$ and $y^{\underline{i}}$ give,
respectively,
\begin{equation}
\partial _2\left(\frac{B}{R}\sqrt{\frac{ R^2+\Pi ^2}{B^2+
  R^2}}\right)=0
 \,, \qquad
 \partial _2\left[2y'^{\underline{i}} \frac{ R^4-\Pi ^2B^2}
 { R^2\sqrt{( R^2+\Pi ^2)( R^2+B^2)}} \right]=0 \,.
 \label{eom}
\end{equation}

The Hamiltonian density   is given by
\begin{equation}
  \mathcal{H}=E\Pi- \mathcal{L}=
\frac{1}{R}\sqrt{(R^2+\Pi ^2)(B^2+R^2)}\,.
 \label{valueH}
\end{equation}
In order to obtain a BPS bound~\cite{Gauntlett:1998ss}, we rewrite the
square of the Hamiltonian density as
\begin{equation}
  \mathcal{H}^2= \left(\Pi \pm B\right)^2+
  \left(\frac{\Pi B}{ R}\mp R\right)^2\,,
 \label{reH}
\end{equation}
from which we obtain the local inequality
\begin{equation} \label{inequality}
  \mathcal{H}\geq |\Pi\pm B|\,,
\end{equation}
which can be saturated only if
\begin{equation} \label{satura}
R^2=y'^{\underline{i}}y'_{\underline{i}}=\pm \Pi B \espai \Leftrightarrow
\espai E^2=1 \,.
\end{equation}
It can be checked that the configurations saturating this bound satisfy
the remaining equations of motion~(\ref{eom}).

Note that the Poynting vector generated by the electromagnetic field is
always tangent to the curve $\CC$ and its modulus is precisely $|\Pi B|$.
We can then use exactly the same arguments as in~\cite{Mateos:2001pi}.
Equation~(\ref{satura}) tells us that, once we set $E^2=1$, and
regardless of the value of $B(\s^2)$, the Poynting vector is
automatically adjusted to provide the required centripetal force that
compensates both the tension and the gravitational effect due to the background
curvature at every point of $\CC$. The
only difference with respect to the original supertubes in flat space is
that the curvature of the background is taken into account in
(\ref{satura}), through the explicit dependence of $R^2$ on the metric of
\eM.

Finally, the integrated version of the BPS bound~(\ref{inequality}) is
\begin{equation}
\tau \geq |q_0 \pm  q_s|\,,
\ee
with
\be \tau\equiv
\int_{\CC} {\rmd \sigma^2} \, \mathcal{H} \,,\qquad
q_0 \equiv \int_{\CC} {\rmd \sigma^2}\, B \,, \qquad
q_s \equiv \int_{\CC} {\rmd \sigma^2} \, \Pi \,.
\end{equation}
and the normalization $0\leq \s^2 < 1$. Similarly, the integrated bound
is saturated when
\begin{equation}
\label{length}
L(\CC)=\int_{\CC} {\rmd \sigma^2} \sqrt{  g_{22}}
 = \int_{\CC} {\rmd \sigma^2} \sqrt{y'^{\underline{i}}y'_{\underline{i}}}
 = \int_{\CC} {\rmd \sigma^2} \sqrt{|\Pi B|}
 =\sqrt{|q_s\,q_0|}\,,
\end{equation}
where $L(\CC)$ is precisely the proper length of the curve $\CC$,
and the last equality is only valid when both $\Pi$ and $B$ are constant,
as will be the case in our examples.

\subsection{Examples}  \label{ch1:sec:examples}

After having discussed the general construction of supertubes in reduced
holonomy manifolds, we shall now present two examples in order to
illustrate some of their physical features.

\subsubsection{Supertubes in ALE spaces: 4 supercharges} \label{ch1:sec:ale}

Let us choose \eM$=\CR^4 \times CY_2$, i.e. the full model being
$\mathbb{R}^{1,5}\times CY_2$. We take the $CY_2$ to be an ALE space
provided with a multi-Eguchi--Hanson metric~\cite{Eguchi:1978xp}
\begin{eqnarray}
&&\rmd s^2_{(4)}=V^{-1}(\yvec) \rmd\yvec \cdot \rmd\yvec +
V(\yvec)\left(\rmd\psi + \vec{A}\cdot \rmd\yvec\right)^2  \,, \nonumber\\
&&V^{-1}(\yvec)=\sum_{r=1}^{N} {Q\over |\yvec-\yvec_r|} \,,
\espai\espai\espai \vec{\nabla}\times \vec{A}=\vec{\nabla}V^{-1}(\yvec)
\,,
 \end{eqnarray}
with $\yvec \in \CR^3$. These metrics describe a $\U(1)$ fibration over
$\CR^3$, the circles being parametrized by $\psi\in [0,1]$. They present
$N$ removable bolt singularities at the points $\yvec_r$, where the
$\U(1)$ fibres contract to a point. Therefore, a segment connecting any
two such points, together with the fibre, form (topologically) an $S^2$.
For simplicity, we will just consider the two-monopoles case which,
without loss of generality, can be placed at $\yvec=\vec{0}$ and
$\yvec=(0,0,b)$. Therefore, the complete IIA background is
\begin{equation} \label{ALEIIA}
\rmd s^2_{IIA}=-(\rmd x^0)^2+(\rmd x^1)^2+...+(\rmd x^5)^2+ \rmd
s^2_{(4)} \,,
\end{equation}
with
\begin{equation}
V^{-1}(\yvec)={ Q \over |\yvec|} +{Q\over
|\yvec-(0,0,b)|} \,.
\end{equation}
Let us embed the D2 supertube in a way such that its longitudinal
direction lies in $\CR^5$ while its compact one wraps and $S^1$ inside
the $S^2$ that connects the two monopoles. More explicitly,
\begin{equation} \label{firstcase}
X^0=\sigma^0\,, \espai X^1=\sigma_1\,,\espai \psi=\sigma^2\,, \qquad
y^3=\textrm{const.} \,, \espai y^1=y^2=0 \,.
\end{equation}

\begin{figure}[here]
\begin{center}
\includegraphics[width=10.5cm,height=4.5cm]{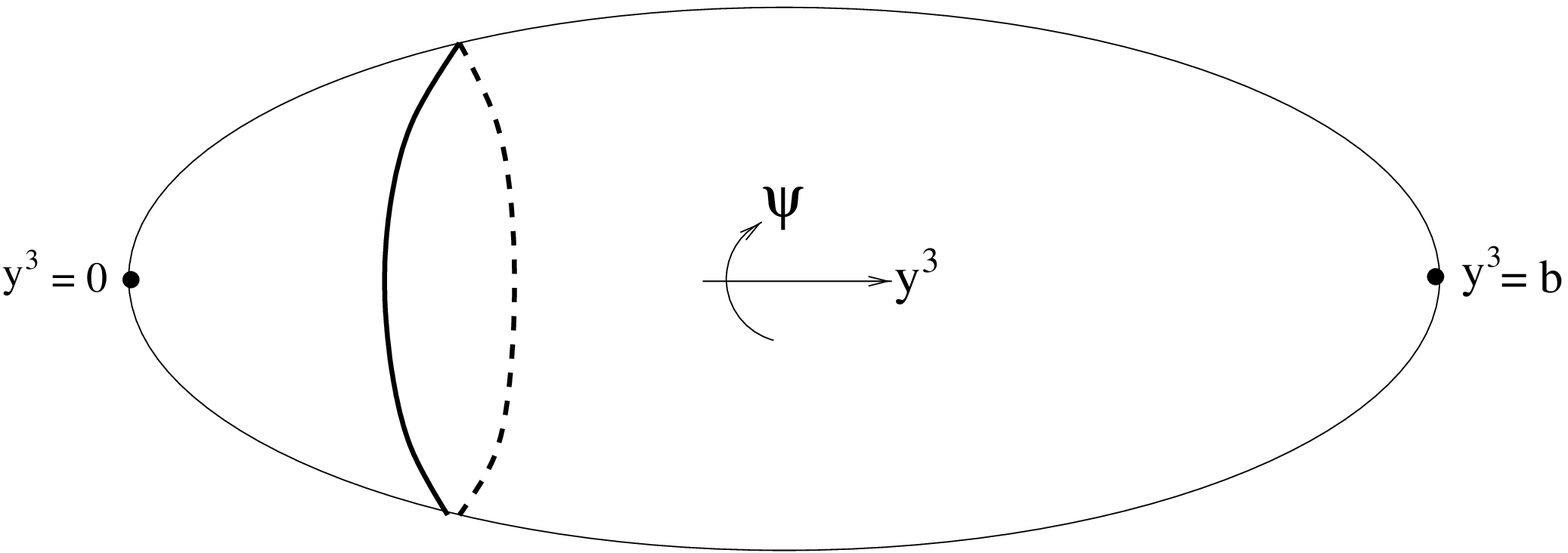}
\end{center}
\end{figure}
Since any $S^1$ is contractible inside an $S^2$, the curved part would
tend collapse to the nearest pole, located at $y^3=0$ or $y^3=b$. As in
flat space, we therefore need to turn on a worldvolume flux as
in~(\ref{fluxes}),
with $E$ and $B$ constant for the moment.

According to our general discussion, this configuration should
preserve $1/4$ of the 16 background supercharges already
preserved by the $ALE$ space. In this case, the $\k$-symmetry equation
is simply
\begin{equation}  \label{ALEkappa}
\left(\G_{\underline{01\psi}}+E \G_{\underline{\psi}}\Gamma
_*+B\G_{\underline{0}}\Gamma _* - \Delta \right)\e=0 \,,
\end{equation}
where $\e$ are the Killing spinors of the background~(\ref{ALEIIA}). They
can easily be computed and shown to be just constant spinors subject to
the projection
\begin{equation} \label{PALE}
\G_{\underline{y^1y^2y^3\psi}}\e=-\e \,.
\end{equation}
Then, the $\kappa $-symmetry equation can be solved by
requiring~(\ref{F1}) and~(\ref{D0}),
which involve the usual D0/F1 projections of the supertube. Since they
commute with~(\ref{PALE}), the configuration preserves a total of $1/8$
of the 32 supercharges.

It is interesting to see what are the consequences of having $E^2=1$ for
this case. Note that, from our general Hamiltonian analysis, we saw that,
for fixed D0 and F1 charges, the energy is minimized for $E^2=1$. When
applied to the present configuration,~(\ref{length}) reads
\begin{equation} \label{selects}
V(y^3)=|q_0 q_s| \,,
\end{equation}
which determines $y^3$, and therefore selects the position of
the $S^1$ inside the $S^2$ that is compatible with supersymmetry.
Since $V(y^3)$ is invariant under $y^3 \leftrightarrow (b-y^3)$,
the solutions always come in mirror pairs with respect to the
equator of the $S^2$. The explicit solutions are indeed
\begin{equation}
y^3_{\pm}={b\over 2} \left(1 \pm \sqrt{1-{4Q\over b}|q_0q_s|}\right)
\;.\end{equation} Note that a solution exists as long as the product of
the charges is bounded from above to
\begin{equation} \label{solu}
|q_0 q_s| \leq {b\over 4Q} \;.
\end{equation}
The point is that this will always happen due to the fact that,
contrary to the flat space case, the $S^1$ cannot grow arbitrarily
within the $S^2$. As a consequence, the angular momentum acquires its
maximum value when the $S^1$ is precisely in the equator. To see
it more explicitly, setting $E^2=1$ and computing
$q_0$ and $q_s$ for our configuration gives
\begin{equation}
|q_0 q_s| = V(y^3) \leq V(y^3 \rightarrow {b\over 2}) ={b\over 4Q} \;,
\end{equation}
which guarantees that~(\ref{solu}) is always satisfied.

Finally, note that we could have perfectly chosen, for instance, a more
sophisticated embedding in which $y^3$ was not constant. This would be
the analogue of taking a non-constant radius in the original flat space
supertube. Again, by the general analysis of the previous sections, this
would just require the Poynting vector to vary accordingly in order to
locally compensate for both the tension and the gravitational effect due
to the background everywhere, and no further supersymmetry
would be broken.

\subsubsection{Supertubes in $CY_4$ spaces: 1 supercharge} \label{ch1:sec:cy4}

The purpose of the next example is to show how one can reach a
configuration with one single surviving supercharge in a concrete
example. One could take any of the $1/8$-preserving backgrounds of the
\eeM Table. Many metrics for these spaces have been recently found in the
context of supergravity duals of non-maximally supersymmetric field
theories. Let us take the $CY_4$ that was found
in~\cite{Gomis:2001vg,Cvetic:2000db} since the Killing spinors have been
already calculated explicitly~\cite{Brugues:2002ff}\footnote{The
construction of this space and its Killing spinors is included
in this thesis, sections~\ref{ch5:sec:d6-solutions} and~\ref{ch6:sec:nc-gm}.}.
This space is a $C^2$ bundle over $S^2\times S^2$, and the metric is
\begin{eqnarray}
ds^2_{(CY_4)}&=&A(r)\left[ \rmd \theta _1^2+\sin^2\theta _1\rmd
\phi _1^2+
\rmd \theta _2^2+\sin^2\theta _2\rmd
\phi _2^2
\right]\nn
&&
  +U^{-1}\rmd r^2 + {r^2 \over 4}\left( \rmd\theta^2+ \sin^2\theta
\rmd\phi^2\right) \nn
&&+ {1\over 4}U r^2\left( \rmd\psi+\cos\theta \rmd\phi
+\cos\theta _1\rmd \phi _1+\cos\theta _2\rmd \phi _2\right) ^2\,,
 \label{metric11}
\end{eqnarray}
where
\begin{equation}
  A(r)={3\over 2}(r^2+l^2)\,, \qquad
  U(r)={3 r^4 + 8 l^2 r^2 + 6 l^4 \over {6(r^2+l^2)^2}} \,,\qquad
  C(r)=\frac14 U\,r^2\,.
 \label{defA}
\end{equation}
By writing the complete IIA background metric as
\begin{equation}
\rmd s^2_{IIA}=-(\rmd x^0)^2 +(\rmd x^1)^2 + \rmd s^2_{(CY_4)}\,,
\end{equation}
and using the obvious vielbeins, with the order
\begin{equation}
   \begin{array}{cccccccc}
    2& 3 & 4 & 5 & 6 & 7 & 8 & 9  \\
     \theta _1 & \theta _2 & \phi _2 & \phi _1 & r & \theta  & \phi  &
     \psi
  \end{array}
 \label{variables}
\end{equation}
the corresponding Killing spinors are
\begin{equation}
  \epsilon =\rme^{-\ft12\psi \Gamma _{\underline{78}}}\epsilon _0\,,
\end{equation}
with $\e_0$ a constant spinor subject to
\begin{equation}
 \Gamma _{\underline{25}}\epsilon_0 =\Gamma _{\underline{34}}\epsilon_0\,,\qquad
 \Gamma _{\underline{25}}\epsilon_0 =\Gamma_{\underline{78}}\epsilon_0 \,,\qquad
 \Gamma _{\underline{67}}\epsilon_0 =\Gamma _{\underline{98}}\epsilon_0 \,.
 \label{projCY4}
\end{equation}
To analyze $\k$-symmetry, let us take
the compact part of the supertube to lie along, say, the $\phi_1$
direction, while setting to constant the rest of the $CY_4$ coordinates.
As in the previous example, this would have the interpretation of
an $S^1$ embedding in one of the two $S^2$ in the base of the $CY_4$. Imposing
$\k$-symmetry:
\begin{equation}  \label{CY4kappa}
\left(\G_{\underline{015}}+E \G_{\underline{5}}\Gamma
_*+B\G_{\underline{0}}\Gamma _* - \Delta \right)\e=0 \,.
\end{equation}
Now, the first projection of~(\ref{projCY4}) happens to anticommute with
the $\gamma_2$ defined in~(\ref{gamma2})
\begin{equation}
\gamma _2=y'^ie_i{}^{\underline{i}}\Gamma _{\underline{i}}\, =\,
A^{1\over 2}(r)\, \sin{\theta_1} \, \Gamma_{\underline{5}} \,.
\end{equation}
In other words, this just illustrates a particular case
of~(\ref{esquematic}) for which the direction $\underline{5}$ plays the
role of $\underline{9}$, and for which $P=\Gamma _{\underline{34}}$ and
$Q=\Gamma _{\underline{25}}$. We can now follow the steps in
section~\ref{ch1:sec:proof} and multiply~(\ref{CY4kappa}) by $P-Q$. This
yields again the usual supertube conditions~(\ref{F1}) and~(\ref{D0}).

Since all the gamma matrices appearing in~(\ref{projCY4}),~(\ref{F1})
and~(\ref{D0}) commute, square to one and are traceless, the
configuration preserves only one of the 32 supercharges of the theory. Of
course, this is not in contradiction with the fact that the minimal
spinors in 2+1 dimensions have 2 components, since the field theory on
the worldvolume of the D2 is not Lorentz invariant because of the
non-vanishing electromagnetic field.

\subsection{Supergravity analysis} \label{ch1:sec:SGanal}

In this section we construct the supergravity family of solutions that
correspond to all the configurations studied above. We start our work
with a generalization of the ansatz used
in~\cite{Emparan:2001ux,Mateos:2001pi} to find the original solutions
in flat space.
Our analysis is performed in eleven dimensional supergravity, mainly
because its field content is much simpler than in IIA supergravity.
Once the eleven-dimensional solution is found,
we reduce back to ten dimensions, obtaining our generalised supertube
configurations.

The first step in finding the solutions is to look for supergravity
configurations with the isometries and supersymmetries suggested by the
worldvolume analysis of the previous sections. Then, we will turn to the
supergravity field equations to find the constraints that the functions
of our ansatz have to satisfy in order that our
configurations correspond to minima of the eleventh dimensional
 action.
Finally, we choose the correct behavior for these functions so that they
correctly describe the supertubes once the reduction to ten dimensions is
carried on.

\subsubsection{Supersymmetry analysis} \label{ch1:sec:supersymmetry}

Our starting point is the supertube ansatz
of~\cite{Emparan:2001ux,Mateos:2001pi}
 \begin{eqnarray}
  \rmd s^2_{\it 10} &=& - U^{-1}
V^{-1/2} \, ( \rmd t - A)^2 + U^{-1} V^{1/2} \, \rmd x^2 + V^{1/2} \,
\delta_{ij}\rmd y^i\rmd y^j \,, \nn B_{\it 2} &=& - U^{-1} \, (\rmd t -
A) \wedge \rmd x + \rmd t\wedge \rmd x\,, \nn C_{\it 1} &=& - V^{-1} \,
(\rmd t - A) + \rmd t \,, \nn C_{\it 3} &=& - U^{-1} \rmd t\wedge \rmd x
\wedge A \,, \nn e^\phi &=& U^{-1/2} V^{3/4} \,, \label{ds10}
 \end{eqnarray}
where the Euclidean space ($\mathbb{E}_8$) coordinates are labelled by
$y^i$, with $i,j,\cdots = (2,\ldots,9)$, $V=1+K$, $A=A_i\,\rmd y^i$ and
$B_{\it 2}$ and $C_{\it p}$ are respectively, the Neveu-Schwarz and
Ramond-Ramond potentials. $V,U,A_i$ depend only on the $\mathbb{E}_8$
coordinates.

To up-lift this ansatz, we use the normal Kaluza-Klein form of the eleven
dimensional metric and three-form,
\begin{eqnarray}
\rmd s^2_{\it 11} &=& e^{-2\phi/3}\rmd s^2_{\it 10}+e^{4\phi/3}(\rmd
z+C_{\it 1})^2 \,,\nonumber\\
N_{\it 3 } &=& C_{\it 3}+B_{\it 2} \wedge \rmd z \,,\label{uplift}
\end{eqnarray}
where $N_{\it 3 }$ is the eleventh dimensional three-form.
The convention for curved indices is $M=(\mu;i)=(t,z,x\,;\,2,3,...9)$ and
for flat ones
$A=(\alpha;a)=(\underline{t},\underline{z},\underline{x}\,;\,\underline{2},\underline{3}...,\underline{9})$.
The explicit form of the eleven-dimensional metric is given by,
\begin{eqnarray}
\rmd s^2_{\it 11} &=& U^{-2/3}\left[ -\rmd t^2 + \rmd z^2 + K(\rmd t+\rmd z)^2
+ 2(\rmd t+\rmd z)A +\rmd x^2\right]+U^{1/3}\rmd s^2_{\it 8}, \nonumber \\
F_{\it 4} &=& \rmd t \wedge \rmd(U^{-1})\wedge \rmd x\wedge \rmd z -
(\rmd t+\rmd z)\wedge \rmd x \wedge \rmd(U^{-1}A)\;, \label{ansatz}
\end{eqnarray}
where $F_{\it 4} = \rmd N_{\it 3}$. This background is a solution of the
equations of motion in eleven dimensions derived from the action
\begin{equation}
S_{11d}=\int \,\left[ R *1 \, - \, {1\over 2} F_{\it 4} \wedge *
 F_{\it 4} \, + \, {1 \over 3}
F_{\it 4} \wedge F_{\it 4} \wedge N_{\it 3} \right] \,,
\end{equation}
when the two functions $K$ and $U$, as well as the one-form $A_{\it 1}$,
are harmonic in $\mathbb{E}_8$, i.e.,
\begin{equation}
(\rmd *_8\rmd)U=0 \,, \zespai (\rmd *_8\rmd)K=0 \,, \zespai (\rmd
*_8\rmd)A_{\it 1}=0 \,, \label{harmonic}
\end{equation}
where $*_8$ is the Hodge dual with respect to the Euclidean flat metric
on $\mathbb{E}^8$. It describes a background with an M2 brane along the
directions $\{t,z,x\}$, together with a wave travelling along $z$, and
angular momentum along $\euc^8$  provided by $A_{\it 1}$.

Next, we generalize the ansatz above by replacing $\euc^8$ by one of the
eight dimensional \eeM manifolds of the table, and by allowing $K$, $U$
and $A_{\it 1}$ to have an arbitrary dependence on the \eeM coordinates
$y^i$. We therefore replace the previously flat metric on $\euc^8$ by a
reduced holonomy metric on \eM, with vielbeins $\te^a$. Hence, in
(\ref{ansatz}), we replace
\begin{equation}
U^{1/3} \delta_{ij} \rmd y^i \rmd y^j \espai \longrightarrow \espai
U^{1/3} \delta_{ab}\te^a \te^b  \,.\label{E8byM8}
\end{equation}

We use a null base of the cotangent space, defined by
\begin{eqnarray}
\label{secondbase}
&& e^+=-U^{-2/3}(\rmd t+\rmd z) \,, \zespai e^-=\undos (\rmd t-\rmd z) -
{K\over 2} (\rmd t+\rmd z) - A \,,\nonumber\\
&& e^x=U^{-1/3} \rmd x  \,, \zespai e^a=U^{1/6} \te^a \,.
\end{eqnarray}
This brings the metric and $F_{\it 4}$ into the form
\begin{equation} \label{flat}
\rmd s^2_{\it 11}=2 e^+ e^- + e^x e^x + \delta_{ab}e^a e^b  \,, \zespai
F_{\it 4}=-U^{-1} \, \rmd U\we e^x \we e^+ \we e^- \, -\, \rmd A \we e^x
\we e^+ \,.
\end{equation}
As customary, the torsion-less condition can be used to determine the
spin connection 1-form $\omega _{AB}$. In our null base, the only
non-zero
components are 
\bea
\omega_{+-}&=&-{U_a \over 3U} e^a \,,\nn
\omega_{+a}&=&\frac{1}{2} U^{1/2}\tilde K_a e^+-{U_a \over 3U}e^- -\frac12a_{ab} e^b\,,\nn
\omega_{-a}&=&-{U_a \over 3U} e^+ \,,\nn
\omega_{xa}&=&-{U_a \over 3U} e^x \,,\nn
\omega_{ab}&=&{U_b \over 6U}e^a -{U^a \over 6U}e^b +\tilde \omega _{ab} +\frac12a_{ab} e^+ \,,
\label{spin2}
\eea
were we have defined various tensor quantities through the relations
\begin{equation}
\rmd U = U_a e^a \,, \zespai  \rmd K=\tilde K_a \tilde e^a \,, \zespai
\rmd A=\ft12a_{ab} e^a \we e^b\,,
\end{equation}
and $\tilde{\omega}^{bc}$ are the spin connection one-forms corresponding
to $\tilde{e}^a$, i.e. $\rmd \tilde e^a+\tilde \omega ^a{}_b\tilde
e^b=0$.

We now want to see under which circumstances our backgrounds preserve
some supersymmetry. Since we are in a bosonic background i.e. all the
fermions are set to zero, we just need to ensure that the variation of
the gravitino vanishes when evaluated on our configurations. In other
words, supersymmetry is preserved if there exist nonzero background
spinors $\e$ such that\footnote{For the components of $p$-forms we use
the notations of~\cite{Candelas:1987is}.}
\begin{equation} \label{gravitino}
\left(\pa_A+{1\over 4}\omega_A{}^{BC}\G_{BC} -{1\over
288}\G_A{}^{BCDE}F_{BCDE}+{1\over 36} F_{ABCD}\G^{BCD}\right)\e=0 \,.
\end{equation}
We will try an ansatz such that the spinor depends only on the
coordinates on \eM. It is straightforward to write down the eleven
equations~(\ref{gravitino}) for each value of $A=\{+,-,x,a\}$. The
equation for $A=x$ is
\begin{equation}
 {U_a \over 6U} \Gamma _a\left(\Gamma_{x}-\Gamma_{+-}\right) \epsilon
 -{a_{ab}\over 12}\Gamma_{ab}\Gamma _-\epsilon =0\,.
 \label{SUSYx}
\end{equation}
Assuming that $a_{ab}$ and $\alpha _a$ are arbitrary and independent we
find
\begin{equation} \label{proj1}
\G_- \, \e=0 \,, \zespai \textrm{and} \zespai \G_{x}\e=-\e \,.
\end{equation}
Using these projections, it is a straightforward algebraic work to see
that the equation for $A=+$ and $A=-$ are automatically satisfied.
Finally, the equations for $A=a$ simplify to
\begin{equation} \label{proj2}
\nabla_{i}\e \, \equiv \, \left(\pa_{i}+{1\over 4}
\tilde{\omega}_i{}^{bc}\G_{bc}\right)\e =0 \,.
\end{equation}

By the same arguments as in the previous sections, the
projections~(\ref{proj1}) preserve 1/4 of the 32 real supercharges. On
the other hand,~(\ref{proj2}) is just the statement that \eeM must admit
covariantly constant spinors. Depending on the choice of \eM, the whole
11d background will preserve the expected total number of supersymmetries
that we indicated in the table on page~\pageref{m8-taula}.

To reduce back to IIA supergravity, we first go to another flat basis
\begin{equation}
  e^+= -U^{-1/3}V^{-1/2}\left( e^0+e^z\right)\,,\qquad
e^-=\ft12 U^{1/3}V^{1/2}\left( e^0-e^z\right)\,,
 \label{e+-tz}
\end{equation}
which implies that
\begin{equation}
  \Gamma _-=U^{-1/3}V^{-1/2}\left( \Gamma _0-\Gamma _z\right).
 \label{Gamma-zt}
\end{equation}
We reduce along $z$, i.e.\ replace $\Gamma _z$ by $\Gamma _*$. The
projections~(\ref{proj1}) become the usual D0/F1 projections, with the
fundamental strings along the $x$-axis.
\begin{equation}
\G_0\Gamma _*\e=-\e  \,, \zespai \textrm{and} \zespai \epsilon =-\Gamma
_x\epsilon =\G_{x0}\Gamma _*\e \,.
\end{equation}

\subsubsection{Equations of motion} \label{ch1:sec:eom}

Now that we have proved that the correct supersymmetry is preserved
(matching the worldvolume analysis), we proceed to determine the
equations that $U$, $K$ and $A_{\it 1}$ have to satisfy in order that our
configurations solve the field equations of eleven-dimensional
supergravity. Instead of checking each of the equations of motion, we use
the analysis of~\cite{Gauntlett:2002fz} that is based on the
integrability condition derived from the supersymmetry variation of the
gravitino~(\ref{gravitino}). The result of this analysis is that when at
least one supersymmetry is preserved, and the Killing vector
$\mathcal{K}_\mu \equiv \bar \epsilon \Gamma_\mu \epsilon$ is null, all
of the second order equations of motion are automatically satisfied,
except for
\begin{enumerate}
  \item The equation of motion for $F_{\it 4}$,
  \item The Einstein equation $E_{++}=T_{++}$,
\end{enumerate}
where $E_{++}$ and $T_{++}$ are the Einstein and stress-energy tensors
along the components $++$ in a base where $\mathcal{K}_\mu=\delta _\mu
^+\mathcal{K}_+$. Let us explain why the above statement is correct. The
integrability conditions give no information about the field equation for
the matter content, therefore the equation of motion for $F_{\it 4}$ has
to be verified by hand. Also, in most cases all of the Einstein equations
are automatically implied by the existence of a non-trivial solution
of~(\ref{gravitino}).

With~(\ref{proj1}) and in the base where the metric takes the
form~(\ref{flat}), and thus $\Gamma _+\Gamma _-+\Gamma _-\Gamma _+=2$, we
have
\begin{equation}
  \mathcal{K}_\mu =\bar \epsilon \Gamma_\mu \epsilon=\ft12\bar \epsilon \Gamma_\mu \Gamma _-\Gamma
  _+\epsilon\,.
 \label{Kmu}
\end{equation}
This vanishes for all $\mu $ except $\mu =+$, implying that our
configuration falls into the classification of those backgrounds that
admit a null Killing spinor and as a consequence the associated Einstein
equations escape the analysis. 
We thus have to check the two items mentioned above.

Let us start with the equation for $F_{\it 4}$, which is
\begin{equation} \label{fourform}
\rmd *F_{\it 4} + F_{\it 4} \we F_{\it 4} = 0 \,.
\end{equation}
Using the fact that the Hodge dual of a p-form
with respect to $e^a$ is related to the one with respect to
$\te^a$ by
\begin{equation}
*_8 C_p = U^{(4-p)/3} \ts_8 C_{p} \,,
\end{equation}
where
\begin{equation}
C_p=\frac{1}{p!}C_{a_1\ldots a_p}\tilde e^{a_1}\wedge \ldots \wedge
\tilde e^{a_p}\ \rightarrow \tilde *_8 C_{p}=
\frac{1}{p!(8-p)!}C_{a_1\ldots a_p} \varepsilon ^{a_1\ldots a_8}\tilde
e^{a_{p+1}}\wedge \ldots \wedge \tilde e^{a_8}\,,
\end{equation}
it is easy to see that~(\ref{fourform}) becomes
\begin{equation}
0= (\rmd\ts_8\rmd) U + (\rmd t+\rmd z)\we (\rmd\ts_8\rmd)A \,.
\end{equation}
This implies that $U$ and $A_{\it 1}$ must be harmonic with respect
to the metric of \eM, i.e.,
\begin{equation}
(\rmd\ts_8\rmd) U = 0 \,, \zespai (\rmd\ts_8\rmd)A_{\it 1}=0 \,.
\label{harmonicUA}
\end{equation}
Finally, using~(\ref{flat}) and~(\ref{spin2}), one can explicitly compute
the $\{++\}$ components of the Einstein and stress-energy tensors, and
obtain
\begin{eqnarray}
E_{++}&=&R_{++}=-\ft12U^{1/3}(\ts_8\rmd\ts_8\rmd)K + \ft12 *_8\left(\rmd
A\we *_8 \rmd A\right) \,,\nonumber\\
T_{++}&=&\ft1{12}F_{+ABC}F_+{}^{ABC} = \ft12 *_8\left(\rmd A\we *_8 \rmd
A\right) \,.
\end{eqnarray}
Therefore, the last non-trivial equation of motion tells us
that also $K$ must be harmonic on \eM,
\begin{equation}
(\rmd\ts_8\rmd)K=0 \,. \label{harmonicK}
\end{equation}

\subsubsection{Constructing the supertube} \label{ch1:sec:construction}

In order to construct the supergravity solutions that properly describe
supertubes in reduced holonomy manifolds, we reduce our
eleven-dimensional background to a ten-dimensional background of type IIA
supergravity, using~(\ref{uplift}) again. We obtain~(\ref{ds10}) with the
replacement~(\ref{E8byM8}), and the constraints~(\ref{harmonicUA})
and~(\ref{harmonicK}).
At this point we have to choose $U$, $K$ and $A_{\it 1}$ so that they
describe a D2-brane with worldvolume $\CR^{1,1} \times \CC$, with $\CC$
an arbitrary curve in \eM. As it was done
in~\cite{Emparan:2001ux,Mateos:2001pi}, one should couple IIA
supergravity to a source with support along $\CR^{1,1} \times \CC$, and
solve the \eeM Laplace equations~(\ref{harmonicUA}) and~(\ref{harmonicK})
with such a source term in the right hand sides. If this has to
correspond to the picture of D0/F1 bound states expanded into a D2 by
rotation, the boundary conditions of the Laplace equations must be such
that the solution carries the right conserved charges. In the appropriate
units,
\begin{equation}
q_0=\int_{\pa \CM_8} \ts_8 \rmd C_{\it 1} \,, \zespai q_s=\int_{\pa \CM_8}
\ts_8 \rmd B_{\it 2} \,, \zespai A_{\it 1}
 \stackrel{\pa \CM_8}{\longrightarrow}L_{ij}y^j \rmd y^i \,.
\end{equation}
Here, as in~\cite{Emparan:2001ux,Mateos:2001pi}, $L_{ij}$ would have to
match with the angular momentum carried by the electromagnetic field that
we considered in the worldvolume approach.

The Laplace problem in a general manifold can be very complicated and, in
most cases, it cannot be solved in terms of ordinary functions. We will
not intend to do so, but rather we will just claim that, once $U$, $K$
and $A_{\it 1}$ have been determined, they can be plugged back
into~(\ref{ds10}), with~(\ref{E8byM8}), and the background will describe
the configurations that we have been discussing here. It will
have the expected isometries, supersymmetries and conserved charges.

\subsection{Conclusions} \label{ch1:sec:concl}
We have shown that the expansion of the D0/F1 system into a D2 can happen
supersymmetrically in all the backgrounds of the form $\CR^{1,1} \times
\CM_8$, with \eeM any manifold of the table, both in the
worldvolume and in the supergravity setting. We would like to stress here that
this is not enough to prove that the system is stable in all regimes,
as we have only worked in the two mentioned approximations. Indeed,
the cross section of the D2-brane can be chosen such that two
pieces of D2 that locally look like a pair D2/anti-D2 are arbitrarily
close to each other.\footnote{See \cite{Hyakutake:2002fk,Bak:2001tt,Bak:2002wy}
for further elaboration on the understanding of supertubes from
various dual configurations.}
 This leads to the suspicion that the system
develops instabilities against annihilation that may have escaped
to our approximations. This led the authors of \cite{Mateos:2001pi}
to explicitly check that no such stability is actually present in
the simplest case of a flat D2/anti-D2 pair in flat space. Although there
is no hope to carry a similar analysis in our more sophisticated backgrounds,
we believe that the same result applies.

We remark that our research is
different from~\cite{Grandi:2002gt}, where it was shown that {\it the
supertube itself}, after some T-dualities, can be described by a special
Lorentzian-holonomy manifold in eleven dimensions.

\chapter{AdS/CFT beyond supergravity and supersymmetry}\label{ch:ads-cft}

This chapter deals with one of the most remarkable
dualities derived from string theory: the AdS/CFT
correspondence~\cite{Gubser:1998bc,Maldacena:1998re,Witten:1998qj}.
First, we describe how it was established
from D-brane considerations. We pay special attention
to  carefully settle the regions where both sides of the duality
are technically tractable. We then move to the difficult
enterprize of trying to find observables that can
be computed on both sides, and then compared.
This will lead us to introduce the work of Berenstein,
Maldacena and Nastase~\cite{Berenstein:2002jq}, and the shortcut
provided by Gubser, Klebanov and Polyakov~\cite{Gubser:2002tv}.

The main part of this chapter is section~\ref{ch2:sec:ads-beyond},
which contains an expanded discussion of the results
presented in~\cite{Mateos:2003de} and~\cite{Mateos:2004rn}.
These papers are devoted to exploiting the ideas
of GKP in order to (try to) test the AdS/CFT
correspondence away from supergravity and supersymmetry.

Finally, section~\ref{ch2:sec:stable-branes} contains unfinished work with
D. Mateos and P. K. Townsend about the possibility of adding stable but
non-supersymmetric
matter to superconformal field theories via the AdS/dCFT correspondence.

\section{The AdS/CFT correspondence} \label{ch2:sec:ads-cft}

Having discussed the two dual pictures of D-branes in chapter~\ref{ch:dbranes},
we are ready to introduce the ideas of
AdS/CFT correspondence, which essentially
consist on taking seriously the equivalence
of both descriptions.\footnote{Again, we
refer the reader to~\cite{{Aharony:1999ti}} for a
deep and careful discussion.}
Our discussion will be now centered in the
case of D3-branes. The generalization to other
Dp-branes will be given in section~\ref{ch5:sec:more-branes}.

\bitem
\item
Let us consider  the open string description.
of a set of $N$ D3 probes in the Minkowsky
vacuum. The world consists of closed strings
oscillating in the 10d space and open strings
oscillating with their endpoints stuck to the probes.
At low energies only the massless excitations matter,
and their dynamics are schematically given by
\be
S = S_{open} + S_{closed} + S_{int} \,,
\ee
where $S_{int}$ governs the open-closed interactions and
all three actions must be understood in the Wilsonian sense,
as having integrated out
the massive modes. We already know that
\bea \label{aa}
S_{open} &=& S\left[\caln=4 \,\, \mbox{ SYM}\right] + \calo(l_s^2) \,,\\
S_{closed} &=& S\left[  \mbox{IIB SUGRA} \right] + \calo(l_s^2) \,,.
\eea
The most important point is whether we can show that
$S_{int}$ is of order $l_s$ or not. What we do know for sure
is that $S_{int} = \calo( l_P)$ where $l_P$ is
the 10d Planck's length
\be
l_P \sim g_s l_s^4 \,.
\ee
This is because the open-closed interaction starts receiving contributions at diagrams
of order $g_s$ (the factor $l_s^4$ must be there on dimensional grounds).
Now the whole subtle point is to realize that \bref{aa} is meaningful
only if we have kept the YM coupling fixed in the $l_s \rightarrow 0$
limit. As for our present case of D3 branes we simply have $\gym^2 \sim g_s$,
there is no obstruction to having simultaneously
\be
l_s \rightarrow 0 \sac \gym^2 = \mbox{fixed} \sac l_P \rightarrow 0 \,.
\ee
We will see in sections~\ref{ch5:sec:d5-diagram} and~\ref{ch5:sec:d6-diagram} the problems that this limit
originates for D5 and D6 branes, respectively.

The conclusion is that in the zero slope limit of this system we obtain
two completely decoupled theories: an \4n SYM theory plus free supergravity
about the flat vacuum.

\item
We now try to do the same in the closed string picture of the D-branes.
The effective action for its massless modes is now just
\be
S = S[\mbox{IIB SUGRA}^{*}] + \calo (l_s^2)\,,
\ee
where by SUGRA$^*$ we mean that one considers supergravity
excitations about the D3 vacuum solution, given by
setting $p=3$ in  \bref{sugrasol1}-\bref{sugrasol3}.
It was realized that in the ${\0a}$ limit this  geometry
becomes disconnected into two regions:
\enub
\item  Near-Horizon region where $ {r/ l_s^2}=\mbox{fixed}$.
\item  Asymptotic  region where ${r/ l_s^2} = \mbox{unbounded} $.
\enue
By 'disconnected' we mean that, in the limit,
no excitation is able to scape from region 1 to region 2
and no excitation from region 2 is able to scatter
with those in region 1. The geometry of both regions can be obtained by
neglecting either the first or the second term
in the function $H=1+{R^4\over r^4}$ of the solution.
The result is
\enub
\item  Near-Horizon metric:
\be
ds^2={R^2\over r^2} dx_{0,3}^2+ {r^2\over R^2} dr^2 + R^2 d\O_5^2 \,,
\ee
which is the metric of \5ads with both factors having the same
radius $R$.
\item  Asymptotic metric: just 10d Minkowsky.
\enue
So we are led again to two decoupled systems, the one being
supergravity excitations about flat space and the other
{\it any closed string} excitation in $AdS_5\times S^5$. The way this decoupling
was obtained makes it clear the we must be careful with
what we mean by {\it any} in the expression in italics. We cannot
allow for almost infinite energy excitations, as they would have never
decoupled from the assymptotics region. One is restricted to consider
those that do not change the assymptotics of \5ads space.\footnote{This
discussion is relevant, for example, in the so-called $AdS/dCFT$ correspondence.
There one introduces an infinite D-brane probe in $AdS$ which intersects
the boundary. There are then extra degrees of freedom that
do not decouple and the dual field theory contains extra matter fields
not present in the \4n supermultiplet.}
\eitem
Summarizing, we end up with the two points of view having split
into two disconnected pieces. As a common factor for both is
just closed string excitations about a flat background, one
is led to conjecture that the remaining two pieces describe
equivalent physics, \ie

\medskip

\caixa{
\cent{
\begin{tabular}{ccc}
\4n SU(N) SYM in 3+1  & = &  IIB string theory on \5ads \\ \\
$\gym^2$ & = & $g_s$ \\ \\
$\l \, \equiv \, \gym^2 N$  & = & $\left(R_{AdS}\,/\,l_s\right)^4$
\end{tabular}}}

\medskip

Note that we wrote $SU(N)$ instead of $U(N)$ as would be expected
from the worldvolume gauge theory considerations of the previous chapter.
This is because a $U(N)$ gauge theory is locally equivalent to a $U(1)$
vector multiplet times an $SU(N)$ theory; the $U(1)$ factor decouples
from the rest of degrees of freedom. There is however no single field
in the supergravity side that does not couple to gravity, so the
$U(1)$ degrees of freedom are not expected to be visible as excitations
in the bulk of $AdS$. They are apparently related to the topological
theory of $B$-fields on $AdS$~\cite{Witten:1998wy}.

This conjecture realizes in an explicit example
the old 't Hooft's idea that
the degrees of freedom of non-abelian gauge theories could be
better described in the non-perturbative phase in
terms of stringy degrees of freedom. He showed that
the perturbative expansion of correlation functions of SYM theories admits
a classification in terms of two parameters: $N$ and
$\l=\gym^2 N$.\footnote{As we said, the \4n gauge Lagrangian
is fixed given $\gym$ and $N$, but it is more convenient
for what follows to consider $\l$ and $N$ as the independent
parameters.}
For example, the partition function can be written as
\be \label{thooft}
\calz = \sum_{g=0}^{\infty} N^{2-2g} f_g(\l) \sac
f_g(\l)=\sum_{n=0}^{\infty} \l^n c_{g,n} \,,
\ee
where $g$ is just the genus of a contributing
diagram written in double-line notation. In other words,
the power of $N$ in each diagram is only controlled by its
topology. This resembles very much the role played by
$g_s$ in perturbative string theory.

't Hooft had in mind the chance of simplifying the
gauge theory by taking $N$ very large. The form of \bref{thooft}
tells us how that, in order to keep interactions, this limit
must be taken such that
\be
\l=\gym^2 N = \mbox{fixed} \sac N \rightarrow \infty \,,
\ee
a limit in which only planar diagrams contribute.

\subsection{Pre-BMN ranges of validity and comparability} \label{ch2:sec:ranges}

The purpose of this subsection is to make clear what are
the regions of the parameter space in both sides of the
duality that are under control, and what should we
expect away from these regions.
\bitem
\item {\bf \4n SYM side.} Although we may not be too familiar
with this fact, the effective coupling of an $SU(N)$ gauge theory
is not $\gym^2$ but $\l=\gym^2 N$ (clearly this distinction does not matter
for the Standard Model!). So we only have access to {\it perturbative
computations} in the field theory side if we take $\l \ll 1$. To
go beyond that we would need non-perturbative computations. We stress
however that the theory is thought to be well-defined for all $\l$,
for example by its path integral or lattice definition. Note as well that
the theory is greatly simplified if we also take $N \gg 1$ since,
by \bref{thooft}, only the planar contributions survive. Summarizing:
\num{
\item Perturbative SYM $\Longleftrightarrow \,\,$ $\l \ll 1$.
\item Perturbative and Planar SYM $\Longleftrightarrow \,\,$ $\l \ll 1, \, N \gg 1$.
}

\begin{figure}[here]
\cent{\epsfig{file=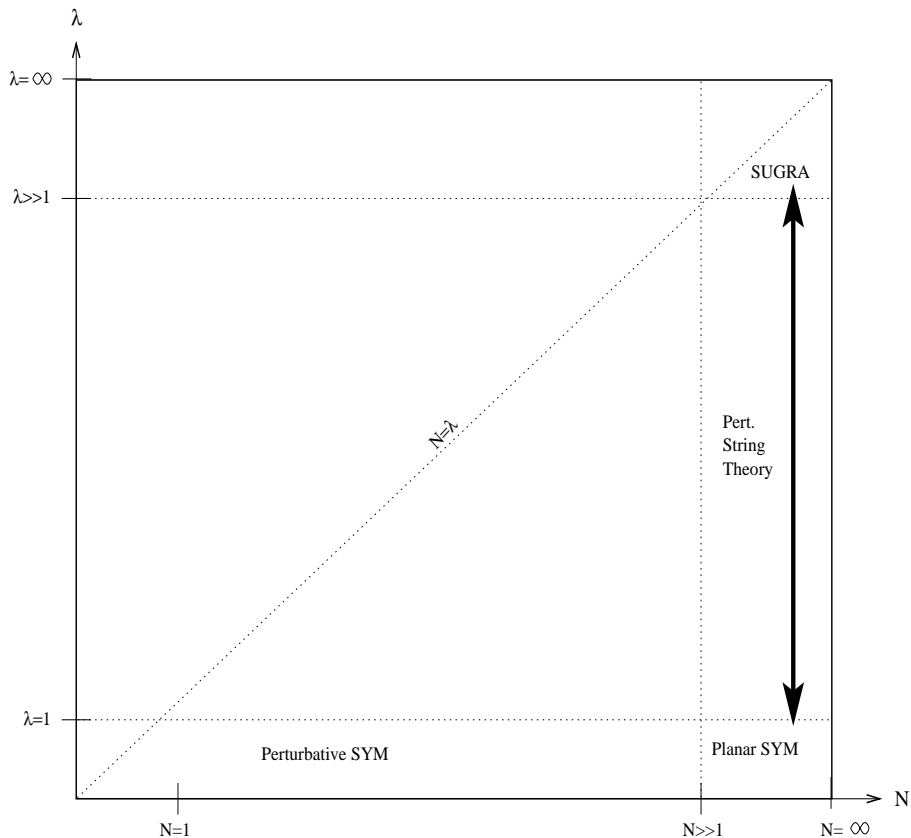, height=11cm, width=12cm}}
\caption{\small{The ranges of validity of the AdS/CFT before BMN. A point
in this $(\l,N)$-plane completely determines the SYM Lagrangian.
The thick line illustrates that the available
{\it tests} of the correspondence involved extrapolation of
SYM observables
whose value does not depend on the coupling $\l$.\label{ads-pre-bmn}}}
\end{figure}

\item {\bf String theory side.} There are three problems one has to
face here. The first one is that we do not even have a non-perturbative
definition of type IIB string theory. Even if we believe in $S$-duality,
there is a whole gap between the weakly and strongly coupled extremes
which is not under control. So the first requirement is more a conceptual
one than a computational one: we need $g_s < 1$; translated to YM variables,
$\l/N <1$. The second problem is that because of the non-linearity of
gravity, the formation of a black hole is typically an inherent non-perturbative
process in string theory. As  the
typical energies of the closed string excitations in \5ads
are of order $E \sim 1/R$, we need to require $1/R \ll 1/l_p$,
\ie $N \gg 1$, in order to prevent black hole formation.
The third problem is that
even under such circumstances we are still unable to quantize the {\it free}
string theory in \5ads due, among other things, to the presence of
RR-fields. Still, if we are ever successful in quantizing it, we know
that its low energy limit will lead to IIB {\it supergravity}, and we
know how to deal with it. This approximation will be valid as long as we
do not reach energies close to the massive states
that we integrated out. This requires  $1/R \ll 1/l_s$ which,
translating to field theory parameters, is equivalent to $\l \gg 1$.
Summarizing:
\num{
\item Perturbative String Theory $\Longleftrightarrow \,\,$ $\l < N \,\,,\,\, N \gg1$.
\item Weakly coupled Supergravity $\Longleftrightarrow \,\,$ $1 \ll \l <N$.
}
\eitem
The important remark is the absolute incompatibility of point 1 in SYM
and point 2 in string theory. This constitutes the main obstacle nowadays
to make comparisons on both sides. Note as well that if we were able to
quantize string theory on \5ads we would have a huge overlap between
perturbative computations on both sides of the duality. We have illustrated
the ranges of validity in figure \ref{ads-pre-bmn}.
Having established the computational ranges of validity do not overlap
at all, it is natural to ask what can be done with this conjecture.
There are essentially two things one may aim to do.
\bitem
\item {\bf Predict.} This is nowadays the less difficult aim. Because
of the reasonably well settled dictionary between observables in
both sides, one can perform the {\it same} calculation in the
two non-overlapping regions. Both answers remain as a prediction
of what the other side should give outside its perturbative domain.
This is what we will do in the following chapters, when we will try
to show issues like confinement or chiral symmetry breaking
for $\caln=1$ and $\caln=2$ gauge theories through their supergravity
duals.
\item {\bf Compare.} It is clear that this is the most difficult
aim as it requires dealing with  non-perturbative physics.
The picture before the BMN work was that most comparisons
had been done under the protection of supersymmetry. For example,
the spectrum of superconformal primary operators and their
conformal dimensions (these concepts are properly defined in the appendix~\ref{ch2:sec:algebra})
are completely fixed by the superconformal
algebra, and one does not need to compute any diagram to determine
them. Thus they are independent of the coupling $\l$ and they provide
answers that can be compared to weakly coupled supergravity computations.
We refer the reader again to~\cite{Aharony:1999ti} for an exhaustive list
of successful comparisons in the literature.

One of the aims of the next sections will be to extend the range of comparability
for some observables and to provide new tests of the correspondence. First
we will need to recall some properties of the \4n SYM theory.
\eitem

\section{The BMN limit of AdS/CFT} \label{ch2:sec:bmn}

There are some good reviews on the BMN limit of the AdS/CFT correspondence,
so we will just describe its basic facts here, specially those that
will be relevant in the next sections.

The work of BMN exploited the fact that \5ads admits a consistent simplification
if we just focus on the geometry in the neighborhood of a null geodesic
on a great circle of the $S^5$. An important and non-trivial
fact about it is that such a limit produces a configuration that
still solves the equations of motion of type IIB supergravity.
Furthermore, the number of supersymmetries is still 32.
The limit is called a Penrose limit and the resulting background
is a maximally supersymmetric pp-wave
\bea \label{bmn-met}
ds^2& =&-4 dx^+ dx^- - x^i x^i (dx^+)^2 + dx^2_{1,8} \sac i=1,...,8 \\
F_5&=&4\mu \, dx^+ \wedge \left[dx^1 \we ... \we dx^4 + dx^5 \we ... \we dx^8 \right]\,.
\eea
Together
with flat space and \5ads, this exhausts the list of maximally supersymmetric
IIB backgrounds (see~\cite{Figueroa-O'Farrill:2002ft} for a proof of this statement).
The last important point, and the one
that made this line of work so powerful, is that the string $\s$-model
in this background is quantizable despite the presence of nonzero
RR fluxes! This opened a new path parallel to the one followed when
the string was quantized in flat space: one can obtain the free spectrum,
construct the vertex operators and analyze interactions.

The quantizability of the $\s$-model by itself would not be so remarkable
if it was not because the pp-wave background is obtained as a limit
of \5ads. After all, there are some few other quantizable backgrounds
in the market which have received no attention at all compared to this
IIB pp-wave. The reason for such a different treatment is that having
the AdS/CFT correspondence allows for an identification of how the limit
acts on the dual CFT; we therefore end up with a quantizable string
theory dual to a sector of observables in the CFT. As we will carefully
analyze, the situation illustrated in figure~\ref{ads-pre-bmn} radically changes and
there will be overlapping regimes of both sides where comparisons can
be performed.

Let us then describe how the limit acts on both sides. On the string
theory side, let us call $E$ the energy of the stringy excitations
and $J$ their angular momentum in the first of the 2-planes of
the $S^5 \subset \CR^2 \times \CR^2 \times \CR^2$.
The metric \bref{bmn-met} is the limit of \5ads where one focuses on a
null geodesic at the origin of the 2nd and 3rd of the $\CR^2$
factors (which makes it a great circle in the $S^5$).
The string states that survive in the limit are those with
\be \label{ads-limit}
J \sim R^2 \sim \sqrt{\l} \sac E - J = \mbox{fixed} \sac \mbox{as }\,\, R / l_s \rightarrow \infty.
\ee
It is important to remark that the limit $R / l_s = \l^{1/4} = (g_s N)^{1/4} \rightarrow \infty$ can
be taken in two different ways. If we want to keep string interactions, we fix $g_s$ and
we let $N \rightarrow \infty$ keeping $J \sim N^{1/2}$. If we prefer to obtain
a free string theory, which can be useful to analyze the spectrum, we
can first take a conventional 't Hooft limit $g_s \rightarrow 0$
keeping $\l$ fixed, and the we perform a large 't Hooft coupling
limit $\l \rightarrow \infty$ keeping $J \sim \sqrt{\l}$.

We now use the AdS/CFT dictionary to translate this limit into the CFT.
The energy was measured in global AdS coordinates and must be identified
with the conformal dimension of its dual YM operator. The angular momentum
is the charge under an $SO(2) \subset SO(6)$ subgroup of the isometries
of the $S^5$, and it must be identified with the $R$-charge of the
dual operator under a $U(1)$ subgroup of $SO(6)$. Finally, using the
relation $R^4 = l_s^4 \gym^2 N$ we can translate \bref{ads-limit}
into
\be
J \sim \sqrt{\l} \sac \Delta-J=\mbox{fixed} \sac \l \rightarrow \infty .
\ee
There are two immediate points to make here.
\bitem
\item
The first one is that
the string states in the pp-wave background will be dual to operators
with very large conformal dimension and $R$-charge. There is a
BPS bound that follows from the $PSU(2,2|4)$ superalgebra\footnote{We will
see in section~\ref{section-bound} an explicit derivation of how to obtain this and other
more general BPS bounds from the algebra.} that implies that all
operators must satisfy
\be
\Delta \ge J \,,
\ee
with equality being valid only for 1/2-BPS operators. There
are a large number of operators that saturate this bound, and they just
differ by the number of traces in the adjoint of the gauge group
(see the appendix~\ref{ch2:sec:algebra} for a proper discussion on chiral operators).
There is an identification in the AdS/CFT correspondence that
relates $n$-trace operators in the YM side to $n$-particle states
in string theory one. This correspondence is based on the
observation that, as states with different number of particles
are orthogonal in supergravity, they should correspond to
'orthogonal' operators in the YM side, where 'orthogonality'
is to be understood as
\be
\langle O_1 O_2 \rangle_{CFT} = 0 \,.
\ee
In general, it is the case that if two operators contain a different
number of traces, then $\langle O_1 O_2 \rangle \sim 1/N^{a}$ with $a >0$,
so that the number of traces is a good quantum number in the large $N$ limit.
However, this result starts to fail when the number of fields inside
each operators start to grow due to large combinatoric factors. Operators
with different number of traces start to mix and the correspondence
to multiparticle states starts to fail.
In our present case we are on the edge to run into this problem.
Although one conventionally takes the ground state of the string
in the pp-wave background to be dual to a single-trace operator
\be \label{singleagain}
\mbox{BMN ground state } \ket{0} \,\, \longrightarrow \,\, \calo_{1,J}=\Tr (X^J) \,,
\ee
it was realized in~\cite{Balasubramanian:2001nh} that as soon as $J \sim N^{2/3}$
the true dual operator involves a linear combination with other\footnote{As
we define in the appendix~\ref{ch2:sec:algebra}, $O_{p,J}$ denotes
an operator with $J$ fields and $p$ traces.} $O_{p,J}$
operators with $p\ge 2$. This kind of phase transition corresponds in
the string theory side to the dual descriptions that a graviton state
admits according to the value of its angular momentum. As $J$ becomes
comparable to $N^{2/3}$ the description is more appropriate in
terms of Giant Gravitons.

The map from the free string theory spectrum to the operators \bref{singleagain}
is still valid, as in the $g_s\rightarrow 0$ limit keeping $J \sim \sqrt{\l}$
we can make $J$ as negligible  as we want in front of $N$. This supports as
well the identification of the first excited modes of the BMN string with
single trace operators obtained from \bref{singleagain} by adding
a few other fields (called impurities). Schematically
\be \label{impurities}
a^{\dagger} ... a^{\dagger} \, \ket{0} \,\, \longrightarrow \,\,
\calo_{1,J}=\Tr (X^J D_{\mu}X \l Y) \, + \, \mbox{perm.}\,,
\ee
where $perm.$ stands for permutations of the impurities inside
the operator with suitably chosen coefficients~\cite{Berenstein:2002jq}. The
operators \bref{impurities} are called BMN or {\it near} 1/2-BPS
operators.

\item
As the BMN limit requires $\l \rightarrow \infty$, it may look like
it renders perturbation theory in the YM side useless.
However, another conspiracy here makes the work of BMN
computationally useful. The point comes from taking into account
combinatoric factors. It was proven that the correlation
functions of BMN operators are not governed by the 't Hooft
coupling but by an effective coupling $\l'$ given by
\be
\l' = {\l \over J^2} = {\gym^2 N \over J^2} \,,
\ee
which is kept finite in the limit. This is one of the highlights
of the BMN limit, as it allows to compute in perturbative
SYM at very large $\l$ (as far as $\l' <1$) where the curvature
of \5ads is very small and supergravity is a good approximation.
Another point to have into account is that also the weight
of non-planar diagrams is modified from
\be
{1\over N} \,\, \longrightarrow \,\,  {J^2\over N}  \,\equiv\, g_2\,.
\ee
Therefore the criteria for performing perturbative field theory
computations with the BMN operators change to
\num{
\item Perturbative SYM $\Longleftrightarrow \,\,$ $\l' \ll 1 \,
\Rightarrow \l \ll J^2$.
\item Perturbative $+$ Planar SYM $\Longleftrightarrow \,\,$ $\l \ll J^2 \ll N$.
}

\eitem

\subsection{Summary} \label{ch2:sec:bmnsummary}

We would like to visually summarize here the BMN limit. The impossibility
of testing the whole AdS/CFT cannot be yet overcome, but a certain
region of strings in AdS do admit comparison to a certain set of
observables in the CFT. The simplification is accomplished by
focusing on observables with large quantum numbers, where quantum
physics typically reduce to classical physics, as we will exploit
in the next sections. Loosely speaking, BMN introduced a new
axis in the $(\l,N)$-plane (which determined the \4n SYM Lagrangian),
an axis of large $R$-charge along which the correlation functions
of some operators simplify.
\be
\epsfig{file=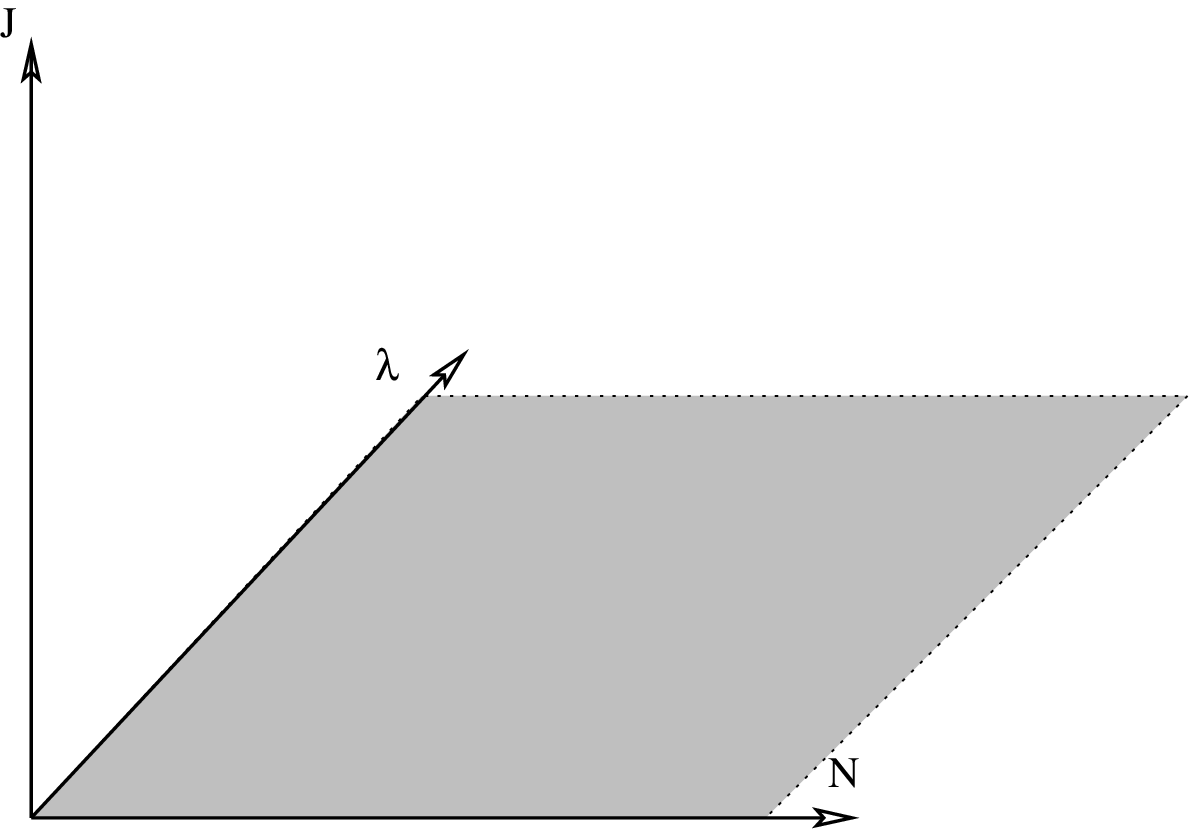, height=5cm, width=7cm}
\ee

\begin{figure}[here]
\cent{\epsfig{file=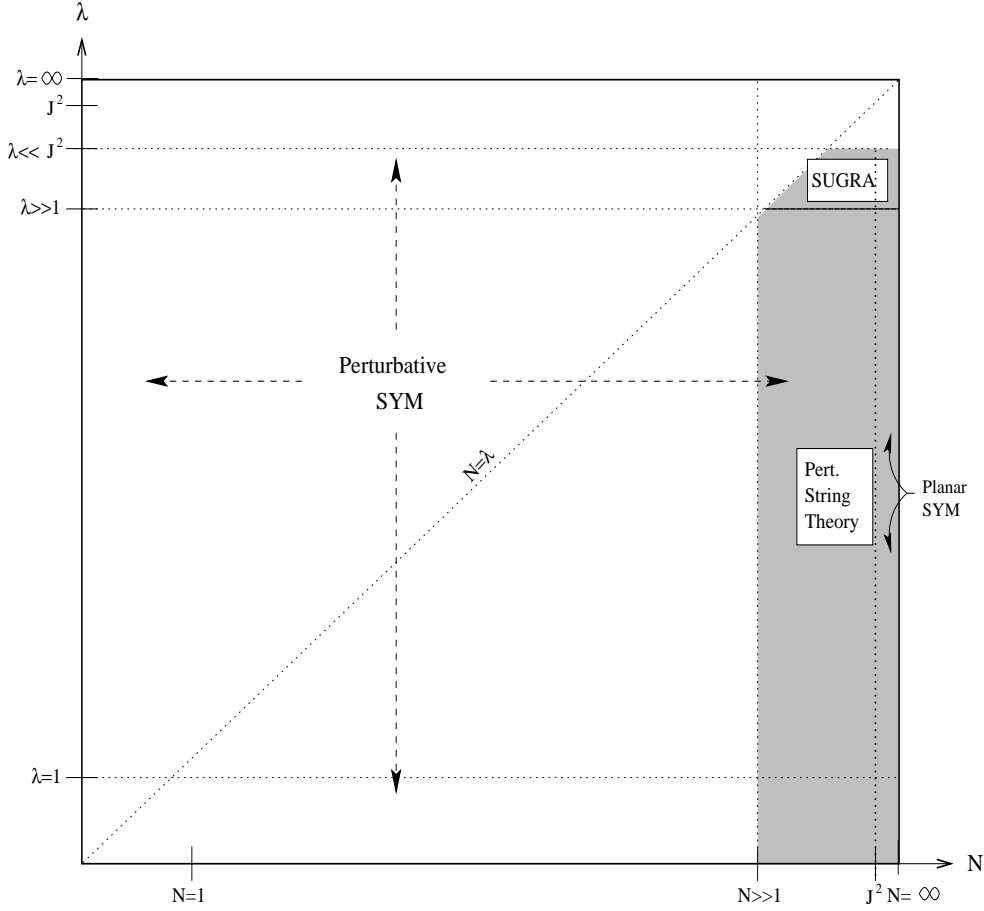, height=12cm, width=13cm}}
\caption{Modification of the AdS/CFT ranges of validity in the BMN
sector ($J$ is taken very large). In the shaded region there is simultaneous validity of
perturbative SYM and perturbative string theory. \label{ads-bmn}}
\end{figure}

We can now modify our plot \bref{ads-pre-bmn} to incorporate
$J$ in both axes and gain a better understanding of the
overlapping region opened in the BMN limit, see figure \bref{ads-bmn}.

\begin{figure}[here]
\cent{\epsfig{file=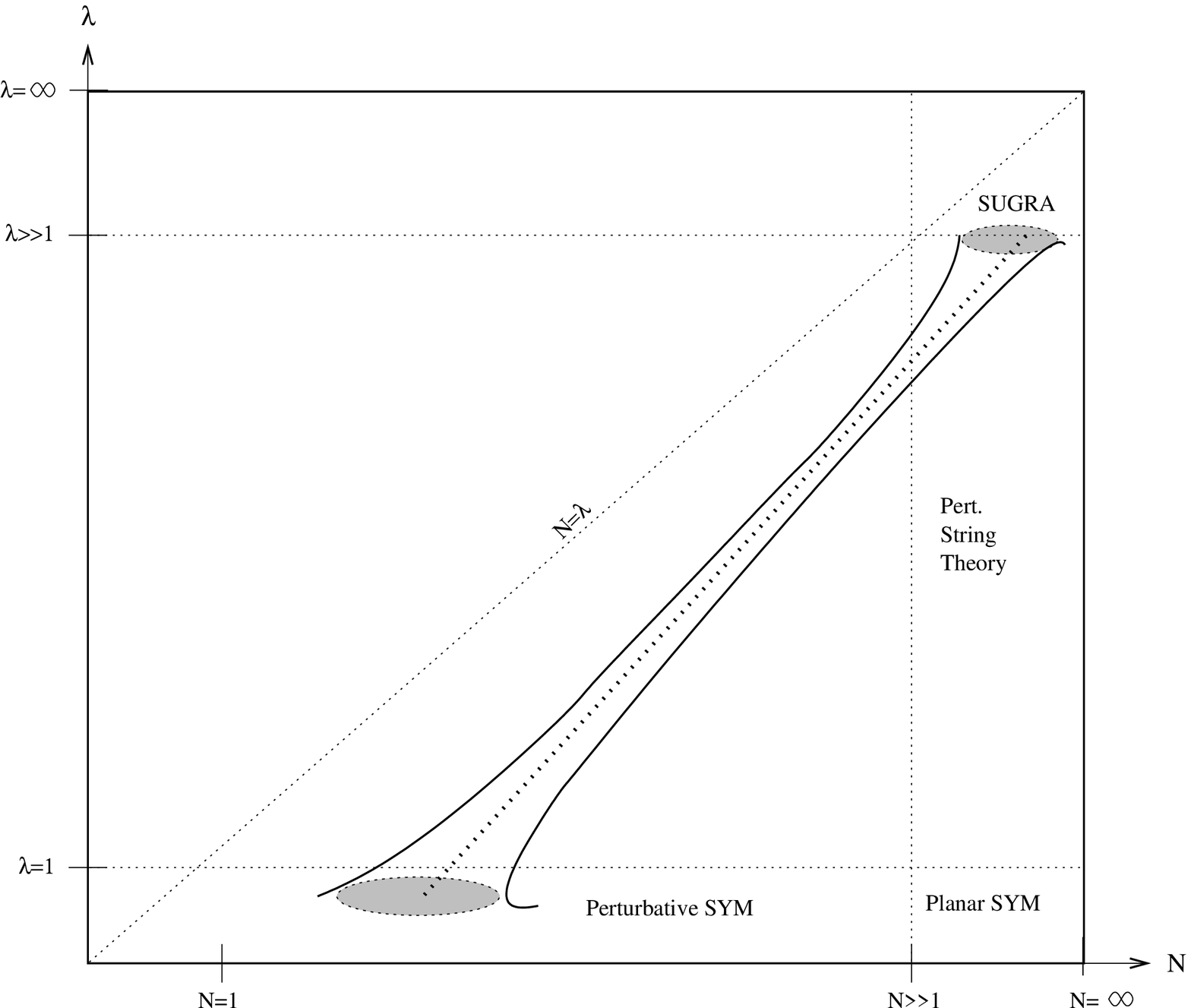, height=12cm, width=13cm}}
\caption{The BMN operators 'dig a safety tunnel' if we imagine
placing the 1/2-BPS protected operators in its middle, and the near
BPS (which receive quantum corrections) about it.\label{tunnel-1}}
\end{figure}

Yet another convenient way of understanding the BMN progress that
will be useful in more complicated cases is to
think of it as follows. The 1/2-BPS operators with one $R$-charge
$J$ have non-renormalized 2 and 3 point functions, so that they
can be safely extrapolated to the string theory region. Operators
a few impurities away from them {\it do receive} radiative
corrections which can be computed in perturbative SYM and,
due to the $J^2$ suppression, they can be extrapolated as well.
Somehow, we have dug a safety tunnel in the $\{\l,N\}$ space;
this tunnel has a nonzero radius, meaning that if we stay close
to its axis (where the BPS operators live), we can
still extrapolate despite the quantum corrections. It is
maybe worth to illustrate this as well, as in figure~\ref{tunnel-1}.

\section{The GKP simplification} \label{ch2:sec:gkp}

We have seen that \4n SYM operators with very large $SO(2)$ $R$-charge
are dual to a simplification of the \5ads background and that,
furthermore, string theory is quantizable in such background.
The observation of Gubser, Klebanov and Polyakov (GKP) is that
one can obtain a part of the BMN results in a much simpler way
which, in turn, allows for an application to many other similar cases.
They proposed that some field theory operators with large quantum
numbers are dual to string theory worldsheet solitons in \5ads.
Recall that the string $\s$-model in \5ads has an effective
coupling $\a \sim 1/\sqrt{\l}$ as can be seen by rescaling the
metric so that
\be
ds^2_{AdS_5\times S^5} = \, R^2 \, d\tilde{s}^2 \,,
\ee
which leads to a bosonic $\s$-model of the style
\be
S_{2d} \sim {1\over l_s^2} \int_{\Sigma_2} d^2\s \sqrt{-g_{AdS^5\times S^5}}
= {R^2 \over l_s^2} \int_{\Sigma_2} d^2\s \sqrt{-\tilde{g}} =
{1 \over \sqrt{\l}} \int_{\Sigma_2} d^2\s \sqrt{-\tilde{g}} \,. \nonumber
\ee
If we find a non-perturbative solution to this $\s$-model, then
its quantum numbers (let us call them generically $Q$) will,
by definition, depend on an inverse power of the coupling
\be
Q \sim {1\over \a^p} \sim \l^p \sac p>0 \,.
\ee
In the region of small curvatures (large $\l$), these charges
are very large and are therefore expected to be dual to operators
with a large number of fields. The expectation is that in the
string side, these quantum numbers can be well approximated
by their {\it classical} value, \ie neglecting quantum
corrections in $\a$.

\medskip

\caixa{Note that the solitons we are talking about
are not topological, in the sense that they are continuously
connected to the vacuum which is typically an almost-collapsed string
(the usual supergravity states). In flat space, where the full
perturbative spectrum is known, one could construct these
solitons by acting with creation operators on the vacuum, and
we would typically obtain a coherent state in the string
Hilbert space. This is not possible to do here because of the
impossibility to quantize the string in \5ads. Expanding
about the classical soliton circumvents this problem.}

\medskip

This is a powerful proposal which, again, allows to make a lot
of predictions but very few tests. The reason is as always that
even though  it extends the AdS/CFT dictionary, one still cannot
extrapolate from SYM computations $\l \ll 1$ to string theory
ones $\l \gg1$. In other words, after computing $Q$ in both sides,
one still needs to answer the following  question
\tem{
\item $Q_{SYM}$ is computed at $\l <<1$ where $Q_{string}$ is
computed at $\l>>1$. Can we extrapolate any of them to the
reciprocal region in order to compare?
}
Let very briefly us discuss two of the most
relevant applications of the GKP ideas.

\subsection{Twist two operators} \label{ch2:sec:twisttwo}

The example originally given
in GKP involved the identification of twist-two operators
with folded spinning strings in the $AdS$ factor of \5ads.
These operators have $\Delta=S+2$, where $S$ is
the charge under one of the $SO(2) \subset SO(2,4)$
and have the form
\be
\Tr \,\, X \nabla_{(\mu_1} ... \nabla_{\mu_n)} X \,.
\ee
They are present in non-supersymmetric theories as well
and, being non-chiral, their conformal dimension
receives quantum corrections. These are all believed (even
non-perturbatively) to have a leading  large $S$ contribution proportional to $\ln S$,
so that
\be
\Delta \underset{S\gg1} {\approx}     S+ f(\l) \ln S \,.
\ee
If $f(\l)$ is computed in perturbation theory, then one gets a power
series in $\l$ and its first term (for QCD) can be checked experimentally!

The identification of GKP with the spinning string allows for a simple
computation of the charges of this soliton and the result was
\be
\Delta \underset{S\gg1}{\approx} S + {\sqrt{\l} \over \pi} \ln (S/\sqrt{\l}) \,.
\ee
As this is a stringy result, it must be understood as valid for very large $\l$,
so that it {\it predicts} the behavior of $f(\l)$ away from perturbation theory.
This is another example of a prediction. There is no hope for comparison since,
on the one hand, the SYM operators are not protected nor their effective
coupling is combinatorially suppressed and, on the other hand, quantum $\s$-model
corrections to the string classical values would become more and more relevant
as we move away from large $\l$.

\subsection{BMN operators} \label{ch2:sec:bmnoperators}

\begin{figure}[here]
\cent{\epsfig{file=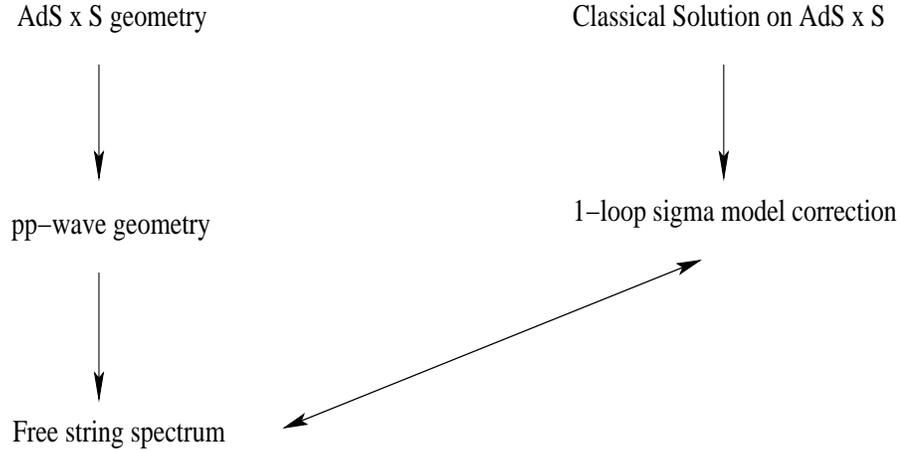, height=6cm, width=12cm}}
\caption{The shortcut provided by GKP.\label{tseytlin}}
\end{figure}

We apply the GKP ideas to the BMN ground state operator \bref{singleagain}.
As it only carries nonzero $\Delta$ and $J$, we must look for a string
state with rotation only in the $S^5$ factor of \5ads, which turns out
to be an almost collapsed closed string.
We will see in detail in the next section how these ideas
are carried out in similar but more sophisticated cases, so here we just cite the results
for the BMN operators. The classical string approximation yields
the relation $E=J$ which is the exact relation $\Delta=J$ for the
1/2-BPS operator. Remarkably, the first $\s$-model correction to
this energy is able to reproduce the whole spectrum of the
string in the pp-wave background! So the diagram~\ref{tseytlin} holds.

This case is essentially different from the previous one (where the
string rotated in the AdS factor) because both the ground
state operator and its dual string solution preserved 1/2 supersymmetry.
This has helped to support the fact that the answers to the questions
we posed are positive: yes, $Q_{string}=Q_{classical}$ at large $J$; and
yes, $Q_{CFT}$ can be extrapolated to the stringy region. Thus supersymmetry
seems to be in the heart of this test of AdS/CFT.

\section{Trying to check AdS/CFT beyond supersymmetry} \label{ch2:sec:ads-beyond}

We have seen that the progress initiated by BMN and then improved by GKP
is able to provide a check of the AdS/CFT in a near-BPS sector of
operators/states. All other quantitative work must be interpreted
as providing predictions. One might suspect that what is behind
the tests described above is supersymmetry rather than AdS/CFT,
although after 7 years of duality most people believe in the
strongest possible version of the AdS/CFT.

So it remains as a major challenge to be able to test (or reject!)
the AdS/CFT conjecture away from supersymmetric sectors. There has been
some recent progress along these lines by an intensive exploitation
of the GKP ideas. Essentially, one would like to follow this route:
\num{
\item Start with a SYM operator with large quantum numbers that is far from any BPS one.
\item Compute a radiative correction, say, to the conformal dimension.
\item Identify the string soliton dual to this operator.
\item Compute the classical energy and relate it to its classical charge.
\item See if both computations can be compared.}
This task will most of the times fail in the last point, the
twist-two operators we discussed above being an example. As
the operators we want to work with are far from BPS, the dual
string states will be far from supersymmetric ones, and it
is then very difficult to answer positively to
\num{
\item is the classical solution stable at all?
\item can we neglect quantum $\s$-model corrections to $E_{classical}$?
\item are the perturbative corrections to $\Delta_{CFT}$  suppressed by $1/J^2$ factors,
so that they can be extrapolated?
}

\subsection{Rotating strings in spheres} \label{ch2:sec:rotating-strings}

We will now describe the attempts to carry on this enterprize which,
despite the  difficulties mentioned above, started around April 2003.
These recent results are based on the
obtention of new $\s$-model solitons. Let us start from the most
basic point: which kind of solutions should we look for? The
BMN ground state operator carries only nonzero $(\Delta, J)$
charges\footnote{We recall that we use the convention
$(\Delta,S_1,S_2,J_1,J_2,J_3)$ to label the charges in
the Cartan subalgebra $SO(1,1)\times SO(2)^5$ of $SO(2,4)\times SO(6)$.
The name {\it spin} refers to $S$-charges and corresponds to motion in $AdS_5$.
The name {\it angular momentum} refers to $J$-charges and corresponds
to motion in the $S^5$.}
and it is dual to an almost-particle (a collapsed string) travelling
on $S^5$. If we want to succeed with the identification of operators/states
we had better look for other kind of operators with definite charges.
The simplest possibility is maybe an operator with $(\Delta,S)$ charges,
but this is similar to the twist-two ones and we saw that they do
not allow for comparisons. Next attempt: consider $(\Delta,S,J)$;
it was seen that they fail again~\cite{Frolov:2002av}. Next attempt: consider
$(\Delta,J_1,J_2)$. This possibility is distinguished from the previous
ones as it must correspond to an extended (as opposite to collapsed)
string rotating in the sphere; a particle could never carry two independent
angular momenta in a sphere.
This means that it must correspond to a truly {\it stringy} state and
if the correspondence worked, it would be a test {\it away from supergravity}.

A string solution with two angular momenta on a sphere had been long known~\cite{Hoppe:1987vv} and its
relevance for AdS/CFT was first introduced by Frolov and Tseytlin (FT) in~\cite{Frolov:2003qc}.
A good way to understand them may start with understanding why are they difficult to find.
An extended object made of mass (so that each of its points attracts the others)
finds it very hard to stabilize in our familiar $\CR^3$ only by rotating. This is because
the rotation of a rigid body is always about one axis (which may rotate as well).
Two points along the axis do not feel a centrifugal force,
so that they just tend to approach by gravitation attraction. If the body was not rigid,
it would just collapse. Flat galaxies are an example; they are more stable
by concentrating all their mass in a plane perpendicular to the axis of rotation.

The situation is dramatically different in $\CR^4$ because $SO(4)$ is a group
of rank 2. This means that a body can have two independent angular momenta
along two different planes. It turns out that this mechanism
can be used to provide the same centrifugal force to each point of a string,
so that gravity is compensated everywhere. This is achieved if we take
the two angular momenta to be equal $J_1=J_2=J$.
\begin{figure}[here]
\cent{\epsfig{file=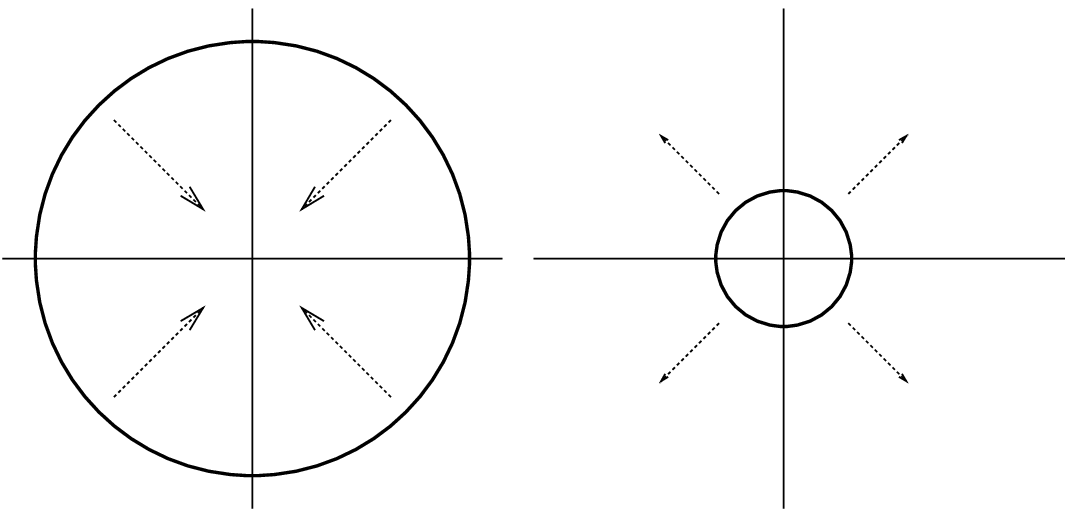, height=2.8cm, width=6cm} ...
\epsfig{file=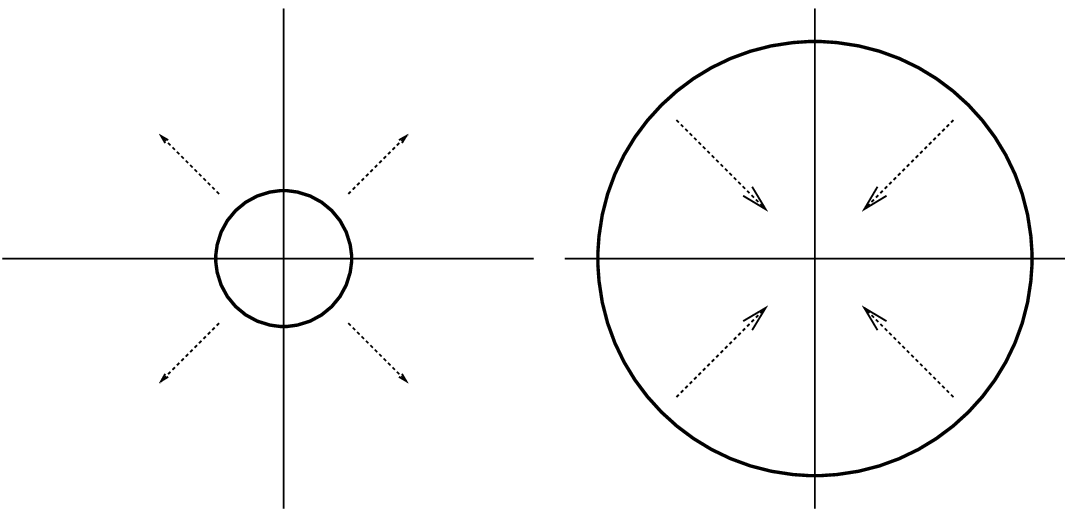, height=2.8cm, width=6cm}}
\caption{On the left, the projections of the string in the $12$ and
$34$ planes of $\CR^4$ at fixed $t_0$; the string contracts in the
first and grows in the latter. On the right, the same but
a bit later; the situation is reversed. Remark: unlike
in $\CR^3$, all points suffer centrifugal acceleration.\label{hoppe}}
\end{figure}

The work of~\cite{Hoppe:1987vv} actually included similar solutions
for general $n$-dimen\-sio\-nal relativistic surfaces, which could be constructed
as long as their ambient space was large enough to allow for enough
independent angular momenta.
A remarkable property of these solutions is that they all actually
happened to lie at all times on a sphere rather than in the whole ambient
space. For example, the string just described happened sweep
an $S^3 \subset \CR^4$  rather than the whole $\CR^4$.
Thus these surfaces are ready to play a game in AdS/CFT by
placing them in the sphere factors of $AdS_{p}\times S^q$.
This is what was done in~\cite{Frolov:2003qc}, where they constructed
a rotating string in the $S^5$ factor of \5ads with
two equal $SO(6)$ momenta.

However, we encounter here the first of the obstacles we mentioned
above. Whereas it was proven~\cite{Demkin:1995td} that all the solutions in flat space
of~\cite{Hoppe:1987vv} are always stable, it was proven that the ones in $S^5$
are always unstable~\cite{Frolov:2003qc,Frolov:2003tu}.
 Both results are not in contradiction because
stability is a property of second order fluctuations about a
given solution, so that it probes the curvature of the background
space.

Should we just stop here our enterprize?
Let us postpone the issue of stability and try to proceed, as
everyone has done in the series of papers that have appeared after
the work of Frolov and
Tseytlin~\cite{Beisert:2003xu,Frolov:2003xy,Arutyunov:2003uj,Beisert:2003ea,Arutyunov:2003rg,Engquist:2003rn,Kristjansen:2004ei}.
We now want to obtain the classical energy as a function of its
classical charges. For that we will need to describe the
solution in more detail.

\subsection{Strings with 3 angular momenta} \label{ch2:sec:strings-with-3}

Here we shall describe in detail the embedding of the
strings we discussed above and obtain the relation
among their charges. Note however that the simplest
string we described lives only in an $S^3\subset S^5$,
so that it still has room for another angular momentum
describing the motion of the $S^3$ inside the $S^5$.
So we will use the notation
\tem{
\item $(\w,J')$ $\,\leftrightarrow  \,$ angular velocity and
momentum due to 'self rotation' in the two planes of $S^3$,
\item $(\nu,J)$ $ \,\leftrightarrow \,$ angular velocity and
momentum due to the $S^3$ motion in $S^5$.
}
Note that both types of motion are of very different physical
origin.

\medskip
\caixa{{\bf Note:} In the following sections we restrict our attention
to strings with two of the three angular momenta being equal.
Although solutions with 3 independent angular momenta
were found afterwards, we prefer to stick to the less
general case because it follows from the intuition we described
above. The generalization to 3 independent $J$'s does not
provide any further insight and will be considered in the appendix
\ref{ch2:sec:generalization}
}
\medskip

We will be considering solutions of the classical equations of motion
derived
from the Nambu-Goto Lagrangian
\be
\call = - \fc{1}{\ap} \,  \sqrt{-\det g} \,,
\ee
where $g$ is the worldsheet metric induced from the \5ads spacetime
metric. The strings we want to consider lie at the origin of \ads{5} and
rotate only in the $S^5$, so effectively they live in a
$\CR{} \times S^5$ subspace of \5ads with metric
\be
ds^2 = R^2 \, \left( - dt^2 + d\Omega_{\it 5}^2 \right) \,,
\ee
where $d\Omega_{\it 5}^2$ is the $SO(6)$-invariant metric on the unit five-sphere
and we recall that $R^2 = \ap \sqrt{\lambda}$.
The $S^5$ may be viewed as the submanifold $|{\bf W}|=1$ of $\CR^{6}$
with Cartesian coordinates $W_i$ ($i=1,\dots,6$). For the
parametrization with
\bea
W_1 + i W_2 &=& \cos \t \, e^{i \chi} \,, \nn
W_3 + i W_4 &=& \sin \t \, \cos \phi
\, e^{i \a} \,, \nn
W_5 + i W_6 &=& \sin \t \, \sin \phi \, e^{i \b} \,,
\label{y}
\eea
this gives
\be
d\Omega_{\it 5}^2 = d\t^2 +
\sin^2 \t \, d\phi^2 + \cos^2 \t \, d\chi^2 + \sin^2 \t \cos^2 \phi
\, d\a^2 +
\sin^2 \t \sin^2 \phi \, d\beta^2  \,.
\label{metric}
\ee

We fix the worldvolume reparametrization invariance by the gauge
choice
\be
t = \tau \sac \a=\s \,,
\ee
where $\tau$ and $\s$ are the worldsheet coordinates.\footnote{
  Note that in our conventions both the spacetime and
  the worldsheet coordinates are dimensionless, which implies that
  the energy of the string, $E$, is also dimensionless. The
  corresponding dimensionful time and energy are obtained by rescaling
  the dimensionless ones by appropriate powers of $R$.}
The string solutions of interest correspond to circular, rotating
strings supported against collapse by their angular momenta;
they are given by
\be
\t = \t_0 \sac \chi = \nu \tau \sac \phi = \omega \tau \sac \b = \s \,,
\label{embedding}
\ee
where $\t_0$ is a constant in the interval $[0,\pi/2]$.
The intuitive picture discussed above is realized here in the fact that,
at any instant, the string is a circle in a two-plane contained within the
3456-space; this plane rotates with angular velocity $\omega$. In turn,
the string's center of mass rotates with angular velocity $\nu$ around a
circle in the 12-plane.

Under the above circumstances, the Nambu-Goto equations are solved if
either of the following relations hold\footnote{Note that in case $(i)$
the string always lies at the origin of the 12-plane, which is a singular
submanifold of the coordinate system we are using; $\nu$ is the angular
velocity in a circle of zero radius. In practice this does not cause
a problem because $\nu$ is irrelevant in this case.}
\bea
(i)&&\qquad \cos \t_0 = 0 \,, \qquad \w^2<1 \,, \nn
(ii)&& \qquad \cos 2\t_0 = {\omega^2 -1 \over \omega^2 -\nu^2} \,,
\qquad \nu^2<1 \,, \qquad 2 \w^2 - \nu^2-1>0 \,. \nonumber
\eea
The restrictions on the angular velocities follow from demanding the
reality of both $\call$ and $\t_0$.
The energy of the rotating string is
\be
E = \rl \, |\sin \t_0| \, \Delta^{-1/2} \,, \qquad \Delta \,\equiv \, 1
-\nu^2 \cos^2 \t_0  -\w^2 \sin^2\t_0  \,,
\label{e}
\ee
while the only non-zero components of the angular momentum two-form,
$J_{ij}$, in the Cartesian coordinates $W_i$, are
\be
J \equiv J_{12} = E\, \nu \, \cos^2 \t_0 \sac
 J' \equiv J_{35} = J_{46} = \fc{1}{2} \, E \, \omega \, \sin^2 \t_0 \,.
\label{j}
\ee
Thus, $J$ and $2J'$ are the momenta conjugate to $\chi$ and $\phi$,
respectively.
Their values for the two possible solutions
of the equations of motion are
\bea
(i) && E = \fc{\rl}{\sqrt{1 - \omega^2}}
            \sac J = 0  \sac  J' = \fc{\rl \,
\omega}{\sqrt{1- \omega^2}} \,,
\\
(ii)     && E = \fc{\rl}{\sqrt{\omega^2 - \nu^2}} \xsac
 J = \rl \, \fc{\left(2\omega^2 -\nu^2-1\right)\, \nu}
{2\left(\omega^2 -\nu^2\right)^{3/2}} \xsac  J' = \rl \,\,
\fc{\omega\left(1-\nu^2\right)}
{2\left(\omega^2 -\nu^2\right)^{3/2}}
 \,. \nonumber
\eea In the first case it is easy to express the energy solely as
a function of the angular momentum, with the result \be (i) \qquad
E = \sqrt{(2J')^2 + \l} = |2J'| \, \left[ 1 + \calo
\left(\fc{\l}{J'^2} \right) \right] \,. \ee The expression for $E$
in terms of $J$ and $J'$ can also be found explicitly in the
second case, but the result is rather messy. However, when
expanded for large $J, J'$, it yields \be (ii) \qquad E =
\left(|J| + |2J'|\right) \, \left[ 1 + \calo \left(\fc{\l}{J^2},
\fc{\l}{J'^2} \right) \right]\,. \ee
{\bf Summary:}

\noindent
The last two equations is what we were looking for. They
provide the string classical result for the relation
between the energy and the angular momentum. There are
a few comments to make here:
\num{
\item Both expressions can be expanded for $J^2 \gg \l$
in such a way that only integer powers of ${\l / J^2}$
appear. This did not happen for the twist-two operators.
Had not it been this way, there would have been no
hope to compare with a perturbative field theory result.
\item The first term, which is $\l$ independent, is
precisely the classical dimensions that the following
operators have
\bea
\calo(J')&=&\Tr \left( Y^{J'} Z^{J'} \right) + \mbox{perm.} \sac \mbox{if } J=0\,, \nn
\calo(J,J')&=&\Tr \left(X^J Y^{J'} Z^{J'} \right) + \mbox{perm.} \sac \mbox{if } J\neq 0\,.
\eea
So it is tempting to relate our string solutions to these operators.
\item
The proposed operators are non-chiral, which agrees with the fact that
these solutions are non-supersymmetric. Indeed, they are a huge
number of impurities away from the BMN ground state operator.
It is tempting to
compare the first terms in the expressions for $E(J,J')$
with a perturbative SYM computation of the anomalous
dimensions of the corresponding operators.
As $E(J,J')$ is a power series in  $\l/J^2$, it seems reasonable to
suspect that the SYM corrections will also be controlled by
$\l'=\l /J^2$ as it happened for BMN operators.
}
Everything leads to the conclusion that it is worth doing
the computation in the SYM side. This process has been
carried out in a number of
papers~\cite{Beisert:2003xu,Frolov:2003xy,Arutyunov:2003uj,Beisert:2003ea,Arutyunov:2003rg,Engquist:2003rn}
and the success
has been spectacular. The SYM computation has been possible
mainly due to the realization that one can map the
problem at 1-loop to different types of spin chains.

We will come back to the discussion of the SYM computation later.
Now, we will spend some time trying to understand what is
actually going on in the string side. Recall that the string solution is not
even stable when $J<2J'/3$ so when Tseytlin and Frolov computed the 1-loop $\s$-model
correction to it in this regime, they were actually doing it {\it about a tachyonic
vacuum!} They used the 1-loop
result to show that it is negligible against the
classical one for $J,J' \gg 1$.
Then arguments
were given that negligibility holds to all orders
in $\s$-model corrections analogously to what
was argued for the BMN case.

Altogether it seems like the comparison is being too successful.
The proposal that we gave in~\cite{Mateos:2003de,Mateos:2004rn} is that actually
we are testing a near BPS sector again. The first indications
are:
\num{
\item All the perturbations that develop tachyonic masses
(and are therefore responsible for instabilities) have the
property that these masses vanish in the large angular momenta
$J,J' \gg \sqrt{\l}$ limit.\footnote{
The relevant tachyonic masses appear in formulas (4.34) of
\cite{Frolov:2003qc} and (2.35), (2.36) of~\cite{Frolov:2003tu}.} This means
that the string becomes asymptotically marginally stable.

\item The leading terms in the energy saturate bounds that
can be derived from the background superalgebra in the
large angular momenta limit. The general
bound leads to preservation of 1/8-supersymmetry; this
is enhanced to 1/4 for the case $J=0$.

\item The last and definitive proof is given below (section
\ref{ch2:sec:k-symmetry}) where we confirm by standard $k$-symmetry arguments
that the solutions preserve the mentioned fractions
of supersymmetry in the limit. In turn, this resolves
the ambiguity posed by the asymptotic marginal stability:
the solution becomes stable in the limit as it carries the
minimum energy for the given charges.
}

We now proceed to show how the BPS bound arises from the superalgebra
and then we proceed to the $\kappa$-symmetry analysis.
A careful discussion of the physical meaning of the limit
considered here is given afterwards in section~\ref{ch2:sec:physics}.

\subsection{BPS Bound from the Superalgebra}  \label{section-bound}

The energy of any supersymmetric  state in \5ads must saturate
a  BPS bound that follows from the $PSU(2,2|4)$ isometry superalgebra of
the \5ads vacuum, and hence its energy can be expressed as a
function of their charges alone. In this section we review the BPS bound
for states that carry the same type of charges as the rotating, circular
strings above, that is, energy and angular momenta on the $S^5$. The BPS
bound we will derive may be equally well understood as a statement about
the supersymmetry properties of operators in the dual CFT
as we will discuss later.

Let $\gamma_m$ ($m=0,\ldots,4$) be the $4\times 4$ five-dimensional
Dirac matrices for $AdS_5$ and let $Q^i$ be the four $AdS_5$ Dirac
spinor charges, transforming as the {\bf 4} of $SU(4)$.
The non-zero anticommutators are
\be\label{5susy}
\{Q^i,Q^\dagger_j\} = \gamma^0 \left[
\left(\gamma_m P^m + {1\over2}\gamma_{mn}M^{mn}\right)\delta^i{}_j
+ 2 \bbi{}\, B^i{}_j  \right] \,,
\ee
where $P, M$ are the $AdS_5$ charges and $B$ is the hermitian
traceless matrix of $SU(4)$  charges. For our spinning string
configurations the only non-zero $AdS$ charge  is the energy $P^0 =E$; in
this case \bref{5susy} reduces to
\be
\{Q^i,Q^\dagger_j\} =\bbi{}\, \delta^i{}_j  E + 2\gamma^0 B^i{}_j \,.
\ee
By means of an $SU(4)$ transformation we may bring $B$ to diagonal form
with diagonal entries $b_i$ ($i=1,2,3,4$) satisfying
\be
\label{trace}
b_1+b_2+b_3+b_4 =0.
\ee
The eigenvalues of the $16\times 16$ matrix $\{Q,Q^\dagger\}$ are
therefore
$E\pm 2b_1, E\pm b_2, E\pm b_3,E\pm b_4$, each being doubly degenerate.
Since this matrix is manifestly positive in any unitary representation,
unitary implies the bound
\be\label{Ebound}
E \ge 2b \qquad b=\mbox{sup}\{ |b_1|,|b_2|,|b_3|,|b_4| \} \,.
\ee
When the bound is saturated the matrix $\{Q,Q^\dagger\}$ will have zero
eigenvalues; the possible multiplicities are $2,4,8,16$. The maximum
number (16) occurs when $b_i=b$ for all $i$, in which case (\ref{trace})
implies $b=0$ and hence $E=0$; this is the $adS_5$ vacuum. Otherwise, one
has preservation of $1/8,1/4,1/2$ supersymmetry when
$\{Q,Q^\dagger\}$ has $2,4,8$ zero eigenvalues, respectively.

The eigenvalues of $B$ are $SU(4)$ invariants and hence determined in
any $SU(4)$ irrep by that irrep's Dynkin labels $(d_1,d_2,d_3)$.
Conversely, the Dynkin labels are determined by the eigenvalues of $B$,
and consideration of the highest weight state leads to the relation
\be
d_1 = b_1-b_2, \qquad d_2=b_2 -b_3,\qquad d_3= b_3-b_4.
\ee
Given the constraint \bref{trace}, this can be inverted to give
\bea
b_1 &=& {1\over 4}\left( 3d_1+2d_2+d_3 \right) \,,\nn b_2
&=& {1\over 4}\left( -d_1+2d_2+d_3 \right) \,, \nn
b_3 &=& {1\over 4}\left( -d_1-2d_2+d_3 \right)\,, \nn
b_4 &=& {1\over 4}\left( -d_1-2d_2-3d_3 \right) \,.
\label{bvalues}
\eea
The $SU(4)$ charges of the spinning strings considered here
correspond to irreps with Dynkin labels~\cite{Frolov:2003qc}
\bea
\label{choiceone}
&& [d_1, d_2, d_3]=[J'-J,0, J + J'] \qquad  \mbox{if}
\qquad J' > J  \,, \\ \label{choicetwo}
&&[d_1, d_2, d_3]=[0,J-J',2 J'] \qquad \,\,\,\,\,\,\,\,
\mbox{if} \qquad J' \le J  \,.
\eea
It follows, for either case, that the four (unordered) eigenvalues of $B$
are
\be \{  J'-\undos J \, , \, \undos J \, , \,
  \undos J \, , \,  - J' - \undos J \} \,.
\label{js}
\ee
Using this in \bref{Ebound}, we deduce that
\be
E \ge |J| + 2|J'|.
\ee
When this bound is saturated the matrix of anticommutators of
supersymmetry charges will have zero eigenvalues, corresponding to the
preservation of some fraction of supersymmetry. Let us determine this
fraction under the assumption that $J'>J$. In this case the Dynkin
labels are given by \bref{choiceone} and hence
\be
b_1 = J' -{1\over2} J, \qquad b_2 = {1\over2} J,\qquad
b_3 = {1\over2} J,\qquad b_4 = -J' - {1\over2} J.
\ee
Generically,  $|b_1| =b$ and all other eigenvalues of $B$ have absolute
value less than $b$ so the supersymmetry fraction is 1/8.
However, this fraction is enhanced to 1/4 if $J=0$ because
then $|b_1|=|b_4|=b$ with $|b_2|,|b_3| <b$.

A similar analysis for $J\ge J'$ again yields the fraction 1/8
generically, with enhancement to 1/2 if $J'=0$; in this case the
string reduces to a point-like string orbiting the $S^5$ along an
equator, as  considered by Berenstein, Maldacena and Nastase (BMN)
\cite{Berenstein:2002jq}. Finally, if $J=J'=0$ then $b_i=0$ for all $i$, $E=0$,
and all supersymmetries are preserved, as expected for the \5ads
vacuum.

Redoing the computation for strings with 3 independent
angular momenta we find that the bound is
$E\ge |J_1| + |J_2| + |J_3|$, with 1/2-, 1/4- or 1/8- preservation
of supersymmetry for one, two or three nonzero angular momenta
respectively.

\subsection{Supersymmetry from $\kappa$-symmetry} \label{ch2:sec:k-symmetry}

The supersymmetries preserved by a IIB string correspond to complex
Killing spinors $\epsilon$ of the background that
satisfy\footnote{Thus,
$\Upsilon/\sqrt{-\det g}$ is the matrix $\G_{\kappa}$ appearing in the
kappa-symmetry transformation of the fermionic variables of the
Green-Schwarz IIB superstring.}
\be\label{kappa}
\Upsilon \, \e =
\sqrt{-\det g} \, \e \sac
\Upsilon = X'{}^M \dot X^N \gamma_{MN}\, K \,,
\ee
where $K$ is the operator of complex conjugation, and $\gamma_M$ are
the (spacetime-dependent) Dirac matrices.

Recall that we are interested in strings that live in the
$\CR \times S^5$ submanifold of \5ads with metric \bref{metric}, and
whose embedding is specified by equation \bref{embedding}.
Under these circumstances\footnote{
As the radius $R$ cancels in the final result we set $R=1$ in this section.}
\be
\sqrt{-\det g} = \sin\t \sqrt{1 - a^2 - b^2} \sac
\dot X \cdot \gamma = \G_t + a \, \G_{\chi} + b \, \G_\phi \,,
\ee
where $\G_\t, \G_\phi, \ldots$ are
ten-dimensional tangent space (\ie constant) Dirac matrices, and
\be \label{aandb}
a= \nu \, \cos \t \sac b = \omega \, \sin \t \,.
\ee
Note that the Lorentzian signature of the induced worldsheet metric
implies that
\be \label{rest}
a^2+b^2 \leq 1 \,.
\ee

The Killing spinors
of \5ads restricted to the relevant submanifold take the form
\be\label{ads5-spinor}
\e =
e^{\fc{t}{2} i \tilde{\G}} \,
     e^{\fc{\t}{2} i \g_* \G_\t} \,
     e^{\fc{\phi}{2} \G_{\t\phi}} \,
     e^{\fc{\chi}{2} i \g \G_{\chi}} \,
     e^{\fc{\a}{2} \G_{\t\a}} \,
     e^{\fc{\b}{2} \G_{\phi\b}} \, \e_0 \,,
\ee
where $\e_0$ is a constant spinor, $\g_* = \G_{\t \phi
\chi \a \b}$, and $\tilde{\G}$ is a constant matrix that commutes with all
other matrices above (its specific form will not be needed).
In our conventions all these matrices are real. For our configurations
$\dot X\cdot X'=0$, so the supersymmetry preservation condition
(\ref{kappa}) can be written as
\be
\label{kappanew}
\left(X'\cdot \gamma\right) \left(\dot X \cdot \gamma \right) \epsilon
=  - \sqrt{-\det g}\ K \, \epsilon \,.
\ee
This must be satisfied for all $\tau, \sigma$, but it is useful to
first consider $\tau=0$, in which case it reduces to
\be
\sin\t \left( \dot X \cdot \gamma \right) \, \G_\a \, K \,
e^{\fc{\t}{2}
i \g_* \G_\t} \,
e^{\fc{\sigma}{2} \left( \G_{\t\a} +  \G_{\phi \b} \right)} \, \e_0 =
\sqrt{-\det g} \,
e^{\fc{\t}{2} i \g_* \G_\t} \, e^{\fc{\sigma}{2} \left( \G_{\t\a} +  \G_{\phi
\b} \right)} \, \e_0 \,. \label{tau0}
\ee
It can be shown that in order for this equation to be satisfied for all $\s$,
one must impose
\be
\G_{\t\a\phi\b} \, \e_0 =  \, \e_0 \,, \label{cond1}
\ee
in which case the equation becomes
\be \label{condi2}
\left[ \G_t + \left( \cos \t - \sin \t \, i \g_*
\G_\t \right) \left( a \, \G_{\chi} + b \, \G_\phi \right) \right] \, \G_\a \,
K \, \e_0 = \sqrt{1-a^2 - b^2} \, \e_0 \,.
\ee
Equations \bref{cond1} and \bref{condi2}  are equivalent to the two equations
\bea
A\, \e_0 & = & \e_0  \sac A \equiv a \, \cos\t \, \G_{t\chi} + b \, \sin\t
\, i \G_{t\chi\a\b}  \,, \\
B\, K \e_0 & = & \sqrt{1-a^2 - b^2}\, \e_0 \sac
B \equiv a \, \sin\t \, i \G_{\phi\b} \,  + b \, \cos\t \,
\G_{\phi\a} \,  \,. \label{B}
\eea
Given that $a$ and $b$ are non-zero, it follows from (\ref{A}) that
\be
i \G_{\a\b} \, \e_0 = s \, \e_0 \sac
\G_{t\chi} \, \e_0 = s \, \e_0 \,,
\ee
and
\be
a \, \cos\t + s \, b \, \sin\t = \tilde{s} \,,
\ee
where $s$ and $\tilde{s}$ are independent signs.
The latter relation is compatible
with the restriction \bref{rest} if and only if
$b\, \cos\t = s \, a \, \sin\t$, and these two relations
for $a$ and $b$ imply, given \bref{aandb}, that
\be
\nu = \tilde{s} \sac \omega = s \, \tilde{s} \,.
\ee
It then follows that the equation \bref{B} is
trivially satisfied, and that $\sqrt{-\det{g}}=0$.
The string worldsheet must therefore be null,
which is only possible for a tensionless string.
Although the IIB superstring is not
tensionless, the energy due to the rotation is much greater
than the energy due to the tension in the limit of
large angular momentum. So in this limit the string is
effectively tensionless.

We continue now by considering only the tensionless string, for which the
supersymmetry preserving condition (\ref{kappanew}) reduces to
\be \label{nullpre}
\left(\dot X \cdot \gamma\right) \, \epsilon =0\, .
\ee
The analysis of this equation for $\tau=0$ reproduces the results already
obtained from an analysis of (\ref{tau0}), which are summarized by the
projections
\be
\label{finalproj}
\G_{\t\a\phi\b} \, \e_0 =  \, \e_0 \sac \G_{t\chi} \, \e_0
= \omega \nu \, \e_0 \sac i \G_{\a\b} \, \e_0 = \omega \nu \, \e_0 \,.
\ee
It is now straightforward to check that a spinor $\e_0$ satisfying these
conditions solves \bref{nullpre} for all $\tau$ and $\sigma$.
It thus follows that the generic null FT string
preserves 1/8 of the 32 supersymmetries of the IIB $AdS_5 \times S^5$
vacuum.

We have assumed above that $a$ and $b$ are non-zero. A solution with
$b=0$ has $J'=0$ and corresponds to a  point-like, collapsed string
moving along a great circle of $S^5$, as considered in~\cite{Berenstein:2002jq},
whereas a solution with $a=0$ has $J=0$. Redoing the analysis it
is easy to see the former preserve 1/2 of the supersymmetry. Similarly,
in the second case one finds that the necessary and sufficient conditions
for preservation of supersymmetry are
\be
\omega^2=1 \sac \cos\t_0=0,
\ee
and that the projections on $\epsilon$ are
\be
\G_{\t\a\phi\b} \, \e_0 \, = \,
\e_0 \sac i \Gamma_{t\chi\a\b} \e_0 = \omega\e_0 \,.
\ee
These projections preserve 1/4 of the thirty-two supersymmetries of the IIB
$AdS_5 \times S^5$ vacuum.

\medskip
\medskip
\caixa{{\bf Note:} In the appendix~\ref{ch2:sec:generalization} we consider the case of
strings carrying 3 different $J$'s directly in the
tensionless case, where the steps are largely simplified.}

\subsection{Physics of the large angular momentum limit} \label{ch2:sec:physics}

It is worth pausing for some time to examine the results just
given. It turns out that in the limit where the comparison with
field theory was being made ($J^2 \gg \l$) the strings become
null. There is a way to understand the inevitability of this. As
we have seen, \tem{ \item the effective string tension in an \5ads
background is $T \sim  \sqrt{\l}$, \item its kinetic energy $K$ is
purely due to  rotation, so that $K \sim J$. } On the other hand,
the expansion parameter on the field theory side had better be
$\l/J^p$ with $p>0$ if we ever want to extrapolate SYM results to
the stringy region. The conclusion is that the effective field
theory coupling constant is precisely \be \l' \sim {\mbox{tension}
\over (\mbox{kinetic energy})^p} \,, \ee and this had better be
small to believe SYM results. In other words, any chance to test
AdS/CFT by these methods will inevitably require a string solution
whose kinetic energy is much larger than its tension: an almost
tensionless string. This happens as well, of course, for the
collapsed string solution dual to the BMN ground state operator.

The understanding of this issue is useful to explain what is
the inherent difference between the apparently so similar
solutions in which the strings rotate in $AdS_5$ or in $S^5$.
Strings start growing in size as we increase
the angular momentum, so that not only the kinetic energy
grows but also the energy due to the tension (which is
proportional to the length). If we were in flat space,
both contributions remain of the same order no matter
how large we take $J$;  the string never becomes tensionless
there. The same applies for strings in $AdS_5$ partly because it is
a non-compact space.

In the $S^5$, however, the string soon reaches a maximum size as it can
not grow larger than the $S^5$. At this point something highly
non-trivial happens because the string is able to absorb more
and more angular momenta by rotating faster, without changing
its length. This is not the case, for instance, of giant
gravitons; they have an upper bound on $J$ beyond which
the solution simply does not  exist. In our case, the phenomenon
of absorbing angular momenta keeping the length fixed finally turns the string
into an almost tensionless one.

The nullity and the supersymmetry properties of these strings
might be intimately related, as it is the case for particles
(where susy implies $p_{\mu}p^{\mu}=0$).
We would like to mention that there have appeared
a number of generalizations of the strings presented
here. In the appendix~\ref{ch2:sec:generalization} we will show that they all
are actually supersymmetric in the same limit in which
they become null, which is also the limit in which
the comparison to field theory is made.
We keep for the discussion the only apparent exception to
this rule.

\subsection{Nearly-BPS Operators} \label{ch2:sec:nearly-bps}

As we mentioned earlier, the macroscopic strings considered here
have been proposed to be dual to operators of the form
\be\label{calop}
\calo (J,J') \, = \, \mbox{Tr} \left(X^J Y^{J'} Z^{J'} \right)
+ \cdots \,.
\ee
Note that these are single-trace operators; their association
to single-string states was discussed in section~\ref{ch2:sec:bmn} to be
valid for $J,J'\ll \sqrt{N}$.\footnote{The assumption that $J,J'\ll \sqrt{N}$ is compatible
with our other assumption that $J,J'\gg \sqrt{\l}$, and the two
together imply that the IIB string theory is weakly coupled.}
Evidence for this correspondence is that the anomalous dimensions of
the $\calo$-type operators have been computed by spin-chain methods
in the one-loop planar approximation~\cite{Beisert:2003xu}, and perfect
agreement has been found with the string prediction in the large
angular momenta limit~\cite{Frolov:2003qc,Frolov:2003tu,Frolov:2003xy}\footnote{See
the review~\cite{Tseytlin:2003ii} and references therein for a more complete list of
work along this direction.}. Note that the spin
chain computation implicitly assumes that $J,J' \ll \sqrt{N}$,
because this condition is needed to justify the restriction to  planar
diagrams; as in the BMN case, non-planar diagrams are expected to be
suppressed by powers of $J^2/N$, $J'^2/N$.

Our results concerning the supersymmetry of the rotating strings dual to
the $\calo$-type operators in the limit of large angular momenta imply
that these operators are `nearly-BPS' in this limit, in a sense that
we now aim to clarify.
Note that these operators are
primary (after diagonalization of the matrix of anomalous dimensions)
but not superconformal primary; for example, the operator with
$J=0$, $J'=2$ is a descendant of the Konishi operator. Moreover, they are
not 1/4-BPS or 1/8-BPS operators, since in $\caln =4$ SYM such BPS
operators are linear combinations of multi-trace operators that involve
at least\footnote{In the free theory, $\l=0$, there exist
purely double-trace 1/4-BPS  and purely triple-trace 1/8-BPS operators
\cite{Andrianopoli:1999vr}. See the
appendix~\ref{ch2:sec:algebra}.} a double-trace or a triple-trace operator, respectively
\cite{Andrianopoli:1999vr,Ryzhov:2001bp,D'Hoker:2003vf}. Therefore, although $\calo$ is a
nearly-BPS operator in the sense that its conformal dimension almost
saturates a BPS bound when $\l/J^2 , \l/J'^2 \gg 0$, it is not
the case that $\calo$ approximates an exact 1/4-BPS or 1/8-BPS
operator in this regime. In this sense the $\calo$-type operators are not
`near-BPS', but they are effectively so for any computation that depends
only on the conformal dimension and R-symmetry quantum numbers. We call
these operators `nearly-BPS'.

We wish to clarify a subtlety that might be confusing when comparing the
two/three angular momenta case to the BMN case. In the latter, the
ground state is BPS, and it is dual to a BPS operator; (most) excitations
about the ground state are non-BPS and they are dual to near-BPS operators.
In the former, the ground state is non-BPS but nearly-BPS; excitations
about it are obviously non-BPS but, to be consistent, they should
be called near-nearly-BPS. We hope that the following tables help to clarify
this issue.

\medskip
\cent{
{\bf BMN $(J,0,0)$}

\begin{tabular}[h]{||c||c||}
\hline
{\bf String Side}  & {\bf SYM side}  \\
\hline
\hline
Ground State & $\calo=\Tr X^J$ \\
is 1/2 susy  & is 1/2 BPS \\
\hline
Excited States  & $\calo=\Tr \left(X X X... D_\mu X...X X\right)$ \\
are near 1/2 susy  & is near 1/2 BPS \\
\hline
\end{tabular}}

\medskip
\cent{
{\bf Frolov-Tseytlin $(J_1,J_2,J_3)$}

\begin{tabular}[h]{||c||c||}
\hline
{\bf String Side}  & {\bf SYM side}  \\
\hline
\hline
Ground State & $\calo=\Tr \left( X^{J_1}Y^{J_2}Z^{J_3} \right) $ \\
is nearly 1/4 or 1/8 susy  & is nearly 1/4 or 1/8 BPS \\
\hline
Excited States  & $\calo=\Tr\left(  XX... D_{\mu}X...X Y^{J_2}Z^{J_3} \right)$ \\
are near nearly 1/4 or 1/8 susy  & is near nearly 1/4 or 1/8 BPS \\
\hline
\end{tabular}}

\medskip

The big difference is that the nearly BPS operators do receive quantum
corrections; they do it however in a controlled way, as they become
negligible as we approach the asymptotic BPS state.
To actually take the limit $\l/J^2 , \l/J'^2\ra 0$ we would need to go to
the free theory, $\l=0$. In this case, the conformal dimension of
$\calo$ does trivially saturate a BPS bound, and therefore must belong to a
shortened supermultiplet. This is possible because operators that are
descendants of superconformal primaries in the interacting theory can
become independent BPS operators in the free-field limit
\cite{Dolan:2002zh,Heslop:2003xu}; put in another way, short BPS states in the free
theory can join into a long one as we turn on the coupling and
receive radiative corrections.
Note that there will be as many of these additional
shortened multiplets as are required to form a long one, so the
shortening provides no  protection against the generation of large
anomalous dimensions: the usual claim that BPS-operators have protected
conformal dimensions is not true without qualification.

What happens when the condition $J,J'\ll \sqrt{N}$ is not satisfied?
On the field theory side, one needs to go beyond the planar
approximation. Moreover, as we described when discussing the BMN limit (see
section~\ref{ch2:sec:bmn})  single-trace operators are no longer
orthogonal to multi-trace operators. On the string theory side,
provided $g_s \ll 1$, single-string states remain orthogonal to
multi-string states. However, the description in terms of elementary
strings is likely to be inadequate. This is indeed the case
for states with $J'=0$, for which the correct semiclassical
description is known to be in terms of non-perturbative, rotating,
spherical D3-branes, the so-called `giant gravitons'~\cite{McGreevy:2000cw}.
The operators dual to these states are not single-trace operators, but
(sub)determinant operators~\cite{Balasubramanian:2001nh}; the latter {\it are}
approximately orthogonal to each other if $J$ is comparable to $N$,
and only those with $J \leq N$ are independent from each other.
A similar situation will presumably hold when both $J$ and $J'$
are non-zero. If this is the case, then the fact that the $\calo$-type
operators are only independent if $2J'+J \leq N$ (since otherwise
they can be expressed as sums of products of operators of the same
type) will be irrelevant to the comparison with string theory, since
these operators will only provide an accurate dual description of the
corresponding string theory states if $J,J'\ll \sqrt{N}$.

\subsection{Discussion} \label{ch2:sec:discussion}

We have discussed why quantitative tests of the AdS/CFT conjecture
that go beyond kinematics are so rare: a weak-coupling computation
on one side generally corresponds to an strong-coupling
computation on the other side. We have seen that an exception to
this state of affairs occurs in the sector of the rotating strings
considered here, for two reasons~\cite{Frolov:2003qc,Frolov:2003tu,Frolov:2003xy,Frolov:2002av}.
First, the energy of the corresponding {\it classical} string
configurations happens to admit an expansion in {\it positive}
powers of $\l/(J+2J')^2$. Second, as argued by FT, partial
cancellations between the vacuum energy of bosons and fermions in
sigma-model quantum corrections imply that all such corrections
containing non-positive powers of $\l$ are suppressed in the limit
$J + 2J' \gg 1$. These two facts allow the comparison of the
string calculation to a perturbative SYM calculation in the regime
in which $J+2J' \gg 1, \sqrt{\l}$.

If $J'=0$ the strings considered here reduce to the BMN strings, that is,
to point-like strings orbiting the $S^5$ around an equator, with angular
momentum $J$~\cite{Berenstein:2002jq}. The dual BMN operators are near-BPS operators,
in the sense that they are `close to' (that is, `a few impurities
away from') an exactly 1/2-supersymmetric operator, the so-called
BMN ground-state; thus, in the BMN case, the agreement tests the
AdS/CFT conjecture in a near-supersymmetric sector, and this fact is
presumably responsible for the partial cancellations of sigma-model
quantum corrections that are essential for the comparison to be possible.

\begin{figure}[here]
\cent{\epsfig{file=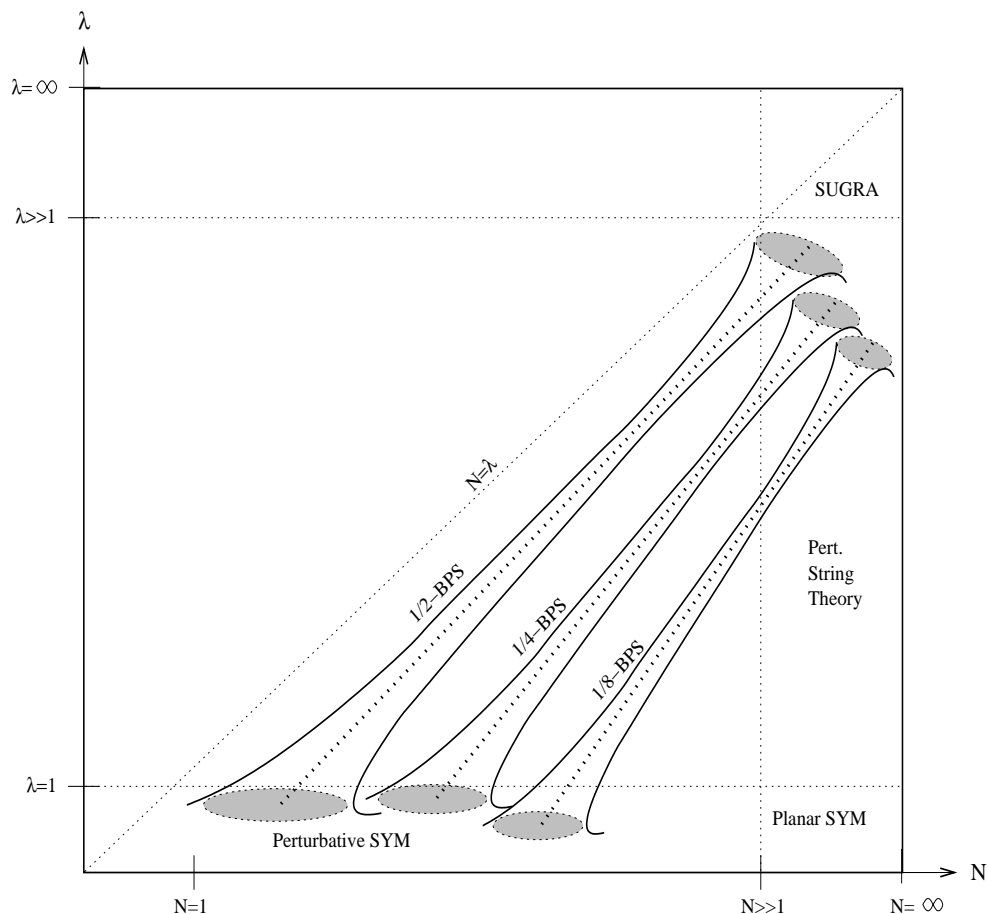, height=12cm, width=13cm}}
\caption{The BMN operators 'dig a safety tunnel' if we imagine
placing the 1/2-BPS protected operators in its middle, and the near
BPS (which receive quantum corrections) about it.\label{tunnel-3}}
\end{figure}

It had not been appreciated previously that the situation is very
similar for the rotating strings discussed here with $J'\neq 0$.
This is implied by the results of our work, since we have shown
that these strings asymptotically become 1/4- or
1/8-supersymmetric in the limit of large angular momenta. A subtle
difference between the extended strings and the BMN collapsed
strings case is, however, that the operators dual to the strings
with $J' \neq 0$ are not near-BPS\footnote{ Except if $J' \ll J$,
in which case they are BMN operators.} but {\it nearly}-BPS, in
the sense that there is no exactly 1/4- or 1/8-BPS operator that
these operators are close to. Supersymmetry could then be
responsible for \num{ \item making sense of computing a 1-loop
correction to a tachyonic $\s$-model vacuum, as this vacuum
becomes supersymmetric and hence stable in the limits considered,
\item suppressing all these $\s$-model loop corrections against
the classical result, \item the replacement $\l \rightarrow
\l/J^2$ in the SYM corrections to the conformal dimensions, as
they must be equal to $J_1+J_2+J_3$ in the limit. } The picture
seems to be that it has been possible to 'dig two other safety
tunnels' in the AdS/CFT parameter space, centered about nearly 1/4
or 1/8 protected operators. Figure~\ref{tunnel-1} should then be
replaced by figure~\ref{tunnel-3}.

We would like to emphasize that tensionless strings arise in our
analysis via an ultra-relativistic approximation in which the energy
due to the string tension is negligible compared to the kinetic energy.
In this sense, we think of the limit $\lambda/J^2 \ra 0$ as a limit
in which $\lambda$ is kept fixed and $J$ is sent to infinity; this
is particularly natural in view of the fact that we also need
$J \gg 1$. However, one can equivalently think of this limit as
fixing $J$ to be much larger than unity and then sending
$\lambda \ra 0$. The latter point of view is needed if one
wanted to obtain the strict limit $\lambda/J^2 = 0$ without
having to consider infinite energies; we then just take $\lambda =0$,
a limit in which the rotating strings
become exactly supersymmetric. This may be relevant to the
correspondence between tensionless strings in
$AdS_5\times S^5$ and free $\caln=4$ SYM theory (see, for example,
\cite{Bianchi:2003wx}). Note that the free field theory has an infinite
number of global symmetries, which could correspond to the gauge
symmetries of massless particles of all spin in the tensionless string
spectrum.\footnote{Tensionless strings could also arise as collapsed D3-branes,
for example, but such configurations could probably not be justified
within a semiclassical approximation.}

All the configurations obtained until now in which the anomalous
dimensions of strings with 2 or more angular have been
successfully compared to field theory results satisfy the two
properties discussed here: they become tensionless and
supersymmetric in the limit in which the comparison is done. We
will prove this in the appendix for all circular strings, but the
same must happen for the other type of strings considered in the
literature: folded strings. This is because they tend to saturate
the same BPS bounds. The only apparent exception that we are aware
of is a pulsating string solution of~\cite{Engquist:2003rn}, for which the classical energy
matches the anomalous dimension of a SYM operator computed in
perturbation theory despite the fact that this energy is not close
to saturating a BPS bound. However, {\it it does become null} in
agreement with the inevitability argument given in section~\ref{ch2:sec:physics}.
This seems to suggest that it might be this last property, and not
supersymmetry the key behind these successful comparisons. The
pulsating string, however, still requires some extra work to be
put into the same footing as the solutions considered here. The
following questions raise suspicions about them: \num{ \item
Having an energy as far as wanted from saturating a BPS bound,
would not it be reasonable to expect them to decay immediately, as
there are states with the same charges but much less energy? \item
It is not clear at all that $\s$-model corrections will be
negligible in this case, as supersymmetry is not restored
asymptotically. If this happened, the classical value would be
meaningless and the comparison would be just a coincidence.}
The second point is a problem of special relevance after the results of
\cite{Bigazzi:2004yt}. In this paper, they consider the two-angular momenta
strings rotating in backgrounds which are dual no non-supersymmetric
field theories (so, obviously, neither the background nor the embedding
of the string preserve any supersymmetry). In a first step, they find
that the energy is a regular expression in the field theory coupling,
so that it looks like we are in a situation were supersymmetry has nothing
to do but still the comparison might be possible. However, they do compute
the 1-loop $\sigma$-model correction and what they find is that, not only
it is not subleading with respect to the classical contribution, but it is
also not regular in the coupling! One might then suspect that it is the generic case
that, in sectors completely unrelated to supersymmetry, the GKP method is not
suitable for truly testing the AdS/CFT correspondence.

We would like to mention that a nice
recent line of research \cite{Kruczenski:2003gt} consists on trying to recover
the $\s$-model Lagrangian directly from the field theory
1-loop operator. This would provide a comparison independent
of the particular solution considered.

\newpage
\section{Stable non-BPS AdS branes}\label{ch2:sec:stable-branes}
\footnote{The
remaining part of this chapter is based on unfinished work with
D. Mateos and P.K. Townsend.}
In this section we conclude our analysis of stringy physics in $AdS\times S$-like backgrounds.
The results that we provide are another good example of the intuition provided by
the dual open/closed string pictures of D-branes.

\medskip

Field theories in anti-de Sitter space have the curious feature that tachyonic modes
do not lead to (perturbative) instabilities as long as they satisfy the
Breitenlohner-Freedman (BF) bound
\cite{Breitenlohner:1982bm,Breitenlohner:1982jf,Mezincescu:1985ev}. This result has an interesting
application to the field theories arising from fluctuations of branes in spacetimes
of the form $AdS_{p+2}\times X_n$ for compact $n$-dimensional manifolds $X_n$.
Given the existence of a `minimal' $n'$-surface $\Sigma$ embedded in $X_n$,
with $n' \leq n$, there exists a minimal $AdS_{p'+2}\times \Sigma$ submanifold of
$AdS_{p+2} \times X_n$, for $p'\leq p$, that is a candidate `AdS vacuum' for a
$(p'+n')$-brane.

\cent{
\begin{tabular}{c|c}
Ambient space & Embedding map \\ \hline \\
$AdS_{p+2} \times X_n$  &  $AdS_{p'+2} \subset AdS_{p+2}$ \\ 
 & $\S \subset X_n$
\end{tabular}}

 Although $\Sigma$ is a `minimal' surface in $X_n$ it need not
have minimal volume within its homology class; it could instead have maximal volume,
as would be the case for a maximal $\Sigma=S^{n'}$ in $X_n=S^n$, and in this case
there exist fluctuations of $\Sigma$ that {\sl decrease} its volume.  Any such fluctuation will lead
to a tachyonic mode on $AdS_{p'+2}$, which would normally lead to an instability,
causing the cycle $\Sigma$ to contract to a lower-dimensional cycle. However, this will
not happen (at least perturbatively) if the tachyonic mode satisfies the BF bound.

The canonical example of this phenomenon, which led to the AdS/dCFT
correspondence~\cite{Karch:2000gx,Karch:2001cw,DeWolfe:2001pq}, is the $AdS_4\times S^2$ embedding of a D5-brane in
the $AdS_5\times S^5$ background  of IIB superstring theory. In this case stability
is guaranteed by the partial preservation of supersymmetry, and this feature
is shared by all previously studied examples of stable `AdS branes'
\cite{Skenderis:2002vf,Mateos:2002bu}.
Any extension of the AdS/dCFT correspondence to non-supersymmetric defect field
theories would presumably require a stable but non-super\-symme\-tric, and hence
`non-BPS', $AdS$-brane. Motivated by this possibility, we examine the stability of a
general class of $AdS$ embeddings of D-branes and M-branes in background spacetimes
of the form $AdS_{p+2}\times X_n$ with $X_n= S^q\times T^r$. We take
$\Sigma = \Sigma_{q'} \times T^{r'}$ and choose $\Sigma_{q'}$ to be a minimal
surface in $S^q$ and $T^{r'}$ to be a minimal surface in $T^r$. This set-up is
sufficiently general to include all previous supersymmetric AdS embeddings and, not
surprisingly, the results of the perturbative analysis confirm the stability
expected on the grounds of supersymmetry. Also included are many non-supersymmetric
AdS embeddings, most of which are shown to violate the BF bound. However, we do find
some new candidates to stable, but non-supersymmetric, AdS embeddings. We cannot
be completely conclusive at this stage however, as these new cases require
some extra work to determine their stability.

In all cases studied here, the $AdS_{p+2}\times X_n$ background can be viewed as
the near-horizon limit of the supergravity fields produced by a large number, $N$,
of other branes. This  allows an interpretation of an AdS embedding as an effective
supergravity  description of a {\sl planar} `probe' brane in the presence of $N$
`background' branes. Consider, for example, the supersymmetric
$(2|D3,D5)$ intersection on a 2-plane of a planar D5-brane with $N$ coincident
planar D3-branes. For large $N$  we may replace the D3-branes by the D3 supergravity
solution, and in the near horizon limit the D5-brane probe is found to be the
$AdS_4\times S^2$ embedding in $AdS_5\times S^5$.  The stability and supersymmetry
of this embedding is thus inherited from the  original flat space $(2|D3,D5)$
intersecting brane configuration. The stability of many  other supersymmetric AdS
embeddings can be understood in this way. Moreover, the fact that these AdS
embeddings are typically members of families of stable and supersymmetric
asymptotically-AdS embeddings can be understood from the fact that the probe is
stable  and supersymmetric at any distance from the background branes~\cite{Mateos:2002bu}.
The {\sl instability} of certain {\sl non-supersymmetric} AdS embeddings can
similarly be understood from the fact that, in the flat space picture, the force on
the probe brane is repulsive.

Conversely, in the flat space picture, an {\sl attractive} force will result in  the
formation of  some bound state (which may or may not be supersymmetric) as the probe
becomes  coincident  with the background branes. In the AdS picture, the probe brane
cannot form a bound state with the background (because the background is fixed in
this approximation) but there must  still exist some minimum energy AdS embedding
corresponding to coincidence of the probe  with the background. In these cases one
expects to find a stable but non-supersymmetric  AdS embedding that, because of
the attractive force, does not belong to a family of  stable asymptotically-AdS
embeddings. We will discuss the existence of several such stable AdS embeddings and
we will confirm that they are not supersymmetric, even when the bound state expected
on the basis of the flat space picture would be. This is similar to the phenomenon of
`supersymmetry without supersymmetry' in which a brane configuration that is
supersymmetric within the full string/M-theory fails to be supersymmetric within
the supergravity approximation, a phenomenon that will be discussed on its
own in sections~\ref{ch5:sec:susy-without-susy} and~\ref{ch6:sec:susy-without-susy}.

However, not all AdS embeddings can be readily understood from a corresponding
flat-space picture. One example discussed here is a non-singular  $AdS_4\times
S^1\times S^1$  embedding of a D5-brane in $AdS_5\times S^5$ that has no flat space
analog in terms of  intersections of planar branes, although it can be considered as
arising from an intersection of the background planar branes with a conical probe.
As might be expected this AdS embedding is unstable. However, the {\sl same}
non-singular  $AdS_4\times S^1\times S^1$ embedding  of an M5-brane in $AdS_4\times
S^7$  (which has a similar interpretation as a conical M5-probe in an M2 background)
turns out to be a candidate for a stable embedding.
In this case, there is no obvious `hidden' supersymmetry that would explain the stability.

We begin with a general analysis of the perturbative stability of $AdS$-branes,
based on the BF bound. Our results are then applied to all known, and some new,
AdS embeddings of branes in $AdS\times S$ backgrounds. Those cases in which we
find candidates for perturbatively stable but non-super\-sym\-me\-tric embeddings
involve either D-branes or the M-branes which are space-filling, so that
their worldvolume gauge fields are actually sourced by the background potentials
and cannot be set to zero. The complete analysis of these cases is still work in progress.

\subsection{Stability of $AdS$-branes} \label{ch2:sec:BBF}

Our concern here is with embeddings of branes in a background of the type
$AdS_{p+2} \times X_n$ with $X_n= S^q \times T^r$. The metric is
\be
ds^2 = R^2\left[r^2 dX\cdot dX + \fc{dr^2}{r^2}\right] + d\Omega_q^2 +
d{\bf Y}\cdot d{\bf Y}\,.
\ee
where ${\bf Y}=(Y^1,\dots,Y^r)$ are cartesian coordinates for $T^r$ and
$d\Omega_q^2$ is the $SO(q+1)$-invariant metric on the unit $q$-sphere.
The $AdS_{p+2}$ metric is in horospherical coordinates with radial
coordinate $r$ and  $(p+1)$-Minkowski coordinates $(X^0,X^1,\dots ,X^p)$. To allow
for an arbitrary ratio of the $AdS_{p+2}$ and $S^q$ radii we  have introduced the
constant $R$ as the $AdS_{(p+2)}$ radius.

We consider a $(p'+ q' +r' +1)$-brane in this background with action
\be
S = -\int_{AdS}\int_{\Sigma} \sqrt{-\det G_{ind}}
\ee
where $G_{ind}$ is the induced metric. The brane vacuum is chosen to be an
$AdS_{p'+2}\times \Sigma_{q'}\times T^{r'}$ embedding. We may take $(\rho,x^0,x^1,\dots,x^m)$
as the worldvolume $AdS$ coordinates and $(y^1,\dots y^{r'})$ as the worldvolume
$T^{r'}$ coordinates, and partially fix the worldvolume diffeomorphism
invariance by the identifications
\bea
r &=&\rho \,,\nn
X^0 &=& x^0, \qquad \dots, \qquad X^m= x^m \,,\nn
Y^1 &=& y^1, \qquad \dots, \qquad Y^{r'}= y^{r'}\,.
\eea
Our choice of brane vacuum is an embedding in $AdS_{p+2}\times X_n$ that is
partially specified by the conditions
\be
X^{p'+1}=\dots = X^p =0, \qquad Y^{r'+1} = \dots = Y^r =0.
\ee
on the worldvolume $X,Y$ scalar fields. The fluctuations of the $Y$ fields cannot lead to instabilities
because the $T^{r'}$ embedded in $T^r$ is stable for topological reasons. However, we should
consider fluctuations of the $X$ fields, which we shall call the `AdS scalars'; they can be
tachyonic as a result of mixing with
modes arising from fluctuations of worldvolume gauge fields~\cite{DeWolfe:2001pq} but it will turn out that
this is not relevant for the cases we consider.

To complete the specification of the brane vacuum we must specify the embedding of $\Sigma_{q'}$
in $S^q$. We must then consider the fluctuations of $\Sigma_{q'}$ in $S^q$, which are fluctuations of what we will call the `sphere scalars'. There are many possible topologies that
$\Sigma_{q'}$ might have. For example, it is known that there exist minimal (maximal-area)
2-surfaces of arbitrary genus in $S^3$~\cite{lawson}, and even ones that divide $S^3$ into two
parts of unequal volume~\cite{math}, but explicit formulas are known (to us) only for
$S^2$ and\footnote{Actually,
an explicit formula is given in~\cite{lawson} for any surface of zero Euler characteristic, which
would include a Klein bottle.} $T^2$. More generally, for $n$-surfaces, there exist explicit formulas for
$S^n$ in any higher-dimension sphere, for $T^n$ in $S^{2n+1}$, and for some products cases like
$S^2 \times S^1$ in $S^5$. These are the cases we will consider here.

\begin{itemize}
\item
$S^{q'} \in S^q$. In this case it is convenient to
parametrize $S^q$ by $q$ angles $(\t_1,\dots,\t_q)$ defined recursively:
\be
d\Omega_q^2 = d\t_q^2 + \sin^2 \t_q \, d\Omega_{q-1}^2.
\ee
Let $\sigma_1,\dots \sigma_{q'}$ be the remaining worldvolume
coordinates. We may complete the gauge fixing by setting
\be
\t_1 = \sigma_1, \qquad \dots , \qquad \t_{q'}= \sigma_{q'}.
\ee
The vacuum configuration, for which the brane is wrapped on a maximal volume $S^{q'}$ in $S^q$,
is then
\be
\t_{q'+1} = \dots = \t_q = {\pi\over2},
\ee
The induced metric is
\bea
ds^2 &=& R^2\left[ \rho^2 dx \cdot dx + {d\rho^2\over \rho^2}\right] + d{\bf y}\cdot d{\bf y}\nn
&& +\ \sin^2\t_q...sin^2\t_{q'+1}\left[ d\t_{q'}^2+
\sin^2 \t_{q'}\left( d\t_{q'-1}^2+... \right)\right] \nonumber
\eea
As we are concerned with fluctuations about the vacuum, we set
\be
\t_{q'+1}=\pi/2+\xi_1 \,, ... \, \t_{q}=\pi/2 +\xi_{q-q'}\,.
\ee
In principle, $\xi$ are fields on $AdS_{p'+2}\times \Sigma$ but the maximally
tachyonic modes arise from fluctuations that are constant on $\Sigma$, so it will be sufficient to establish
stability for these modes. We thus take $\xi$ to be constant on $\Sigma$. In this case the action for small fluctuations is
\be
S= vol(\Sigma)\int_{AdS} \sqrt{-\det g} \left[-1 + {\cal L}\right] \,,
\ee
where $g$ is the $AdS_{p'+2}$ metric of radius $R$, and
\be
{\cal L} = -{1\over2} g^{\alpha\beta}\partial_\alpha \xi \cdot \partial_\beta \xi +
{q'\over2} \xi\cdot \xi.
\ee
The product of the $q-q'$ fluctuation fields is the Euclidean product.
From this result, we see that fluctuations are indeed tachyonic, with mass squared $-q'$.
As the $AdS$ space has radius $R$, the BF bound requires
\be \label{gen1}
q' \, \leq {(p'+1)^2 \over 4 R^2} \,.
\ee
If this bound is not satisfied then the brane vacuum is unstable.

\item
$T^{q'} \in S^q$. We assume here that $q\ge 2q'-1$. We take the metric on $S^q$ to be
\be
d\Omega_q^2 = d\t_q^2 + \sin^2\t_q d\t_{q-1}^2 + \dots + \sin^2\t_q \cdots \sin^2\t_{2q'}
d\Omega_{2q'-1}^2\,,
\ee
with $d\Omega_{2q'-1}^2$ defined recursively by the formula
\be
d\Omega_{2q'-1}^2 = d\psi_{q'-1}^2 + \cos^2\psi_{q'-1} d\phi_{q'}^2 +
\sin^2\psi_{q'-1} d\Omega_{2q'-3}^2.
\ee
Let $(\sigma_1, \dots ,\sigma_{q'})$ be the remaining worldvolume coordinates. We fix the remaining
worldvolume diffeomorphisms by the identifications
\be
\phi_k = \sigma_k , \qquad (k=1,\dots, q') \,,
\ee
so we have a $T^{q'}$ embedding into an $S^{2q'-1}$ in $S^q$. The volume of the embedded torus is
\be
(2\pi)^{q'} \prod_{i=2q'}^q \sin^{q'}\t_i \prod_{k=1}^{q'-1} \cos\psi_k \sin^k\psi_k
\,.
\ee
This is maximal when
\bea
\t_i &=& {\pi\over2}, \qquad (i= 2q',\dots, q)\,, \nn
\psi_k &=& {\rm arctan} \sqrt{k}, \qquad (k=1,\dots,q'-1)\,.
\eea
To allow for fluctuations about this vacuum solution, we set
\bea
\t_i &=& {\pi\over2} + \xi_i,  \qquad (i= 2q',\dots, q)\,, \nn
\qquad \psi_k  &=& {\rm arctan} \sqrt{k} + \sqrt{q'\over 1+ k}\, \zeta_k \,,
\eea
where the factors ensure canonical normalization of the kinetic terms for the fluctuation fields.
Proceeding as before, we then find the following Lagrangian on $AdS_{p'+2}$ for these fields:
\be
{\cal L} = -{1\over 2} g^{\alpha\beta}\left[\partial_\alpha \xi \cdot \partial_\beta \xi +
\partial_\alpha \zeta \cdot \partial_\beta \zeta\right] +
{1\over 2} \left[\xi\cdot \xi + 2 \zeta\cdot \zeta \right].
\ee
Note that the $\zeta$ fields are twice as tachyonic as the $\xi$ fields. Their BF bound is
\be \label{gen2}
2 q' \, \leq {(p'+1)^2 \over 4 R^2},
\ee
which is more restrictive than the corresponding bound for the $\xi$ fluctuations.
Thus stable AdS branes with toroidal embeddings in $S^q$ are likely to be
rarer than those with spherical embeddings.

\item
$S^2 \times S^1 \in S^5$. We consider this case as an illustration of the general
`product' case. The metric on $S^5$ can be written as
\be
d\Omega_5^2 = d\psi^2 + \cos^2\psi d \phi^2 +
\sin^2\psi \left[ d\t^2 + \sin^2\t d\Omega_2^2 \right] \,.
\ee
We identify $\phi$ and the coordinates parametrizing the 2-sphere with the three
worldvolume coordinates. This embeds $S^2 \times S^1$ in $S^5$. The volume of this
embedded manifold is maximal when $\t= \pi/2$ and $\psi = {\rm arctan} \sqrt{2}$,
so we set
\be
\t= \pi/2 + c_1 \xi, \qquad \psi = {\rm arctan} \sqrt{2} +  c_2 \zeta \,,
\ee
where the factors $c_1,c_2$ must be chosen again to ensure canonical
normalization of the fluctuation fields.
Proceeding as before, we find that the squares of the masses of these fields are
\be
m^2_\xi = -{9\sqrt{3}\over 2}, \quad m^2_\zeta = - 6.
\ee
If $S^5$ is itself embedded in a higher-dimensional sphere then there will be
additional tachyonic modes with $m^2= -3$, which is the value of $m^2$ for an
$S^3$ embedding. The value of $m^2$ for a $T^3$ embedding is $-6$, so there is
a fluctuation of the  $S^2\times S^1$ embedding (the $\xi$ fluctuation) that
is more tachyonic than any of the fluctuations of either the $S^3$ or $T^3$
embeddings.

\end{itemize}

This last case illustrates the pitfalls of some simple generalizations that might
be suggested by the first two cases. Even though $S^2$ is more stable than $T^2$,
the product $S^2\times S^1$ is less stable than $T^2\times S^1 \cong T^3$. Therefore,
one can say that the most stable surface of a given dimension is a sphere,
followed by a torus. We will now apply these results to specific string/M-theory
brane configurations. In doing so we will need to use the appropriate value of
the ratio $R$, which depends on the type of background brane, as given in the following
table:

\begin{center}

\otaula {|c|c|c|} 
\hline
 Brane Configuration  & Background  & R    \\ \hline \hline
M5 & \adss{4}{7} & 2 \\ \hline
M2 & \adss{7}{4} & 1/2 \\ \hline
D5 & \adss{5}{5} & 1 \\ \hline
D5/D1 & \adss{3}{3}$\times T^4$ & 1 \\ \hline
\ctaula

\end{center}

\subsection{Applications to string/M-theory} \label{ch2:sec:strings}

There are a variety of AdS embeddings of string/M-theory branes in
$AdS\times S$ backgrounds to which we can apply the stability results
just obtained. This method has been previously used to establish the stability
of some AdS embeddings but here we shall consider a much larger class. In
general we should also consider fluctuations of worldvolume gauge fields
and the AdS scalars,
but we will deal with this issue on a case by case basis. The cases we consider
are conveniently divided into three categories.

\begin{itemize}
\item
Supersymmetric embeddings. As explained in the introduction, these arise from
standard supersymmetric intersections of planar D-branes or M-branes.
Because of the WZ term in the worldvolume action, the derivatives of the AdS scalar
mix with the gauge fields, and the mass squared of their fluctuations is obtained
by diagonalization of an infinite matrix~\cite{DeWolfe:2001pq}. Typically some
eigenvalues are tachyonic but they
satisfy the BF bound and thus do not lead to any instability. This is expected from
supersymmetry; indeed supersymmetry typically {\sl requires} that some tachyonic modes
arise in this way in order that all fluctuations form complete supermultiplets.

\begin{center}

\petit{\otaula {|c|c|c|c|c|c|} 
\hline

 Brane conf.   & Embedding & $(m R)^2$ & Bound
 & Stability & Susy    \\ \hline \hline
 (2$|$D3,D5)          & \adss{4}{2} & -2 &
 -9/4 & stable & yes     \\ \hline

 (1$|$M2,M5)          & \adss{3}{7} & -3/4
 & -1 & stable  & yes    \\ \hline

 (1$|$D3,D3)          & \adss{3}{1} &-1
& -1 & stable  & yes    \\ \hline

 (3$|$D3,D7)          & \adss{5}{3} & -3
& -4 &  stable  & yes    \\ \hline

\ctaula}

\end{center}

\item
Non-supersymmetric spherical embeddings. These arise from non-supersymmetric intersections
of planar branes. If the probe only overlaps the background branes, rather than intersecting them,
then there will be a force on the probe (and the corresponding embedding will be only asymptotically AdS).
This force may be attractive or repulsive.

If the force is repulsive then the AdS embedding will be unstable; one would expect the $q'$-sphere to
collapse to a point. In all cases of this type one may verify that the BF bound is violated
for at least one of the sphere scalars considered in the previous section. In the
$(0|D3,D3)$ case, there is a potential additional contribution to the kinetic terms of the fluctuations
coming from the WZ term in the D3-brane action which could change the conclusion, but
inspection shows that it contributes only to the cubic couplings. Also, there are other
scalars (as discussed above), but their fluctuations do
not mix with the fluctuations of sphere scalars (to quadratic order) so their presence cannot
affect the conclusion either. Thus, instability is verified in all these cases.

If the force is attractive then one expects stability. In all of the cases of this type listed in the
table, \ie $(3|D3,D5)$ and $(2|M2,M5)$, the probe brane completely fills the background AdS space, so
there are no AdS scalars. However, there is the additional problem that the worldvolume
gauge fields cannot be set to zero because they are sourced by the background potentials.
Under these circumstances, the determination of the embedding stability becomes more involved
and it requires a careful examination of the resulting equations of motion. We cannot
give a definite answer yet, but we hope to report on this in the near future.
In the table we include the masses that fluctuations in these two cases would have
if we could set the gauge fields to zero, and we leave the stability column with a
question mark.

\begin{center}
\petit{\otaula {|c|c|c|c|c|c|} 
\hline
 Brane conf.   & Embedding & $(m R)^2$ & Bound
 & Stability & Force    \\ \hline \hline
 (3$|$D3,D5)          & \adss{5}{1} & -1 &
 -4 & ?  & attract.  \\ \hline

 (0$|$D3,D3)          & \adss{2}{2} & -2
 & -1 & unstable  & repuls.    \\ \hline

 (2$|$M2,M5)          & \adss{4}{2} &-1/2
& -9/4 & ?  & attract.    \\ \hline

 (0$|$M2,M5)          & \adss{2}{4} & -1
& -1/4 &  unstable  & repuls.    \\ \hline

 (3$|$D5/D1,D5)          & \adss{2}{1}$\times T^3$ & -1
& -1/4 &  unstable  & repuls.    \\ \hline

\ctaula}

\end{center}

\item
Non-spherical embeddings. The main case to consider here is $T^{q'}$ for $q'>1$.
If one interprets the background AdS geometry
in terms of background branes then the probe brane must be asymptotically conical, with $T^{q'}$ as the base
of the cone. The vertex of this cone would be singular in flat space, but the singularity is removed
by the `back-reaction' of the background branes on the geometry, in a way that leads to a completely
non-singular $AdS_{p'+2}\times T^{q'}$ embedding of the probe in $AdS_{p + 2}\times X_n$. There is no
obvious reason why such an embedding should be supersymmetric, so generically we should expect instability.
Indeed, in all cases that we have considered, but one, the sphere scalar fluctuations violate the BF bound,
and this is sufficient (by the argument given before) to prove instability. The same argument applies
to the non-toroidal $S^2\times S^1$ embedding listed in the table.

The one non-spherical case for which the BF bound is satisfied is the $AdS_4 \times T^2$ embedding of
an M5-probe in the $AdS_4\times S^7$ near-horizon geometry of the supergravity M2-brane.
As this is $AdS$ space-filling, there are no AdS scalars,
and hence no tachyonic modes arising from the worldvolume 2-form gauge potential.
The sphere scalars can be expanded in harmonics on $T^2$. The constant harmonic
yields the maximally-tachyonic modes, with $(mR)^2 = -2$; this correspond to conformal coupling
of scalars in $AdS_4$ and hence stability. If the embedding were supersymmetric there would be an
{\sl even} number of these scalars. As the number is odd, the embedding is not supersymmetric.
However, this case suffers from the same problem as the attractive cases considered
above: it is not a consistent truncation to set the worldvolume gauge field to zero
because it is sourced by the background potential. A more careful analysis is
in progress.

\begin{center}

\petit{\otaula {|c|c|c|c|c|c|} 
\hline

 Brane conf.   & Embedding & $(m R)^2$ & Bound
 & Stability & Force    \\ \hline \hline
 (2$|$D3,D5)          & $AdS_4\times T^2$ & -4 &
 -4/9 & unstable & repuls.  \\ \hline

 (1$|$M2,M5)          & $AdS_3\times T^3$ & -3/2
 & -1 & unstable  & repuls.    \\ \hline

 (1$|$M2,M5)          & \adss{3}{2}$\times S^1$ &-${9 \sqrt{3}\over 8}$
& -1 & unstable  & repuls.    \\ \hline

 (2$|$M2,M5)          & $AdS_4\times T^2$  & -1
& -9/4 &  ?  & attract.   \\ \hline

\ctaula}

\end{center}

\end{itemize}

\subsection{Discussion} \label{ch2:sec:stable}

Recall that the $AdS_5\times S^3$ embedding of a D7-brane in $AdS_5\times S^5$
has been used, via the AdS/dCFT correspondence, to couple N=2 quark multiplets
to N=4 SYM theory. This is made possible by the fact that the D7 fills the
$AdS_5$ space. A non-supersymmetric stable AdS embedding of this type would
similarly enable us to make progress, via the AdS/dCFT correspondence, in
understanding non-supersymmetric gauge theories, and this was one motivation
for the work reported here. We have shown that there is indeed a candidate
for an embedding of this type in which a D5-brane
fills the $AdS_5$ and wraps a maximal $S^1$ in
$S^5$. However, in the full string theory, which involves consideration of
open strings connecting the probe D5 to the background D3-branes, we expect
to find tachyons that imply an instability in which the D-branes dissolve
into flux on the D5-brane, leading to a supersymmetric bound state. Of course,
this is not seen in our approximation.

According to the AdS/CFT correspondence, M-theory on $AdS_4\times S^7$ is
equivalent a (2+1) CFT with $O(8)$ symmetry and 16 supersymmetries. According
to the AdS/dCFT correspondence, AdS-embedded M5-brane probes are dual to
defects in this CFT. The (possibly) stable non-supersymmetric $AdS_4\times S^2$ embedding
of an M5 in an M2-background that we have found might be used to add non-supersymmetric
matter to the CFT, but this suggestion presumably suffers from the same type of problem
alluded to above because in the context of M-theory we know that the M2-branes
will dissolve in the M5-brane. However, we found one other example of a (possibly) stable
non-supersymmetric AdS-filling embedding of an M5-brane in $AdS_4\times S^7$ which
merits further investigation in this respect. Note that {\sl all} of the candidates to
stable non-supersymmetric AdS embeddings that we have found are
$AdS$-space-filing.

We hope to report soon about the fate of these three configurations. What is clear
at this stage is that the fact that the gauge fields do not vanish forces the
static embedded brane's worldvolume to wiggle, and this could be related to the
breaking of the $R$-symmetry of the dual theory by the addition of non-supersymmetric
matter.

\chapter{Engineering the gauge/string duality}\label{ch:wrapped}

This chapter is devoted to extending the AdS/CFT duality
to more realistic field theories with less than 16 supercharges.
The departure from flat embeddings of D-branes in flat space
is presented as a set of logical steps that lead one to
consider D-branes wrapping calibrated cycles of special
holonomy manifolds.
The chapter is quite self-contained and it discusses in some
detail all the steps of the sequence:
\tem{
\item the twisting of supersymmetric gauge theories from a purely
field-theoretical point of view (section~\ref{ch5:sec:twisting-field-theories}),
\item  the mathematical background concerning special
holonomy manifolds (section~\ref{ch5:sec:special-holonomy})
and calibrated cycles (section~\ref{ch5:sec:calibrations}),
\item the relation between the worldvolume theories on wrapped D-branes
and the field-theoretical twisting (section~\ref{ch5:sec:geometrical-twist}),
\item the technical progress in trying to find
the closed string duals provided by gauged supergravities
(section~\ref{ch5:sec:gauged-sugras}).
}

We then illustrate the whole procedure with the explicit
construction of the supergravity dual (in the IR) of an $SU(N)$ $\caln=2$
susy field theory in 2+1 dimensions, as reported in~\cite{Gomis:2001vg}.
Having obtained the dual, we study its non-perturbative moduli space
as we did in~\cite{Brugues:2002pm}. During this process we will encounter
a difficulty which goes under the name of 'supersymmetry without
supersymmetry'. That this was a general phenomenon in the duals
of non-maximally susy field theories was reported in~\cite{Brugues:2002ff} and we
carefully discuss it here in section~\ref{ch5:sec:susy-without-susy}
and in the chapter~\ref{ch:nc-sugra} (section~\ref{ch6:sec:susy-without-susy}).

\section{More general dualities involving flat D-branes} \label{ch5:sec:more-branes}

Even if the general purpose of the chapter is to find
more AdS/CFT-like dualities for more realistic field
theories than \4n in 3+1, we first need to understand
what happens if we consider Dp-branes with $p \neq 3$
and we repeat the decoupling process. Naively
we would expect to obtain a duality between
\bea
\begin{array}{|ccc|}
\hline
&&\\
& & \mbox{IIA/IIB in the near horizon limit} \nn
\mbox{SYM with 16 susys in D=p+1} & \leftrightarrow & \nn
& & \mbox{of the (p+1)-brane SUGRA solution} \nn
&&\\
\hline
\end{array}
\eea

There are however new problems that enter as soon as we
abandon $p=3$. As we mentioned in section~\ref{ch1:sec:closed-picture}, the
p-brane solutions have non-constant curvature for $p\neq 3$. The dilaton
is not constant either, so the coupling controlling
the loop corrections of string theory becomes a function
of the transverse distance to the origin. Therefore the ranges
of validity become more complicated here, as they are not
the same for all points of the background. The limit
in the supergravity side is always taken in such a way that
we retain excitations in the throat that are
dual to finite energy configurations of the field theory.
We keep  finite, for example, the mass of the $W$-bosons
\be
M_{W} = {r\over l_s^2} \equiv U = \mbox{fixed} \,.
\ee
It is worth writing down in one equation the near-horizon limit of
a general p-brane supergravity solution (we skip the RR forms)
\bea \label{nh-forall}
ds^2&=&\ap \left( {U^{(7-p)/2} \over \gym \sqrt{d_p N}} \,dx_{0,p}+
{\gym \sqrt{d_p N} \over {U^{(7-p)/2}}} \left[ dU^2 + U^2 d\Omega_{8-p}^2 \right] \right)
\nn
e^{\phi}&=& (2\pi)^{2-p} \gym^2 \left( {\gym^2 d_p N  \over U^{7-p}} \right)^{{3-p \over 4}} \,,
\eea
All quantities were defined in section~\ref{ch1:sec:closed-picture}, but we recall here that
\be
\gym^2 = (2\pi)^{2-p} g_s l_s^{p-3} \label{gym-again} \,.
\ee
The curvature scalar for these backgrounds can be easily computed and will be needed below
\be
l_s^2 \, \calr \sim {U^{3-p \over 2} \over \sqrt{\gym^2 N}}\,.
\ee
The supergravity solution will be valid in the regions where we can
simultaneously keep $l_s^2 \, \calr, e^{\phi} \ll 1$.

\medskip

How do we take the limit in the open string picture of the system?
As in the D3 analysis that led to the AdS/CFT, we should
\num{
\item decouple the open and closed string massive modes, so $l_s\rightarrow 0$,
\item decouple the open/closed interactions, so $l_P = g_s^{1/4} l_s\rightarrow 0$,
\item obtain a finite $\gym$ in the low energy effective action of $S_{open}$.
}
The third condition can be used together with \bref{gym-again}
 to obtain that $g_s \sim l_s^{3-p}$ in the limit,
so that it diverges for $p>3$. Using this scaling in the second
condition, we obtain that open/closed string interactions decouple
only if \be l_P \sim l_s^{7-p \over 4} \rightarrow 0 ,\,\,
\mbox{as }\, l_s \go0 \sothat p<7\,. \ee The naive conclusion is
then that it is not possible to decouple the closed strings for
Dp-branes with $p\ge 7$ and that we will need
 dual strong coupling descriptions for D4,
D5 and D6 branes.
As actually the string coupling constant
is $U$-dependent, the validity of the various descriptions will
require more care as they will be energy-dependent. Having kept
fixed $\gym$, we obtain a low energy description of the system in
terms of a $(p+1)$-dimensional SYM theory with 16 supercharges and
gauge group $SU(N)$. The coupling
constant is then dimensionful, and the dimensionless coupling that
truly controls the perturbative expansion is \be \label{geff}
g^2_{eff} \sim \gym^2 N U^{p-3}\,. \ee

\noindent
{\bf Summary:}

\noindent
The picture is that one has one single system. Depending on the parameters that
one can freely choose, say $N$ and $g_s$, {\it and} the energy scale $U$ at which we probe it,
 the system is best described in terms of perturbative SYM, supergravity
in the backgrounds \bref{nh-forall} or its strong coupling duals.

\section{Phase diagrams for flat D5 and D6 branes} \label{ch5:sec:d5-d6-diagrams}

In developing the gauge/string duality we will need to make extensive
use of D5 and D6 branes in complicated target manifolds; it
is therefore instructive to first understand their properties
in flat space. In this section we study their phase diagrams
by applying the conditions obtained above.

\subsection{Flat D5 Branes} \label{ch5:sec:d5-diagram}

The decoupling limit for $N$ D5 branes is
\be
U={r\over l_s^2}=\mbox{fixed} \sac \gym^2=(2\pi)^3 g_s l_s^2 =\mbox{fixed}
\sac l_s \go0 \,.
\ee
Perturbative SYM is valid when
\be
g^2_{eff}=\gym^2 N U^2 \ll 1 \sothat \gym U \ll {1\over \sqrt{N}} \,,
\ee
which is the deep IR region. As we increase the energy we will enter
in the realm of supergravity, which is valid when
\bea \left. \begin{array}{ccc}
l_s^2 \calr \ll 1 &\sothat&  \gym U \gg {1\over \sqrt{N}}  \\ \\
e^{\phi} \ll 1 &\sothat&  \gym U \ll \sqrt{N}  \\ \end{array} \right\}
\sothat
 {1\over \sqrt{N}} \ll \gym U \ll \sqrt{N} \,.
\eea
It is in that region that we can trust the supergravity solution.
As we increase the energy so that $\gym U \gg \sqrt{N}$, the string coupling
becomes large and we need to perform an S-duality, where the $N$
D5-branes are turned into $N$ IIB NS5-branes. The dual background is then
\bea \label{ns5-nh}
ds^2_{NS5}&=&dx_{0,5}^2 + g_s N \ap \left({dU^2 \over U^2} + d\Omega_3^2 \right) \,, \\
e^{\phi} &=& \left( {(2\pi)^3 N \over \gym^2 U^2 }\right)^{1/2} \,.
\eea
This is the familiar near horizon region of the full $N$ NS5-branes solution,
which would be
\bea \label{ns5-full}
ds^2_{NS5, full}&=&dx_{0,5}^2 + e^{2\phi} \left(dU^2  +  U^2 d\Omega_3^2 \right) \,, \\
e^{2 \phi_{full}} &=& e^{2\phi(\infty)} + {g_s N \ap \over U^2} \,.
\eea
This system presents a problem that we had not encountered in the D-brane backgrounds before.
The metric \bref{ns5-full} exhibits an infinite throat as $U \go0$, a limit
in which the radius of the $S^3$ remains finite and where we recover
the $\CR^{1,6} \times S^3$ geometry \bref{ns5-nh}. However, it is
readily checked that massive geodesics can propagate in a finite proper
time along the throat, which means that they do not decouple and that
extra degrees of freedom should be added to a description in terms
of \bref{ns5-nh}. The best way to deal with this case is to go back to
the string $\s$-model in this background and realize that it is actually
an exact CFT. The problem of geodesics escaping along the throat is then seen
as the lack of decoupling of some stringy states in the low energy limit,
where one obtains what has been named a 'little string theory'.
The following diagram (extracted from~\cite{Itzhaki:1998dd}) summarizes the
phases of the D5-branes,
\medskip

\be
\includegraphics[width=9.3 cm,height=7 cm]{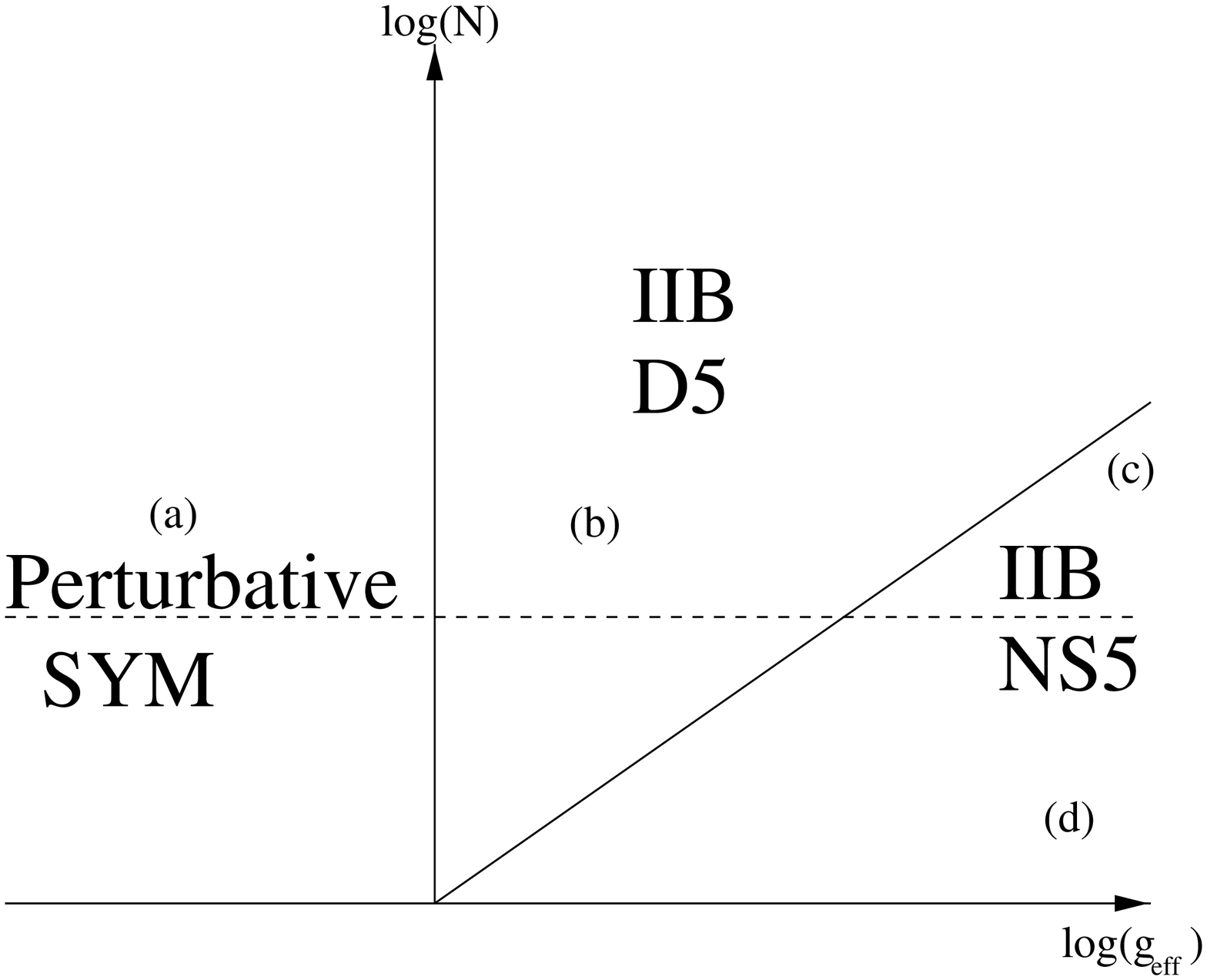}
\ee

\bigskip

\subsection{Flat D6 branes} \label{ch5:sec:d6-diagram}

We now repeat a similar analysis for the $N$ D6-branes system.
The near horizon limit  is
\be
U={r\over l_s^2}=\mbox{fixed} \sac \gym^2=(2\pi)^4 g_s l_s^3 =\mbox{fixed}
\sac l_s \go0 \,.
\ee
In this limit, the radius of the M-theory circle diverges
\be \label{r11}
R_{11}=g_s^{2/3} l_s \sim l_s^{-1} \rightarrow \infty \,,
\ee
and the system should then be described by going to 11 dimensions.
Once again, this the naive expectation as it depends on the energy at which
we probe the system. Let us be more careful; repeating the steps above we find that
perturbative SYM is valid when
\be
g^2_{eff}=\gym^2 N U^3 \ll 1 \sothat \gym U \ll {1\over (\gym^2 N)^{1/3}} \,,
\ee
which is the deep IR region. As we increase the energy we will enter
in the realm of supergravity, which is valid when
\bea \left. \begin{array}{ccc}
l_s^2 \calr \ll 1 &\sothat&  \gym U \gg {1\over (\gym^2 N)^{1/3}}  \\ \\
e^{\phi} \ll 1 &\sothat&  \gym U \ll {N \over \gym^{2/3}}  \\ \end{array} \right\}
\sothat
 {1\over (\gym^2 N)^{1/3}} \ll U \ll {N \over \gym^{2/3}} \,. \nonumber
\eea
It is in that region that we can trust the IIA supergravity solution.
As we increase the energy so that $\gym U \gg \sqrt{N}$, the string coupling
becomes large and the correct version of equation \bref{r11} becomes
\be
R_{11}(U)=e^{{2\over 3}\phi} \,l_s \gg l_s \,,
\ee
so that we have to uplift the IIA solution to M-theory.
As we will thoroughly discuss in section~\ref{ch5:sec:d6-solutions} this
leads to a solution of 11d supergravity which is
purely gravitational (it does not involve any gauge field)
and describes an ALE space with an $SU(N)$ singularity. Defining
 $y^2=2N \gym^2 U /(2\pi)^4$ we obtain
\bea \label{ale-space}
ds^2_{11d}&=&dx_{0,6}^2 + dy^2 + y^2 \left(d \theta ^2+\sin^2 \theta d\vp^2 + \cos^2\theta d\phi^2 \right) \,\\
(\vp,\phi) & \sim & \left( \vp+ {2\pi \over N},\phi +{2\pi \over N}\right) \sac 0\le \theta < {\pi\over 2}
,\,\, 0\le \vp,\phi < 2\pi \,. \nonumber
\eea
The identification in the angles makes the $S^3$ metric a $U(1)$ bundle over $S^2$ with
monopole charge $N$. Note that the metric \bref{ale-space} is locally flat, so that
the curvature vanishes everywhere except at the singularities. However, at very large values
of $y$ (in the far UV), the proper radius of these circles becomes very large and the 11d solution
can be trusted everywhere. As the proper length of the circles is of order $y/N \sim
\gym N^{-1/2} U^{1/2}$, and the 11d Planck length is $l_P^{(11)} = g_s^{1/3} l_s$, we will
have an everywhere flat 11d background as long as
\be
R_{circles} \ll l_P^{(11)} \sothat U \gg {N\over \gym^{2/3}} \,,
\ee
which means that we should trust that 11d supergravity is a good
description in the UV for any N. Thus we encounter a similar problem
to the one we found for the $NS5$-branes. When a massive radial geodesic
in the IIA near-horizon $D6$ background runs away from the
small $U$ region, it starts seeing the extra 11th dimension and
the geometry becomes flat, so that it can easily escape to infinity.
The proper description should then be the whole M-theory (and not
just supergravity excitations) in the ALE space background.
The phase diagram is then
\medskip

\be
\includegraphics[width=9.3 cm,height=7 cm]{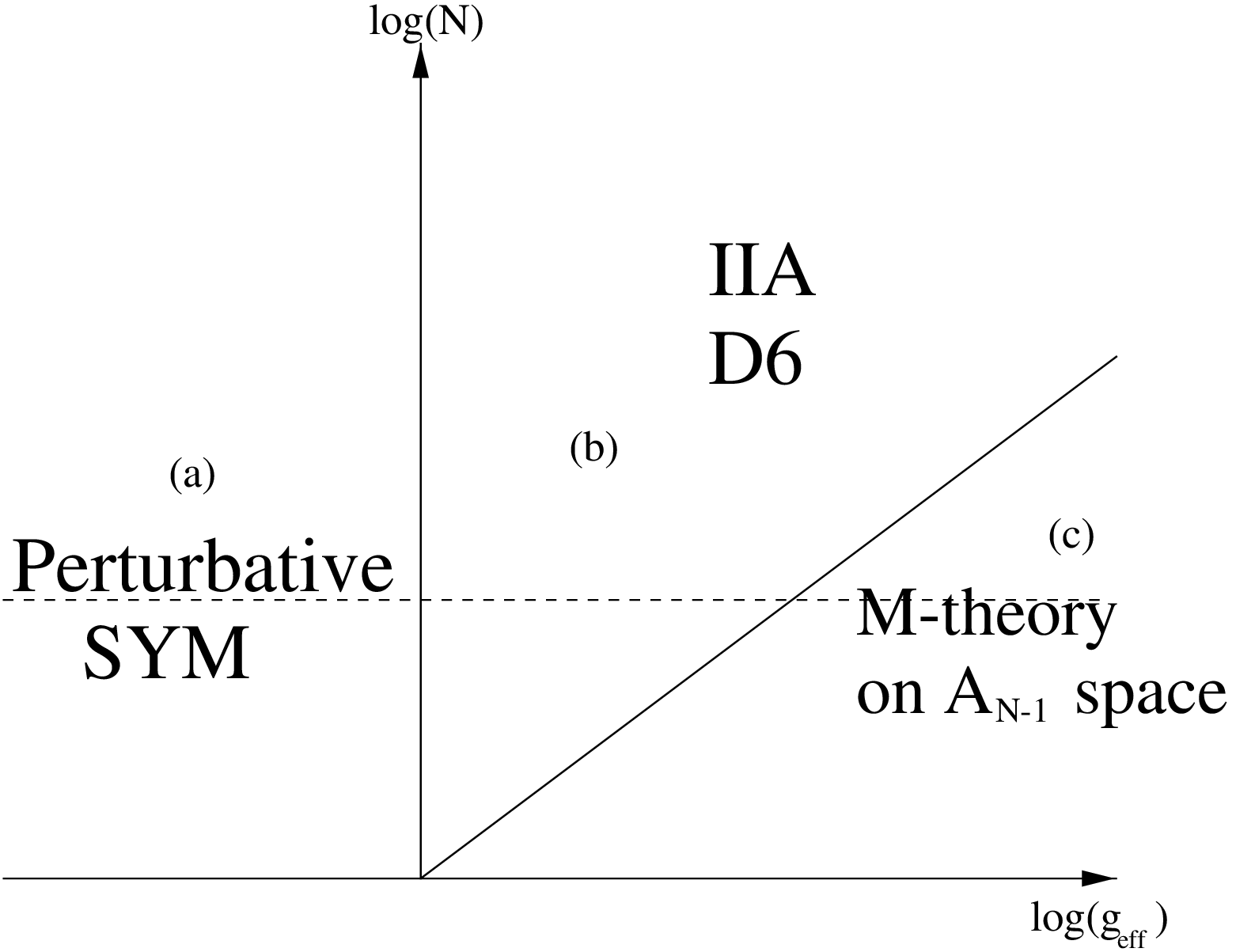}
\ee

\bigskip

\section{Moving away from flatness} \label{ch5:sec:away-from-flatness}

At this point we have essentially exhausted the possibility of
obtaining AdS/CFT-like dualities by means of flat D-branes
in flat space, and we encountered that they all involved
unrealistic field theories with maximal supersymmetry.
The enterprize of extending these dualities to less or
non supersymmetric cases is a difficult one, but it
has received a lot of attention after the original
AdS/CFT appeared. The process necessarily goes through
a modification of the flatness of the D-brane, of the
flatness of the target space, or both simultaneously.
Let us summarize the attempts performed until now\footnote{This list is not exhaustive,
see~\cite{Bigazzi:2003ui} and references therein for a more
exhaustive list.}
\tem{
\item The first possibility is to replace the $S^5$ factor by
another 5d manifold $X_5$.
As the brane is still flat we still have an $AdS$ factor;
this means that {\it the field theory will still be conformal}. There are two
well-known ways to modify $S^5 \rightarrow X_5$:
\tem{
\item The easiest is to keep the transverse space to
the brane flat except for singularities at some points.
This is easily done by replacing the transverse $\CR^6
\dreta \CR^6/\Gamma$, with $\G \subset SO(6)$. As the
radial distance is $SO(6)$-invariant, we simply obtain
a near-horizon limit $AdS_5 \times (S^5/\G)$. The number
of supersymmetries correspond to the fraction of the
original 16 supercharges which are left invariant by $\G$.
The smaller $\G$ is the more susy is preserved. The result is
\num{
\item $\caln=2$ if $\G \subset SU(2)$,
\item $\caln=1$ if $\G \subset SU(3)$,
\item $\caln=0$ if $\G \subset SU(4)$.}
\item A not so straightforward procedure consists on replacing
the transverse space $\CR^6$ by some other complicated
manifold $X_6$. If one puts the metric in the standard conical (radial$\times$compact)
form,
\be
ds^2_{X_6}=dr^2 + r^2 ds^2_{X_5} \,,
\ee
then any choice of an Einstein metric on $X_5$ leads to a 6d Calabi-Yau
space which, as we will see below, does not destroy all supersymmetries.
All one has to do to obtain the supergravity description is
to replace $d\Omega_5^2 \dreta ds^2_{X_5}$ in the corresponding
D3-brane solution. If we just restrict our attention to
cases in which $X_5$ is a quotient group $G/\G$, then there
are only 2  possibilities
\num{
\item $X_5 = SO(6)/SO(5)=S^5$, which leads to 16 supersymmetries (+16 special
conformal ones),
\item $X_5 = {SU(2)\times SU(2)\over U(1)}=T^{1,1}$, which leads to 8+8
supersymmetries.}
}
\item The second possibility is to replace the $AdS$ factor, which
corresponds to having a non-flat D-brane, and leads to the
{\it breaking of conformal invariance}. There are two completely
different ways of attacking this problem.
\tem{
\item One can consider small deformations of the D-brane embedding
which do not change its flat asymptotics. The near-horizon is
then still asymptotic to $AdS_5$. This means that in the UV,
where the energies are much larger than those
of the deformation, the theory must flow to a conformal fixed
point. The dual field theory is then typically an \4n SYM
plus some mass terms.

This method suffers from a very basic problem if what one would
like is to obtain something similar to QCD. To that aim,
one must require that the masses $M$ of the unwanted degrees of
freedom decouple from the theory before the strong-coupling
phase is reached. However, the renormalization
group invariant scale is \label{problema-1}
\be\Lambda_{QCD} \sim M e^{{-1 / N\gym^2 }} \,,\ee
which means that this decoupling can only be obtained if
we let $M \dreta \infty$ together with $\gym^2 N \go0$, so
that $\L_{QCD}$ is kept fixed. Because of $\gym^2 N \go0$,
there is little hope to apply this 'deformation method' to  study
strongly coupled pure QCD via supergravity. Insisting with
supergravity is still justified, but one must keep in mind
that what is really being studied is a $QCD$-like theory
with a (typically infinite) set of unwanted degrees of
freedom.}
\item
The remaining part of this chapter will deal with
another possibility which completely breaks conformal invariance
by abandoning the D-brane flatness property, even asymptotically.
Stability will be achieved by wrapping it in minimal cycles and
supersymmetry will be preserved by a beautiful
mechanism called twisting. The supergravity description
will then deal with spaces which have nothing to
do with $AdS$.
These models will show
similar problems to the 'deformation method' ones,
although some proposed way outs have been more or less
successful in this respect.
}
From now on we will concentrate only on the last mentioned
possibility.

\section{Twisting gauge theories} \label{ch5:sec:twisting-field-theories}

If our aim is to preserve less supersymmetry by considering curved
branes in curved manifolds, the first step is to understand how
to deal with a supersymmetric {\it field} theory in a curved space.
This section is purely field theoretical and hopes to provide
an understanding of what will come next, when we wrap branes
in complicated spaces.

The problem of formulating susy theories in curved spaces is a difficult one.
We are used to having supersymmetric
theories in curved backgrounds, but these are {\it supergravity} theories
in which the presence of spin two particles is a consequence of global
supersymmetry being promoted to a local one. On the other hand we are
used to formulating all kind of supersymmetric field theories (with all spins
$\leq 1$
and number susys $\leq 16$), but they are always in flat space (or at most
in spaces of constant curvature like $AdS$ or $dS$).
One naive way of illustrating the difficulties of changing flat
space by a curved manifold is that
as soon as one replaces ordinary derivatives by covariant ones,
there are new terms appearing in the transformation of the
Lagrangian which are proportional to the curvature. Maybe a more
clear way to say the same is that if the supersymmetry is to be
realized globally (locally would require supergravity), we need to
be able to deal with covariantly constant spinors, \ie \be
\label{cov-spi} D_{\mu} \e = (\pa_{\mu} + {1\over 4}\, \w_{\mu})\e
= 0 \,, \ee where $\w_{\mu}$ is the spin connection on the
manifold in the spinorial representation. The large majority of
manifolds we can think of do not admit any non-zero solution to
\bref{cov-spi}, $S^2$ being a simple example. We should stress
here that we are just working with a quantum field theory on a
curved {\it fixed} space; the equation \bref{cov-spi} will
reappear in the following sections, but the meaning will be
absolutely different, as then we will be in supergravity, where
$\w_{\mu}$ is dynamical.

So, by now we have to face the problem that we cannot formulate a
supersymmetric theory in a generic curved background
due to the obstruction \bref{cov-spi}. The first
breakthrough was due to Witten~\cite{Witten:1988ze}, who realized that one needs
to modify the various irreps in which the standard fields transform.
Supersymmetric theories typically involve a global $R$-symmetry that
rotates the susy charges and, hence, the various fields in a multiplet.
Witten discovered that by redefining  the Lorentz group as a mixture
of the usual one with the global $R$-symmetry it was possible to prevent
all curvature terms to appear in the susy variation of the action.
This was because under the new Lorentz group, all the curvature terms
appeared 'hitting' fields that are now scalars, so that
$[\nabla_{\mu},\nabla_{\nu}] \phi = 0$. The change of Lorentz irreps
of the fields gave the name of {\it twisting} to this procedure.

We will not go any deeper into Witten's point of view because it was realized
some time later~\cite{Eguchi:1990vz,Labastida:1997rg} that there was an easier way to mimic the twisting.
It was shown that it was all equivalent to choosing some of the conserved $R$-symmetry
currents and coupling them to a new non-dynamical gauge connection $A_{\mu}$ in such a way that
the modified version of the equation \bref{cov-spi}
\be \label{cov-spi-2}
D_{\mu} \e = (\pa_{\mu} + {1\over 4}\, \w_{\mu} + A_{\mu} )\e = 0 \,,
\ee
admits solutions. In other words, one is making {\it local} a part of
the originally {\it global} $R$-symmetry. Given a curved manifold $\calm_d$
in which we want to work, the spin connection is fixed and
it transforms in representations  of $G \subseteq SO(d)$.
If the manifold is not generic, then $G$ can be smaller than $SO(d)$; for
example, for $\calm_6=\CR^4 \times S^2$, $G=SO(2)$.
If we want $A_{\mu}$ to cancel a part of this spin connection, we first
need to set it equal (up to a sign) to $\w_{\mu}$ as 1-forms. Then we must
hope that the spinors transformed  in an irrep of the $R$-symmetry
with the  charges fine-tuned to cancel with the spin-connection.
There are indeed very few ways to gauge part of the $R$-symmetry
so that it all finally works,
and each way gives rise to a different number of preserved spinors.
The name {\it twisting} is here justified by the way the covariant
derivatives effectively act on the various fields, which is modified
by the presence of $A_{\mu}$.
The final result is a topological field theory, in the sense that
the curvature drops out of all computations, as it is effectively
cancelled by $A_{\mu}$.
We will give in section~\ref{ch5:sec:geometrical-twist}
a nice geometrical meaning to the apparently artificially
introduced $A_{\mu}$ and to the few various ways to gauge a
part of the gauge connection.

The best way to finish this section is by giving a concrete example.
Let us consider a 6d gauge theory with 16 supercharges. The minimal
supermultiplet consists of a vector $V_\mu$, four scalars $\phi^i$
and two complex Weyl spinors of opposite chirality $\psi^{\pm}$.
The Lorentz group is $SO(1,5)$ and the $R$-symmetry is $SO(4)_R
\cong SU(2)_L \times SU(2)_R$.
Consider now replacing $\CR^{1,5} \dreta \CR^{1,3} \times S^2$.
The spin connection is only nontrivial in the tangent bundle
to $S^2$, so it is valued in $SO(2)_{spin} \cong U(1)_{spin}$.
Before doing any twist, let us recall the precise irreps
of these groups in which the fields transform.\footnote{The
subscripts $\{0,\pm 1,\pm 2,...\}$ always denote $U(1)$ charges. In this
case there is only one $U(1)$ group present, so there is no room for confusion.
Even when there will be more $U(1)$'s are present, it is often clear to which
one they refer.}

\bea
\begin{array}{c|c|c}
& SO(1,5) \times SO(4) & SO(1,3)\times U(1)_{spin} \times SU(2)_L \times SU(2)_R \nn
\hline
&&\\
V^{\mu} & (6,1) & (4_0,1,1) \oplus (1_{+1},1,1) \oplus (1_{-1},1,1) \\
\phi^i & (1,4) & (1_0,2,2) \\
\psi^+ & (4,2) & (2_{+1},2,1) \oplus (\bar{2}_{-1},2,1) \\
\psi^- & (4',2') & (\bar{2}_{+1},1,2) \oplus (2_{-1},1,2)
\end{array}
\eea

\medskip
\medskip

Now the twist begins.
To cancel partially the transformation of the fields due to
the $S^2$ curvature, we gauge a $U(1)_R$ subgroup of the $SO(4)_R$, \ie
we couple this $U(1)_R$ current to an external non-dynamical
gauge field $A_{\mu}$ which we set equal to $\w_{\mu}$. There are only
two topologically non-equivalent embeddings of $U(1)_R \subset SU(2)_L \times
SU(2)_R$ which we consider separately.
\tem{
\item $U(1)_R \subset SU(2)_D$, where $SU(2)_D$ is the diagonal subgroup of
$SU(2)_L\times SU(2)_R$. So we need to understand how the  $SU(2)_L\times SU(2)_R$
irreps appearing in the table above decompose under this $U(1)_R$:

\bea
\begin{array}{c|c|c}
 SU(2)_L \times SU(2)_R & SU(2)_D & U(1)_R \subset SU(2)_D \nn
\hline
&&\\
(1,1) & 1 & 0 \\
(2,2) & 1 \oplus 3 & 0 \oplus 0 \oplus +1 \oplus -1 \\
(2,1) & 2 & +1 \oplus -1 \\
(1,2) & 2 & + 1\oplus -1
\end{array}
\eea

Finally, we need to identify the $U(1)_{spin}$ with $U(1)_R$,
\ie define $U(1)_{final}$ as their diagonal subgroup,
and retain only the fields which are invariant under $U(1)_{final}$.
The final set of charges under the relevant groups are then:

\bea
\begin{array}{c|c}
& SO(1,3)\times U(1)_{final} \nn
\hline
&\\
V^{\mu} &  4_0 \op 1_{+1}  \op 1_{-1} \\
\phi^i & 1_0 \oplus 1_0 \oplus 1_{+1} \oplus 1_{-1} \\
\psi^+ & 2_2 \op 2_0 \op \bar{2}_0 \op \bar{2}_{-2} \\
\psi^- & \bar{2}_2 \op \bar{2}_0 \op 2_0 \op 2_{-2}
\end{array}
\eea

We see that the fields which are invariant under $U(1)_{final}$
give, from a 4d point of view, one vector $4_0$, two scalars $1_0 \op 1_0$,
and two Majorana spinors $(2_0+\bar{2}_0) \op (2_0+\bar{2}_0)$.
This is precisely the $\caln=2$ gauge supermultiplet in four dimensions,
so we expect the whole theory on $\CR^{1,3}\times S^2$ to preserve
$1/2$ of the original 16 supercharges in $\CR^{1,5}$.

This result could have been anticipated by performing the same
decomposition on the supersymmetry parameters. In the original
$\CR^{1,5}$ theory, the four supercharges transform exactly
in the same irrep as $\psi^+$ and $\psi^-$, which means
that we can read directly from the last table that there
will only remain two Majorana supersymmetry generators,
\ie $\caln=2$.

\item $U(1)_R \subset SU(2)_L \subset SU(2)_L \times SU(2)_R$.
We just need to repeat the same steps. The decomposition
of the various $SU(2)_L \times SU(2)_R$ irreps under this choice
of $U(1)_R$ are

\bea
\begin{array}{c|c|c}
 SU(2)_L \times SU(2)_R & SU(2)_L & U(1)_R \subset SU(2)_L \nn
\hline
&&\\
(1,1) & 1 & 0 \\
(2,2) & 2 \oplus 2 & +1 \oplus -1 \oplus +1 \oplus -1 \\
(2,1) & 2 & +1 \oplus -1 \\
(1,2) & 1 \op 1  & 0 \oplus 0
\end{array}
\eea

We now identify $U(1)_{spin}$ with this $U(1)_R$ and consider the
decomposition under the diagonal $U(1)_{final}$

\bea
\begin{array}{c|c}
& SO(1,3)\times U(1)_{final} \nn
\hline
&\\
V^{\mu} &  4_0 \op 1_{+1}  \op 1_{-1} \\
\phi^i & 1_{+1} \oplus 1_{-1} \oplus 1_{+1} \oplus 1_{-1} \\
\psi^+ & 2_2 \op 2_0 \op \bar{2}_0 \op \bar{2}_{-2} \\
\psi^- & \bar{2}_{+1} \op \bar{2}_{+1} \op 2_{-1} \op 2_{-1}
\end{array}
\eea

We see that the fields which are invariant under $U(1)_{final}$
give, from a 4d point of view, one vector $4_0$
and one Majorana spinor $(2_0+\bar{2}_0)$.
This is precisely the $\caln=1$ gauge supermultiplet in four dimensions,
so we expect the whole theory on $\CR^{1,3}\times S^2$ to preserve
$1/4$ of the original 16 supercharges in $\CR^{1,5}$.

Again, recalling that the susy parameters in the
$\CR^{1,5}$ theory transform exactly
in the same irrep as $\psi^+$ and $\psi^-$, we can read directly from the
last table that there will only remain one Majorana spinor supersymmetry parameter,
\ie $\caln=1$.
}

We have learnt that the twisting mechanism allows us to put
a supersymmetric field theory in $\CR^{1,3}\times S^2$
at the price of preserving only $1/2$ or $1/4$ supersymmetry.
At distances much larger than the radius of the $S^2$ (in the IR) the
theory is expected to be well described by its truncation
to the massless KK modes, leading to a 4d $\caln=1,2$ gauge field theory.

\newpage

\section{D-branes wrapping cycles in special holonomy manifolds} \label{ch5:sec:wrapping-branes}

At this point we have understood from a purely field-theoretical
point of view how to put a supersymmetric field theory in
a curved background. This section describes one of the most
amazing geometrical interpretations that D-branes in string theory
have provided of a field theory phenomenon. We will
see that the extra gauge connection introduced in the previous
section can be interpreted as a usual spin connection on the
normal bundle to the D-branes.
First, however, we need to understand which type of manifolds are candidates
to accept cycles where D-branes can be supersymmetrically wrapped.

\subsection{Special holonomy manifolds} \label{ch5:sec:special-holonomy}

If the worldvolume theory of a D-brane is to be supersymmetric, the
first condition is that it must be put in a background that
preserves at least 1 supersymmetry. Let us then look for
supersymmetric solutions of the low energy supergravity theories of
IIA, IIB or M-theory\footnote{Although we have been
mainly concerned about D-branes, the discussion that follows
applies to M-theory backgrounds and M-branes wrapped on
their cycles as well.}  which preserve supersymmetry.
Looking for all such possible solutions is a huge task,
and a complete characterization has only been achieved
for maximally supersymmetric solutions\footnote{By
'characterization' we understand giving the explicit
solution, up to coordinate transformations.}~\cite{Figueroa-O'Farrill:2002ft}. Luckily, we
will only need a small subset of these solutions, characterized
by being {\it purely gravitational} and of the form
\be
M_D = \CR^{1,d-1} \times X_{d'} \sac \mbox{with $D$=10,11 and
d+d'=D} \,.
\ee
We will use an index notation such that
\bea
M_D & \dreta & M,N=0,1,...,D-1 \,, \nn
\CR^{1,d-1} & \dreta  & \mu,\nu=0,1,...,d-1 \,, \nn
X_{d'} & \dreta & i,j=1,...,d' \,. \nonumber
\eea
The solutions we are looking for are also purely bosonic,
which means that all the fermions are set to zero. This
statement is not supersymmetric invariant in general, because
fermions and bosons mix under a supersymmetry transformation.
So if we look for backgrounds which do not spontaneously
break supersymmetry we need to require, schematically,
\bea
\d_{susy} ~ \mbox{bosons} |_{sol} &\sim& \mbox{fermions} |_{sol} = 0 \\
\d_{susy} ~ \mbox{fermions} |_{sol} &\sim& \mbox{bosons} |_{sol} = 0 \,,
\eea
where the subscript '$|_{sol}$' indicates that a expression must be
evaluated on the solution. As fermions$|_{sol}$$=0$, the first
equation is always satisfied. The second, however, is a non-trivial
requirement.
Being purely gravitational backgrounds makes it possible to
treat IIA/IIB/M-theory simultaneously, as their only non-trivial equations of motion
are the Einstein equations in the vacuum, and the only non-trivial
variation of the fermions is that of the gravitino $\d \Psi_M=D_M \e$.\footnote{The
variation of the dilatini in type IIA/IIB is proportional to the gauge fields
and derivatives of the dilaton, which are all zero.} Therefore,
supersymmetric backgrounds of IIA/IIB/M-theory are solutions of
\bea
\mbox{Supergravity e.o.m.:} &&R_{MN} = 0 \sothat R_{ij}=0 \\
\mbox{Killing spinor equation:} && D_M \e = 0 \sothat
D_i\e= \left(\pa_i+ {1\over 4} w_i\right) \e =0 
\nn \label{kil-spi}
\eea
Together, they imply that $X_{d'}$ must be a Ricci-flat manifold
with covariantly constant spinors. Having a spinor that is
parallel transported along any curve must imply a restriction
on the holonomy of the manifold. In particular, by
considering a closed curve, we find that the spinor is unchanged,
so that the holonomy group $H$ of the manifold must admit an invariant
subspace, thus it can not be as large as $SO(d')$.
This is seen explicitly by taking the commutator of \bref{kil-spi},
which gives the change of an object under an infinitesimal
closed path,
\be 0=
[D_i,D_j]\e={1\over 4} R_{ij\a\b} \G^{\a\b} \e \,,
\ee
and recognizing $R_{ij\a\b} \G^{\a\b}$ as the generators of
the holonomy group. So the condition of supersymmetry
preservation can be rephrased in terms of $H$ by stating that
$SO(d')$ must admit at least one singlet under
the decomposition of its spinorial representation in irreps
of $H \subset SO(d')$. Such a manifold is called a {\it reduced}
or {\it special holonomy manifold}.

One important result which will be used repeatedly is that it can
be shown that the connection $\nabla$ and the curvature tensor $\calr$ in such manifolds
are restricted
\bea
\nabla \, \in \, End(TM) \, \otimes \, \L^1 T^* M & \rightarrow &
                  \nabla \, \in \, Hol(M) \, \otimes \, \L^1 T^* M \,,\nn
\calr \, \in \, End(TM) \, \otimes \, \L^2 T^* M & \rightarrow &
                  \calr \, \in \, Hol(M) \, \otimes \, \L^2 T^* M \nonumber \,,
\eea
which essentially means that the comparison of tensors/spinors of tangent
spaces at different points can be performed by $Hol(M)$-rotations, not
with the whole most general $SO(d')$. This means that the decomposition
of fields in different irreps of $SO(d')$ under $H \subset SO(d')$
provides irreps that do not mix, and can be considered as different
fields on the manifold.

We will discuss the general classification of these manifolds below.
Let us first note some nice properties which are common to all of them
and which resemble very much the isomorphism between homology and cohomology
group, in the sense that they relate global-topological with local-differential
properties of a manifold.

Let us consider the spectra of three of the most important differential
operators that may be defined on a manifold:
\num{\item the Hodge-de Rham operator $(d+\d)^2$ acting on forms,
\item the Dirac operator $i\G^a D_a$ acting on spinors,
\item the Lichnerowicz operator $\nabla_L$ acting on symmetric traceless tensors.}
In a general manifold, the spectra of these operators will be unrelated.
However, on a manifold with reduced holonomy we can use $H$ instead of $SO(d')$
to classify the fields over which these operators act, as the generators
of $H$ commute with the 3 operators. This is because $H$ is obtained
by parallel-transporting the fields and the 3 operators can be written
in terms of covariant derivatives. The conclusion is that the spectra
of these differential operators must form a representation of $H$
at each level.

\tem{ \item
{\bf Example: }Let us consider a 7d manifold $X_7$ with $G_2$ holonomy
to illustrate what has been said in this section.
That this preserves supersymmetry is readily seen from the decomposition
of the spinorial representation under the holonomy group
\bea
\ba{ccc}{
SO(7) & \dreta & G_2 \subset SO(7) \\ &&\\
8 & \dreta & 1+7
}
\eea
so that there is one spinor left-invariant by $H$, which is the Killing spinor.
In such a manifold, if we have a scalar eigenstate $\phi$ of the Hodge-de Rham operator,
\ie
\be
-\nabla^i \nabla_i \phi = \l \phi \,,
\ee
then we can immediately obtain two eigenstates of the Dirac operator out of $\phi$ by
\be
\chi_{\pm} = \left[\phi \pm i \l^{-1/2} (\nabla_i \phi)\G^i \right]\eta \sothat
i\G^i\nabla_i ~\chi_{\pm} = \pm \l^{1/2} \,\chi_{\pm} \,,
\ee
with $\eta$ a constant spinor. Similarly, if $V^i$ is an eigenstate of
the Hodge-de Rham operator
\be
(d+\d)^2 V_i =  \nabla^j \nabla_j V_i + R_{ij}V^j = \l V_i\,,
\ee
then we can immediately obtain two eigenstates of the Dirac operator out of $V^i$ by
\be
\chi_{\pm} = \left[i V^i \G_i \pm \l^{-1/2} (\nabla_i V_j)\G^{ij} \right]\eta \sothat
i\G^i\nabla_i ~\chi_{\pm} = \pm \l^{1/2} \,\chi_{\pm} \,.
\ee
One could also give the inverse transformations from eigenspinors to eigenscalars
and eigenvectors.}

Having seen how the spectra are related, let us focus on the zero modes.
Out of a Killing spinor $\e$ we can construct
forms of any degree $\w_n$ by contracting them with gamma matrices,
\be \label{nforms}
\w_n ~=~ {1\over n!} ~\bar{\e} \,\G_{i_1 ... i_n} \e ~\, dx^{i_1} \we ... \we dx^{i_n} \,.
\ee
We are not being careful with the conventions here, but once they are taken
into account, it can be seen that there are a few $n$ such that $\w_n = 0$
identically. However, those $\w_n \neq 0$ constructed this way are
automatically zero modes of the Hodge-de Rham operator, so that they
lie in a non-trivial class in cohomology. Their
corresponding homology $n$-cycles are therefore non-trivial neither.

\tem{
\item Back to the $G_2$ example, it can be checked that the
only non-zero forms that can be constructed by \bref{nforms} are an $\w_3$
and an $\w_4$, both related by Hodge duality.
}

This is as far as we will need to go with our study of reduced holonomy
manifolds for the moment. The classification of the possible holonomy
groups that lead to supersymmetry and their non-zero harmonic forms
are given in the table below. We also indicate how many supersymmetries
would be preserved if the total manifold was an M-theory background
of the style $\CR^{1,10-d'}\times X_{d'}$.

\cent{
\begin{tabular}{|c|c|c|c|c|}
\hline
dim($X_{d'}$) & Holonomy group & Susy & Forms & Name \\
\hline
\hline
4 & SU(2) & 16 & $\w_2$ & Calabi-Yau \\
6 & SU(3) & 8  & $\w_2, \w_3, \w_4$ & Calabi-Yau \\
7 & $G_2$ & 4  & $\w_3, \w_4$ & $G_2$-manifold \\
8 & SU(2) $\times$ SU(2) & 8  & $\w_2, \w_4, \w_6$ & Calabi-Yau\\
8 & Sp(2) & 6  & $\w_2, \w_4, \w_6$ & Hyper-K\"{a}hler \\
8 & SU(4) & 4  & $\w_2, \w_4, \w_6$ & Calabi-Yau \\
8 & Spin(7) & 2  & $\w_4$ & Spin(7)-manifold \\
10 & SU(3) $\times$ SU(2) & 2  & $\w_2, \w_3, \w_4,\w_6,\w_8$ & Calabi-Yau \\
10 & SU(5) & 2  & $\w_2, \w_4,\w_5, \w_6,\w_8$ & Calabi-Yau \\
\hline
\end{tabular}
}
\label{special-hol-table}

It is instructive to work out the number of preserved supersymmetries
from a simple group theory argument. Let us examine the cases of
8d manifolds. The $Spin(1,10)$ structure group is broken by the background
to $Spin(1,2) \times Spin(8)$. The Majorana representation splits
into irreps of the latter
\bea
\ba{ccc}{
 Spin(1,10) & \dreta & Spin(1,2) \times Spin(8) \\ &&\\
32 & \dreta & (2,8_+) + (2,8_-)
}
\eea
The $Spin(8)$ factor is further reduced because $X_8$ has
special holonomy.
\tem{
\item If $X_8$ is a $Spin(7)$-manifold, then we need to further
decompose
\bea
\ba{ccc}{
 Spin(8) & \dreta & Spin(7) \subset Spin(8) \\ &&\\
8_+ & \dreta & 1 + 7 \\
8_- & \dreta & 8
}
\eea
so that we only find one singlet. There are then only 2 Killing spinors
that form a doublet of $Spin(1,2)$ and a singlet of $Spin(7)$. A
$2+1$ physicist would call it $\caln =1$.
\item If $X_8$ is a $CY_4$-manifold, then we need to further
decompose
\bea
\ba{ccc}{
 Spin(8) & \dreta & SU(4) \subset Spin(8) \\ &&\\
8_+ & \dreta & 1 + 1 + 6  \\
8_- & \dreta & 4 + \bar{4}
}
\eea
so that we find two singlets. There are then 4 Killing spinors
that form two doublets of $Spin(1,2)$.  A
$2+1$ physicist would call it $\caln =2$.
\item If $X_8$ is a $HK_4$-manifold, then we need to further
decompose
\bea
\ba{ccc}{
 Spin(8) & \dreta & Sp(2) \subset Spin(8) \\ &&\\
8_+ & \dreta & 1 + 1 + 1 + 5  \\
8_- & \dreta & 4 + \bar{4}
}
\eea
so that we find three singlets. There are then 6 Killing spinors
that form three doublets of $Spin(1,2)$.  A
$2+1$ physicist would call it $\caln =3$.
}

\subsection{Calibrations} \label{ch5:sec:calibrations}

Now that we know in which backgrounds we should place the D-branes
if we want to preserve supersymmetry, let us classify the kind
of cycles that they can wrap. For that we need some mathematical
background first. The plan of this section is to
\num{
\item define the notion of calibration and calibrated cycles,
\item discuss which calibrations admit the special holonomy manifolds,
\item prove the isomorphism between certain calibrated cycles and supersymmetric cycles.
}

\subsubsection{Definitions and properties of calibrations} \label{ch5:sec:definition-calibration}

{\it Definition.} A calibration on a Riemannian manifold $X_{d'}$
is a $p$-form $\w$ satisfying
\bea
d\w &=& 0 \,, \\
\label{calibrated}
\w|_{\xi^p} &\leq& vol|_{\xi^p} \sac \forall \xi^p \,,
\eea
where $\xi^p$ is any tangent $p$-plane and $vol$ is the volume
form on the cycle induced from the metric on $X_{d'}$. A $p$-cycle
$\Sigma_p$ is calibrated by $\w$ if the inequality \bref{calibrated}
is saturated for all tangent planes to $\Sigma_p$.
The main physical intuition of calibrated cycles comes from
the fact that they minimize the volume within their homology class.
This is easily proven by considering a calibrated cycle $\Sigma_p$
and any other cycle in the same homology class $\Sigma'_p$ so that
$\S_p -\S'_p = \pa \Xi_{p+1}$. Then
\be
\mbox{Vol}(\S_p)=\int_{\S_p} \w = \int_{\Xi_{p+1}} d\w + \int_{\S'_p} \w
= \int_{\S'_p} \w \leq \, \mbox{Vol}(\S'_p) \,.
\ee
The first equality follows from $\S$ being calibrated. The second uses
Stokes theorem. The third follows from the closure of $\w$ and the fourth
from \bref{calibrated}.

This is a rather deep result. The problem of finding minimal surfaces
in a given space has been much studied in the mathematical literature and its
non-simplicity arises from the fact that it requires solving a
second order differential system. The problem of finding calibrated
cycles is a first order one, as determining whether $\S$ is calibrated
or not depends only on the embedding map and the tangent spaces
to $\S$. Calibrated geometry is a fertile source of examples
of minimal submanifolds. It will come as no surprise that
the equations  we will have to solve when wrapping a brane
on a calibrated cycle will also be of first order.

\subsubsection{Calibrations of special holonomy manifolds} \label{ch5:sec:calibrations-specials-hol}

It turns out that all the special holonomy manifolds discussed in the
previous section happen to admit calibrations. The harmonic
forms whose existence could be derived from the existence
of Killing spinors turn out to satisfy the axioms of
a calibration. Being harmonic, they are closed; the second
axiom can be verified case-by-case. Let us analyze these
calibrations in more detail.
\tem{
\item On a $Spin(7)$-manifold, there exists a harmonic 4-form which
is called a Cayley calibration; its corresponding calibrated
four-cycles are called Cayley cycles.
It can always be written in a given orthonormal frame as
\bea
\w_4 &=& e^{1234}+e^{1256}+e^{1278}+e^{3456}+e^{3478}+e^{5678} + e^{1357} \nn
&-&e^{1368} - e^{1458}-e^{1467}-e^{2358}-e^{2367}-e^{2457}+e^{2468} \,,
\label{cayley}
\eea
where $e^{i_1...i_n}=e^{i_1} \we ... \we e^{i_n}$.

\item On a $G_2$-holonomy manifold there exist calibrations of degree 3 and 4
related by Hodge duality. Cycles calibrated by the first are called associative
and by the latter co-associative.
In an orthonormal frame we can write
\be
\w_3=e^{246}-e^{235}-e^{145}-e^{136}+e^{127}+e^{347}+e^{567}\,.
\ee

\item Calabi-Yau $n$-folds, where $n$ is the complex dimension, admit two classes
of calibrations. The first one is given by its \kah 2-form $J$ and powers of it,
\be
\w_2 = J \sac \w_4 = \undos J\we J \sac ... \,\,, \, \w_{2p}= {1\over p!} J^{p} \,.
\ee
The \kah form can always be written in an orthonormal frame as
\be
J=e^{12}+e^{34} +...+e^{(2n-1)(2n)} \,.
\ee
Cycles calibrated by these forms are called holomorphic because it can be
proven that their embedding in the CY manifold can be given in terms
of holomorphic maps.

The other type of calibrations are given by the real part of a certain
holomorphic $n$-form. This form is always fixed up to a phase.
In an orthonormal frame we can write,
\bea \label{slag}
\w_n &=& \mbox{Re}\left[e^{i\theta}\Omega_n\right]  \,,
\eea
with $\theta \in S^1$ and
\bea
\Omega_n &=& (e^1+i e^2)(e^3+i e^4)...(e^{2n-1}+i e^{2n}) \,. \nonumber
\eea
Cycles calibrated by this form are called special Lagrangian (SLAG).
For a four-dimensional Calabi-Yau, \ie for a $CY_2$,
both types of calibrations have the same degree, although they do not
necessarily coincide. Indeed in this case there is a third 2-form
calibration. The reason is that $SU(2)=Sp(1)$, which means that the
Calabi-Yau is also a Hyper-\kah manifold. We study this case below.
The next coincidence of degree appears for a $CY_4$. This is because
$SU(4)\subset Spin(7)$, so that a $CY_4$ is a particular case of $Spin(7)$.
In such case, the $4$-form \bref{cayley} can be written as
\be
\w_4 = \undos J\wedge J + \mbox{Re}\left[e^{i\theta} \Omega_4\right] \,.
\ee

\item Hyper-\kah manifolds with real dimension less than 10 exist
only in $d'=4$ and $d'=8$. As mentioned above, for $d'=4$ they
coincide with a $CY_2$. For both $d'=4,8$ they admit three different
\kah 2-forms which are also calibrations. For $d'=8$, the fauna
of calibrated cycles is quite large. Apart from these 3 \kah
calibrations (and their wedge-products), and having into
account the sequence $Sp(2)\subset SU(4) \subset Spin(7)$,
there also exist SLAG and Cayley calibrations. A cycle can be
calibrated with respect to more than one of these calibrations.

In an orthonormal frame, we can always write the 3 \kah forms of a
$Sp(2)$-manifold as
\bea
J^1&=&e^{12} + e^{34} + e^{56} + e^{78} \nn
J^2&=&e^{14} + e^{23} + e^{58} + e^{67} \nn
J^3&=&e^{13} + e^{42} + e^{57} + e^{86} \nonumber \,,
\eea
and its corresponding holomorphic 4-forms
\bea
\Omega^1 &=& \undos\, J^3 \we J^3 - \undos \, J^2 \we J^2 + i \, J^2 \we J^3 \nn
\Omega^2 &=& \undos\, J^1 \we J^1 - \undos \, J^3 \we J^3 + i \, J^3 \we J^1 \nn
\Omega^3 &=& \undos\, J^2 \we J^2 - \undos \, J^1 \we J^1 + i \, J^1 \we J^2 \nonumber \,.
\eea
}

\subsubsection{Calibrated cycles are supersymmetric cycles} \label{ch5:sec:calibrated-is-susy}

With the technology of calibrations one can easily classify the
cycles that probe D- or M-branes can wrap preserving supersymmetry.
Such cycles are called {\it supersymmetric cycles}.

We first try to develop some intuition, and for the sake of simplicity
we will start considering an M2 brane in an 11d supersymmetric background
of the form discussed above $\CR^{1,d}\times X_{d'}$.
Let us single out time and write the metric as
\be
ds^2_{11}=-dt^2 + g_{ij} dx^i dx^j \sac i,j=1,...,10.
\ee
In the absence of a background $3$-form, the action for the
membrane is just of a Nambu-Goto type, so that the dynamics
will be such that the worldvolume tries to minimize its
spacetime volume. For static configurations we can fix the
gauge $t=\s^0$ and let $x^i$ be independent of $\s^0$.
Doing so we can easily compute the energy of the M2 brane
\be
E= T_2 \int_{\S_2} d^2\s \, \sqrt{\det{m_{\a\b}}} \,,
\ee
where $m_{\a\b}$ is the induced metric from the background
to the spatial part of the worldvolume $\S_2$. Therefore
static configurations minimize the spatial volume of the
membrane. As supersymmetry requires that the energy of a state
is minimum for its given charges, we obtain that a necessary
condition for preservation of supersymmetry is that
the M2 wraps a calibrated cycle. To prove that it is
sufficient we should check the $\k$-symmetry
condition for worldvolume preservation of susy, which
for an M2 reads
\bea \label{ka}
\G_{\k} \e &=& \e \,, \\
\G_{\k} &=& -{1\over \sqrt{\det m}} {1\over 2!} \e^{ab} \pa_a x^i \pa_b x^j \G_{0ij}\,,
\eea
where $\e$ is an 11d Majorana spinor and $\G_i$ the corresponding curved
$\G$-matrices, \ie $\{\G_i,\G_j\}=2g_{ij}$. Equation \bref{ka} can be thought
of as an equation for the embedding map $x(\s)$; its solutions preserve
worldvolume supersymmetry. There is a nice way to characterize its solutions.
Let us consider the positive definite quantity
\be \label{almostcal}
\e^{\dagger} {(1-\G_{\k})\over 2} \e =  \e^{\dagger}  {(1-\G_{\k})\over 2} {(1-\G_{\k})\over 2}
 \e = ||  {(1-\G_{\k})\over 2} \e ||^2 \, \ge 0 \,.
\ee
We conclude that $\e^{\dagger}\e \ge \e^{\dagger} \G_{\k} \e$, with equality
if and only if \bref{ka} is satisfied. Let us rewrite the inequality
as
\be \label{popoloco}
\sqrt{\det{m}} \ge \e^{\dagger} {1\over 2!} \e^{ab} \pa_a x^i \pa_b x^j \G_{0ij} \e \,.
\ee
Defining $\bar{\e}=\e^{\dagger} \G_0$, and a 2-form $\w_2$ by
\be \label{cali}
\w_2 = - {1\over 2!} \, \bar{\e} \,\G_{ij}\e \,\, dx^i\we dx^j \,,
\ee
we see that \bref{popoloco} becomes the second condition \bref{calibrated} for $\w_2$ to be a calibration.
It can also be proven~\cite{Gauntlett:2003di} that $\w_2$ is close.

The conclusion is that the
cycle is supersymmetric if and only if it is calibrated by the 2-form \bref{cali}
constructed out of the Killing spinors. This statement can actually be
generalized to all the other D- and M-branes:
\medskip

\caixa{
An M- or D-brane can only wrap supersymmetrically a cycle that is calibrated
by a $p$-form which can be constructed out of the background Killing spinors.
}
\medskip

I am not aware of any example of a calibration that cannot be constructed out of Killing
spinors, but if examples exist, the discussion above shows that they would not lead
to  supersymmetric cycles.

\subsubsection{A caveat on homology and homotopy} \label{ch5:sec:homotopy}

Let us just comment on a little subtlety that can be confusing. Our
intuition of why D-branes are stabilized when they wrap minimal cycles
is that they do it because there is simply no way to contract any further.
So our intuition would lead us to conclude that it is homotopy what
matters; we just need to find a non-contractible cycle and wrap the
brane around it. However, all the discussion of the previous sections
is based on homology and cohomology. Indeed, it is homology what really
matters for D-branes, as dictated by their WZ couplings to the
various background forms. And we know that homology and homotopy
form classes of equivalence which are, in general, different
(see figure~\ref{homotopy}).
\clfigu{9}{6}{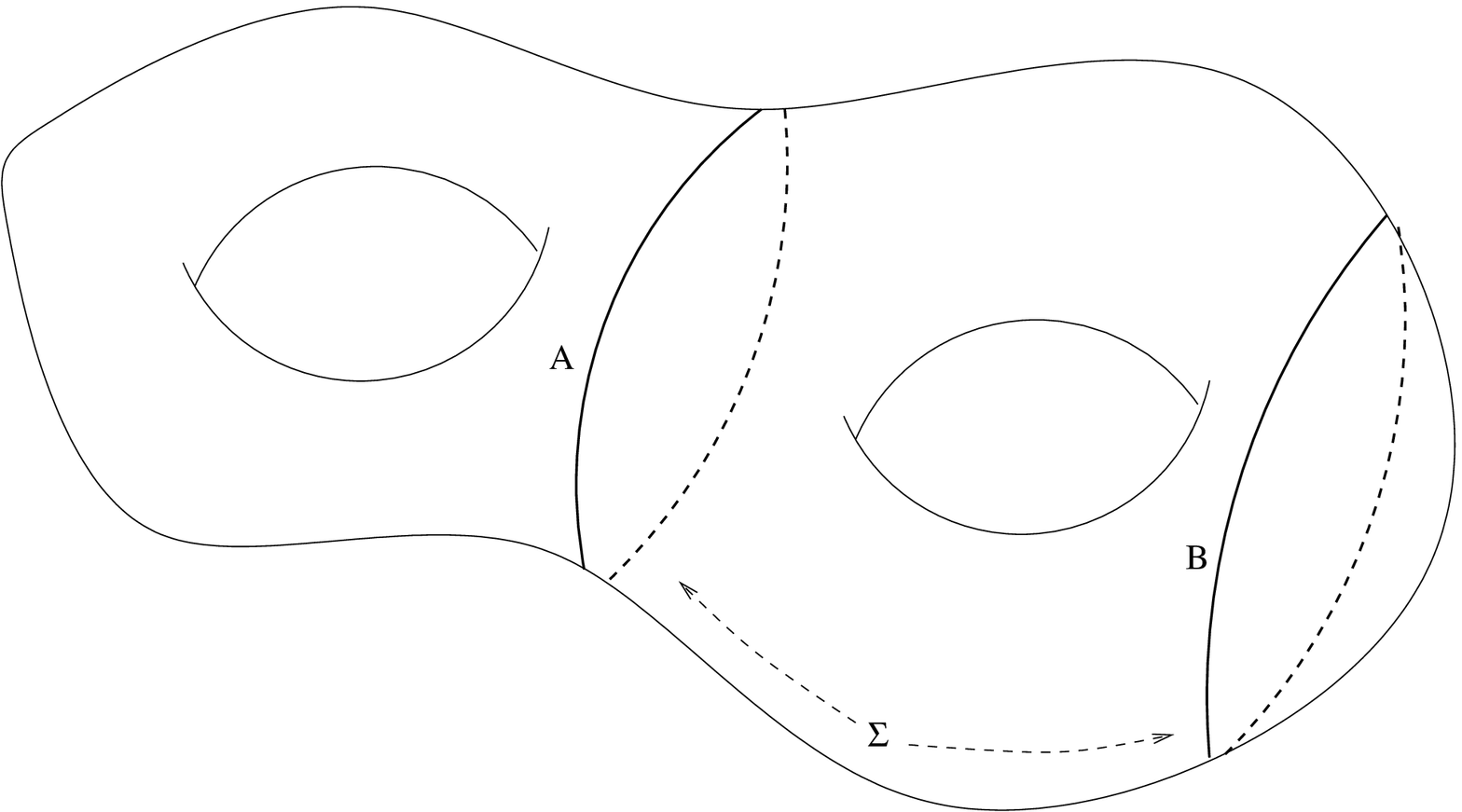}{The 1-cycles $A$ and $B$ are homologous
as they are the boundary of the 2-surface $\Sigma$, but not homotopic
as we must break $A$ to deform it to $B$. In particular $A$ and $B$ are trivial
in homology as $B$ is a boundary, but $A$ is not contractible.}{homotopy}

The reason why our intuition did not fail yet is that it is true
that if $A_n$ and $B_n$ are any two cycles of equal dimension $n$, then
\be \label{theo}
A_n \mbox{ homotopic to } B_n ~~\Rightarrow~~ A_n  \mbox{ homologous to }  B_n \,,
\ee
as can be understood by imagining $A$ moving continuously towards $B$
while tracing an $(n+1)$ dimensional submanifold with boundary $A \cup B$
(the theorem does not work in the inverse direction, though).
In particular, we can consider $B_n$ to be homologically trivial and
then the negation of \bref{theo} reads
\be \label{theo2}
A_n \mbox{ not trivial in homology}  ~~\Rightarrow~~ A_n  \mbox{ not trivial in homotopy}  \,.
\ee
So a brane that wraps a calibrated cycle (which is homologically non-trivial)
is also non-contractible as our intuition required. Nevertheless, as \bref{theo2}
is not a both-ways implication, a brane can wrap a non-contractible cycle
and be non-BPS, unstable and decay.

\section{The geometrical twisting} \label{ch5:sec:geometrical-twist}

We have introduced a good amount of mathematical machinery to end up saying
that branes must wrap calibrated cycles in special holonomy manifolds.
As the worldvolume of D-branes carry a gauge theory, we are saying
that it is possible to construct a supersymmetric gauge field theory
on a curved manifold just by letting the D-brane wrap a calibrated cycle.
However, we saw in section~\ref{ch5:sec:twisting-field-theories} that the
inherent difficulties of formulating
non-gravitational supersymmetric theories in curved spaces forced us to
make use of the twisting mechanism. This involved the introduction of
an external gauge field $A_{\mu}$ without a physical interpretation, which
was needed in order to make local (a part of) the $R$-symmetry.
The purpose of this section is to exploit the geometrical
understanding that D-branes provide and show that the twisting mechanism
is naturally realized by wrapping the D-branes in calibrated cycles.

Let us start with the general picture which will be illustrated below
with the example of M5-branes wrapping SLAG 3-cycles. Consider a Dp-brane
with worldvolume $\CR^{1,n-1} \times \S_{n'}$ embedded in IIA/IIB
background of the form $\CR^{1,d-1} \times X_{d'}$, with $X_{d'}$
a special holonomy manifold so that

\cent{
\begin{tabular}{c|c|c}
Worldvolume & Target space & Embedding \\
\hline
&&\\
$\CR^{1,n} \times \Sigma_{n'}$ & $\CR^{1,d-1} \times X_{d'}$ &
$\CR^{1,n} \subset \CR^{1,d-1} \sac\S_{n'}  \subset  X_{d'}$
\end{tabular}}

\noindent
with $p=n+n'$ and $10=d+d'$. As we know, the low energy theory of
the brane worldvolume contains a set of scalars that can be
interpreted as the fluctuations of the brane in the transverse space.
In this case, the transverse space consists of a piece with
$d-n-1~$ directions along $\CR^{1,d-1}$ and $d'-n'$ along $X_{d'}$.
The scalars referring to the former must remain massless
as they are the Goldstone bosons for the broken translational
invariance along the flat part. The scalars referring to the
latter require more work. It is useful first to decompose
the tangent bundle of $X_{d'}$ as
\be
\calt_{X_{d'}} = \calt_{\S_{n'}} + \caln_{\S} \,,
\ee
where $\calt_{\S_{n'}}$ is the tangent bundle to $\S_{n'}$
and $\caln_{\S}$ its normal bundle. The latter has dimension
$d'-n'$, which is the correct one to give the transverse
scalars inside $X_{d'}$ the character of sections of
the normal bundle.

\begin{figure}[here] \begin{center}
\includegraphics[width=5 cm,height=4 cm]{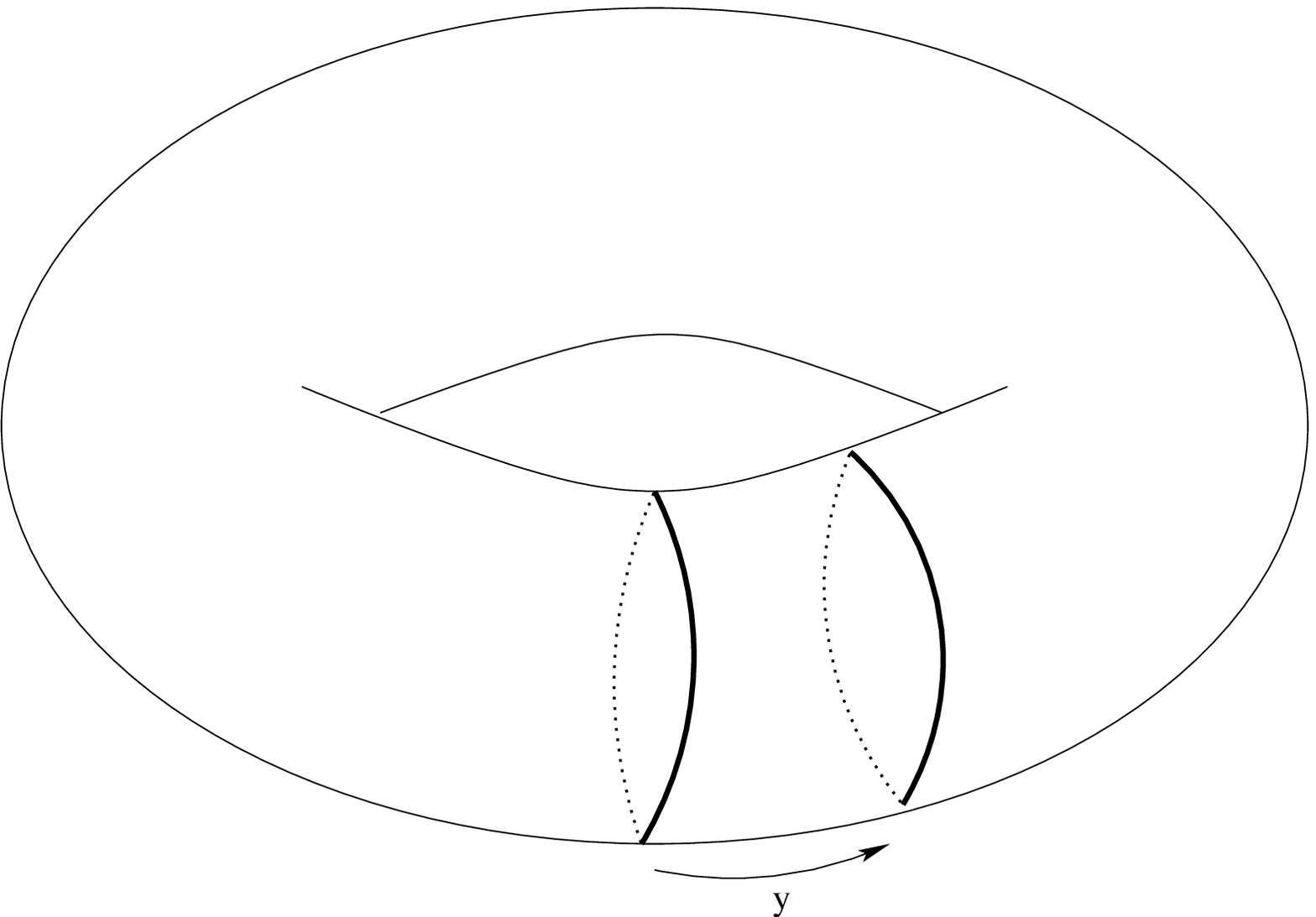} ~~~~
\includegraphics[width=5 cm,height=4 cm]{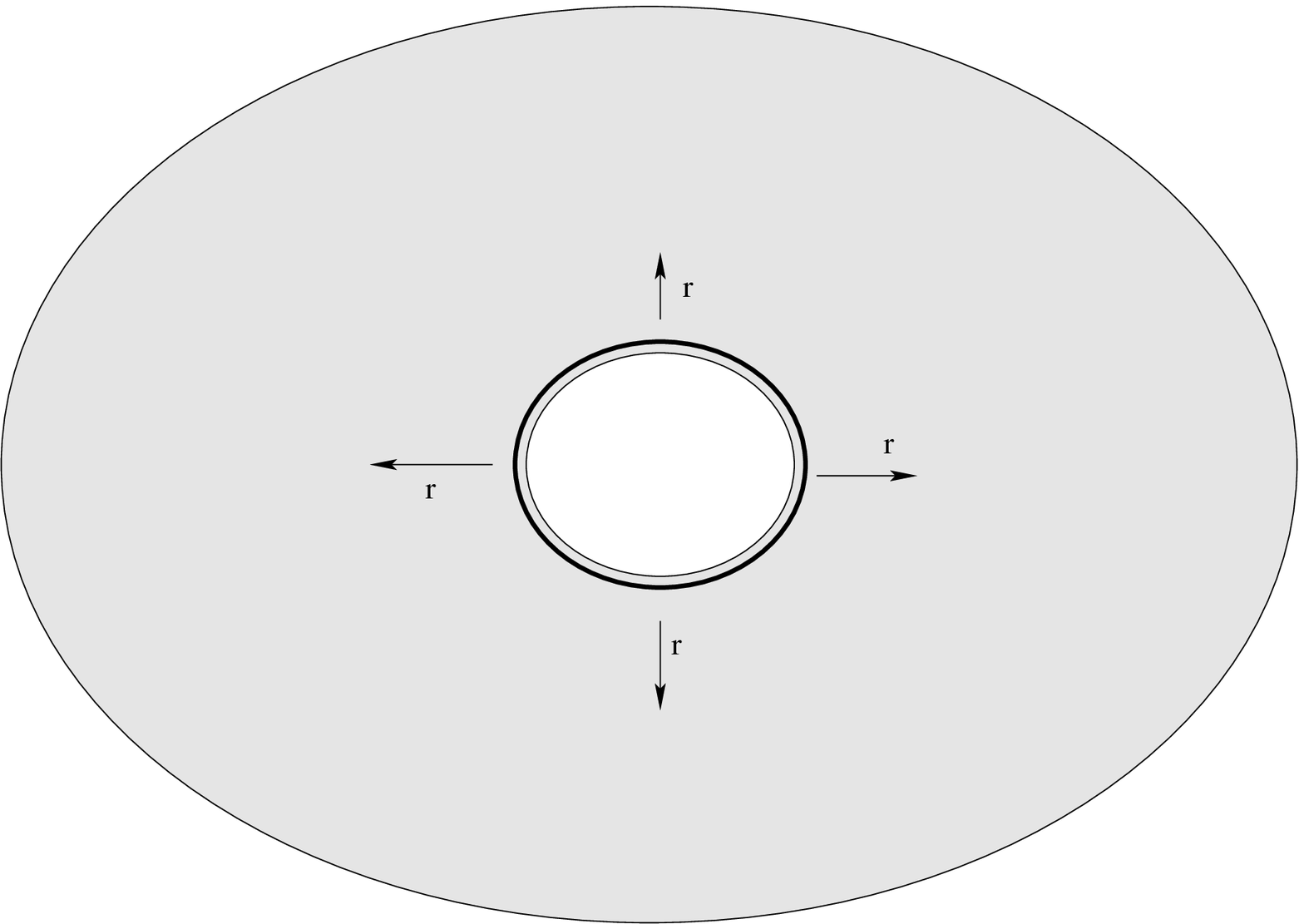}
\caption{The figure on the left shows a brane (thick line) wrapping a
non-rigid cycle of a torus. Perturbations described by the scalar $y$
are massless. On the right, a 2d cut of a manifold which has a hole
in the center. If a brane wraps it, there is no other neighboring supersymmetric
cycle and perturbations of the scalar $r$ are massive.\label{rigid}}
\end{center} \end{figure}

This interpretation allows, among other things, to answer the relevant question:
do we see any of these scalars in the low energy description of the D-brane?
The geometric answer is simple: a scalar will remain massless, and
thus will have to be included in the low energy action, if the supersymmetric
cycle is not rigid. By rigid, we mean that there are no other supersymmetric
cycles in its homology class which are continuously connected to it.
Not being rigid means that we can move the brane away from the cycle through
a set of supersymmetric cycles. Figures in~\ref{rigid} exemplify this concept.

In other words, to determine the number of scalars inside $X_{d'}$ that
will be present in the effective description, we need to count the
number of vector fields in $\caln_{\S_{n'}}$ that give deformations
along a family of supersymmetric cycles. This can be usually translated into
a purely topological  argument by mapping the normal bundle to the bundle
of forms of some type. Let us give two examples which may help to clarify
this issue.

\tem{
\item
Let $X_{d'}=CY_{d'/2}$ and
consider a SLAG $d'/2$-cycle on it, \ie a cycle
parametrized by the real calibration $\w_{d'/2}$ of \bref{slag}.
Let us characterize the normal bundle.
It is easily seen that the restriction of the \kah form $J$ to the cycle is zero
everywhere. Therefore, for any vector
field $V$ on the cycle, we can form a 1-form $J^ij V^j$ which is
normal to all vectors on $\S$, and therefore belongs to $\caln_{\S}$.
This shows that for SLAG cycles
$\calt_{\S} \cong \caln_{\S}$. On the other hand, it shows that one
can think of $\caln_{\S}$ as a frame bundle of 1-forms.
A result due to Mclean~\cite{Mclean} is that a deformation
is through a set of SLAG cycles if and only if the 1-form $J_{ij}V^j$
is harmonic. So we have a topological argument in which the first Betti number
of the cycle counts the number of scalars that remain massless.
\item Let $X_{d'}=CY_{d'/2}$ and consider \kah cycles on it,
which as we said are calibrated by $J$ or powers of it. An
interesting result is that, as the first Chern class of a
Calabi-Yau is zero, and this number is additive under direct
sum of bundles,
\be
c_1[\caln_{\S}] = -c_1[\calt_{\S}] \,. \label{chern}
\ee
We will later deal with a particular case in which the
cycle is co-dimension two (called a {\it divisor}).
For such cases it can be shown that the normal bundle
is isomorphic to a complex line bundle.
It is a standard result that complex line bundles are
classified by their first Chern class, so that \bref{chern}
completely specifies the normal bundle of divisor cycles.
The notation is $\calo(p)$ for a complex line bundle with
Chern class $p$.
}

These examples show that the normal bundle to the cycle
can be rather complicated depending on how
the cycle is embedded in the curved manifold. We now make contact
with the twisting procedure. We want to take the point of
view of an observer sitting in the Dp-brane. He would interpret
$X_{d'}$ not just a $d'$-dimensional manifold, but rather
as bundle over the cycle where he is living in, the fibers being
the normal vector space at each point. The connection on the
normal bundle would be seen by him as an external field. Furthermore,
if the observer was good at compactifying field theories, he could
check that his lower dimensional SYM action can be derived from
reducing 10d SYM down to his world, just as the SYM theories
on Dp-branes are obtained from reduction of 10d SYM.
He would then find that the surviving fields couple naturally
to the normal bundle connection in a precise way dictated by
how the cycle is embedded in $X_{d'}$. The action he would write
down would be equivalent to the twisted actions we discussed in
section~\ref{ch5:sec:twisting-field-theories}. For example, the fact that for SLAG cycles
we had $\calt_{\S} \cong \caln_{\S}$ is a way of rephrasing
that the external gauge connection $A_{\mu}$ is equal to the spin connection.

We would then wonder what the possible non-equivalent twits that he
can do correspond to in terms of branes wrapping cycles. The answer is
that each twist corresponds to the brane wrapping different cycles
in (possibly) different special holonomy manifolds. Let us reinterpret
the examples we gave in section~\ref{ch5:sec:twisting-field-theories},
which corresponded to
6d field theories in $\CR^{1,3}\times S^2$. We associate them to the
field theory on a $D5$-brane in a 10d IIB background of the
type $\CR^{1,3} \times X_6$ and wrapping a cycle of $X_6$. The
precise matching is given in the following table:

\cent{
\otaula{c|c|c}
Theory & Twisting & Geometry \\
\hline
&&\\
4d $\caln=2$ SYM & $U(1)_{spin}=U(1)_R$ & D5 brane wrapped on \\
 & $ \subset \left[\su2 \right]_D$ & $S^2 \subset CY_2\times \CR^2$ \\
&&\\
4d $\caln=1$ SYM & $U(1)_{spin}=U(1)_R$ & D5 brane wrapped on \\
 & $ \subset SU(2)_L \subset \su2 $ & $S^2 \subset CY_3$

\ctaula
}

The counting of preserved supersymmetries is even easier in this
picture. As we saw in section~\ref{ch5:sec:special-holonomy},
a $CY_2, CY_3$ destroys $1/2,1/4$ of
the 32 supersymmetries of Minkowsky vacuum, whereas the D5 typically
breaks 1/2 more. This makes $8,4$ remaining supersymmetries, which
from a 4d point of view is $\caln=2,1$.

The counting of scalars that survive is also easier. The transverse
directions to the $D5$ are all inside $X_6$. As the $S^2$ is known
to be rigid in a $CY$, no scalars survive when $X_6=CY_3$ and
only two survive when $X_6=CY_2 \times \CR^2$;
this is in agreement with the expected number of dimensions
of their corresponding moduli spaces.

Finally, if we put $N$ of these branes on top of each other,
the gauge group of the dual theory is expected to grow to $SU(N)$
and, at distances much larger than the $S^2$ (in the IR) the theory
becomes effectively 4d.

\subsection{A problem common to (almost all) supergravity solutions}\label{ch5:sec:problem}

In page \pageref{problema-1} we briefly discussed the impossibility
of studying less than maximally supersymmetric field theories
by adding deformations to the $\caln=4$ Lagrangian. Essentially,
supergravity was valid only in a region of the parameter space
in which the added degrees of freedom do not decouple before
the non-perturbative phase is reached.

Although of a qualitatively different nature, the same problem reappears
in the wrapped branes arena. The intuitive understanding is clear, although
one ultimately needs to check it in the final solution. The problem is
that one is looking for solutions describing branes wrapping minimal
cycles of the ambient space {\it in the limit in which these cycles
are very small}, as required in order that the field theory effectively
lives in the unwrapped part of the brane.
The supergravity requirement that the curvatures be small typically
leads to the opposite limit in which the radii of the non-contractible
spheres is large.

Again one hopes that some qualitative (and, with some luck, even
quantitative) features of the wanted field theory are still captured
by the supergravity approximation. Indeed, we will present some
supergravity results that fit really well with the field theory expectations
for some non-perturbative observables.

And finally, we remark that an exception to the rule will be provided by
the closed string background dual to a noncommutative $\caln=1$ SYM
in 3+1. We will see that the newly introduced NC scale $\t$ can
be fine-tuned in order to decouple the KK states. The reason why
this scale leads to such qualitatively different results from the
mass-deformations of the $\caln=4$ might be due to the fact that,
as extensively discussed in this thesis, a NC deformation is not
just a deformation through some finite set of operators but,
at best, by an infinite set of them. For example, a magnetic
NC deformation introduces spatial non-locality and it radically
changes the classical and specially quantum properties.

Unfortunately, the limit in which $\t$ can be used to decouple
the KK modes is a limit of very large $\t$, so that one is
left with a theory at least as unrealistic as the 16-supersymmetric
one!

\section{How to find supergravity solutions of wrapped branes} \label{ch5:sec:sugra-wrapped}

\subsection{Motivation} \label{ch5:sec:motivation}

What we have learnt in this chapter is how to deal with D/M-brane
proves with curved embeddings in curved manifolds. This is the open
string description that we discussed in chapter~\ref{ch:dbranes}.
We can place more and more probes on top of each other in a
supersymmetric way, as they all preserve the same type of
Killing spinors. As the number of probes $N$ increases the
backreaction cannot be neglected and we expect that
a closed string description arises, just like in the
case of flat branes in flat space.

Once again, the same remark of section~\ref{ch1:sec:closed-picture} applies here:
in the closed string description of branes in special holonomy manifolds
we will not see any special holonomy manifold and, sometimes, not even
branes. Maybe the simplest way to see it is that the backreaction
will always involve the supergravity gauge field-strength $F$ that couples
to the brane, and this will modify the background Killing spinor
equation. Schematically
\be \label{co-sp}
D_{\mu} \e \sim \left( \pa_{\mu} + \, \w_\mu \, + \, F_{\mu \a_1...} \G^{\a_1 ...} \right) \e =0 \,.
\ee
So it is not ordinary covariantly constant spinors what we need, but covariant
spinors in the sense of \bref{co-sp}. There has been recent progress in the
geometric understanding of these spinors and in the whole supergravity solution
of wrapped branes. The geometrical description is in terms $G$-structures, which
are the proper generalization to manifolds with background fluxes
of the concept of holonomy. This is however beyond the scope of this thesis.

If we were able to find the supergravity solution of one of these  wrapped branes,
we could try to take the near-horizon limit an expect to find an AdS/CFT-like
duality relating

\bea
\begin{array}{|ccccc|}
\hline
&&&&\\
&& & \mbox{IIA/IIB in the near horizon limit}& \nn
&\mbox{IR region of SYM with} & \leftrightarrow &
\mbox{of the SUGRA solution describing} &\nn
&\mbox{$\leq$16 susys in $\CR^{1,p}\times \S_{p'}$}& & \mbox{(p+p')-branes wrapping $\S_{p'}$ inside} &\nn
&& & \mbox{a special holonomy manifold} &\\
&&&&\\
\hline
\end{array}
\eea

We will see that these dualities are hard but possible to obtain. However, if
the aim is to end up with a field theory only in the $\CR^{1,p}$ factor,
we must be able to take the limit in which $vol(\S_{p'}) \go0$ on
both sides. This is what will not be possible to achieve, as we discussed in
section~\ref{ch5:sec:problem}, being at present the main drawback against
this line of research.

\subsection{Using gauged supergravities to find the solutions} \label{ch5:sec:gauged-sugras}

The purpose of finding the supergravity solutions is as hard as
finding exact solutions to general relativity. Despite the
fact that restricting to those solutions that preserve supersymmetry
turns most of the second order differential equations of motion
into first order,\footnote{Not all of them, as we described in
section~\ref{ch1:sec:supersymmetry}.}
the enterprize is still a hard one. The success in this respect during the last years
is partly due to the observation of Maldacena and \nun that one
may use gauged supergravities to find them.

Let us discuss why gauge supergravities work. First of all, when we focus on the
geometry of a brane wrapped on a compact cycle on a compact special holonomy manifold,
it turns out that its normal bundle looks like if it was non-compact. This is because
the limit in which the worldvolume theory of the $N$ probes is the twisted
SYM theory in $\CR^{1,p}\times \S_{p'}$ requires $l_s \go0$ keeping fixed
the volume of the cycle and of the special holonomy manifold. This means that
the theory on the worldvolume is a large volume compactification
in terms of the string length (see figure~\ref{wrappedbrane}).
If it had not been this way, the enterprize would have been simply impossible; for example,
we still lack one single metric for a compact $CY_2$! So if our aim is
just to describe the near horizon region, the answer must be in terms
of a metric for a noncompact space.

\clfigu{12}{5.5}{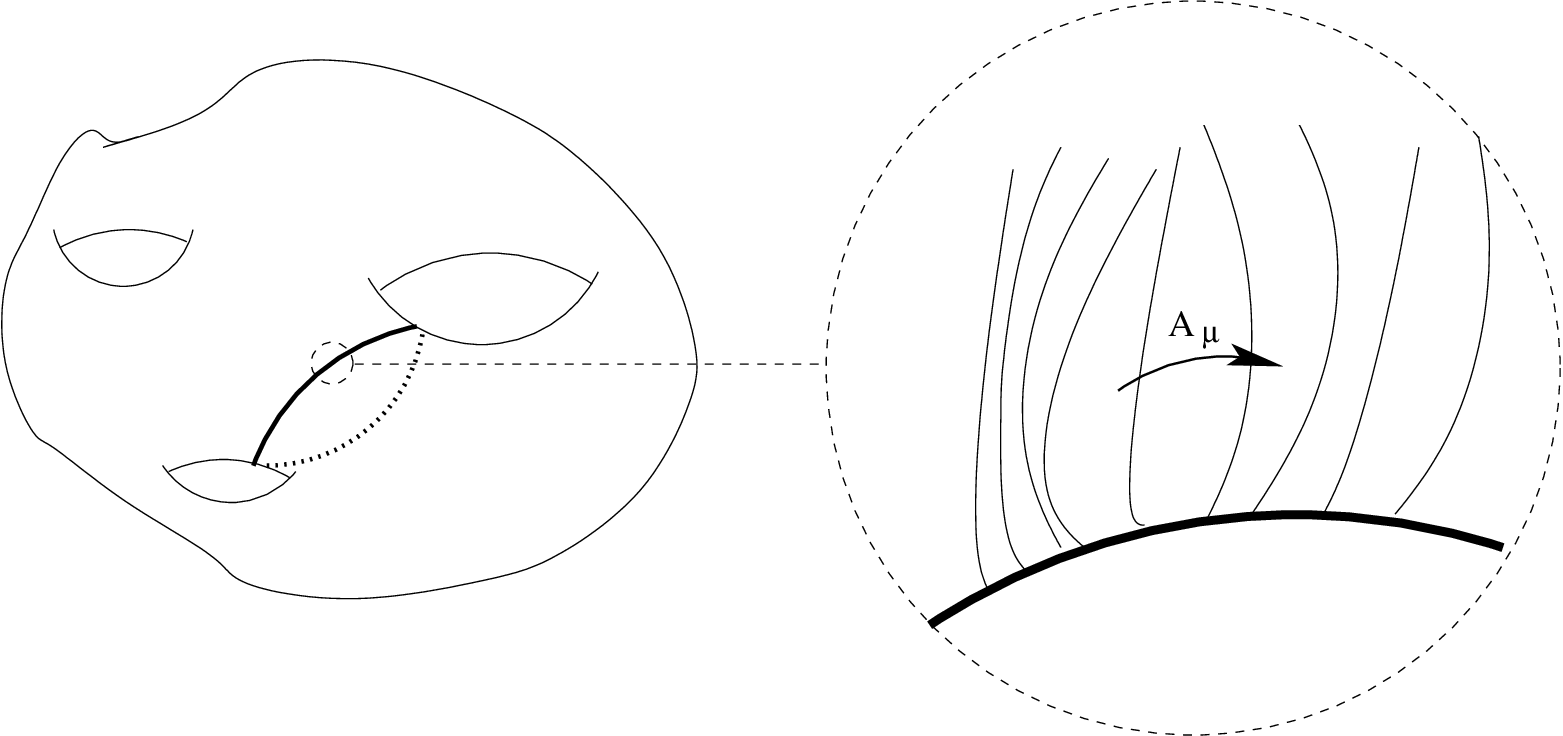}{On the left, a brane (thick line) is wrapping
a nontrivial compact cycle of a complicated compact special holonomy manifold.
By focusing on its worldvolume, normal bundle becomes a set of noncompact
fibers; the way they curved with respect to its neighbors is controlled
by the connection $A_{\mu}$.}{wrappedbrane}

Having said this, let us guess the kind of near horizon limit that we expect.
The boundary conditions on the metric must be such that it approaches
$\CR^{1,p}\times \S_{p'}$ at the boundary instead of $AdS$. Similarly,
the $R$-symmetry gauge fields must approach its field theory values
needed in order to perform the twist. Recall that if the transverse
space to the brane is $n$-dimensional, the $R$-symmetry
is the isometry group the transverse $S^{n-1}$.
It turns out that for all the cases that lead to twisted field
theories~\cite{Maldacena:2000mw},
the supergravity fields excited by their couplings to the gauge
theory belong to the multiplet which is massless upon compactification
of the 10d or 11d supergravity on $S^{n-1}$.
The conclusion is that it is much easier to give an ansatz in a supergravity
theory where all the modes that are massive upon such compactification
are truncated, and these are precisely the gauged supergravities. Indeed,
we need much less than a lower dimensional supergravity in which the
whole $SO(n)$ is gauged, because we know that the twist requires
only a part of the $R$-symmetry to be made local. In such a reduced
supergravity, we know exactly how to perform the ansatz because
the brane looks like a domain-wall there, and because the
understanding of the twist from a field theoretical point of view
tells us exactly which fields should be turned on.
It is possibly better to explain this by finding an explicit
solution, and this is the purpose of the next sections.

\section{Supergravity duals using D6 Branes} \label{ch5:sec:d6-solutions}

The purpose of this section is to find the near-horizon supergravity description
of D6-branes wrapping \kah four-cycles inside $CY_3$ manifolds, as reported
in~\cite{Gomis:2001vg}.
This should provide the AdS/CFT-like dual in the IR of
a gauge theory with $\caln=2$ in 3d.

\subsection{D6 branes and M-theory} \label{ch5:sec:d6-and-11d}

Before going to technicalities, it
is worth mentioning a remarkable property that D6-branes have. The point
is that their IIA supergravity solutions typically involve only the
metric, the dilaton and the $C_1$ RR-potential, which are all fields
that directly descend from the degrees of freedom of the 11d metric.
So, even if their IIA supergravity solutions do not provide metrics
for noncompact special holonomy manifolds, their uplift to 11d
does~\cite{Gomis:2001vk}.
We will repeatedly make use of this uplift, so it is good to keep
the ansatz for doing so in mind:
\bea
 \label{reduction}
 ds^2_{(11)}&=& e^{-{2\phi\over 3}} ds^2_{IIA}
+ e^{4\phi\over 3}\left( dx^{T}+C_{[1]}\right)^2, \\
A_{[3]}&=&- C_{[3]} + dx^{T} \wedge B_{[2]}, \eea
where $A_3$ is the 11d 3-form potential and $C_3$ the type IIA RR one.

We already used the simplest example when discussing supertubes:
the uplift of the flat D6 supergravity solution in flat space
(given by \bref{sugrasol1}-\bref{sugrasol3} with $p=6$). Using \bref{reduction} we obtain
\be
\label{flatcase}
ds^2_{(11)}= dx^2_{0,6}+H \left(dr^2+{r^2}
[d\t^2+\sin^2 \t d\phi^2]\right)+ R^2 H^{-1} \left( d\psi+\cos\t d\phi
\right)^2,
\ee
where $N$ is the number of D6-branes and we recall that
\be
H(r)=1+{R\over r}, \espai\espai R=g_s N \sqrt{\ap} \,.
\ee
This is a purely gravitational solution with metric $\CR^{1,6}\times CY_2$,
where the particular $CY_2$ is a Euclidean Taub-Nut space.
Its near horizon limit is the ALE space discussed in section~\ref{ch1:sec:ale}.

This means that the problem of finding sugra solutions of wrapped D6 branes
is doubly motivated. For a physicist, they provide non-perturbative data
of SYM theories with a low degree of supersymmetry; for a mathematician
they provide explicit metrics for special holonomy manifolds. The latter
is a point of special relevance for special holonomy manifolds
other than Calabi-Yau spaces, like $G_2$ and $Spin(7)$. This is because
for such manifolds we do not have the analogous of Yau's theorem which
states the existence and uniqueness of a Ricci-flat metric in each
\kah class. One has had to prove that several such metrics exist
by brutal force until now: just going and finding them. For instance,
there were only 3 examples of $G_2$-holonomy metrics until 2000
which were explicitly constructed in~\cite{Salamon,Gibbons:1990er}.
After the use of wrapped
D6 branes, many other explicit metrics
appeared~\cite{Brandhuber:2001yi,Cvetic:2001zx,Cvetic:2001ih}.

\subsection{Twisting to get $\caln=2$ in 2+1 dimensions} \label{ch5:sec:twist-gm}

Let us show that the low energy effective theory of $D6$-branes wrapping a general
\kah four-cycle inside a Calabi-Yau three-fold $CY_3$
is an $\caln=2$ SYM theory in 2+1 dimensions.
We just need to repeat the steps described in section~\ref{ch5:sec:twisting-field-theories}.

A configuration with a D6 in flat
space would have an $SO(1,6)\times SO(3)_R$ symmetry, the
last group corresponding to the transverse directions to
the worldvolume. The number of linearly realized supersymmetries would be 16.
Consider now that our target space is instead $\CR^{1,3}\times CY_3$,
and that we wrap the $D6$ in a \kah four-cycle inside the
$CY_3$ in such a way that its flat directions fill
an $\CR^{1,2}\subset \CR^{1,3}$, \ie

\cent{
\begin{tabular}{c|c|c}
Worldvolume & Target space & Embedding \\
\hline
&&\\
$\CR^{1,2} \times \Sigma_{4}$ & $\CR^{1,3} \times CY_6$ &
$\CR^{1,2} \subset \CR^{1,3} \sac \S_{4}  \subset  CY_6$
\end{tabular}}

The worldvolume symmetry is broken to
$SO(1,2)\times SO(4) \cong SO(1,2)\times SU(2)_1 \times
SU(2)_2$. Being a \kah four-cycle, its holonomy is only
$U(2)$, which we identify with $SU(2)_2\times U(1)_1$,
the latter being a subgroup of $SU(2)_1$.

On the other hand, the  $R$-symmetry
will be broken to a $U(1)_R \times \CR$, with $U(1)_R$
corresponding to the two normal directions to the $D6$ that are inside
the $CY_3$ and $\CR$ to the one which is in $\CR^{1,3}$. The latter
gives a massless scalar, as we can put the brane supersymmetrically
anywhere in that direction. We summarize the way the various fields
transform in the  original and final symmetry groups
in the following table. As always, we indicate the $U(1)$ charges in subscripts.

\begin{center}
\otaula {c||c|c|}
 & $SO(1,6)\times SO(3)_R$
          & $SO(1,2)\times \left[SU(2)_2\times U(1)_1\right]\times U(1)_R$
                        \\ \hline \hline
&&\\
Scalars & ({\bf 1},{\bf 3}) & ({\bf 1},{\bf 1})$_{(0,0)}\oplus$
({\bf 1},{\bf 1})$_{(0,1)}\oplus$
({\bf 1},{\bf 1})$_{(0,-1)}$ \\ &&\\ \hline

&&\\

Spinors & ({\bf 8},{\bf 2}) & ({\bf 2},{\bf 1})$_{(\undos,\undos)}\oplus$
({\bf 2},{\bf 1})$_{(-\undos,\undos)}\oplus$
({\bf 2},{\bf 1})$_{(\undos,-\undos)}$ \\
& &$\oplus$({\bf 2},{\bf 1})$_{(-\undos,-\undos)}$ $\oplus$
({\bf 2},{\bf 2})$_{(0,-\undos)}\oplus$
({\bf 2},{\bf 2})$_{(0,\undos)}$ \\
&&\\
 \hline
&&\\

Vectors & ({\bf 7},{\bf 1}) & ({\bf 3},{\bf 1})$_{(0,0)}\oplus$
({\bf 1},{\bf 2})$_{(\undos,0)}\oplus$
({\bf 1},{\bf 2})$_{(-\undos,0)}$   \\
&&\\
\hline
\ctaula

\end{center}

The twisting can now be understood as an identification of both
$U(1)$ groups, so that only those states neutral under $U(1)_D=
\left[ U(1)_1 \times U(1)_R\right]$ survive. These are those
irreps with opposite charge with respect to both $U(1)$'s in the
table. The resulting field content consists of two Weyl
fermions, one scalar and one vector, which is precisely the
field content of an ${\cal N}=2$ $D=3$ SUSY theory. Later, from a supergravity
point of view, we will see that these are the spinors naturally
selected from the requirement that our solutions be supersymmetric.

\subsection{BPS equations in D=8 gauged supergravity} \label{ch5:sec:bps-8d}

The aim of this section is to construct a supergravity
solution describing the
aforementioned $D6$-brane configurations making use
of the gauged supergravities, as described in section~\ref{ch5:sec:gauged-sugras}.
For this particular case, we need to work
with eight dimensional supergravity since this is the
theory that results from reducing type IIA on the $S^3$
transverse to the D6-branes. To be more explicit,
our framework will be maximal $D=8$ gauged supergravity,
obtained in~\cite{Salam:1985ft} by dimensional reduction of
$D=11$ on an $SU(2) \cong S^3$ manifold. We proceed to
very briefly mention their results and explain our notations.

Following the usual conventions, we will use the Greek alphabet
to denote curved indices and Latin ones to denote
flat ones. We split the D=11 indices in $(\mu,\alpha)$
or $(a,i)$, the first ones in the D=8 space while
the second ones in the $SU(2)\cong S^3$.
The bosonic field content consists of the usual metric $\met$
and dilaton $\Phi$, a number of forms that we will set to zero,
an $SU(2)$ gauge potential $\amu^i$, and five scalars
parametrizing the coset $SL(3,R)/SO(3)$ through the unimodular
matrix $L^i_{\a}$. Finally, the fermionic content consists
of a 32-components gaugino $\psi_{\mu}$ and a dilatino $\chi_i$.

We will need to make use of the susy transformations for
the fermions
\petit{
\bea \label{susy1}
\d\psi_{\rho}&=&D_{\rho}\e+{1\over 24}e^{\Phi}
F_{\mu\nu}^i\G_i\left(\G_{\rho}^{~\mu\nu}-
10 \d_\rho^{~\mu}\G^\nu\right)\e-
{g\over 288}e^{-\Phi}\e_{ijk}\G^{ijk}\G_{\rho}T\e \,\,\,
\\
\d\chi_i&=&\undos\left(P_{\mu ij}+{2\over 3}\d_{ij}
\pamu\Phi\right)\G^j\G^\mu\e-{1\over 4}e^{\Phi} F_{\mu\nu
i}\G^{\mu\nu}\e \nn
 && -{g\over 8}\left( T_{ij}-\undos
\d_{ij}T\right)\e^{jkl}\G_{kl}\e
\label{susy2}
\eea }
The definitions used in this formulae are \bea D_{\mu}\e &=&\left(\pamu+{1\over
4}w_{\mu}^{ab} \G_{ab} + {1\over 4}Q_{\mu ij}\G^{ij}\right)\e \,,
\label{key}
\\
P_{\mu ij}+Q_{\mu ij} &\equiv& L_i^\a\left(\d_\a^{~\b}
\pamu - g \e_{\a\b\g} A^{\g}_\mu\right)L_{\b j}\,,
\\
T^{ij}&=&L^i_\a L^j_\b \d^{\a\b}\,, \\ T&=&\d_{ij}T^{ij} \,.
\eea
Notice that $SU(2)$ indices are raised and lowered with $L_i^\g$,
\eg $\amu^{\g}=L_i^\g\amu^i$. Finally we choose the usual
$\g$-matrices representation given by \be \G^a=\g^a \otimes I \sac
\G^i=\g_9 \otimes \s^i \,, \ee where $\g^a$ are any representation
of the $D=8$ Clifford algebra, $\g_9=i\g^0\cdots \g^7$, and $\s^i$
are the usual $SU(2)$ Pauli matrices.

We now proceed to obtain the solution.
Since we look for purely bosonic SUSY backgrounds,
we must make sure that the susy transformation
of the fermions \bref{susy1}\bref{susy2} vanishes.
One of the ingredients that we put by hand is that
the background Killing spinor is required to satisfy
the same equation as the spinor in the twisted field theory.
In other words, we impose that the first term in \bref{susy1} vanishes by itself,
\ie $D_{\mu}\e=0$.
The first immediate condition that we get
is that the metric in the four cycle must
necessarily be Einstein~\cite{Gauntlett:2000ng}, so that
\be
R_{ab}=\L g_{ab} \espai \L=cte \label{einstein} \,.
\ee
Inspired by the
case in which the four-cycle is $CP_2$,
we take the  metric normalized in such a way
that \footnote{See next section for
a discussion about the case $\L<0$.} $\L=6$.
We then make the following domain-wall ansatz for the $D=8$ metric
\be
ds^2_{(8)}=e^{2f(r)}dx^2_{(1,2)}+
e^{2h(r)}ds^2_{\S_4}+dr^2 \,,
\ee
where $ds^2_{\S_4}$ is any Einstein metric on the
\kah 4-cycle  that we want to choose.

Now, guided by the identifications between the normal
bundle and the spin connection that we discussed
in the last section, we complete our ansatz by
switching on only one of the $SU(2)_R$ gauge fields, $\amu^3$, so that we
break $R$-symmetry to $U(1)_R$, and one of the
scalars in $L^i_\a$. This matrix can therefore
be brought to~\cite{Edelstein:2001pu}
\be
L^i_\a=diag(e^\l,e^\l,e^{-2\l}) \,.
\ee
Indeed, $\l$ parametrizes the Coulomb branch of the gauge theory,
as we discuss below.
We choose vielbeins for the
four-cycle such that the \kah structure takes the form
$J=e^0\wedge e^3+ e^1\wedge e^2$.
In this basis, $D_{\mu}\e=0$ further implies
the following identification
between the $R$-symmetry gauge field and the four-cycle
spin connection
\be
A^3=-{1\over 2g}w_{ab}J^{ab} \espai \Rightarrow \espai
F^3=dA^3=-{6\over g}\,J \,, \label{aas}
\ee
and the following projections on the
supersymmetry spinor \footnote{Every time we
write down a concrete index, we will underline it
 only if it is flat. Therefore, indices in
\bref{aas} are curved while those in (\ref{b},\ref{c})
are flat. Also, $\{0,1,2,3\}$ label coordinates
in the four-cycle.}
\bea
\g^{\ur}\e&=&\e \label{b} \,,
\\
\g^{\uu\ud}\e &=& \g^{\uz\ut} \e ~=~\G^{\uu\ud}\label{c}\e \,.
\eea
The projections that survive to these projections
form a 4d vector space, which means that we are breaking
1/8 of the 32 background supersymmetries as expected. Finally,
the remaining information that  we can extract
from our BPS equations is in the following set of coupled first-order
differential equations for the functions of our ansatz $f(r)$,
$h(r)$, for the dilaton $\Phi(r)$ and for the excited scalar $\l(r)$
\bea
3f' ~=~\Phi'&=&{g\over 8}e^{-\Phi}(e^{-4\l}+2e^{2\l})
-{6 \over g}e^{\Phi-2h-2\l} \label{bps1}\,,
\\
h'&=&{g\over 24}e^{-\Phi}(e^{-4\l}+2e^{2\l})
+{4 \over g}e^{\Phi-2h-2\l}\label{bps2}\,,
\\
\l'&=&{g\over 6}e^{-\Phi}(e^{-4\l}-e^{2\l})
+{4 \over g}e^{\Phi-2h-2\l}\label{bps3}\,.
\eea

\subsection{Solutions of the BPS equations} \label{ch5:sec:bps-solution}

For the case in which the scalar $\l$
is constant, we could obtain the following
exact solution
of the BPS equations (\ref{bps1},\ref{bps2},\ref{bps3})
\be
e^{2\Phi}={9 g^2 \over 2^{1\over 3} 128} \, r^2 \sac
e^{2f}=C \, r^{{2\over 3}} \sac e^{2h}={27\over 16}r^2
\sac e^{6\l}=2 \,.
\ee
There are two arbitrary integration constants. One of them is not
shown explicitly, since it just amounts to a shift in
the coordinate $r$. The other one is $C$, appearing
in the solution for $f(r)$.

Note that if we had taken a negative value for $\Lambda$
in \bref{einstein}, the only difference would have been
a change of sign in all last terms containing $1/g$.
This translates into a change of sign in the solution for
$\l$ to $e^{6\l}=-2$. Hence, there is no supersymmetric
solution for the cases $\Lambda<0$.

One can now lift this solution
to the original $D=11$ supergravity by using the ansatz \bref{reduction}.
After performing a suitable redefinition of the radial variable, we obtain
\be
ds^2_{(11)}=dx^2_{0,2}+2 dr^2 +
{1\over 4}r^2(d\theta^2+sin^2\theta d\phi^2)+
{3\over 2}r^2 ds^2_{\S_4} + \undos r^2 \sigma^2
\label{singular} \,,
\ee
where \footnote{These metrics were obtained in~\cite{Cvetic:2000db}
in a completely different approach. Here we follow their
notation.}
\be
\sigma = d\psi -\undos \cos\theta d\phi+ \tilde{A}_{[1]}
\label{sigma} \,.
\ee
Here we have defined
$\tilde{A}_{[1]}={g\over 2} A^3_{[1]}$, so that we have
$d\tilde{A}_{[1]}=3J$. The periodicities of the Euler
angles are $0\leq \t \leq \pi$, $0\leq \phi \leq 2\pi$,
whereas the periodicity of $\psi$ depends on which
particular four-cycle we choose, and we leave this issue
for the particular examples.

This M-theory solution has
the topology of $R^{1,2}\times CY_4$, the Calabi-Yau
four-fold being a $C^2/Z_n$ bundle over the \kah four-cycle
(again, $n$ depends on the particular four-cycle chosen).
Everything matches. As discussed in the introduction of
this section, the uplift to $M$-theory had to provide
an explicit metric for a special holonomy manifold.
Looking at the table of special holonomy manifolds (page \pageref{special-hol-table})
 we see that the only possibility
that preserves four supercharges is a $CY_4$.

Our metric describes a cone, with $r=cte$ hypersurfaces
being a $U(1)$ bundle over the base
$S^2\times \S_4$. The particular fibration will
depend again on the four-cycle chosen. As a good $CY_4$,
the eight-dimensional metric is \kah and Ricci-flat,
thus it automatically provides vacuum solution of the D=11 equations.

Note that the metric has a conical singularity at $r=0$,
where the fiber, the $S^2$ and the four-cycle collapse to
a point.
One can now try to resolve this singularity by
obtaining solutions in which at least one of the
factor spaces in the base of the cone remains finite
for $r\rightarrow 0$. This can be done here
by dropping the assumption that the scalar $\l$ is constant.
We could find a more general solution to the BPS equations,
which is best described by first changing the radial variable
from the old $r$ to $R$ by
\be
{dr\over dR}=\left({gR\over 4}\right)^{\undos} U^{-{5\over 12}}(R) \,,
\label{change}
\ee
where
\be
U(R)= {3 R^4 + 8 l^2 R^2 + 6 l^4 \over 6(R^2 +l^2)^2} \,.
\ee
There exists a whole family of solutions parametrized by the constant
$l$ given by
\bea \label{sol1}
e^{6\l(R)}&=&U^{-1}(R) \,, \\ e^{4f(R)}&=&{g^2\over 16} R^2 U^{1\over3}(R)
\,, \\
 e^{2\Phi(R)}&=&\left({g R\over 4}\right)^3 U^{\undos}(R) \,, \\
e^{2h(R)}&=&{3g\over 8} R U^{1\over6}(R) (R^2+l^2) \,.
\eea
Repeating the lifting process to M-theory, we obtain the
following 11d metric
\bea
ds^2_{11}&=&dx^2_{(1,2)} + ds^2_{(8)} \label{gib1} \,,
\\
ds^2_{(8)}&=&U^{-1}(R) dR^2 + {1\over 4}R^2(d\theta^2+sin^2\theta d\phi^2)+
{3\over 2}(R^2+l^2) ds^2_{\S_4} \nn &&+ \,\,U(R) R^2 \sigma^2 \label{probeta} \,.
\eea
Note that for $l=0$ this collapses to the original singular solution \bref{singular}.
On the other hand, for $l\neq 0$ the four-cycle has blown-up, and its
size remains finite at $R\rightarrow 0$, although the $S^2$ and
the $U(1)$ fiber still collapse.
Nevertheless, recall~\cite{Cvetic:2001ma} that
the condition for local regularity in this limit implies that
{\it at most one} of the factors in the base of the $U(1)$ fiber
can collapse. Our manifold is therefore locally
regular. Globally, it will depend on the four-cycle chosen,
as the following examples show.

\avall

\tem{
\item
{\bf Example I:} Consider the choice of
a $\S_4=CP_2$. This a \kah holomorphic cycle of codimension two
in a Calabi-Yau manifold, \ie a divisor. We described in
section~\ref{ch5:sec:geometrical-twist} how the normal bundle is for such cases and
found that it must form a  holomorphic line
bundle with opposite Chern class with respect to the cycle.
Given that in our conventions $c_1[CP_2]=3$
the normal bundle must be an $O(-3)$ bundle.

After this discussion about the global structure, we aim
to make explicit all the functions that were left unspecified
for being cycle-dependent.
We provide the $CP_2$ base with the standard
Fubini-Study unit metric, \ie
\be
ds^2_{CP_2}={1\over (1+\rho^2)^2}d\rho^2 +
{\rho^2\over (1+\rho^2)^2}\sigma_3^{~2}
+ {\rho^2\over 1+\rho^2}\sigma_1^{~2} +{\rho^2\over 1+\rho^2}\sigma_2^{~2}
\label{vielbeins} \,,
\ee
where $\s_i$ are the $SU(2)$ left-invariant one forms
normalized such that $d\s_i=\e_{ijk}\s_j\s_k$.
This metric is Einstein, with $R_{ab}=6\,g_{ab}$ as required
by our conventions.
When we plug this metric in our M-theory solution \bref{probeta},
we obtain that $\tilde{A}_{[1]}=-{3\over 2}\rho e_3$. We substitute
this in \bref{sigma} and, applying the arguments in~\cite{Cvetic:2001ma},
we see that the maximum range of the $U(1)$ fiber angle must
be restricted to $(\Delta \psi)_{max}=\pi$ instead of the
normal $2\pi$. We have a $CP_2$ bolt at the origin.
This is why the $U(1)$ fibers over $S^2$ do not describe an
$S^3$ (viewed as a Hopf fibration), but an $S^3/Z_2$.

\avall
\item
{\bf Example II:} We give now an example in which the
four-cycle is taken an $S^2\times S^2$. As the metric had to
be Einstein both spheres need to have the same radius.
Finally, in order to normalize them such that
$R_{ab}=6\,g_{ab}$, their radii must be $r^2=1/6$, so that
\be \label{s2xs2}
ds^2_{~S^2\times S^2}={1\over 6} (d\t_1^{~2}+sin^2\t_1 d\phi_1^{~2})
+ {1\over 6} (d\t_2^{~2}+sin^2\t_2 d\phi_2^{~2}) \,.
\ee
Now $\tilde{A}_{[1]}=\undos\left[ \cos\t_1 d\phi_1+cos\t_2 d\phi_2\right]$
so, unlike before,
this allows a standard range  $(\Delta \psi)_{max}=2\pi$. Hence, topologically,
the manifold is a regular $C^2$ bundle over $S^2 \times S^2$.
}

\section{Non-perturbative physics of $\caln = 2$ in 2+1 from its
supergravity dual}

There is a good amount of non-perturbative
qualitative (and sometimes quantitative) physics of gauge theories
that can be extracted from the string duals. In the next chapter
we will analyze issues like confinement or chiral symmetry breaking
in both ordinary and NC $\caln=1$ $SYM$ in four dimensions.
Here we will devote our attention to the above obtained supergravity dual
of its cousin $\caln=2$ in 3d, which corresponds to its reduction
on an $S^1$. We will just explore the data that this dual provides
about its moduli space. As discussed in sections~\ref{ch1:sec:multiple-D} and
\ref{ch1:sec:4n-sym}, this
can be studied by introducing a probe brane in the background
created by the others and computing its effective action.
As our solution has been found in 11d, the first task is to reduce it
back to type IIA and then put a probe.
However, in the paper where the solution \bref{gib1}-\bref{probeta} was originally
constructed~\cite{Gomis:2001vg} we did not have into account that a problem usually
referred to as {\it supersymmetry without supersymmetry} was going
to be relevant in our case. Let us summarize what this problem consists on.
We will then show how it affected our reduction and the incorrect
conclusions that were originally derived. We then perform the
correct reduction and discuss the non-perturbative moduli space
of the $\caln=2$ SYM in 2+1.

\subsection{Supersymmetry without supersymmetry} \label{ch5:sec:susy-without-susy}

The fact that the radius of the eleventh dimensions
is proportional to the IIA string coupling constant,
$R_{11} = g_s^{2/3} l_s$, has deep consequences
on the  physics  felt by observers in 10 or 11 dimensions.
It is well known that the compactification from 11d to 10d
is a consistent one for any 11d vacuum that can be
viewed as a $U(1)$ bundle over some base manifold $\calm$.
The point is that $M$-theory states, even those of its
massless supergravity sector, which are charged under $U(1)$
rotations will look massive from a IIA point of view, with
masses
\be
m_{IIA} \sim {1\over g_s^{1/3} R_{11}} \sim {1\over g_s l_s} \,,
\ee
where the extra factor $g_s^{1/3}$ appears when we measure distances
in terms of the string metric in 10d.
An example of such states are excitations of the 11d metric which are not $U(1)$ invariant;
they become IIA states which can be identified with D0-branes. These are
invisible if we just do perturbative string theory. However, from
an 11d point of view there is no such distinction between perturbative
and non-perturbative states, which means that a 10d observer
may feel like his supersymmetry multiplets are shorter as
a consequence of describing the world perturbatively.

This phenomenon was named 'supersymmetry without supersymmetry' in
\cite{Duff:1997qz} as supersymmetry is actually present but in a nonperturbative
way for observers in the compactified theory. They provided a very
nice example in which 11d supergravity was reduced on $\adss{4}{7}$
following two routes.
\tem{
\item When compactified directly from 11d to 4d, the massless sector
falls into $SO(8)$ $\caln=8$ supermultiplets, and it is described
by the corresponding $\caln=8$ supergravity in 4d.
\item As $S^7$ can be seen as a $U(1)$ bundle of $CP^3$, one can
first go to IIA and then reduce on $CP^3$. The point is that in
the first step, some states will disappear due to their nonperturbative
condition, the consequence being that the final 4d result will
appreciate less than $\caln=8$ supercharges. Depending on the
orientation of the $S^7$ it was shown that the 4d observer would
measure either $\caln=6$ or $\caln=0$.
}

A  more subtle point arises when one reduces from 11d
to 10d on a bosonic vacuum that is completely $U(1)$ invariant.
Of course, in the quantum theory one needs to consider the whole
tower of 11d excitations, some of which will correspond to D0-branes,
but here we want to concentrate on the vacuum only.
Being a bosonic background, it reduces down to another bosonic background;
being $U(1)$ invariant means that all its bosonic fields fit into the
reduction ansatz \bref{reduction} and no 'non-perturbative' states are
generated. But apart from the 11d supermultiplet,
the reduction ansatz involves
the background Killing spinors as well. The danger is that it
can happen that only the Killing spinors fail to be $U(1)$ invariant.
What are the consequences then? As all bosonic
fields do fit in the ansatz, the 10d configuration will perfectly
solve the IIA equations of motion. However, the configuration
will be more or less supersymmetric depending on how many
11d Killing spinors are singlets under the $U(1)$.
This problem is difficult to avoid, the safest way possibly
being the brute force explicit computation of the 11 Killing
spinors.

Needless to say, this problem extends to any reduction
of supersymmetric bosonic solutions of any theory. The
next subsections focus on the particular 11d-to-10d case,
and we will see many new examples in chapter~\ref{ch:nc-sugra} arising in
more general $11d$ compactifications.

\subsection{A non-supersymmetric compactification
and a zero-di\-men\-sio\-nal moduli space} \label{ch5:sec:zero-d}

We now go back to our study of the $\caln=2$ SYM in
2+1 via the supergravity solution. In order
to analyze its moduli space, we first need to
reduce the solution \bref{gib1}-\bref{probeta}
from 11d to IIA, and we will do it here along the simplest
possible $S^1$.
Since the metric \bref{probeta}
has a $U(1)$ isometry, with Killing vector $\partial_{\psi}$,
we choose that direction as the M-theory circle.
Using the KK ansatz \bref{reduction} we
obtain a bosonic type $IIA$ solution with
the following values for the metric, the dilaton and the
$RR$ one-form
\bea
ds^2_{IIA}&=&e^{2\Phi/3}\left[ dx^2_{1,2}+U^{-1}dr^2+{r^2\over 4} (d\t^2+
sin^2\t d\phi^2) + {3\over 2}(r^2+l^2)ds^2_{\S_4} \right] \,,
 \nn \label{tenmetric} \\
e^{4\Phi/3}&=&U(r) r^2  \,, \\
C_{[1]}&=&A_{[1]}-\undos \cos\t d\phi \,.
\label{dilaton}
\eea
Notice that the dilaton vanishes at $r\rightarrow 0$ and
diverges at infinity, which means that one expects a good
description with classical string theory only for small
values of $r$. Essentially, this problem comes from the
fact that our $U(1)$ fiber radius
in the eleven-dimensional metric already diverged. Obtaining
solutions with a finite circle at infinity would probably
require an analysis beyond gauged supergravity.
A different approach, based on imposing directly the
required symmetries in the whole D=11 supergravity,
enabled the authors of~\cite{Brandhuber:2001yi}
to construct such kind of solutions.

Our metric is clearly singular at $r\rightarrow 0$.
In order to apply the criteria for good/bad singularities
of~\cite{Maldacena:2000yy}, one needs to put the metric \bref{tenmetric}
in the Einstein frame, which just amounts to multiplying by
$e^{-{\Phi\over 2}}$. It can be seen that $g_{00}$ decreases (and
it is bounded) as we approach the singularity, which
means that excitations of fixed proper energy are seen with
lower and lower energy from an
observer at infinity as we approach the origin. Thus we
conclude that it is a {\it good} one, properly describing
the $IR$ behavior of the dual theory.

Let us now put a probe brane in
the background of the wrapped $D6$ that we have obtained.
We consider a probe wrapping the same cycle but at some
distance in the $\CR^{1,3}$ factor, so that one can think of it as
pulling one of the $D6$ apart from the others.
The effective action for such a probe in the case that $\S_4=CP_2$  is,
from \bref{DBI}-\bref{WZ},
\bea
S&=&-\mu_6\int_{R^{1,2}\times CP_2}d^7\xi \,\, e^{-\Phi}
\sqrt{-det[G+B_{[2]}+2\pi\a'F_{[2]}]} \nn
&&+ \mu_6 \int_{R^{1,2}\times CP_2}
[exp(2\pi\a'F+B)\wedge \oplus_n C_{[n]}] \label{dbi}
\eea

In our solution \bref{tenmetric}\bref{dilaton} we have
$B_{[2]}=0$ and only $C_{[1]}\neq 0$. In order to
pull back our fields we choose a static gauge, in which
we identify the worldvolume coordinates $\{\xi^i,\, i=0,...,6\}$
with the space time coordinates $\{x^0,x^1,x^2,\rho,\tth,
\tphi,\tpsi\}$, the first three parametrizing $R^{1,2}$, and
the other four the $CP_2$. We will look for the vacuum configuration
and so we will set to constant the three space time coordinates
normal to the brane $\{r,\t,\phi\}$. With these choices, our
formula \bref{dbi} becomes
\be
S=-\mu_6 \,\, Vol \left[ R^{1,2} \right] \,\, \int_{CP_2}
d\rho d\tth d\tphi d\tpsi \,\, {a^{3/2} \rho^3
(a+b\rho^2)^{1/2} \sin\tth \over
8 (1+\rho^2)^3} \label{vacuum} \,,
\ee
where $a$ and $b$ are the following functions of $r$
\be
a(r)={3\over 2}r U(r)^{\undos}(r^2+l^2) \sac
b(r)={9\over 4} r^3 U(r)^{3\over 2} \,.
\ee
Looking at the integrand, which is always positive,
we already see that its minimum is at $r=0$ where,
indeed, $S = 0$.

The dimension of the moduli space
can be determined by looking at the kinetic terms
arising from the DBI action when one allows for the
transverse coordinates $\{r,\t,\phi\}$ to depend
on the flat worldvolume ones $\{\xi^0,\xi^1,\xi^2\}$.
The exact expression one obtains is identical to that
in \bref{vacuum} but replacing
\be Vol \left[ R^{1,2} \right] \,\, \rightarrow \,\,
\int d\xi_1 d\xi_2 d\xi_3 \sqrt{det\left(  \d_{ij} +  \partial_i r \partial_jr+
{1\over 4} \partial_i \t \partial_j \t +
{1\over 4}sin^2\t \partial_i \phi \partial_j \phi\right)}
\ee
Here $\{\partial_i = {\partial \over \partial \xi^i},\, i=0,1,2\}$.
Clearly, evaluating this at the minimum $r=0$ still
makes the whole expression vanish. Hence,
In this approximation we find that the moduli space is zero-dimensional!

\medskip

\noindent
{\bf What went wrong?}

\noindent
We will discuss in detail below what is known about the moduli space
of the $\caln=2$ SYM theory and what our supergravity analysis
was expected to give. Even without that knowledge, there is simple
way to see that something is going wrong and to explain
why we got a zero-dimensional moduli space. Despite
the fact that we reduced along a $U(1)$ isometry, \ie all
our bosonic fields are $U(1)$ invariant, the Killing spinors
of the 11d solution are all charged under this $U(1)$.
Their explicit computation will be performed in section~\ref{ch6:sec:11d-solution-wrapped}
and their relevant expression is given in formula~\ref{bigspin}.
For our purposes here we just need to note that they all
depend on $\psi$, which means that they all acquire
masses in 10d.\footnote{The reader might complain
that not having checked the 11d supersymmetry variations yet
we should not claim that the 11d
solution \bref{gib1}-\bref{probeta} is supersymmetric.
However, this is not necessary as 11d supersymmetry
is guaranteed by the fact that the solution
is the uplift of a supersymmetric 8d one.}
So, as explained in the previous subsection, we should
expect this 10d solution to solve the IIA equations
of motion but to preserve no supersymmetry at all. This
is indeed the case as we explicitly checked.
To conclude this caveat, the lack of supersymmetry in
the background prevents one of the branes to be
pulled away from the others in a supersymmetric way;
in field-theoretical terms, the moduli space must be
zero-dimensional.
The conclusion is simply that {\it this IIA background
is not dual to the 2+1 $\caln=2$ SYM gauge theory.}

\subsection{A supersymmetric compactification and an
all-loops perturbative moduli space} \label{ch5:sec:two-d}

\subsubsection{The IIA solution} \label{ch5:sec:IIA-non-susy}

As the reduction along the $\psi$-circle destroyed all
supersymmetries, it is logical to wonder whether we can
find any other $U(1)$ isometry such that the Killing
spinors are invariant. Again we need to advance results that
will be properly obtained and discussed in section~\ref{ch6:sec:susy-without-susy}.
It turns out that the correct reduction must be performed
along the isometry generated by the Killing vector $\pa_{\phi}$
of the 11d metric  \bref{gib1}-\bref{probeta}. Note that
both $\psi$ and $\phi$ are angles on the $S^3$ transverse to
the 4-cycle inside the $CY_4$. As this $S^3$ is squashed, it
makes physical difference to distinguish between both angles.
For the sake of simplicity we will restrict here to the case
when the 4-cycle is an $S^2\times S^2$ with equal radii, so
that as in \bref{s2xs2},
\be
ds^2_{\S_4}={1\over 6} \left( d\t_1^2 +\sin^2\t_1 d\phi_1^2 +
d\t_2^2 +\sin^2\t_2 d\phi_2^2 \right) \,.
\ee

Reducing along $\pa_{\phi}$ we obtain
\bea
ds^2_{IIA}&=&\left( {4 g_s N \over r^2 f}\right) ^{-\undos}
 dx_{0,2}^2 + \left(g_s N \right)^{\undos} \left\{ {3\over 2}(r^2+l^2)ds^2_{S^2\times S^2}
\right. \nn
&& \espai \left. +U^{-1}dr^2+
{r^2 \over 4} [ d\t^2 + m B_{[1]}^2]\right\} \,,
\label{IIAgood1} \\
e^{4\Phi/3}&=&{g_s^{1\over 3}r^2 f \over 4N}\,,
\\
C_{[1]}&=&-N\, Uf^{-1}\cos\t\,B_{[1]} \,.
\label{IIAgood3}
\eea
We have defined various quantities in order to make
the above expressions more compact. These are
\bea \label{petercond}
f(r,\t)&=&\sin^2\t+U \cos^2\t \sac  m(r,\t)=(U^{-1}+\cot^2\t)^{-1}
\,, \nn
B_{[1]}&=&d\psi+\cos\t_1 d\phi_1+\cos\t_2 d\phi_2 \,.
\eea
Note that the metric has cohomogeneity two as it depends on
the two coordinates $r$ and $\t$. This is natural from the fact
that the background \bref{IIAgood1}-\bref{IIAgood3} describes
$N$ D6-branes wrapped on a \kah 4-cycle inside a $CY_3$; there
is one transverse coordinate to the D6 inside the $CY_3$ and
one which is in $\CR^{1,3}$.

Before entering into the analysis of the moduli space, let us conclude
by studying the range of validity of this solution. The supergravity approximation
is valid where the curvature and the dilaton remain small.
In this case, these restrictions imply
\be \label{validity}
{1\over (g_s N)^{1/6}}\, \ll \, r \ll \,{N^{1/4} \over g_s^{1/12}} \,\,\,.
\ee
Within this range, this is the type IIA background which is proposed to
be dual to the  $\caln=2$ $SU(N)$ SYM in 2+1, without any matter
multiplet.

\subsubsection{The  moduli space from supergravity} \label{ch5:sec:moduli-space}

In this section we repeat the steps that led us to a zero-dimensional
moduli space in the non-supersymmetric reduction considered
above.
We will  analyze the Coulomb branch of this theory by giving
a non-zero vacuum expectation value to the scalars in a $U(1)$ subgroup
of $SU(N)$. As is well known, this is easily implemented in the supergravity
side by pulling one of the $N$ D6-branes away from the others.
The $U(1)$ degrees of freedom on the probe brane can be effectively described
by the DBI action, where the rest of the branes are substituted by the
background that they create. 

If we want to break the gauge group without breaking supersymmetry, we
must make sure that no potential is generated.
So the first thing to look at is the vacuum configuration of the probe brane.
With this purpose, we take the static gauge where the first seven space-time
coordinates are identified with the worldvolume ones, and all the rest,
\ie $\{r,\t,\psi\}$, are taken to be constant. In this way, only the
potential is left in the DBI action. It is not possible to give a
closed analytic
expression for it but, numerically, it is easy to see that it vanishes
only at $\t=0$ and $\t=\pi$, independently of $r$ and $\psi$.

We therefore locate the probe brane at such values of $\t$ and look at
the low energy effective action for its massless degrees of freedom.
This is accomplished by allowing $\{r,\psi\}$ and the worldvolume field-strength
$F_{[2]}$ to slowly depend on the worldvolume coordinates, so that only
the terms quadratic in their derivatives are kept in the expansion of the DBI action.
Indeed, in the limit in which the four-cycle is taken to be small and
physics are $(2+1)$-dimensional, one can
simply consider excitations about the flat non-compact part of the worldvolume.
Both locus $\t=0,\pi$ give the same effective action:
\be
-S_{probe}=\int d^3x
\left[ a^2 N (g_sN)^{3/2}C^2(r) r^2 \,(\partial r)^2 \, \, +\, \,
{1\over g_s N^2}{1\over 4 C^2(r)} \, (\partial y)^2 \right] \,,
\ee
where $a^2=2 \pi^2\mu_6$, $C(r)={1\over 4}(r^2+l^2)$
and $y$ is the compact scalar of period $2\pi$ that
one obtains after dualizing the gauge field $V_{[1]}$.

The moduli space is therefore two-dimensional and, after gluing the two
locus at the origin, it turns out to have the topology of a cylinder.
The metric is just
\bea \label{sugra-metric}
ds^2_{moduli}
&=&a^2 N (g_sN)^{3/2} C^2(r)\,
 r^2 \,dr^2 \, \, +\, \, {1\over g_s N^2}{1\over 4 C^2(r)}  \, dy^2 \nn
&=&  d\rho^2+{4 a  \over a g_s N^2 l^4 + 16 (g_s N^3)^{1/4} \,\rho}\, dy^2
\eea
In the last step we redefined the radial coordinate
$\rho={a g^{3/ 4}N^{5/ 4}\over 16}\, r^2\left(r^2+2l^2\right)$
in order to put the metric in a more standard form. It is easy to prove
that this metric is \kah by explicitly constructing the \kah potential.
In order to do so, first define complex coordinates
\be
z=y+i\chi(r)\,,
\ee
with
\be
\chi(r):={a\over 48}N(g_sN)^{5/4}
\left(r^6+3l^2 r^4+3l^4 r^2\right) \,.
\ee
One can then show that
\be
ds^2_{{\rm moduli}}\,=\,2\, g_{z\oz}\, dz \otimes d\oz\,,
\ee
with $g_{z\oz}=\partial_z \partial_{\oz} \vp$ and
\be
\vp
 = {a^{2/3}\over 128}g_s^{1/2}N^{5/6}\left( a g_s^{3/4} N^{5/4}l^6
+{48 \over g_s^{1/2}N} \,\, {z-\oz \over 2i} \right)^{4\over 3}
={a^2 N (g_sN)^{3/2}\over 128}(r^2+l^2)^4 \,.
\ee
The fact that $\vp$ is a real function completes the proof that the metric
is \2ka (see~\cite{Gomis:2001aa,Gauntlett:2001ur} for similar results using different
branes).

\subsection{Comparison with the field theory results} \label{ch5:sec:comparison}

We shall now compare the results obtained using supergravity with the
ones that are known from the field theory.
The first immediate comment is that in the absence of matter multiplets,
instantons of non-abelian gauge theories with $\caln=2$ in 2+1 dimensions
develop a superpotential that completely lifts the Coulomb branch
\cite{Affleck:1982as}.
This is not in contradiction with our result because these contributions
are exponentially suppressed with $N$, so they are not expected to
be visible in the supergravity side.

On the other hand, $\caln=2$ supersymmetry implies that the moduli space
must be a 2d \kah manifold, in agreement with what we have seen from supergravity. Furthermore,
it must have the topology of a cylinder, with the compact direction
coming from the dualized scalar, and the non-compact one coming
from the vacuum expectation value of the other scalar in the multiplet.
The one loop corrected metric for an $SU(2)$ theory~\cite{deBoer:1997kr} is
\be \label{1loop}
ds^2={1\over 4}\left({{1\over e^2}-{2\over r}}\right) dr^2+
\left({{1\over e^2}-{2\over r}}\right)^{-1}dy^2 \,,
\ee
and it is valid for $r \gg 2e^2$. Asymptotically, it tends to
the classical prediction which, after generalizing to $SU(N)$,
is just a flat cylinder with metric
\be
ds^2={1\over 4 e^2} dr^2+{e^2 \over N} dy^2 \,.
\ee

\begin{figure}
\begin{center}
\includegraphics[width=6cm,height=4cm]{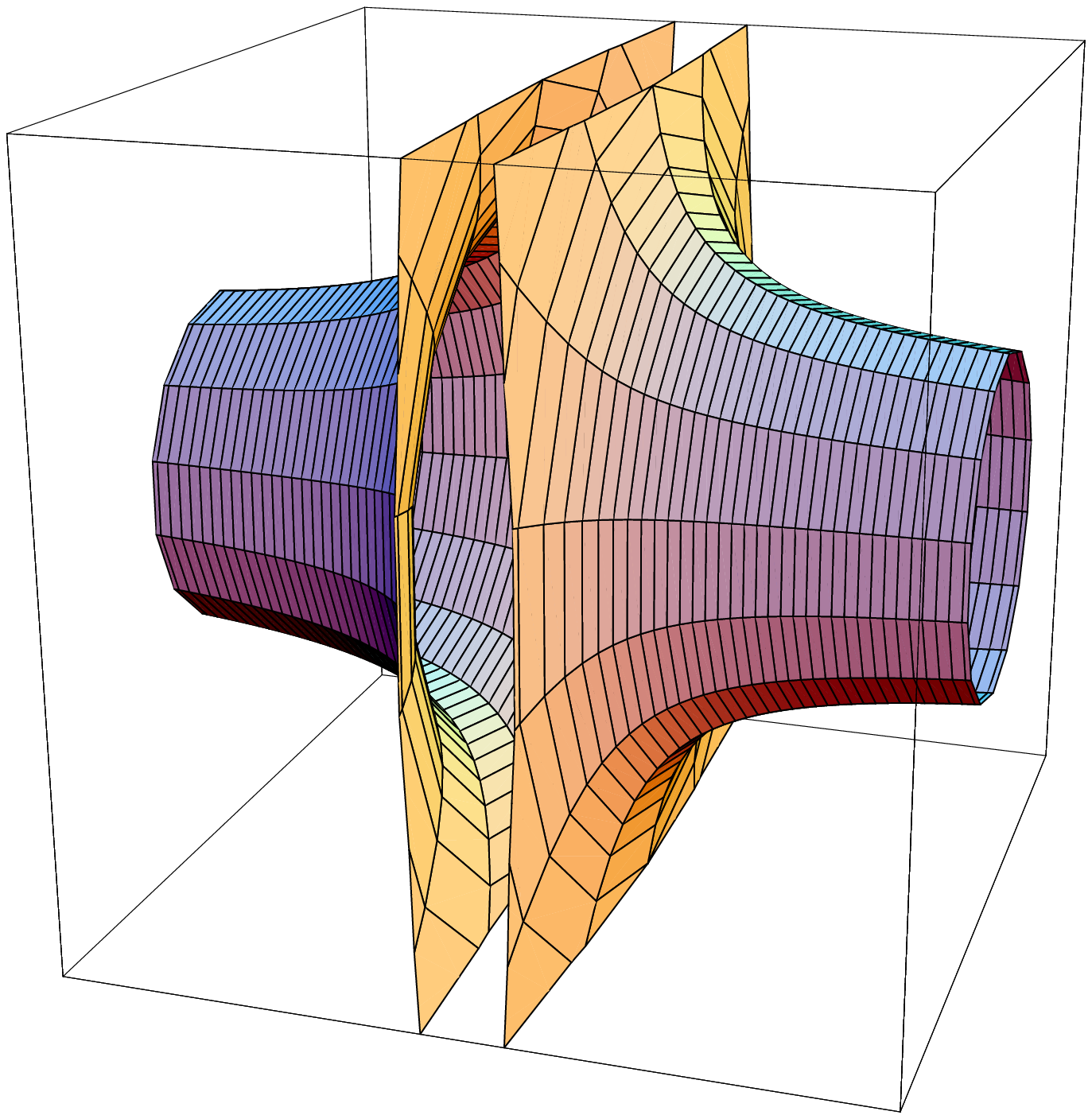} 
\includegraphics[width=6cm,height=4cm]{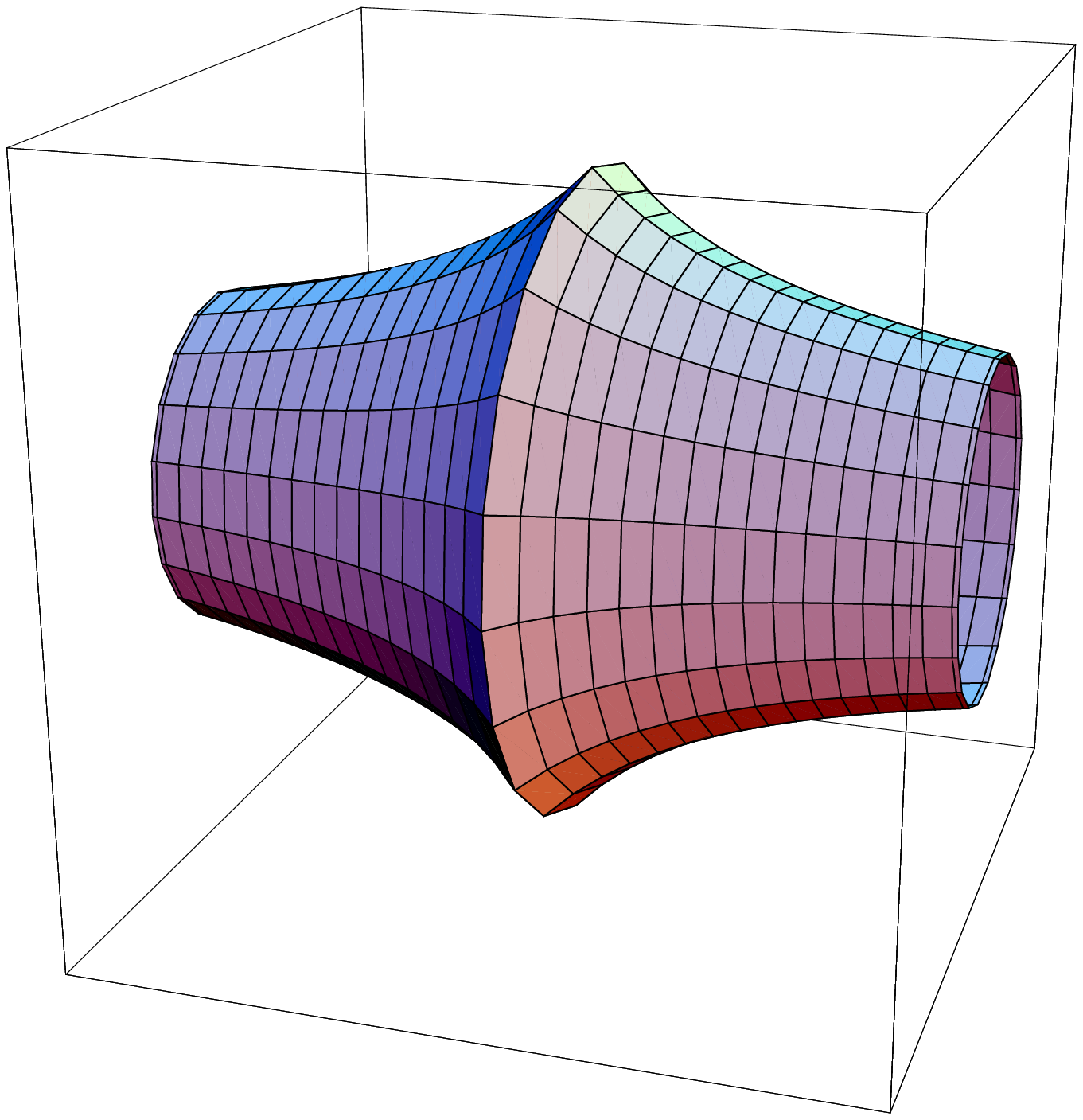}
\end{center}
\caption{Moduli spaces from 1-loop field theory (left) and supergravity (right).\label{radi}}
\end{figure}

In order to compare these metrics with our supergravity result \bref{sugra-metric}
we shall perform a change of variables in \bref{1loop} so that the metric is in
the standard form $ds^2=d\rho^2+f(\rho) dy^2$. Unfortunately, the change of
variables is not expressible in terms of elementary functions. Anyway, we can
solve numerically for $f(\rho)$ and plot the two moduli spaces, as we have
done in figure~\ref{radi}.
The plot on the left shows the one-loop corrected moduli space predicted by
field theory calculations. At very large values of the non-compact scalar,
it  tends to flat cylinder with radius proportional to $|e|/\sqrt{N}$.
As this $vev$ decreases higher loop corrections are needed. In particular,
the one loop calculation diverges at $r=2 e^2$.

On the other hand, the figure on the right shows the moduli space predicted
by supergravity. It also tends to a cylinder with vanishing radius at large
values of the non-compact scalar, so it qualitatively agrees with the $N \rightarrow \infty$
limit of the field theory. It also smooths the divergence of the one loop
calculation, which could maybe  correspond to a resummation
of infinite loops contributions. Strictly speaking, we see from \bref{validity}
that the supergravity approximation is not valid at $r=0$, where the curvature
of our background blows up. In any case, we can still use it as close to the
origin as needed by taking $g_s N$ large enough.

Finally, we shall make more explicit the
relation between the parameters in supergravity ($g_s$, $N$ and $l$)
and in the field theory ($e$ and $N$).
As usual, the number $N$ of D6-branes is the rank of the gauge group.
On the other hand, in the supergravity side,
a non-zero value for $l$ prevents the radius from diverging
as we approach the origin.
Nevertheless, it is difficult
to make the dictionary more precise. In any case, one can read the
gauge coupling for the $U(1)$ degrees of freedom at a certain
point of the moduli space by identifying the coefficient in front
of the $F^2$ term in the DBI action of probe. The result is
\be
{1 \over g_{U(1)}^2}=4 \pi^2 \mu_6 \, g_s N^2 \, (r^2+l^2) \,.
\ee

\chapter{From D-branes to NC Field Theories}\label{ch:nc-theories}

This chapter is devoted to the study of noncommutative field theories.
We review the original motivation that led physicists to consider them,
which dates back to more than 50 years. We take some time to review
the Landau problem, and show that this is essentially the way that NC theories
ultimately arise in String Theory. We then analyze the most remarkable
classical and quantum properties that these theories have, paying
special attention to the Seiberg-Witten map, the UV/IR mixing and
the lack of unitarity in the electric cases.
This last issue is most clearly discussed in section~\ref{ch3:sec:unitarity-non-rel},
which contains some detailed one-loop computations that
were reported in~\cite{Mateos:2001tj}.

We leave for chapter~\ref{ch:hamiltonian} the Hamiltonian analysis
of non-local field theories and its application to the
electric NC theories. The construction of string
duals of magnetic NC theories and the non-perturbative
physics than can be extracted from them is left for
chapter~\ref{ch:nc-sugra}.

\section{The interest of NC field theories {\it per s\'e} \label{ch3:sec:interest}}

Despite the recent interest for NC theories originated by their
appearance in string theory, the idea of extending the canonical
position-momentum noncommutativity to the spacetime variables,
represented by Hermitian operators on a Hilbert space such that
\be \label{nc-algebra}
[x^{\mu},x^{\nu}]=i \theta^{\mu\nu}(x) \,,
\ee
is quite an old one. Apparently the first proposal along these
lines is due to Heisenberg and dates back to the late 30s. He
was hoping that such a modification could help curing the UV
divergences typical of field theories. He probably mentioned this
to Peierles who used it in a phenomenological
approach to the study of electrons in external fields~\cite{Peierls}.
Then Peierles told about it to Pauli, who told Oppenheimer, who told
to his student Snyder, who published the first paper with a systematic
analysis on the subject~\cite{Snyder:1947qz}.

Commutation relations like \bref{nc-algebra} can be thought of an
effective approach to whatever theory of quantum gravity rules
Nature at the highest energies. For the simplest case of constant $\theta^{\mu\nu}$,
this matrix plays a universal  role analog to $\hbar$ for the spacetime
coordinates and, in particular, it leads to uncertainty relations of the style
\be \label{uncertainty}
\Delta x^{\mu} \Delta x^{\nu} \ge | \,\theta^{\mu\nu} | \,.
\ee
It therefore predicts a minimal spacetime area that can be possibly
probed, its size being of order $\theta$. This strongly reminds
the properties of some other aimed-to-be fundamental theories like:
\bitem
\item
{\bf String theory.} It is well-known
that the effective Heisenberg principle that is relevant for strings takes the form
\be
\Delta x \ge {\hbar \over 2} \left( {1\over \Delta p}+l_s^2  \Delta p \right) \,.
\ee
This implies, by minimization of the RHS with respect to the momentum, that the
minimal spacetime length that can be measured is $\Delta x \sim l_s$.
\item {\bf $\kappa$-Minkowsky spacetimes.} This is an approach to Quantum Gravity
in which two fundamental invariant scales are introduced. The first one is the usual
speed of light $c$, which leads to the standard \Poin algebra. The second one is the
Planck length $l_P$ which, as mentioned, is promoted to a length that is invariant
for all inertial observers. This clearly does not affect the translational
sector of the \Poin subalgebra, but it completely modifies the boost sector, since
the Lorentz contraction effect would spoil the invariance of $l_P$. Altogether,
it requires a modification of the commutation relations among the spacetime coordinates
to what is known as the algebra of $\kappa$-Minkowsky spacetime
\be
[x^i,x^0]=i \, l_P \, x^i \sac [x^i,x^j]= 0 \,.
\ee
\eitem

Any theory of quantum gravity has to propose how the spacetime notion must be
modified at the shortest scales, and it seems hard to make compatible the
principles of Quantum Mechanics with our familiar description of spacetime
via differential manifolds. It is however not clear whether a more proper
description will arise in terms of some completely new mathematical construction,
or whether the concept of a manifold will survive, but with modification
of the standard commutation rules. In any case, the more humble aim of
studying noncommutative spacetimes as toy models for an effective description
seems reasonable and, as we will see,
even the simplest infinitesimal deviation of $\theta$ from zero radically
changes the quantum properties of the theory. This in turn will force us to deal
with new unexpected physics.

Before finishing this section, we would like to review a familiar problem in
standard physics in which space noncommutativity appears: the Landau problem.
Apart from showing that noncommutativity is not a bizarre property
of sophisticated limits in string theory, it will set up an intuitive understanding
that will be needed in the following sections.

\subsection{The Landau Problem}  \label{ch3:sec:Landau}

The Landau problem is possibly one of the simplest setups to describe the
fundamental notions of noncommutativity. Consider the motion of $N$ interacting non-relativistic
electrons in a plane with a constant transverse magnetic field. Denote their positions
and velocities by
\be
\vec{r}_a=(x_a,y_a) \sac \vec{v}_a=\dot{\vec{r}}_a \sac a=1,...,N \,.
\ee
Pick a gauge where the vector potential is just $\vec{A}(\rvec_a)=(0,B x_a)$,
with $\vec{B}=\nabla \times \Avec=B \hat{z}$. The Lagrangian for the system is
\be \label{lag-landau}
L=\sum_a \left(\undos m_e \vvec_a^2+{e\over c} \vvec_a \dot \Avec(\rvec_a) - V(\rvec_a)\right)
- \sum_{a<b}U(\rvec_{a}-\rvec_{b}) \,.
\ee
We have included a term $V$ accounting for the possible interaction of the electron with
the medium, and a sum of $U$-terms accounting for the pair-interaction among the electrons.

Let us perform the canonical quantization of the system. The canonical momenta are
\be \label{cano}
\pvec_a=m_e \vvec_a +{e\over c} \Avec(\rvec_a) \,,
\ee
and they satisfy the usual commutation relations
\be
[x_a,p^x_b]=i\hbar \delta_{ab} = [y_a,p^y_b] \sac
[x_a,y_b]=0 = [p^x_a,p^y_b] \,.
\ee
Recall that the canonical momenta are not gauge invariant quantities, as it
is obvious from \bref{cano}. The gauge invariant, and therefore physical,
quantities are $\pivec_a=m_e \vvec_a$, and these momenta satisfy a noncommutative
algebra
\be
[\pi^x_a,\pi^y_b]=i\hbar {e B \over c} \delta_{ab} \,.
\ee
The Hamiltonian for the system is
\be \label{ham-landau}
H=L=\sum_a \left({\pivec_a^2 \over 2 m_e}+ V(\rvec_a)\right)
+ \sum_{a<b}U(\rvec_{a}-\rvec_{b}) \,.
\ee
For $U=V=0$, we can write the physical momenta in terms of annihilation/creation operators
and one finds that the energy levels are
\be
E=\sum_a \hbar \w_c (n_a + \undos) \sac n_a=0,1,2,...
\ee
where
\be
\w_c={e B\over m_e c} \,,
\ee
is the cyclotronic frequency.
This is the famous expression for the {\it Landau levels}.
Note for future reference that the energy gap $\Delta$ between the ground state and the first
excited ones is $\Delta=\hbar \w_c/2$.

\subsection{Projecting to the first Landau level} \label{ch3:sec:landau-project}

We would now like to take a limit in the parameter space of the Landau
problem that will reappear (disguised) later when we deal with
string theory. The limit can be thought either as a limit of {\it strong
magnetic field} or of {\it small electron mass}. The dimensionless
quantity controlling it is the quotient $B/m_e$ which we take
to be very large. In this limit, the Lagrangian \bref{lag-landau}
simplifies to
\be
L \approx \sum_a \left({e B \over c} x_a \dot{y}_a - V(\rvec_a)\right)
- \sum_{a<b}U(\rvec_{a}-\rvec_{b}) \,.
\ee
Note that we are dealing now with a first order Lagrangian,
which is therefore singular. This means that we will get primary
constraints that will not allow us to isolate all velocities in terms
of the momenta. For example, the canonically computed momenta are now
\be
p^x=0 \sac p^y={e B\over c} x \,,
\ee
which do not involve the velocities and should therefore be treated
as constraints in the Hamiltonian formalism.

One can now run the
whole machinery of Dirac formalism for quantization
in the presence of constraints. First, since the Poisson bracket
of the constraints is non-zero they are second-class constraints.
This allows one to solve them once and for all if one replaces
all Poisson brackets by Dirac brackets, which are the ones
that are finally promoted to commutators in the quantized
theory.
Having done this, one finds that the (reduced) phase space of the theory
is two-dimensional; both momenta have been solved, and one is left
with the following canonical structure for the remaining coordinates
\be
[x_a,y_b]=i {\hbar c \over e B}  \delta_{ab}\,,
\ee
which can be written in the standard form
\be
[x^i_a,x^j_b]=i\theta^{ij} \delta_{ab} \sac \theta^{ij}={\hbar c \over e B} \e^{ij} \,.
\ee

Can we understand what is going on in the limit just taken?
The first point to notice is that in this limit the energy gap $\Delta$ is
scaled to infinity, so that the ground state is decoupled from the rest
of excited states. This limit is such that the phase space is reduced, as
we have actually seen; indeed, the constraints on the momenta project
the allowed states of the system to its ground state. Moreover, the theory
becomes topological in the sense that the Hamiltonian \bref{ham-landau}
reduces to
\be \label{final-h}
H_0 \approx \sum_a  V(\rvec_a)
+ \sum_{a<b}U(\rvec_{a}-\rvec_{b}) \,,
\ee
and it vanishes in the absence of potentials, showing that there are no
propagating degrees of freedom.

\subsection{Weyl Quantization}  \label{ch3:sec:Weyl}

Having established the first simple example of an algebra of quantum operators
in which coordinates do not commute, it is useful to switch back to
the classical mechanics phase space and look for an alternative formalism
that takes noncommutativity into account. The method known as Weyl quantization
will do the work for us.  The general idea is to define a map from
the algebra of quantum operators on a Hilbert space to the algebra of
{\it functions} on the classical phase space.
In what follows, in order to make clear the distinction
between classical functions and operators, we will use the standard convention of putting hats
over the latter.

The  first ingredient we need
a proposal for a map between such functions and operators. Let us
consider the so-called Weyl map defined by
\be \label{Weyl-map}
W: F(x,p) \longrightarrow \hO_F=W[F] := {1\over 2 \pi}\int d\a d\b
f(\a,\b) e^{i(\a \hx+\b \hp)} \,,
\ee
where $f(\a,\b)$ is the Fourier transform of $F(x,p)$.
The second ingredient needed is a set of commutation relations among the
quantum coordinate and momentum operators. This is necessary in order to define
the product of two functions of the quantum phase-space operators, as
can be seen by computing
\be \label{product}
W[F] W[G] =
{1\over (2 \pi)^2}\int d\a d\b \int d\a' d\b'
 f(\a,\b) g(\a',\b') e^{i(\a \hx+\b \hp)} e^{i(\a' \hx+\b' \hp)} \,.
\ee
We need to know how to evaluate the product of the exponentials in
order to find whether the RHS is the $W$-image of some function $H(p,q)$.
What is clear is that unless we choose all commutators among coordinate
and momenta to be zero, it will follow that $W[F]W[G]\neq W[FG]$ and
therefore the Weyl map will not preserve the usual product of functions
on phase space. Let us show two of the main examples:

\bitem
\item If we take the usual quantum mechanics relations
\be \label{cano-com}
[\hx,\hx]=0 \sac [\hp,\hp]=0 \sac [\hx,\hp]=i\hbar \,,
\ee
then we can use in \bref{product} the Baker-Campbell-Hausdorff (BCH) expansion
\be
e^A e^B=exp\left(A+B+\undos [A,B]+{1\over 12}[[A,B],B]+{1\over 12}
[[B,A],A]+...\right) \,,
\ee
and obtain
\bea
W[F] W[G] &= &
{1\over (2 \pi)^2}\int d\a d \b \int d\a' d \b'
 f(\a',\b') g(\a'-\a,\b'-\b) \times \nn
 & \times &  e^{i(\a \hx+\b \hp)} e^{{i\over \hbar}(\a\b'-\ap \b)}
= {1\over 2 \pi}\int d\a d \b \, a(\a,\b) \, e^{i(\a \hx+\b \hp)}
\,, \nonumber
\eea
where
\be \label{int1}
a(\a,\b)={1\over 2 \pi}\int d\a' d \b' f(\a',\b')g(\a'-\a,\b'-\b)
e^{{i\over \hbar}(\a\b'-\ap \b)}
\,.
\ee
This immediately forces us to define a new product $*$ for functions
in the classical phase space by requiring that
\be
(F * G)(x,p)=W^{-1} \left(W(F)W(G) \right)={1\over
2 \pi} \int d\a d \b \, a(\a,\b) e^{i(\a x+\b p)} \,.
\ee
So all we need to know is to inverse-Fourier transform of \bref{int1},
which is easily verified to give the following differential expression
\be
(F * G)(x,p) = F(x,p) \exp{{i\over 2} \hbar \left( \overleftarrow{\partial_x}
\overrightarrow{\partial_p} -\overleftarrow{\partial_p} \overrightarrow{\partial_x}\right)}
G(x,p) \,.
\ee
It can be verified as well that this product is associative but noncommutative.
As a final check note that, as expected,
\be
[x,p]_* \equiv x*p-p*x = i\hbar \,.
\ee

\item
We now wish to consider how the Weyl map works for
the commutation relations that we obtained in the
Landau problem (and that we will also obtain from the string theory low energy
dynamics). This will lead us to the concept of {\it classical} noncommutative spaces
over which fields will be defined. We therefore require a deformation of
\bref{cano-com} that includes a spacetime noncommutativity with constant $\t$:
\be \label{nc-rel}
[\hx^{\mu},\hx^{\nu}]=i\t^{\mu\nu} \sac [\hp^{\mu},\hp^{\nu}] =0 \sac
[\hx^{\mu},\hp^{\nu}]=i \hbar \, \eta^{\mu\nu} \,.
\ee
The strategy here is to first define the concept of {\it noncommutative spacetime}
by Weyl-mapping only the coordinate sector, \ie not including the momenta.
One then has the capability of defining functions or fields in such spacetimes
and, in particular, one can write down actions for such fields by means
of whatever noncommutative product we obtain. At the end, if wanted, one
can proceed by the usual quantization of such classical theories which,
after all, takes into account the usual coordinate-momentum noncommutativity.

So, we set $\beta=0$ in the definition of the Weyl map \bref{Weyl-map}
\be 
W: F(x) \longrightarrow W_{\t}[F] := {1\over (2 \pi)^4}\int d^4\a f(\a)
e^{i\a_{\mu} \hx^{\mu}} \,,
\ee
and we compute the product of
operators using the first commutator in \bref{nc-rel}. The steps are
analog, and indeed simpler, to the previous case considered.
The final answer is then
\be
W_\t[F] W_\t [G] = W_\t [F*G] \,,
\ee
with
\be \label{Moyal}
(F * G)(x) = F(x) \exp{{i\over 2} \theta^{\mu\nu} \left( \overleftarrow{\partial_{x^{\mu}}}
\overrightarrow{\partial_{x^{\nu}}} -\overleftarrow{\partial_{x^{\nu}}} \overrightarrow{\partial_{x^{\mu}}}
\right)} G(x) \,.
\ee
This is the so-called Weyl-Moyal product that we will be using very often
throughout the first part of this thesis.
Again one can check that this product is associative but noncommutative, and that
\be
[x^{\mu},x^{\nu}]_* = i\t^{\mu\nu} \,.
\ee

\eitem

\subsection{A few properties of the Weyl-Moyal product} \label{ch3:sec:Weyl-properties}

Because of the relevance of the Moyal product \bref{Moyal}, it is worth
making a pause and collecting some of the main properties that
will be needed when dealing with NC field theories.

Most properties follow from examining the expansion of the $*$-product
of any two functions
\be \label{expa}
f*g =fg + {i \over 2} \theta^{\mu\nu} \pa_{\mu} f \pa_\nu g -
{1\over 4} \theta^{\mu\nu}\t^{\a\b} \pa_\mu \pa_\a f \pa_\nu \pa_\b g \,+\, ... \,.
\ee
Because of the antisymmetry and the constancy of the matrix $\t$, all terms but
the first one can be written as a total derivative. For example, for the terms
considered in \bref{expa},
\be
f*g =fg + \pa_\mu \left({i \over 2} \theta^{\mu\nu} f \pa_\nu g -
{1\over 4} \theta^{\mu\nu}\t^{\a\b}  \pa_\a f \pa_\nu \pa_\b g \right) \,+\, ... \,.
\ee
Immediate consequences affect integrals of $*$-products with boundary conditions
such that surface terms vanish:
\bea
\int d^4 x F * G &=& \int d^4x \, FG \,, \label{ch3:integration} \\
\label{ch3:cyclicity}
\int d^4 x F_1 * F_2 * ... * F_n &=& \int d^4 x \, F_n * F_1 * ... * F_{n-1} \,.
\eea
The first property will imply that the free part of the NC field theory actions will
coincide with the commutative one: propagators will not change. The second cyclic
property will imply that vertices in Feynman diagrams for NC theories
will be invariant under a cyclic rotation of the incoming legs.

The last useful property we will review has to do with the way that plane waves
are treated under $*$-products. It is not hard to check that
\be \label{prod-exp}
e^{ik_1 x} * e^{ik_2 x} * ... * e^{ik_n x} = e^{ix \sum_i k_i - {i\over 2}
\sum_{i<j} k_i \t k_j } \,,
\ee
which will allow us to write the Fourier transform of interactions by means of
\be \label{prod-fun}
\int d^4 x F_1 * F_2 * ... * F_n =  {1\over (2\pi)^{n-1}}
\int \prod_{i=1}^{n} \left[d^4 p_i f_i(p_i) \right]
   \, e^{- {i\over 2} \sum_{j<l} k_j \t k_l } \,.
\ee

\newpage
\section{From D-branes to NC theories} \label{ch3:sec:D-nc}

We now wish to explain how NC theories arise in string theory as
originally discovered by Seiberg and Witten in~\cite{Seiberg:1999vs}. The final
goal is to show that the low energy worldvolume effective action of D-branes
in the presence of a constant background B-field is a NC theory.

Since D-brane dynamics are described in perturbation theory by the excitations
of the open strings that lie on them, let us start by considering the
worldsheet 2d CFT for such strings. The bosonic part of the
action in conformal gauge is
\be \label{nc-action}
S={1\over 4\pi \ap} \int_{\Sigma_2} \left( g_{MN} \partial_{\a} X^{M}
\partial^{\a} X^{N} - 2\pi i \ap B_{MN} \e^{\a\b} \partial_{\a} X^{M}
\partial_{\b} X^{N} \right) \,.
\ee
Some clarification on the conventions:
\bitem
\item Spacetime $\zespai \longleftrightarrow \zespai$ Capital indices $M,N=0,...,D-1$.
\item Worldsheet $\zespai \longleftrightarrow \zespai$ Greek indices $\a,\b=\{ \tau, \sigma \}$.
\item The worldsheet metric is Euclidean, and hence the $i$ in front of the second term.
\item We will divide the spacetime indices among indices parallel $(\mu,\nu...)$ and transverse
$(i,j,...)$ to the brane.

\eitem
Note that since we take the background $B$-field to be constant, the second term
of the action can be written as a total derivative
\be
S[B] = i \int_{\Sigma_2} P[B_{2}] = {i\over 2} \int_{\Sigma_2} P\left[
d( B_{\mu\nu} X^{\mu} dX^{\nu} )\right]
= {i\over 2} \int_{\pa \Sigma_2} B_{\mu\nu} X^{\mu} \pa_t X^{\nu} \,,
\ee
where we recall that $P$ is the pullback operator and $\pa_t$ stands for the derivative tangent
to the boundary $\pa \Sigma_2$.

The background we want to deal with has
\be
g_{MN}=\eta_{MN} \sac B_{\mu\nu}=\mbox{ct.}\, , \,\, B_{ij}=0=B_{\mu i} \sac \Phi=\mbox{ct.}
\ee

It is very important to remark here that a {\it constant  $B$-field would be pure gauge
if it was not for the presence of D-branes.} When branes are present, trying to
gauge away a B-field with components along the brane
induces a worldvolume field-strength $F_2$ since the truly invariant
quantity is then $\calf_2 = B_2 + 2\pi \ap F_2$. In other words, we can
always go to a gauge where all the components of the constant $B$-field transverse
to the brane directions are zero. On the other hand, the gauge ambiguity
on how to distribute the value of $\calf_2$ among $F_2$ and $B_2$ is
commonly resolved by the adhoc boundary condition that $F_2$ vanishes
at infinity.
This relation between $B_2$ and $F_2$ motivates the name {\it electric} for
the $B_{0\mu}$ components and {\it magnetic} for the $B_{ij}$ ones.

Having fixed the background we vary the action to obtain the equations of
motion and the boundary conditions. Since the $B$-term in the Lagrangian
is a total derivative, the equations of motion are the usual ones;
however the requirement that the surface terms vanish is modified to
\bea \label{nc-bcs}
\pa_n X_\mu + 2\pi i \ap B_{\mu \nu} \pa_t X^{\nu} |_{\pa \Sigma_2}& =& 0 \sac
\mu=0,...,p\,, \\
\pa_t X^i |_{\pa \Sigma_2}& =& 0 \sac i=p+1,...,9 \,.
\eea
Note that the effect of the $B$-field is to interpolate between Neumann
BC's at weak field and Dirichlet BC's at strong field. In the latter limit,
which is the one we will take in a while, the endpoints of the strings are fixed
at one point of the worldvolume of the brane, as if they were attached to a D0
brane.

We will now move on an try to compute S-matrix elements with this CFT with
the aim to extract its low energy physics and see what kind of field theory
approximates them.
As we are interested in the physics of the D-branes, we will mostly
be concerned with open string diagrams. Let us start by tree level
diagrams with the topology of a disc and map it to the upper half plane
by a conformal transformation. The BC's \bref{nc-bcs} read in the
usual complex coordinates
\be
(\pa-\cpa) X_\mu + 2\pi i \ap B_{\mu \nu} (\pa+\cpa) X^{\nu} |_{z=\cz} = 0 \,.
\ee
The propagator satisfying these boundary conditions is
\bea \label{nc-propagator}
\langle X^{\mu}(z) X^{\nu}(z') \rangle & =& -\ap  \big[  \eta^{\mu\nu} \log|z-z'|- \eta^{\mu\nu} \log|z-\cz'| \\
& & + G^{\mu\nu} \log|z-\cz'|^2 + {1\over 2\pi \ap} \t^{\mu\nu} \log{ z-\cz' \over \cz-z'}+D^{\mu\nu}
\big] \,, \nonumber
\eea
where
\bea \label{canvi-1}
G^{\mu\nu} &=& 
\left({1\over
\eta +2\pi\alpha' B} \eta {1\over \eta-2\pi\alpha' B}\right)^{\mu\nu} \,, \\
\label{canvi-2}
G_{\mu\nu} &=&\eta_{\mu\nu}-(2\pi\alpha')^2 \big(B \eta^{-1} B\big)_{\mu\nu} \,, \\
\label{canvi-3}
\theta^{\mu\nu}&=& 
 -(2\pi \alpha')^2 \left({1 \over \eta+ 2\pi\alpha'
B} B {1 \over \eta- 2\pi\alpha'
B}\right)^{\mu\nu} \,.
\eea
and the constants $D^{ij}$ can depend on $B$ but not no $z$ and $z'$,
and therefore play no essential role.

If we are only interested in the interactions of open strings with other open strings,
all vertex operators that we need to insert in the path integrals must lie in
the boundary of $\Sigma_2$. So, effectively, we just need to know the propagator \bref{nc-propagator}
restricted to the boundary. Naming $\tau=\mbox{Re}[z]$, we get
\be \label{open-st-prop}
\langle X^{\mu}(\tau) X^{\nu}(\tau') \rangle  =
-\ap  G^{\mu\nu} \log(\tau-\tau')^2 + {i\over 2} \t^{\mu\nu} \e(\tau-\tau')
\,. \nonumber
\ee

We are ready to give a physical interpretation to the two metrics that
have appeared at this point. Whereas the short distance behavior of
the propagator \bref{nc-propagator} is controlled by the term
$-\ap \eta^{\mu\nu} \log|z-z'|$, the corresponding leading term
for points in the boundary is control\-led by
$-\ap G^{\mu\nu} \-\log|z-z'|$.
Recall that one way to obtain the mass-shell condition for the string
states is to impose that its corresponding vertex operator has dimension
one. This in turn can be calculated via its OPE with the energy momentum
tensor, and it is easy to see that the relevant singular terms are
precisely the ones containing the logarithms; the metric multiplying
them is the one that has to be used when computing the mass.
Therefore we can say that the closed strings (whose vertex operators
are inserted inside the disc) still feel the presence of a flat background
metric, whereas open strings effectively see $G_{\mu\nu}$. Henceforth,
the former will be called the {\it closed string metric} and the latter,
the {\it open string metric}. As a consequence, the correct vertex operators
for the tachyon and the massless gauge field of the open string spectrum are
\be \label{v-ops}
V_{{tachyon}}(k) \sim \, :e^{i k \punt X}: \sac
V_{{gauge}}(k,\xi) \sim \, :  \xi \punt \pa \, e^{i k \punt X}:
\ee
where the dot-contractions are taken with respect to the open string metric $G$.
In particular, the gauge boson vertex operator with polarization $\xi$ is
physical when
\be
G^{\mu\nu} \xi_\mu k_\nu = 0 = G^{\mu\nu} k_\mu k_\nu \,.
\ee

The tensor $\t^{\mu\nu}$ in \bref{open-st-prop} has also an interpretation.
By the usual calculus in CFT, commutators of operators are translated
into time-ordered products, so that
\be \label{nc-string}
[X^{\mu}(\tau),X^{\nu}(\tau)]=\langle X^{\mu}(\tau) X^{\nu}(\tau-\e)
- X^{\mu}(\tau) X^{\nu}(\tau+\e) \rangle = i \t^{\mu\nu} \,.
\ee
Therefore, the open strings feel that they are effectively living on
a noncommutative space with parameters $\t^{\mu\nu}$.

\subsection{The low energy limit for magnetic backgrounds} \label{ch3:sec:magnetic-limit}

Our aim now is to establish what the low energy effective theory is.
We want to decouple all the massive states, which as usual
should be accomplished by taking $\ap \rightarrow 0$. However we will see
that this is not possible to do in the cases for which there
are non-zero electric components $B_{0\mu} \neq 0$. We therefore restrict
in this section to magnetic backgrounds and postpone the discussion of
electric ones until section~\ref{ch3:sec:electric-case}.

The limit $\ap \rightarrow 0$ must be taken with care if we want to
remain with a sensible theory for open strings; we need to keep
finite the parameters that control their dynamics, \ie $G_{\mu\nu}$
and $\t^{\mu\nu}$. Taking a look at the formulas \bref{canvi-1}-\bref{canvi-3}
we see that this can be achieved by taking $\ap B \gg 1$, which
is a {\it strong magnetic field limit}.
In the original paper of Seiberg and Witten, instead of sending the magnetic
field to infinity, they equivalently  scaled to zero the closed string metric $\eta$,
keeping $B$ fixed.
The exact way in which the limit is taken is then
\bea \label{nc-limit}
\ap \sim \e^{\undos} \rightarrow 0 \sac
\eta_{\mu\nu} \sim \e \rightarrow 0 \,.
\eea
Then, the relations \bref{canvi-1}-\bref{canvi-3} reduce to
\bea \label{canvi-21}
G^{\mu\nu} &=&
- {1\over (2\pi\ap)^2} \left({1\over B} \eta {1\over B}\right)^{\mu\nu} \,, \\
\label{canvi-22}
G_{\mu\nu} &=&-(2\pi \ap)^2(B\eta^{-1} B)_{\mu\nu} \,, \\
\label{canvi-23}
\theta^{\mu\nu}&=& \left({1\over B}\right)^{\mu\nu}  \,.
\eea
Using \bref{nc-limit} we see that $G$ and $\t$ are finite in the limit.
Let us analyze how other quantities look like in the limit. The
propagator \bref{open-st-prop} becomes
\be
\langle X^{\mu}(\tau) X^{\nu}(0) \rangle ={i\over 2} \t^{ij} \e(\tau) \,.
\ee
Note that it looses its $\ap$-dependence; a sign that we are decoupling
all the massive modes of the string. Indeed, all its dependence on the
worldsheet coordinates is through $\e(\tau)$, which is a constant function
everywhere except at one point. In a sense, the CFT looses all its propagating
degrees of freedom. This is clearly seen by taking a look at the action \bref{nc-action}
in this limit
\be
S \rightarrow -{i\over 2} \int_{\pa \Sigma_2} B_{\mu\nu} X^{\mu} \pa_t X^{\nu} \,.
\ee
We see that the kinetic term in the bulk has become negligible, and that we are
left with a one-dimensional action for two oppositely charged particles in
a large constant magnetic field. This is precisely the Landau problem
that we considered in \ref{ch3:sec:Landau}! We can therefore understand better
why the noncommutativity \bref{nc-string} appears, since position operators
are now canonical pairs describing   electrons forced to remain in its Landau
ground state. It is worth mentioning that the fact that our two electrons are indeed
the endpoints of a string makes a little difference in this guise, specially
when one compactifies some direction, say, on a torus; there we can see global
effects arising. However, for the most part of what follows, the discussed
interpretation remains a useful guide.

\subsection{The effective action from the S-matrix} \label{ch3:sec:leff}
\label{s-matrix}

Let us setup the technicalities that we will need in order to compute
S-matrix elements and deduce the effective action.
The analog of the $*$-product that we found by Weyl mapping in \bref{ch3:sec:Weyl}
is provided here by normal ordering.
By using repeatedly the operator product expansion of two exponentials
one can prove the following properties:
\begin{enumerate}
\item
$
:e^{i p \punt X(\tau)}::e^{i q \punt X(0)}: =
e^{-{i\over 2}\e(\tau) (p\t q)} :e^{i p \punt X(\tau)+iq \punt X(0)  }: \,,
$
\item
$
:f(X(\tau)): :g(X(0)): = :e^{{i\over 2} \e(\tau) \t^{\mu\nu} {\pa \over \pa
X^{\mu}(\tau)}{\pa \over \pa X^{\nu}(0)}} f(X(\tau))g(X(0)):
$
\item
$
\lim_{\tau \rightarrow 0^+} :f(X(\tau)):::g(X(0)): = : (f * g)(X(0)) : \,,
$
\end{enumerate}
with the $*$-product defined in \bref{Moyal}.
From these properties it follows that the correlation functions of
open string tachyon vertex operators \bref{v-ops} on $\pa\Sigma_2$ are given by
\be \label{general-tachyon}
\langle \prod_n :e^{i p^n \punt X(\tau_n)}: \rangle=
e^{-{i\over 2} \sum_{n> m}p^n \punt \theta \punt p^m}
\delta (\sum_n p^n) \,.
\ee
It is worth pausing a minute to examine this equation.
There a two main difference with respect to the same
disc computations made in a background without $B$-fields.
The first one is, as mentioned, the appearance of the
open string metric $G$ in the various scalar products.
The second one is the appearance of a momentum dependent
phase. By the $*$-product cyclic property \bref{ch3:cyclicity}, we see these correlation
functions are invariant under cyclic permutations of the operators
on the boundary, although {\it they are not invariant under non-cyclic
permutations.} This resembles very much the non-abelian properties of
open strings with Chan-Paton factors, although we remark that
we have not yet introduced any. As we will see, abelian noncommutative
actions\footnote{We recall that in the whole thesis we use
'abelian' to refer to gauge groups and 'commutative' to
refer to spacetime commutativity.} share
a lot of properties of standard non-abelian theories. In particular,
the lost of cyclicity will lead to the classification of diagrams
in terms of planar and non-planar ones.

Equation \bref{general-tachyon} can be rewritten in terms
of $*$-products since they are the natural operation
for the problem. Indeed, using the property \bref{prod-exp}
in \bref{general-tachyon} we get
\be
\langle \prod_n :e^{i p^n \punt X(\tau_n)}: \rangle=
\int dx  \,  e^{i p^1 \punt x} * e^{i p^2 \punt x} *...*
e^{i p^n \punt x} \,.
\ee
More generally, all expectation values made of the scalar
fields (with no derivatives) can be shown to give
\be \label{general-vev}
\left\langle \prod_n :f_n(X(\tau_n)):\right\rangle  = \int
dx (f_1*f_2*\dots * f_n)(x) \,.
\ee

Equations \bref{general-tachyon}-\bref{general-vev}, together
with their extension to include derivatives of the scalar fields
(which just follow by the usual procedure) are all we need to
compute our first S-matrix element: the 3 gauge bosons interaction.
An intermediate result is
\bea \label{first-s}
&
\big\langle \xi^1\cdot \partial x  e^{ip^1 \cdot x(\tau
_1)}~\xi^2\cdot \partial x e^{ip^2 \cdot x(\tau _2)}~\xi^3\cdot
\partial x e^{ip^3 \cdot x(\tau _3)}
\big\rangle
\sim
{1 \over (\tau
_1-\tau _2)(\tau _2-\tau _3)(\tau _3-\tau _1)} \nn &
\cdot \left( \xi^1 \cdot\xi^2 p^2 \cdot \xi^3 +  \xi^1 \cdot \xi^3p^1
\cdot \xi^2 +  \xi^2 \cdot \xi^3 p^3 \cdot \xi^1
\right) \nn &
\cdot   e^{-{i\over 2}\left(p^1 \punt \theta \punt p^2 \epsilon (\tau
_1-\tau _2) +p^2 \punt \theta \punt p^3\epsilon (\tau _2-\tau _3)
+p^3 \punt \theta \punt p^1\epsilon (\tau _3-\tau _1) \right)} \,,
\eea
where we are not writing an explicit delta-function for the momentum
conservation.
To compute the S-matrix element, we should integrate over all possible
positions of the vertex operators in $\pa \Sigma_2$, and then divide
over the volume of the conformal group of the disc $SL(2,R)$.
Equivalently, for this 3-point amplitude $SL(2,R)$ is large enough
as to allow us to place the vertex operators wherever we like
(typically, at $\{0,1,\infty\}$). In any
way we make it, the result is just equation \bref{first-s} removing
the $\tau$-dependent denominator, and with a little modification
of the exponential
\be
\left( \xi^1 \cdot\xi^2 p^2 \cdot \xi^3 +  \xi^1 \cdot \xi^3p^1
\cdot \xi^2 +  \xi^2 \cdot \xi^3 p^3 \cdot \xi^1
\right)
e^{-{i\over 2}(p^1 \punt \theta \punt p^2)} \,.
\ee
This amplitude can be reproduced by a computation of the
3-point function evaluated with the field theoretical action
\be \label{nc-ym}
{(\alpha')^{3-p\over 2}\over 4(2\pi)^{p-2}G_s } \int d^{p+1} x \,
\sqrt{\det G} G^{\mu\mu'}G^{\nu\nu'}\mbox{Tr} \ \hat F_{\mu\nu}*\hat F_{\mu'\nu'} \,,
\ee
where $G_s$ will is what we will call the open string coupling (we
will fix its value later) and
\be      \label{nc-f}
\hat F_{\mu\nu}=\partial_\mu\hat A_\nu -
\partial_\nu \hat A_\mu-i\hat A_\mu*\hat A_\nu+i\hat A_\nu*\hat A_\mu \,.
\ee
We are advancing what will follow a little bit by
putting hats over all NC-fields, not to confuse them with the commutative ones
(both concepts,  will be defined shortly).

We finish this section by generalizing the result to include any number
of gauge fields. It can be seen that
\be
\left\langle
\prod_n \xi^n\cdot \partial x e^{ip^n\cdot
x(\tau_n)}\right\rangle_{G,\theta}
=
e^{-{i\over 2} \sum_{n> m}p^n \punt
\theta \punt p^m \epsilon(\tau_n-\tau_m)}\left\langle \prod_n \xi^n
\cdot  \partial x e^{ip^n\cdot x(\tau_n)}\right\rangle_{G,\theta=0} .
\ee
In other words, if the effective action is expressed in terms
of the open string quantities, all the $\t$-dependence is only
in the $*$-product.

\subsection{A look at the NC Yang-Mills action and NC gauge invariance} \label{ch3:sec:nc-gauge}

The action \bref{nc-ym} is the first NC field theory that we
encounter. A little bit expectedly, it is not obtained from an  ordinary U(1)
YM action simply  by replacing standard products by $*$-products; the
field strength \bref{nc-f} has been non-linearized as well. Indeed,
equation \bref{nc-f} defines a NC field strength exactly in the same way
as we would define a non-abelian one. Further more, the action \label{bc-ym}
is not invariant under usual gauge transformations $\delta A_{\mu}=\pa_\mu \lambda$
but rather under a NC version of them
\be \label{nc-gauge}
\delta \hA_{\mu}=\pa_\mu \lambda +i [\l,\hA_{\mu}]_* \,.
\ee
The need for this gauge transformation can be understood
directly from string theory. In the presence of a $B$-field,
the coupling of the gauge field $A_\mu(X)$ to the worldsheet
\be \label{cou}
-i \int d\tau A_{\mu}(x) \pa_\tau X^{\mu}\,,
\ee
is no longer invariant under $\delta A_{\mu}=\pa_\mu \lambda$
at the quantum level. This is because the gauge transformation
of \bref{cou}, which is a total derivative $\int d\tau \pa_\tau \l$,
can produce divergences when meeting other insertions in
the path integral. For example, a term like
\be
\int d\tau A_{\mu}(x) \pa_\tau X^{\mu} \int d\tau' \pa_{\tau'} \l \,,
\ee
appears in the variation of the path integral, and needs to be regularized
at points where the operators are very close.
Seiberg and Witten showed that point splitting regularization
produces a finite contribution which can only be cancelled if
the full gauge transformation contains the extra piece in \bref{nc-gauge}.
Due to the fact that point splitting explicitly breaks gauge invariance,
a modification was to be expected.
However, what if we use a regularization compatible with gauge invariance?
Had we used Pauli-Villars, the original transformation would have remained
an invariance. Conclusion: getting a NC field $\hA$
or an ordinary one $A$ (the adjective referring to their kind of gauge
transformation) is a matter of choice of regularization method.

Are we then just being masochists by choosing $*$-products and NC fields?
The answer to this question is arguably {\it no}. The freedom
in choosing commutative or NC fields corresponds to the well-known ambiguity
in low energy Lagrangians derived from string S-matrices. The S-matrix
is unchanged under field redefinitions of the effective Lagrangian \bref{nc-ym}.
We could have well written an analog Lagrangian in terms of the
closed string quantities $\{\eta_{\mu\nu},g_s,B_2\}$ and the commutative fields
$A_\mu$ reproducing the same S-matrix element. For example, had we used
Pauli-Villars in the worldsheet, the commutative effective action
would have been an ordinary U(1) Maxwell Lagrangian
replacing $F_2$ by $F_2+B_2$ everywhere, as the latter is the only
gauge-invariant quantity and PV preserves gauge invariance.
However, computations are carried more easily with the NC action \bref{nc-ym}
because all the dependence on $\t$ is hidden
inside the $*$-product, which is quite a manageable structure.

Let us finish this section by fixing the value of $G_s$. This can
be done by comparing the DBI actions that one would obtain using
closed or open string variables.
As already discussed in sec. \ref{ch1:sec:multiple-D} the DBI action in the
presence of B-fields (which we now reinterpret as being written
in the closed string variables) is
\be \label{c-dbi}
\call_{DBI}(F)={1 \over g_s (2\pi)^p(\alpha')^{p+1\over
2}}\sqrt{\det (g+2\pi\alpha'(B+F))}
\ee
whereas the NC counterpart must have its $\t$-dependence
in the $*$-product and it is thus expected to be
\be \label{nc-dbi}
\call_{DBI}(\hF)={1 \over G_s (2\pi)^p(\alpha')^{p+1\over
2}}\sqrt{\det (G+2\pi\alpha' \hF)_*} \,.
\ee
All products in any expansion of \bref{nc-dbi} are understood
to be $*$-products. Comparing the coefficients of both
Lagrangians for constant gauge fields
\be
\call_{DBI}(F=0) = \call_{DBI}(\hF=0) \,,
\ee
we obtain the required value for the open string coupling $G_s$
before and after the zero slope limit:
\bea \label{Gs}
G_s &=&g_s\left({\det G\over \det{(g+2\pi
\alpha'B)}}\right)^{1\over 2}=
g_s\left({\det G \over \det g}\right)^{1\over 4}  \nn
&=&
g_s\left({\det (g+2\pi \alpha' B) \over \det g}\right)^\undos  \,,\\
G_s |_{\ap \rightarrow 0} & = & g_s\mbox{det'} (2\pi \alpha' Bg^{-1})^\undos \,,
\eea
where $\mbox{det'}$ stands for the determinant of the non-vanishing block of
$B$.

We can now obtain the NC Yang-Mills coupling that originates from expanding
the NC DBI action \bref{nc-dbi} and picking the coefficient of the $\hF^2$ term:
\be
{1 \over g_{YM}^2} = {(\alpha')^{3-p\over 2}\over
(2\pi)^{p-2}G_s } \,.
\ee
In order to keep it finite in the zero slope limit, we need
to scale $G_s$ exactly in the same way as one would
scale $g_s$ in the $B=0$ cases~\cite{Itzhaki:1998dd}, \ie
$G_s \sim \e^{{3-p}\over 4}$. The complete
zero slope limit is then, in terms of closed string quantities,
\be
\ap \sim \e^{\undos} \rightarrow 0 \sac
\eta_{\mu\nu} \sim \e \rightarrow 0 \sac g_s \sim \e^{{3-p+r}\over 4} \,,
\ee
where $r$ is the rank of $B_2$.

Before finishing this section it is worth noticing a subtle fact.
Whereas we know that the group $U(N)$ is actually isomorphic
to $U(1)\times SU(N)/Z_N$, so that the $U(1)$ photon decouples
from the rest of degrees of freedom in a $U(N)$ gauge theory,
this does not apply to NC theories. If $f$ and $g$ are two matrix
valued fields, we have that in general $\det(f*g)\neq \det (f)
* \det (g)$ so that $SU(N)$ does not give rise to any gauge group
on a NC $\CR^D$. In other words, the $U(1)$ photon of NC $U(N)$
gauge theories couples to the rest of fields in the gauge multiplet.
\label{nc-photon}

\subsection{The Seiberg-Witten map} \label{ch3:sec:sw-map}

As we have seen, different regularizations in the worldsheet lead to
different effective YM theories, a commutative and a NC one.
But in a QFT, different regularizations differ always by a
choice of contact terms and so, theories defined with
different regularizations are related by coupling constant
redefinitions. As coupling constants in the worldsheet
Lagrangian are precisely the spacetime fields,
we conclude that there must be a map between  the commutative
and the NC fields which maps the ordinary to the NC gauge transformations.

Even before trying to find such a map, it is worth guessing how it
will look like. The naive attempt to look for a map among
fields and gauge parameters of the form
\be \label{naive-map}
\hA= \hA (A,\pa A,...;\t) \sac
\hl=\hl(\l,\pa \l,...;\t) \,,
\ee
would never work. The reason is that this would imply that there is
an isomorphism between the ordinary and the NC YM groups; this is impossible
since even in the simple U(1) YM case, the one is an abelian group and
the other is non-abelian.

The way that Seiberg and Witten managed to find such a map was by
relaxing the aim. Instead of looking for a map between gauge groups,
one must look for a map between gauge orbits of the groups. Two
field configurations related by a gauge transformation must
be mapped to two other field configurations related by a gauge transformation.
The infinitesimal version of this statement requires then that
\be \label{gauge-class}
\hA(A)+\hat{\d}_{\hl} \hA(A) = \hA (A+\d_\l A) \,.
\ee
Under a modification of \bref{naive-map} to include a field-dependent transformation
for the gauge parameters $\hl=\hl(\l,\pa \l,...;\t;A)$, Seiberg and Witten
managed to solve exactly the equation \bref{gauge-class} for the $U(1)$ case:
\be 
\hF={1\over 1+F\t} F \,,
\ee
or written in terms of the $B$-field 
\be
\hF=B{1\over B+F} F \,.
\ee
This shows that a NC description is not possible in the case that $\calf=F+B=0$
as was to be expected\footnote{It would otherwise be really shocking if
one could describe a usual commutative YM QFT by means of a NC one. This
would mean that the various phenomena that we will encounter later on, like
UV/IR mixing, are nothing but an artifact of our description.}.

\subsection{Electric Backgrounds} \label{ch3:sec:e-bckg}

In section \ref{ch3:sec:magnetic-limit}, when trying to find an $\ap \rightarrow 0$
limit of string theory, we restricted ourselves to the cases
in which $B_{0\mu}=0$ and postponed the discussion of the
electric backgrounds. It is time to justify this separation.

Let us put the electric field in the direction of $X^1$, so that we take
$B_{01}=E \neq 0$. In this subsection, we restrict the range of
the indices $\mu,\nu$ to $\{0,1\}$ to focus on the electric sector.
Writing explicitly for our case the relations \bref{canvi-2}, \bref{canvi-3}
and \bref{Gs}  between closed
and open string quantities, we find
\be G_{\mu\nu} = G \eta_{\mu\nu} \sac
\t^{\mu\nu} = \Theta \left( \begin{array}{cc} 0 & 1 \\ -1 & 0 \end{array} \right)\sac
G_s=g_s \sqrt{ 1-\tE^2  } \,,
\ee
where we have defined the constants
\be
\tE={E\over E_{cr}} \sac E_{cr}={g\over 2\pi \ap} \sac \Theta={\tE \over E_{cr}}
{1\over 1-\tE^2}\,,
\ee
and $G$ is introduced to later rescale the metric.

We see that the expressions become singular as we increase the value of
the electric field until $E\rightarrow E_{cr}$. This singularity is
indeed related to the fact that the gaussian string vacuum becomes
unstable for $E > E_{cr}$. We first give the qualitative picture of what
is happening and then the quantitative one. Physically, the strings may be
understood as a lace with two oppositely charged particles, one at each
endpoint, and both endpoints are forced to lie on the brane.
If we turn on an electric field in one of the brane directions, the lace
will like to minimize its energy by aligning along the electric field,
and each point will pull in opposite directions. The lace does not break
because of the tension. However, as the electric field
increases, it may happen that the tension is not enough to keep the system
stable, and there appears an instability against breaking and creation
of new strings. To make it quantitative, one can use the
DBI action (which we recall that it is exact for constant field-strengths)
in this background. From \bref{DBI} we get
\be \label{im-lag}
\call_{DBI} \sim \sqrt{1-(2\pi \ap E)^2} \,,
\ee
which becomes imaginary precisely at $E=E_{cr}$.
Yet another way to understand this is by performing a T-duality along the
direction of the electric field. Having into account only the $(01)$
coordinates, the resulting configurations describes a relativistic
particle moving at speed $v= (E/E_{cr}) c$, so that \bref{im-lag}
becomes
\be 
\call \sim \sqrt{1-{v\over c}^2} \,.
\ee

The conclusion we draw from this is that we cannot take a
large electric field limit as we did in the magnetic case,
which required $B \gg 1/\ap$. In our present case, however,
the electric field must be kept below $1/\ap$.
It is not hard to convince oneself that there is now way
to send $\ap\rightarrow 0$ keeping finite the open string
quantities, from which we must conclude that  NC {\it field} theories
with electric $\t$ do not arise as low energy descriptions of
string theories.

There is however an interesting limit that can be taken, now
that we have a new parameter ($E$) to play with. The limit is
a near-tensionless limit in which $E\rightarrow E_{cr}$
and the closed string metric is scaled to infinity. The only way
to keep the interactions in this limit ($G_s$ finite) is to
put a large number $N$ of D-branes on top of each other, so
that effective coupling is actually $G_s^{eff} = G_s N$.
One can then let
\be \label{ncos-limit}
E \rightarrow E_{cr} \sac -g_{00} =g_{11} \sim {1\over 1-\tE^2}
\sac N \sim {1\over \sqrt{1-\tE^2}}
\,,
\ee
and verify that all open string quantities remain finite.
It is important to remark that {\it we are not sending $\ap\rightarrow 0$},
so that we are not decoupling the massive string states, and
{\it the resulting theory is not a field theory, but a stringy one}.

One of the main properties of the resulting theory is that these open strings
do not couple to the closed strings, in apparent contradiction to the
general expectation that all open strings contain closed strings.
Can't we just bend an open string topology to form
a closed one? The answer is that this is not forbidden by any
defining property of the theory, but by its dynamics.
The intuitive way to understand it is that, being close to
its critical value, the electric field keeps that open strings
completely rigid, and it takes an infinite energy to bend it
to the point of bringing together the endpoints to form
a closed string.

Summing up, the limit \bref{ncos-limit} provides us with a new non-critical interacting
string theory in which
\bitem
\item open string massive modes still survive,
\item the brane becomes invisible to the bulk closed string physics,
\item the spacetime seen by the open strings is noncommutative.
\eitem
This theory has been called NC Open String theory (NCOS)\cite{Seiberg:2000ms,Gopakumar:2000na}.

\newpage

\section{Quantum NC Field Theories} \label{ch3:sec:quantum-NC}

Having discussed how NC field theories arise as a low energy
description of D-branes in backgrounds with $B_2 \neq 0$,
we now move to quantize these theories.
The analysis of the quantum properties that follows for the rest of
this chapter is purely perturbative. We will see that some of the
phenomena we will encounter are rather difficult to understand
within our present knowledge of quantum field theories. A
non-perturbative analysis is then extra motivated, and this
will be the subject of chapter~\ref{ch:nc-sugra}.

\subsection{Perturbative NC $\phi^4$} \label{ch3:sec:phi-4}

Let us illustrate the main new phenomena that occur
in the quantum NC-field theories by using the $\phi^4$ theory in 4 dimensions.
The action is
\begin{equation}
S=\int dx^{4}\left\{
\frac{1}{2}\partial_{\mu}\phi\partial^{\mu}\phi+\frac
{1}{2}m^{2}\phi^{2}+\frac{\lambda}{4!}\phi * \phi * \phi * \phi\right\}
 \label{action4} \,.
\end{equation}
Note that because of the property \bref{ch3:integration}, it is a matter of choice whether
to put $*$-products or ordinary ones in the first two terms of the action,
and it is obviously easier to leave it as it is.

The equations of motion are obtained as usual by varying the action,
and one gets
\begin{equation}
(\square+m^{2})\phi=\frac{\lambda}{6}\phi * \phi * \phi.
\end{equation}
Before going on, let us briefly mention that this equation admits
some solutions qualitatively different from its commutative
counterpart. In particular, it admits solitonic solutions.
This does not violate Derrick's theorem because of the loss
of \Poin invariance due to the presence of the new scale $\t$.

Let us proceed to obtain the Feynman rules for this theory.
Our choice of not including star products in the quadratic
part of the action leads to the usual Feynman propagator.
We just need to obtain the interaction vertex. In momentum
space, we can use the property \bref{prod-fun} to write
\bea
S_{int}& =& \frac{\lambda}{4!}\int d^4x \underset{4}{\underbrace{\phi
 * ... * \phi}} \nn
& = & \frac{\lambda}{4!}\int\frac{d^4p_{1}...d^4p_{4}%
}{\left(  2\pi\right)  ^{3}}\tilde{\phi}(p_{1})...\tilde{\phi}(p_{n}%
)\delta^{4}(p_{1}+...+p_{4})\exp(-\frac{i}{2}\sum_{i,j=1,i<j}^{4}p_{i}\theta
p_{j}). \nn \label{sint}%
\eea
From here we read that the only effect of the noncommutativity is the
appearance of the global phase. Momentum is still conserved, but now
the vertex is invariant only under cyclic permutations of its 4 legs.
An immediate consequence is the classification of diagrams
in terms of {\it planar} and {\it non-planar} ones, according
to whether they can be drawn in a plane or not.

Let us illustrate this in a simple example. Consider a 1-loop
diagram in which we have two external legs with 4-momenta
$p$ and $k$. Let us start by attaching the $p$-leg to
one of the four legs of the vertex. Due to cyclic invariance,
all choices are equivalent, so we will connect it to leg $1$.
Now we want to attach the $k$-leg. If we attach it to leg 2 or
to leg 4, we will be able to complete the diagram by pairing
the two neighboring legs that remain,

\be
\includegraphics[width=4in]{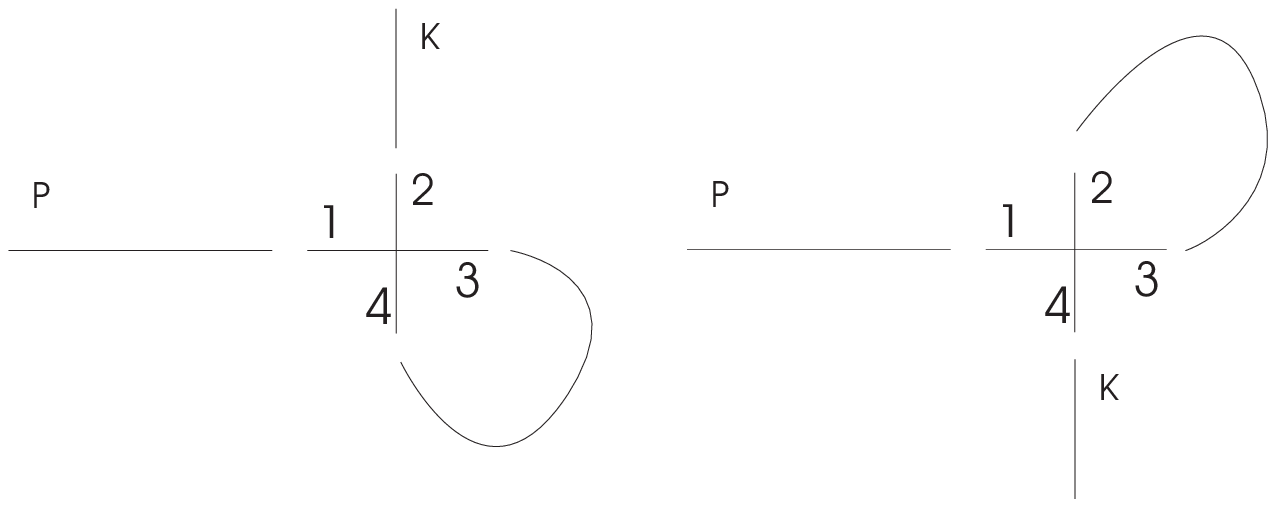} \nonumber
\ee
This can be drawn
in a plane and it therefore corresponds to a planar diagram.
Its phase factor is
\be
\exp{\left[-{i\over 2}  p \,\t k \right]} \,,
\ee
which would be trivial if we impose momentum conservation
on the external legs, since $p \,\t p=0$.
Consider now the other option, \ie linking the $k$-leg to
leg 3. This forces us to pair 2 with 4, which cannot
be done in a plane, and produces a non-planar diagram,

\be
\includegraphics[width=3in]{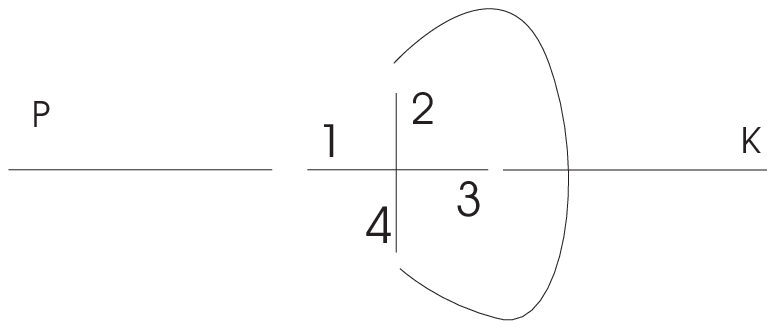} \nonumber
\ee
The important point here is that a phase remains
which {\it depends on the internal momentum $l$}
\be
\exp{\left[-{i\over 2}  p \,\t k \right]}
\exp{\left[-i \,   l \,\t k \right]} \,.
\ee
There is a quick way to see that this extra phase will completely
change the UV behavior of the diagram. The integration
over high $l$-momentum is now moduled by an infinitely fast
oscillating factor which, as we know from distribution theory,
tends to the zero distribution and it therefore effectively
removes the UV part of the loop.

This example illustrates a rather general fact. It can be shown that
all planar diagrams differ from the usual commutative ones
by a global phase\footnote{And, of course, a symmetry factor if we compare
planar NC diagrams to all commutative ones.\label{foot-factor}} which depends
only on the external momenta. Their divergences are then
identical to those in the commutative theory and they do not add
any qualitatively new phenomena. Non-planar diagrams, however,
are in general self-regulated in the UV by the parameter $\t$
and they require more care. As we will see, they will be
responsible for a non-habitual UV/IR mixing.

\subsection{The 1-loop correction to the self energy and
UV/IR mixing}\label{uv-ir}

Let us exactly compute the  1-loop
correction to the self energy for this NC $\phi^4$ theory.
{\it We will only deal with magnetic backgrounds} in this subsection because,
as we will see later, NC field theories in electric ones suffer from the
problem of lack of unitarity. We will skip most of the
explicit calculations here because in section \bref{ch3:sec:unitarity-non-rel}
we will see in all detail how this works for a similar
theory.

As discussed above, the 0th-order contribution is just the usual
\be
\Gamma_{(0)}^2=p^2+m^2 \,,
\ee
and the 1st-order one is given by the sum of a planar
and a non-planar diagram,
\be
\includegraphics[width=4in]{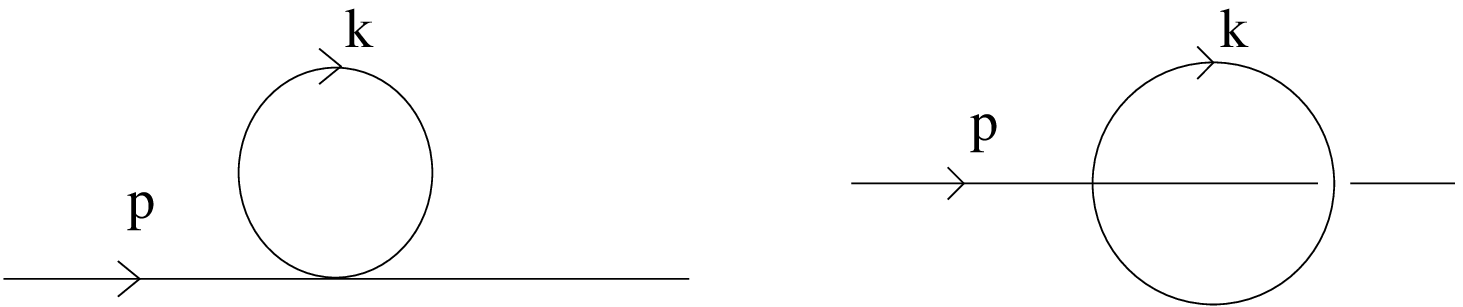}
\ee
The integrals we need to evaluate are then
\begin{align} \label{gp}
\Gamma_{\text{pl}}^{2}(p)  &  =\frac{\lambda}{3}\int\frac{d^4k}{(2\pi)^{2}%
}\frac{1}{k^{2}+m^{2}},\\
\label{gnp}
\Gamma_{\text{npl}}^{2}(p)  &  =\frac{\lambda}{6}\int\frac{d^4k}{(2\pi)^{2}%
}\frac{\exp[ip \, \theta k]}{k^{2}+m^{2}}. %
\end{align}
We skip the intermediate steps, but after introducing a UV cutoff $\L$, one finds
\begin{align}
\Gamma_{\text{pl}}^{2}(p)  &  =\frac{\lambda}{48\pi^{2}}\left[\Lambda^{2}-m^{2}%
\ln\left(\frac{\Lambda^{2}}{m^{2}}\right)\right] +\mbox{finite},\label{ft}\\
\Gamma_{\text{npl}}^{2}(p)  &
=\frac{\lambda}{96\pi^{2}}m \Lambda_{\text{eff}} K_{1}\left(
\frac{m}{\Lambda
_{\text{eff}}^{2}}\right)  =\frac{\lambda}{96\pi^{2}}\left[\Lambda_{\text{eff}}%
^{2}-m^{2}\ln\left(\frac{\Lambda_{\text{eff}}^{2}}{m^{2}}\right)\right]+\mbox{finite},
\end{align}
where
\begin{equation} \label{eff-cutoff}
\Lambda_{\text{eff}}^{2}=\frac{1}{1/\Lambda^{2}+\pop} \,.
\end{equation}
Due to its often appearance in this type of computations it is worth
to define the following $\circ$-product
\be \label{pop}
k \, \circ p \equiv - k^{\mu} \t^2_{\mu\nu} p^{\nu} \,,
\ee
from which it is easy to show that $\pop \ge 0$ for magnetic
backgrounds.

As promised, the planar diagram is the same as in the commutative theory
(this time, even without any external phase) and it has the
usual quadratic plus logarithmic divergences in the UV. The non-planar
diagram, however, is best expressed in terms of the effective
cutoff \bref{eff-cutoff}. If we just let $\L \rightarrow \infty$,
then the effective cutoff remains finite
\be \label{limit-1}
\L_{eff} \underset{\L \rightarrow \infty}{\longrightarrow} {1 \over \pop} \,.
\ee
As explained above, this finiteness is due to the fact that
the internal phase in \bref{gnp} acts as a regulator.

Notice however that \bref{limit-1} is IR divergent after this limit, but
it was perfectly IR finite before the limit. One is not even allowed
to set $\t$ back to zero after all the UV modes are included!
It is clear that the IR physics depend on the exact UV physics of
the theory. This is in frontal clash with the Wilsonian picture
of renormalization, where field theories should
be thought of as coming with an explicit UV cut-off. Renormalization
flow is then a flow towards the IR governed by equations that
impose that the long distance physics do not depend on
the specific way we used to regularize the theory at the shortest
distances.

This point of view seems to be not applicable to NC theories. Despite
being causal and unitary (as we will shortly discuss),
the physics that we would observe at energy scales of the
order of $1$ $eV$ are radically different if our theory
is given a cut-off at $100$ $GeV$ or at $10^{10}$ $GeV$! It turns out
that precisely the modes between the latter two scales
can cause long distance divergences.

One possible reason why this mixing occurs is to interpret the
uncertainty relation of NC spacetimes
\be
\Delta x \, \Delta y \ge \t \,,
\ee
as telling us that specifying the theory at shorter and shorter
distances in one direction affects more and more the long distance
properties in the other direction.

{\bf On $\b$-functions:}

Here comes an important issue that should be made clear
before proceeding: how do we define the $\b$-functions
in such theories? Do we consider non-planar diagrams as divergent or
as finite? As far as I understand, before discussing $\b$-functions,
one should first give a clearer meaning to the renormalization, \ie
to the issue of dealing with infinities in these theories (if possible
at all!). Nonetheless, the standard rule is to proceed by considering
that non-planar diagrams are auto-regulated by their phases\footnote{In
some more complex Feynman diagrams the auto-regulation may require
a case-by-case check.}, so that all divergences come from the
planar ones. As their divergent structure coincides with that in
commutative theories, the $\b$-functions typically coincide (again,
up to symmetry factors discussed in footnote \bref{foot-factor}).
Let us mention in support of this way of proceeding that the $\b$-functions
that we will extract from the string duals of these theories
match with the ones obtained in this way.

\subsection{Optical theorem and unitarity} \label{ch3:sec:optical}

Having established the peculiar behavior of the NC $\phi^4$
perturbative expansion, one could wonder whether it is all an
artifact of having been dealing with an ill-defined quantum
theory. In this section we show that whereas magnetic theories
are unitary at 1-loop, electric ones are not. This
is in complete agreement with the fact that the magnetic ones
arise as the low energy description of string theory (and
hence inherit unitarity) whereas electric ones do not.
Non-unitarity is normally a sign of not having dealt with
a complete set of degrees of freedom, which can be interpreted
in this case as the string modes that we showed not to decouple
in the electric cases.

The optical theorem follows from demanding unitarity of the S-matrix
and yields an exact (non-perturbative) requirement. The main power
of it is that it provides a non-linear relation from which it follows
that the imaginary part of the 1 to 1 forward scattering
of a field is equal to the probability of decaying {\it into an
arbitrary final state} (made of an arbitrary number of particles).
When the S-matrix is considered in perturbation theory, it
yields for a $\phi^3$ theory
\be \label{optical-1}
\includegraphics[width=5in]{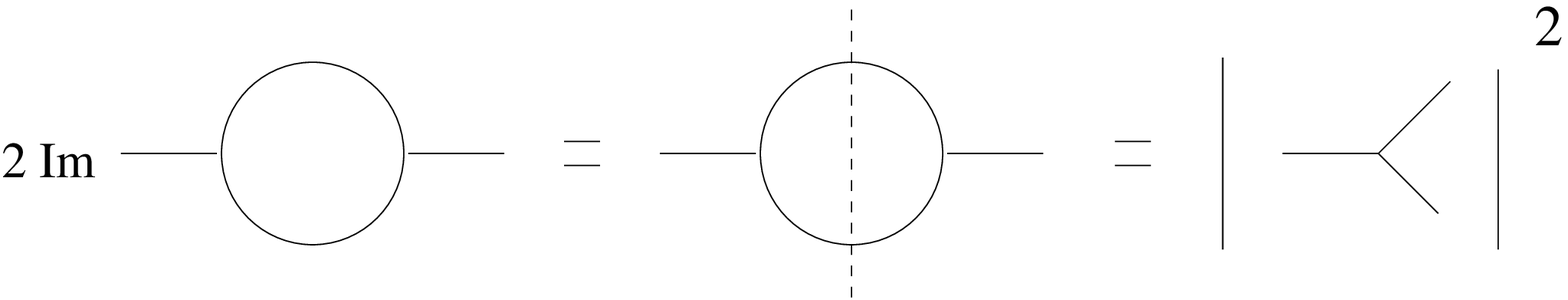}
\ee
The first non-trivial condition that follows is at order
$\l^2$. Again we do not give details on this computation
since a similar one will be carefully considered in section
\ref{ch3:sec:unitarity-non-rel}. We just mention that the result is~\cite{Gomis:2000zz}
\be \label{lhs1}
2 \, \mbox{Im} M  =\left\{ \begin{array}{cc}
{\lambda^2 \over 32\sqrt{p^2}} J_0\left[{\sqrt{1-4m^2/p^2}\sqrt{p^2 \, p\circ p} \over
2}\right]  & \mbox{, if  } p \circ p>0  \,,\\ \\
{\lambda^2 \over 32 \pi} \int_0^1 dx \,J_0 \left[ \sqrt{|p
\circ p|(m^2 +|p^2| x(1-x))}\right] & \mbox{, if  } p\circ p <0 \,,
\end{array} \right.
\ee
whereas
\be \label{rhs1}
\sum|M|^2 =\left\{ \begin{array}{cc}
{\lambda^2 \over 32\sqrt{p^2}} J_0\left[{\sqrt{1-4m^2/p^2}\sqrt{p^2 \, p\circ p} \over
2}\right]  & \mbox{, if  } p \circ p>0  \,,\\ \\
0 & \mbox{, if  }p\circ p <0 \,.
\end{array} \right.
\ee
We see that the optical theorem is verified or not depending only on the sign
of $\pop$ (the $\circ$-product was defined in \bref{pop}.
Let us reexamine the correspondence of this sign with magnetic
or electric backgrounds.
\bitem
\item For magnetic backgrounds we can box-diagonalize the matrix $\t$
so that its only nonzero entries are $\t^{23}=-\t^{32}=\t$. Then
\be
\pop = \t^2 \left[ (p^2)^2 + (p^3)^2 \right] \ge 0 \,.
\ee
So $\circ$ provides a definite positive inner product. If any of the
components $p^2$ and $p^3$ is nonzero, then $\pop>0$ and we are in the first
rows of \bref{lhs1} and \bref{rhs1}. The optical theorem is verified at order
$\l^2$. When both momenta are zero, the S-matrix is ill-defined because
of the IR singularities mentioned in the previous section.

\item For electric backgrounds, we can box-diagonalize the matrix $\t$
so that its only nonzero entries are $\t^{01}=-\t^{10}=\t$. Then
\be
\pop = \t^2 \left[ (p^0)^2 - (p^1)^2 \right] \,,
\ee
and the sign becomes negative as soon as $p^0 < |p^1|$. We are then
in the second rows of \bref{lhs1} and \bref{rhs1}, so that the
optical theorem is not verified. The way to understand the zero
appearing in the second row of \bref{rhs1} is by noticing
that the condition $p^0 > |p^1|$ requires a space-like initial
momentum, which makes it kinematically impossible to decay into two massive
on-shell particles.
\eitem
The result is exactly analogous in the NC $\phi^4$ theory.

\subsection{Trying to restore unitarity. The $\chi$-particles.} \label{ch3:sec:chi-particles}

In the previous section, we illustrated the loss of unitarity
in a $\l^2$ computation for a NC $\phi^3$ theory. This is the
way this issue was originally discussed in the literature~\cite{Gomis:2000zz}.
There is however a much simpler diagram in the massless $\phi^4$ theory,
which we will use now to study a possible restoration of unitarity.
This diagram arises from the condition of order $\l$ imposed by
the optical theorem. Since there is no way to produce an order $\l$
contribution in the RHS of the optical theorem (because it is
always the square of something), we deduce that the imaginary
part of the following diagram should vanish
\be \epsfig {file=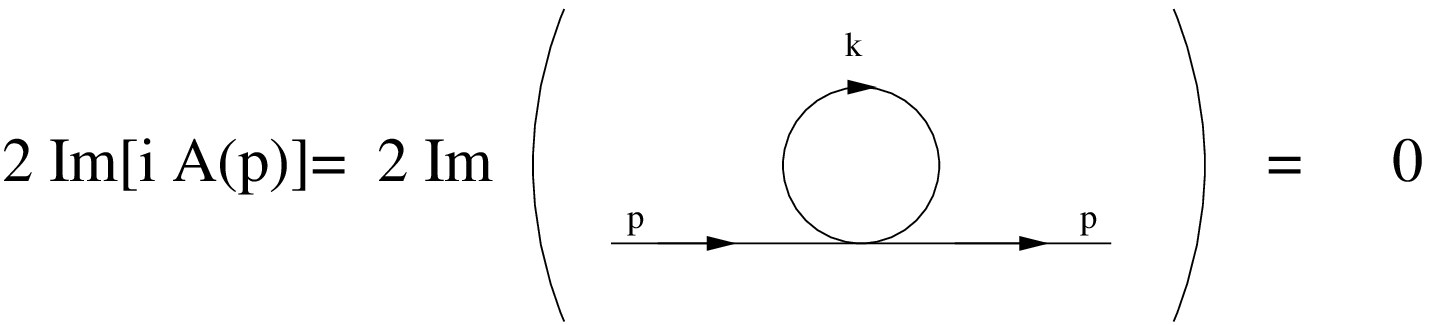, height=2cm, width=9cm} \ee
In purely electric backgrounds, a quick computation yields
\bea \label{Alvarez-Gaume:2001ka}
2\, \mbox{Im} \,[A(p)] &=& \int
{d^4k \over 2 (2\pi)^3} \,  \rho(\l,\t)  \,\delta^{(4)}(p-k) \,, \\
\rho(\l,\t) & \sim &  { \l \over  \t^2} \d\left[(p^0)^2-(p^1)^2\right]\,,
\label{gaume2}
\eea
which is clearly nonzero and it therefore shows the loss of unitarity.
The point of writing the result in the form \bref{Alvarez-Gaume:2001ka} is that
it immediately allows us to reinterpret its RHS as the contribution
we would have gotten had we included in the original Lagrangian
an extra field $\chi$ with coupling $\l_{\phi\chi}=\rho^{1/2}(\l,\t)$ to
our original field $\phi$. Due to the delta function in \bref{gaume2},
this particle must have the rather strange dispersion relation
$k^0=|k^1|$. Note as well that the coupling diverges in the $\t \rightarrow
0$ limit, a region affected by the discussed UV/IR problems.

However, this interpretation is jeopardized by the fact that if
one repeats our analysis for higher powers of $\l$,
the masses of the new required $\chi$-particles are
tachyonic.
This somehow transforms the problem of perturbative  unitarity
into an inconsistency of perturbation theory about an
unstable vacuum.
The conclusion is that although we seem to have a tempting
interpretation of the loss of unitarity, and of the possible
way of restoring it, it remains almost impossible to verify
it quantitatively, at least in perturbation theory.

\section{Unitarity of non-relativistic NC theories} \label{ch3:sec:unitarity-non-rel}

The purpose of this section is to summarize what we have learnt until
now about NC field theories and their quantum properties, and to
illustrate it in detail for a particular interesting model.
Let us summarize
\begin{enumerate}
\item
We have explicitly showed that
NC $\phi^4$ and $\phi^3$ field theories
are non-unitary when the non-commutativity
involves the time coordinate, a case in which
they also exhibit an acausal behavior~\cite{Seiberg:2000gc}.
This is related to the fact that only magnetic theories
arise as decoupling limits of string theory.

\item
We have also seen
that order by order in $\l$ one can try
to restore unitarity by adding extra degrees
of freedom ($\chi$-particles), although
their masses are typically tachyonic. They
are proposed to correspond to the instability
of the string vacuum in the zero slope limit
of electric backgrounds.

\end{enumerate}

These properties will now be reexamined for
a {\it non-relativistic} and NC field theory.
It is not straightforward to extend properties $1$ and
$2$ (above) to non-relativistic theories,
where the treatment of space and time is completely
different, and time non-locality may not lead to the
same consequences.

The particular model we will study is
a non-relativistic $NC$ $\phi^4$ theory in 2+1 dimensions, which
can be nicely viewed as the realization of the
Galileo group with two central extensions (the mass
and the non-commutativity parameter $\t$)\cite{Lukierski:1997br}.
Our conclusions will be that property $1$ is still
valid, whereas property $2$ fails: there is no way
to restore unitarity by the addition of new states
even at first non-trivial order in $\l$.
This casts even more doubts on the mentioned interpretation
of the $\chi$-particles in the discussed relativistic models.

\subsection{Four Points Function and Unitarity} \label{ch3:sec:4-points}

To set up our framework, we define a non-relativistic
NC scalar field theory in $D=2+1$ with quartic interactions
by the Lagrange density  \footnote{Some perturbative properties of
this model in the magnetic case have been studied in~\cite{Gomis:2000pf}
and some exact results can be found in~\cite{Bak:2000dx}.} $^,$ \footnote{
It can be seen that having taken the other possible
ordering of the vertex, \ie $\dphi*\phi*\dphi*\phi$,
would have led to exactly the same unitarity problems.}
\be
\call_{nr}=\dphi \left(i\pa_t + {\vec{\nabla}^2 \over 2} \right)
\phi - {\lambda \over 4} \, \dphi*\dphi*\phi*\phi.
\label{lnr}
\ee

Following the steps described in section \ref{ch3:sec:optical},
we will study the unitarity of the theory by checking whether the
Optical Theorem is fulfilled at the level of two particles scattering.
The analog of figure \ref{optical-1} in our case is then
\be  \epsfig {file=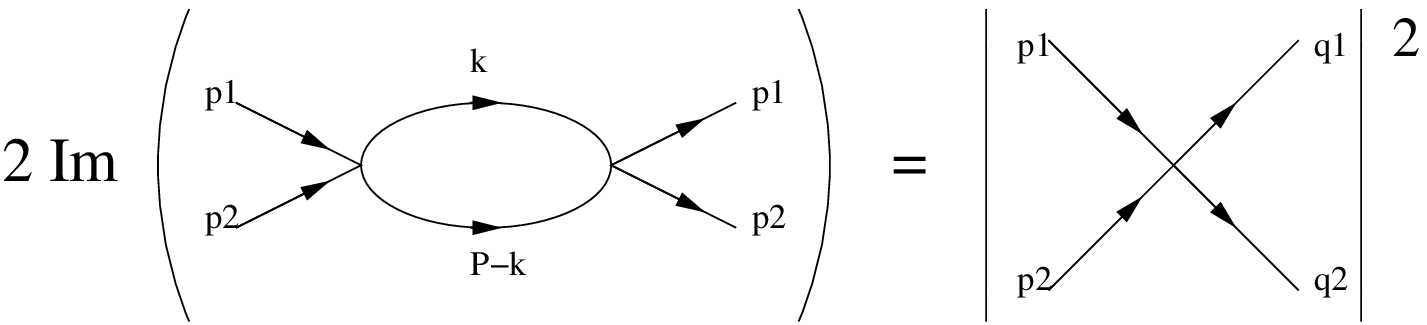, height=2cm, width=9cm} \ee
The left hand side (LHS) and the right hand side (RHS)
can be written as
\bea
\mbox{LHS} &\equiv& 2\, \mbox{Im} \bl
-i \frac{\lambda^2}{2} \, \cos^2 {\otp_1  p_2 \over2} \,
\nn && \times \int {d^2k dk^0 \over (2\pi)^3}\,
{cos^2 { \tp  k \over 2} \over
[k^0 - \frac{\kvec^2}{2} +i\e]
 [p^0-k^0 - \frac{(\pvec-\kvec)^2}{2} +i\e ]} \br \,,
\label{LHS} \\
\mbox{RHS} & \equiv & {\lambda ^2\over 4\pi} \cos ^2 { \otp_1  p_2\over 2}
 \int d^3 q_1 \int d^3 q_2 \,
 \delta (q^0 _2 - {{\vec {q_2}}^2
\over {2}})
\delta (q^0 _1 - {{\vec {q_1}}^2 \over {2}}) \nn
&  & \times \,\,\del3 (p_1+p_2-q_1-q_2)
 \cos ^2 {\tilde{q}_1 q_2 \over 2} \,,
\label{preRHS}
\eea
where $\otp^{\mu} \equiv p_{\nu}\t^{\nu \mu}$,
$P^{\mu}=p_1^{\mu}+p_2^{\mu}$ and the products
are defined by $pk \equiv p^0k^0-\pvec\cdot\kvec$.
Using the identity $2cos^2x=1 + \cos2x$ for the
cosine involving integrated momenta, both
sides can be written as a sum of a planar integral
plus a non-planar one. It is straightforward to
show that the planar parts are identical in both
sides. Therefore, the only job left is to check
for the non-planar ones. The RHS can be written
as
\be \mbox{RHS}|_{\mbox{npl}}=
\frac{\lambda^2}{4}
cos^2 \otp_1p_2 \int \frac {d^3 k}{2\pi}
\delta (P^0-k^0-\frac{(\opvec - \kvec )^2}{2})
\delta (k^0-\frac{\kvec^2}{2}) \cos \tp k \,,
\label{RHS}
\ee
irrespective of whether the background is electric
of magnetic. The LHS requires considering both cases
separately.

\vskip 4mm

\subsubsection{Magnetic Case} \label{ch3:sec:magnetic-case}

Take the non-commutativity only in the two spatial
coordinates $[x,y]=i\t$. In this case we have $\tp^0=0$
and so we can take the cosine of \bref{LHS}
out of the $k^0$ integral. Therefore, we can perform the $k^0$ integral using
Cauchy's theorem. We are left with
\be
\mbox{LHS}|_{\mbox{npl}}\,=\,
- {\lambda^2 \over 2(2\pi)^2} \cos^2{\otp_1 p_2 \over 2} \,
\mbox{Im} \int d^2 k {cos\tp k \over P^0 - \frac{\kvec^2}{2}
- \frac{(\opvec-\kvec)^2}{2}+i\e}.
\ee

The imaginary part is extracted by using
that $(x+i\epsilon)^{-1}=
{\mathcal{P}} \frac {1}{x}-i\pi \delta (x)$, and it is
then straightforward to show that we obtain exactly \bref{RHS}.
It can be easily seen that these last two steps are equivalent to
replacing the internal propagators by delta functions.
Indeed, this is nothing but a proof that
the cutting rules are valid for the magnetic case.
Notice that we have been able to check the Optical Theorem
to all orders in $\t$.

\vskip 6mm

\subsubsection{Electric Case} \label{ch3:sec:electric-case}

Now, take non-commutativity to affect space and time,
\ie $[t,x]=i\t$. The main difference with respect to the
magnetic case is that now $\tp^0 \neq 0$ and, therefore,
the cosine factor in \bref{LHS} cannot be taken out of the
$k^0$ integral.
We will find that the order zero $\t$-term is different
from the one we obtain in expanding the RHS \bref{RHS}. Furthermore,
a linear term arises, in contrast with the RHS, where the
first $\t$ term is quadratic.
Here one needs to go through Feynman parameters and residue integrals.
We arrive to:
\be \mbox{LHS}|_{np}\, =\,
{i \lambda^2 \over 16\pi}
\int_0 ^1 dx {1\over |1-2x|} \,\, \big\{ e^{i f(P,\t,x)} \Omega(P^0,\t)
+ e^{i{f(P,-\t,x)}} \Omega(P^0,-\t) \big\} \,,
\label{intx}
\ee
with
\bea
f(P,\t,x) &\equiv&
\frac{|\tp_0|}{2|1-2x|}
\bl 2P^0(1-x)-\opvec^2x(1-x)) \nn &&+
\frac{\vec{\tp}^2}{2|\tp_0|} |1-2x|
-\vec{\tp} \cdot \vec{P} (1-x) \br \label{def} \,,\\
\Omega(P^0,\t) &\equiv& \Theta(\tp^0) \Theta(x-\undos)
+ \Theta( \tp^0)\Theta(\undos-x)\,,
\eea
where we have chosen the symbol $\Theta(x)$ to name the
step function, not to be confused with the
non-commutative parameter $\t$.

The integral \bref{intx} cannot be solved exactly.
However, every term in
\bref{def} is linear in the non-commutativity
parameter $\t$, since $\tp^0 = \t P^1$ and
$\vec{\tp}=(\t P^0 ,0)$. Therefore,
we can expand the exponentials of \bref{intx}
in order to obtain a power series in $\t$ in the LHS.
Some care is needed due to the singular behavior of \bref{def}
about $x=\undos$, so we will only expand the exponential of
the non-singular terms. Taking all this into account we finally obtain
\be
\mbox{LHS}|_{np} \, =\, \,
{\lambda^2 \over 16} \,\,
\,\,
+ \,|\t| \,{\lambda^2 \over 32\pi} \, \left(
{|P^1|\; \opvec\;^2} + {2(P^0)^2 \over  |P^1|} \right)
+ \, \, \l^2 \calo(\t^2) \,.
\label{result}
\ee
The first term arises from expanding a gamma
function with imaginary argument, in contrast with
the logarithms one finds in the relativistic case
\cite{Gomis:2000zz}. Its value is exactly half of its
RHS counterpart \bref{RHS}, and so unitarity
is violated. The linear term is not present in
\bref{RHS} either.
Notice that only the absolute value of $\t$ appears in
\bref{result}, in agreement with the original symmetry $\t
\rightarrow -\t$ in \bref{LHS}.

\subsection{Two Points Function and the failure of $\chi$-particles.} \label{ch3:sec:2-points}

Does the method of adding new fine-tuned degrees of freedom
to restore unitarity work in this non-relativistic case?
Let us try to reproduce the analysis of section \ref{ch3:sec:chi-particles}
and apply the optical theorem to the one-to-one scattering
amplitude, which implies
\be \epsfig {file=Gaume.eps, height=2cm, width=9cm} \ee
A  short calculation shows that
\be
A(p) =
-i \lambda \int {dk^0 d^2k \over (2\pi)^3}
{cos^2 {\otp k \over 2} \over k^0 - \frac{\kvec^2}{2}
+i\e} =
{-i \l \over 16 \pi} \, \Lambda^2 \, \, + \, \,
i {\lambda \over 8\pi}
{exp\left({ \vec{\otp}^2 \over 2 \otp^0}\right) \over |\otp^0|}.
\label{two3}
\ee
In obtaining this result, we have introduced a hard cutoff
for the planar integral (it diverges as in any
commutative theory).
For magnetic cases, we have $\otp^0=0$. If we take this limit
in our result \bref{two3}, we recover the result of
\cite{Gomis:2000pf}, \ie $A(p) = \l
\delta^{(2)}(\pvec) / 4\t^2$. It has no imaginary part,
and so unitarity is preserved. On the contrary,
for electric cases $\otp^0$ is finite, and
there is always an imaginary contribution
\be
2\, Im \,[A(p)] \,=\, {\l \over 4 \pi} \,
{cos {\vec{\otp}^2 \over 2 \otp^0} \over |\otp^0|} \,,
\label{cos-a}
\ee
which should be compared to the relativistic formula \bref{Alvarez-Gaume:2001ka}.
Our result \bref{cos-a} can not be interpreted as coming from
new particles that couple to our original field, not even
if  we allow for
the coupling to depend on the momenta.
This is due to the fact that our expression
is a smooth function of the momenta, and so it
can never be written as a delta function times
a coupling. There is no way to have momentum conservation
in the vertices then.

\chapter{Supergravity duals of Noncommutative field theories}\label{ch:nc-sugra}

In this chapter we link the three major subjects of this thesis:
the AdS/CFT duality, its extension to less than maximally
supersymmetric cases, and the noncommutative field theories.
We develop the technical tools to construct the closed string
duals of maximal and less than maximal NC field theories
and apply it to obtain the duals of
\tem{
\item a $U(N)$ NC $\caln=1$ SYM in 3+1 (section~\ref{ch6:sec:nc-mn},
reported in~\cite{Mateos:2002rx}),
\item a $U(N)$ NC $\caln=2$ SYM in 2+1 (section~\ref{ch6:sec:nc-gm},
reported in~\cite{Brugues:2002ff,Brugues:2002pm}).
}
For the first theory we discuss a good amount of nonperturbative
properties derived from the closed string dual: the presence of
UV/IR mixing, confinement, the $\b$-function and chiral-symmetry
breaking. We will see an interesting property which is absent
from the commutative counterpart: the new scale introduced by
the noncommutativity can be fine-tuned so that it allows for
a decoupling of the KK modes.
In constructing the dual of the second theory,
we make some precise general remarks about the effect of the
'susy-without-susy' phenomenon in supergravity solutions
of NC theories (section~\ref{ch6:sec:susy-without-susy})
and we analyze its moduli space.

Needless to mention, all cases discussed here involve
only magnetic NC theories which, unlike the electric ones,
do not suffer from unitarity of causality problems.

\section{Introduction} \label{ch6:sec:intro}

The analysis of NC theories performed until now was
purely field-theoretical and perturbative. We saw that one of their most
amazing properties is the UV/IR mixing in the non-planar Feynman diagrams
of the theory, a property that frontally clashed with the Wilsonian
interpretation of renormalization.

The possibility of studying these theories by means of AdS/CFT-like
dualities has shed new light on the subject. The supergravity dual
is supposed to capture non-perturbative properties of these theories
(at least at large $N$)
and renormalization flow has typically the simple interpretation
of flowing towards the horizon. The first proposed duals of
NC theories are due to Maldacena and Russo~\cite{Maldacena:1999mh} and
Hashimoto and Itzhaki~\cite{Hashimoto:1999ut}.
In particular, one of these backgrounds is dual to the NC deformation
of the usual $\caln=4$ SYM in 3+1. We will later discuss how such backgrounds
may be constructed and concentrate now on their physical consequences.

\tem{
\item In the first place the NC solution reduces to \5ads very close
to the horizon. By the usual radius/energy relation this implies
that in the deep IR the NC field theory reduces to the commutative
one. This result is far from trivial as the UV/IR mixing showed
that the physics at distances much larger than the NC scale $d\gg \sqrt{\t}$
did not decouple from those at $d\ll \sqrt{\t}$. The supergravity result
would seem more trustable as it is not based on any perturbative
(within supergravity, of course) artifact like Feynman diagrams.
They found however that the solution started deviating from \5ads
at scales of order $d \sim R \sqrt{\t}/l_s$, which in the
supergravity approximation ($R \gg l_s$) is much larger
than the expected $d \sim \sqrt{\t}$.
The UV/IR mixing could be responsible for modifying the physics
until such large distance scales.

\item There were some extra difficulties in setting up
the computation of Wilson loops, as we will show in a particular
case below. They found however that the very deep IR behavior is
the same as in the commutative $\caln=4$, \ie the energy
is proportional to the inverse distance between quarks.
}
There are two reasons why these results should be taken with care.
The first one is that, as we discussed in section~\ref{uv-ir},
the UV/IR mixing is mostly present in the non-planar sector of the theory.
We know however that all such diagrams are suppressed against
planar ones by factors of $N$. This means that, as supergravity
typically requires $N\gg 1$, we may be dealing with backgrounds
which are dual to NC theories collapsed to the planar sector only.
This could explain the coincidence in the deep IR between the
commutative and the NC supergravity solutions.\footnote{See
\cite{Cai:1999aw,Cai:2000hn} for an extension of the results in
\cite{Maldacena:1999mh} to various other maximally supersymmetric
NC theories via supergravity, which indicate
that commutative and NC field theories may have the same
number of degrees of freedom.}

The second one is that the $\caln=4$ theory is supposed to be finite as
a quantum theory. If this is so, then there is no ambiguity
on how to compute even the non-planar diagrams, so there is
simply no UV/IR mixing. The earlier than expected deviation
from the commutative solution would be then due to a UV/IR mixing
at strong coupling which would not be visible in perturbation
theory.

This last comment makes it more interesting to extend the duality
to non-conformal NC theories which are plagued with UV/IR mixing
in almost all observables, and see what supergravity can tell us
about it. We will do so in the remaining sections and show that,
at least in the case of NC $\caln=1$ SYM in 4d, there seems
to be a strong UV/IR mixing that renders the commutative
and the NC backgrounds different at all scales.

\section{Constructing solutions dual to NC theories with
less than 16 supercharges}

The aim of this section is to describe the two techniques
that have been used to construct supergravity duals
of NC field theories with less than 16 supercharges.
Essentially we need to find IIA/IIB backgrounds
with a nonzero $B$-field. As we want to study only
magnetic noncommutativity, only the space/space components
of $B$ are turned on, and they must be constant along the
directions of the brane. The requirement that we have
less than 16 supercharges leads us to consider D-branes
which wrap cycles of special holonomy manifolds.

Note that we do not want to put the $B$-field along the
wrapped directions but along the flat noncompact ones
(see figure~\ref{ncbrane}).
This is because we want to end up with a NC-theory in the
flat directions at distances much larger than the cycle.
This means that {\it we do not introduce any flux along
the special holonomy manifold}, which implies that all the
discussion about covariantly constant spinors, special holonomy, etc.,
is unchanged. The only effect of the $B$-field will be
a modification of the dilaton and the necessity of turning on
another $RR$-potential. This will be understood when we discuss
the second method.

\clfigu{10}{5}{nc-brane-c}{The magnetic B-field is turned on
only in the noncompact part of the branes worldvolume,
leaving untouched the special holonomy manifold.}{ncbrane}

\subsection{Method one: brute force} \label{ch6:sec:method-1}

The more pedestrian method consists on making a bosonic ansatz for
all the fields that one thinks that should be turned on,
then solving the supersymmetry variation of the fermions,
and then checking explicitly that they solve the equations
of motion. Even if one imposes all the isometries in the
ansatz, the process can be 'extremely painful'. It can be however carried out and we
will apply it later to construct the NC deformation of
the wrapped D6 branes that we discussed in the previous
chapter. We were probably successful because the D6 branes
have a simple description in 11d supergravity.

Despite being technically complicated, this method has two
advantages. First, the construction is very transparent as
it is just a matter of making an ansatz for what we look for
and solving it. Second, and most important, it necessarily
provides the Killing spinors of the background. This can be
useful for a number of reasons. First, as we saw in section
\ref{ch5:sec:calibrated-is-susy}, one can construct most of
the covariantly constant tensors
in the special holonomy manifold; in particular, the calibrations.
Second, it allows one to understand which kind of compactifications
will be free from 'supersymmetry without supersymmetry' problems
and which ones will not. A remarkable result that we obtained
from this method is that, in general, {\it there is no way to
find the supergravity duals of NC theories in the
corresponding gauged supergravity.} We will study this in
section~\ref{ch6:sec:susy-without-susy}.

\subsection{Method two: T-dualities} \label{ch6:sec:method-2}

The method which finally turns out to be easier to implement
exploits the T-duality that is believed to exist between
type IIA and IIB compactified on a circle; in particular
we will use the fact that a T-duality along a diagonal direction
to a Dp-brane (\ie along a line with nonzero projection along
one direction tangent to the brane and one transverse to it)
produces Dp-D(p-2) bounds states with a background
$B$-field. This idea was proposed in~\cite{Breckenridge:1997tt, Costa:1998zd} before the
understanding of noncommutative field theories as low-energy limits of string
theory, and the technique has been greatly improved in the past two years.
Let us briefly review how the original and the improved methods work, and
why they are equivalent.

Suppose we have a Dp-brane in flat space along the
directions $\{x^0,...,x^p\}$. We would like to perform
a T-duality along a diagonal axis in the plane $(x^p,x^{p+1})$. Equivalently,
we can rotate the brane in that plane and simply T-dualize along $x^{p+1}$.
In the last picture, the originally tilted brane had coordinates satisfying
\be \label{orig-bcs}
\partial_{n}\left(x^p + \tan\t \,  x^{p+1}\right)=0 \sac
\partial_{t}\left(x^p - \cot \t \, x^{p+1}\right)=0 \,,
\ee
where $\partial_n$ and $\partial_t$ are normal and tangent derivatives with
respect to the string worldsheet's boundary, and $\t$ is the angle of rotation. Now,
T-duality along $x^{p+1}$ exchanges Neumann and Dirichlet conditions,
so it transforms \bref{orig-bcs} into
\be
\partial_{n} x^p + \tan\t \, \partial_{t} x^{p+1}=0\sac
\partial_{n} x^{p+1} - \tan \t \, \partial_t x^{p}=0\,.
\ee
This mixed boundary conditions can be interpreted as those of a
string attached to a $D(p+1)$ brane in the presence of a $B$-field
\be
\partial_n x^{\mu}\, - \,\calf ^{\mu}~_{\nu} \, \partial_t x^{\nu}\,=\,0 \,,
\ee
where $\calf _{[2]}=B_{[2]}+2\pi \a' F_{[2]}$ and, in this case, we
have induced $\calf _{12}=-\tan\t$. Such gauge invariant field strength
produces D(p-1) charge in the world-volume of the D(p+1) through
the Wess-Zumino term.

This is, {\it grosso modo}, the original method proposed in~\cite{Breckenridge:1997tt,Costa:1998zd},
where it was applied to several cases of branes in flat space to
produce various Dp-D(p-2) bound states. What we have seen now is the
open string picture of the method, which is a rather simple one.
When moving to the closed string picture, the method
still had some technicalities that made it difficult to
implement in cases other than the description of flat branes in
flat backgrounds.
Maybe the most relevant difficulty was that T-dualities
need to be performed along isometries. In our case, the T-duality was
performed along a diagonal direction involving one coordinate along the
brane and one transverse to it, and this is not
an isometry of the supergravity solution. Originally, this was solved by
delocalizing the Dp branes along
the $x^{p+1}$ axis before applying the T-duality,
for example by adding an infinite number of parallel branes.
In the supergravity solution of flat p-branes, this just amounts to
changing slightly the form of the harmonic function $H(r)$: instead of being
harmonic in the whole transverse space of dimension $10-p-1$, one can choose it
to be harmonic in one dimension less, \ie in a $10-p-2$ space. Schematically,
\bea {\rm Dp \,\,localized:} \espai
H(r)=1+{1\over r^{7-p}}\, , \espai r^2=\sum_{i=p+1}^{10} (x^i)^2\,.\\
{\rm Dp \,\,delocalized \,\, in \,\, x^{p+1}:} \espai
H(\tilde{r})=1+{1\over \tilde{r}^{6-p}}\, , \espai \tilde{r}^2=\sum_{i=p+2}^{10} (x^i)^2\,.
\label{replace}
\eea
As can be seen, delocalizing a brane is fairly simple when we are in flat
space and we know the whole geometry solution. The difficulty would increase
if we were only given the near horizon region. There, the  harmonic function
can be very hard to recognize depending on the coordinates
we are given. Indeed, if we also abandon flat space backgrounds,
the transverse space to the brane is typically a sophisticated fibre bundle, and
a better method to delocalize the brane is needed.

The way this can be achieved is just by starting with a brane of one dimension
higher, say a D(p+1) along $\{x^0,...,x^{p+1}\}$ and by T-dualizing along $x^{p+1}$.
In the supergravity dual, one just needs to use the T-duality
rules to transform the closed string background. In flat space, it is easy to
check that this is equivalent to the replacement \bref{replace}, no matter if we
started with the whole geometry or just the near-horizon.

The last refinement of the original method consists on substituting the rotation
of the delocalized brane by a more mechanical algorithm. It just exploits the fact
that rotating the brane is equivalent to: first T-dualizing one of the world-volume
directions, then turning on a constant $B$-field, and then T-dualizing back.

Therefore, the improved method for producing the noncommutative configurations can
be summarized, from a supergravity point of view, as follows (see also figure~\ref{tdualfig})

\avall

{\bf (i)} Start with a supergravity solution of a Dp along $\{x^0,...,x^p\}$. We require
that at least two of these directions, say $\{x^1,x^2\}$,  are flat, while the others
may or may not be wrapped along any compact cycle. We compactify $x^1$ and $x^2$ on
a torus so that $\partial_{x^1}$ and $\partial_{x^2}$ generate circle isometries.

{\bf (ii)} T-dualize along $x^2$. This produces a D(p-1) brane delocalized along $x^2$.

{\bf (iii)} Rotate the D(p-1) along the $(x^1,x^2)$ plane by T-dualizing along $x^1$,
turning on a constant $B$-field $B=\T\, dx^1 \wedge dx^2$, and T-dualizing along $x^1$ again.
The introduction of $B$-field does not modify the equations of
motion because its field strength is zero, and the Chern-Simon's term of the
corresponding supergravity Lagrangian is a total derivative.

{\bf (iv)} T-dualize back on $x^2$. This is the diagonal T-duality
of a delocalized and rotated brane that we mentioned. It produces
a bound state of Dp-D(p-2) in the
background of a non-trivial $B$-field. Finally, uncompactify $\{x^1,x^2\}$
by sending the radii of the torus to infinity.

\cltfigu{14}{8}{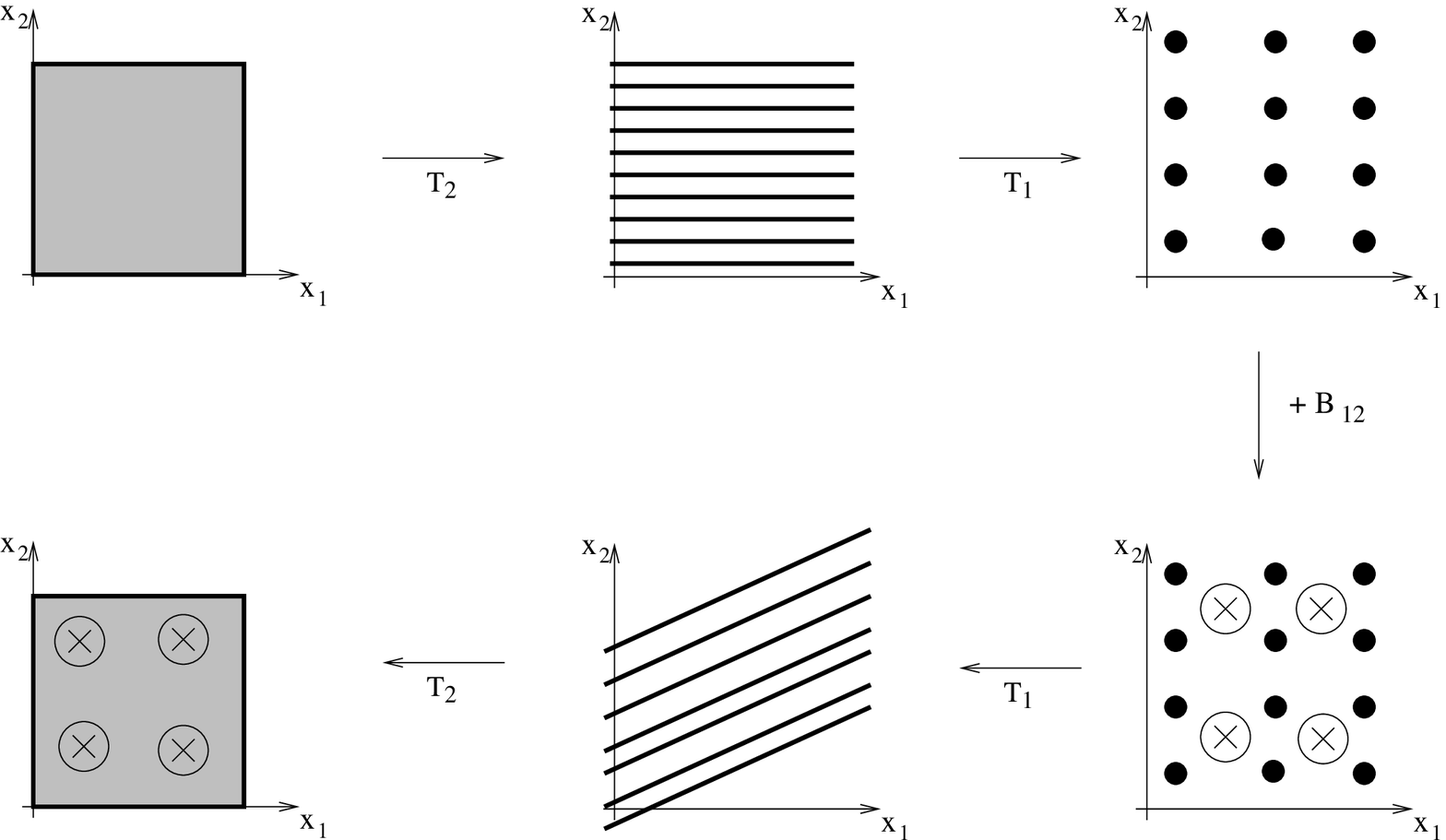}{The procedure of consistently introducing
a magnetic $B$-field in a D2-brane. Thick lines are the worldvolume of
the branes, circles-with-crosses are magnetic fields and $T_i$ refers
to a T-duality along the axis $x^i$.}{tdualfig}

\avall

Supersymmetry is preserved throughout this procedure if the spinors originally
did not depend on $x^1$ and $x^2$~\cite{Bergshoeff:1995cb}, as is typically the case. The introduction
of the $B$-field in step $(iii)$ does not break supersymmetry either, since only
$H_{[3]}=dB_{[2]}=0$ appears in the supersymmetry variations of supergravity.

We conclude this section by making a few remarks about the improved method.
The first is that it generalizes easily to include $B$-fields with rank
higher than two. The second is the non-trivial fact that, as pointed out
in~\cite{Sundell:2000jx}, when the $B$-fields are magnetic,
the following diagram holds.
\be
\includegraphics[width=12 cm,height= 10cm]{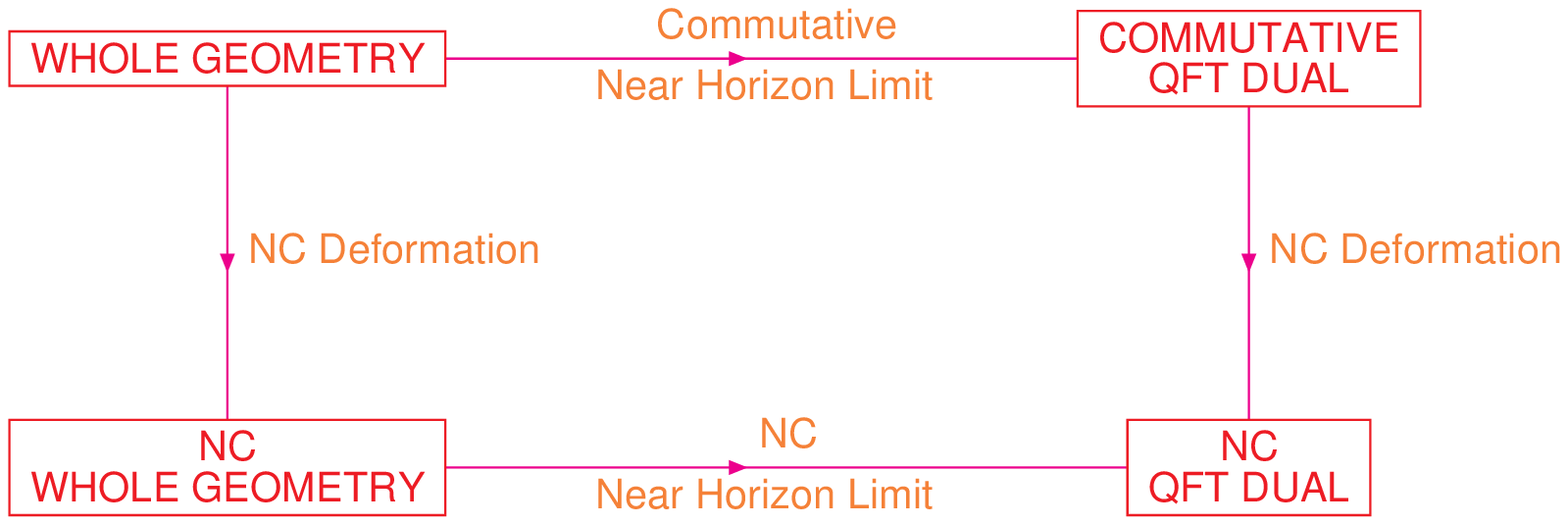}
\ee

This is crucial for our purposes, since in the  supergravity duals of
wrapped branes one only knows the near-horizon region. The third and last remark is that
the improved method has been widely used to obtain duals of maximally supersymmetric
field theories, as in~\cite{Maldacena:1999mh,Hashimoto:1999ut,Larsson:2001wt,Berman:2000jw},
and of $\caln=2$  as in~\cite{Buchel:2000cn}.

\section{The supergravity dual of the NC $\caln=1$ SYM in 3+1} \label{ch6:sec:nc-mn}

The purpose of this section is to apply the second method discussed
above to construct  the supergravity dual of the NC version of
the 'pure superglue' theory: an $\caln=1$ $SU(N)$ SYM in 3+1 without
matter supermultiplets. The commutative version
has one of the most interesting phenomenologies among all the supersymmetric
field theories, mainly because of the possibility of adding chiral matter.
This is not possible in theories with more supercharges as the multiplets
are too large and left/right matter cannot lie in different susy irreps.
The $\caln=1$ also enjoys other interesting non-perturbative phenomena like
chiral symmetry breaking by fermion bi-linears condensation and confinement.

The outline of what follows in this section is:
\tem{
\item
A brief discussion of one of the most successful supergravity duals
constructed until now, due to Maldacena and \nun and dual to the
commutative $\caln=1$ theory.
\item
An application of the second method to construct its NC deformation.
It will be shown that UV/IR effects seem to persist even in the deep
IR, unlike in the dual of a NC $\caln=4$. It is also shown that the
new scale introduced by the noncommutativity can be used to decouple
the KK modes and to a true 3+1 dimensional theory in the IR.
\item
A computation of the area of the basic Wilson loop in the
string theory side. The short distance behavior is different
from the commutative theory, but the confining phase is
also reached in the IR, with the same string tension
as in the commutative theory.
\item
A computation of the $\b$-function in the supergravity side
in both the commutative and the NC versions of this $\caln=1$.
}

\subsection{The NC deformation of the Maldacena-\nun background} \label{ch6:sec:nc-mn-deformation}

We studied in chapter~\ref{ch:wrapped} how to construct the supergravity
duals of field theories with less than maximal supersymmetry.
The explicit example we discussed involved D6-branes wrapping
\kah 4-cycles in $CY_3$ manifolds.
In the examples of
section~\ref{ch5:sec:twisting-field-theories} we discussed from a purely field-theoretical
point of view how to perform a twist in a 6d SYM theory
in order to put it in $\CR^{1,3}\times S^2$. We saw that
there were two possible twists, one of them preserving only
4 supercharges. At very low energies, or distances much larger than
the $S^2$, the theory is effectively the $\caln=1$ SYM in 4d.

The understanding of the twist allowed Maldacena and \nun to construct
the supergravity solution. Being a 6d theory, the natural branes to
look at are D5 branes in type IIB. As their transverse space is
$\CR \times S^3$, the natural supergravity to construct the solution
is 7d gauged supergravity, which appears upon reduction of IIB on
$S^3$. We skip the details here because all the steps are exactly parallel to
those of section~\ref{ch5:sec:d6-solutions}.

The final solution represents a stack of $N$ D5 branes wrapping
an $S^2$ inside a Calabi-Yau three-fold, and reads\footnote{We will set $l_s=1$
in this chapter.}
\bea
\label{metrica}
ds^2_{IIB}&=&e^{\Phi} \left[  dx^2_{0,3}+
N \, \left( d \rho^2+ e^{2g(\rho)} d\Omega_2 + {1 \over 4} (w^a-A^a)^2 \right)
\right]\,,
\\
F_{[3]}&=&dC_{[2]}= {N\over 4} \biggl[ -(w^1-A^1)\wedge(w^2-A^2)\wedge(w^3-A^3)  \nn
&& \espai \espai ~~~+  \sum_{a=1}^3 F^a\wedge(w^a-A^a)\biggl]\,,
\\
\label{com-dilato}
e^{2\Phi}&=&e^{2\Phi_0}\,\,{\sinh 2\rho \over 2 e^{g(\rho)}} \,,
\eea
where the definitions of the quantities appearing above are written
in the appendix~\ref{ch6:sec:conventions}.
To construct the NC deformation we just use the second method explained
in the previous section. We skip the intermediate steps and give the result
for the case of a magnetic $B$-field along the $\{x^2,x^3\}$ plane,
\bea
ds^2_{IIB}&=&e^{\Phi} \left[ dx^2_{0,1}+
 h^{-1} dx_{2,3}^2 +
N \, \left( d \rho^2+ e^{2g(\rho)} d\Omega_2 + {1 \over 4} (w^a-A^a)^2 \right)
\right]\,,
\nn  \label{nc-mn}
\\ \label{nc-dilato}
F_{[3]}&=& dC_{[2]}= {\rm unchanged}\,,
\\
e^{2\tPhi}&=& e^{2\Phi} h^{-1}\,,
\\
B_{[2]}&=&-\, \Theta \, {e^{2\Phi} \over h} \,dx^2 \wedge dx^3\,,
\\
C_{[4]}&=&\Theta \, {e^{2\Phi}\over 2h} \, C_{[2]} \wedge dx^2 \wedge dx^3 \,,
\eea
where we defined
\be \label{last}
h(\rho)=1+\Theta^2 e^{2\Phi}\,.
\ee
Notice that we use $\tPhi$ for the new value of the dilaton and $\Phi$ for the
one appearing in \bref{com-dilato}. Also, $\T$ is the noncommutative
parameter\footnote{In this chapter we use a capital $\T$ to denote
noncommutativity not to confuse it with the angles of the spheres
we will have to deal with.}, while $C_{[2]}$ and $C_{[4]}$ are the type IIB
Ramond-Ramond potentials, with field strengths $F_{[3]}$ and $F_{[5]}$
respectively. Note as well that the $B$-field is not trivial
(its field strength is non-zero) but it is constant along the directions
of the brane.

A few remarks concerning (\ref{nc-mn}-\ref{last}) are in order. First of all, notice
that in the commutative limit $\T \rightarrow 0$ we have $h(\rho)\rightarrow 1$
and hence we smoothly recover the whole commutative background of MN.
Second, the solution describes a bound state of D5-D3 branes,
with the D3 smeared in the world-volume of the D5, and partially
wrapped in the two-sphere. If we denote by
$(\theta,\phi)$ the coordinates of the $S^2$ in \bref{nc-mn}, and
by $(\t_1,\phi_1,\psi)$ the ones of the transverse $S^3$, we can
summarize the configuration in the following array
\begin{center}
\[
\begin{array}{c | c c c c c c c c c c}
{\rm IIB} &x^0 & x^1 & x^2 & x^3 & \t & \phi & \rho & \t_1 &\phi_1 & \psi \\ \hline
{\rm D5}& -&-&-&-&-&-&&&& \\
{\rm D3}&-&- & &&-&-&&&&\\
B_{[2]} & & &- &-&&&&&&\\
\end{array}\]
\end{center}
Third, as in the MN solution, the metric is completely
regular at the origin.

\subsubsection{Validity of Supergravity and KK states} \label{ch6:sec:validity}
\label{sugra}

\begin{figure}[t]
\begin{center}
\includegraphics[width=6cm,height=4cm]{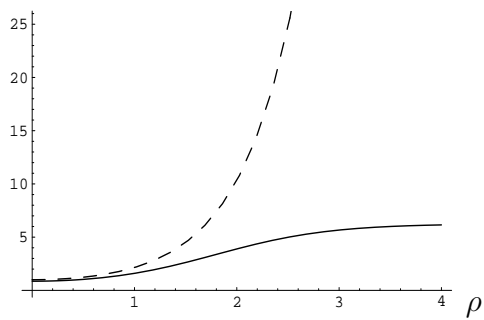}
$\rho$
\end{center}
\caption{Dilaton behavior as a function of the transverse coordinate.
The full line corresponds to $e^{2\hat{\Phi}(\rho)}$ (NC case)
and the dashed line to $e^{2\Phi(\rho)}$ (commutative case). While the former
remains finite at any value of the variable $\rho$, the latter blows up at infinity.
\label{figdila}}
\end{figure}

Before continuing with our discussion,
let us analyze the conditions for the NC-MN solution to be a valid approximation
of string theory. The main difference with respect to the commutative solution
is that the dilaton does not diverge at the boundary, due to the factor $h^{-1}$ in
\bref{nc-dilato}. It acquires its maximum value at infinity -see fig. \bref{figdila}-,
where $e^{\hat\Phi}\rightarrow \T^{-1}$.
So if we want to keep small everywhere the corrections coming from
higher order diagrams of string theory, we just need to demand
\be \label{condi}
\T\gg 1\,.
\ee

The second validity requirement comes from the curvature. In the noncommutative
geometry \bref{nc-mn}, the scalar of curvature ${\cal R}$ vanishes at infinity and it
acquires its maximum value at the origin. Requiring the curvature to be small everywhere
implies explicitly
\be \label{curva}
\vert {\cal R}\vert_{\rm max}\, =\, \vert {\cal R}(\rho = 0)\vert \,
= \, \frac{32}{3N}\,\frac{e^{-\Phi_0}}{ \left(1+ \T^2e^{2\Phi_0}\right)}
\, \ll \, 1\,.
\ee
In order to obtain a truly pure $\caln=1$  NC-SYM at low energies,
conditions \bref{condi} and \bref{curva} should be compatible with the
decoupling of the massive Kaluza-Klein modes of the wrapped $S^2$.
Since the only change in the metric with respect to the commutative
one is in the $(x^2,x^3)$-plane, the KK modes decoupling condition
is exactly the same as in~\cite{Maldacena:2000yy}, namely
\be \label{kk-res}
N e^{\Phi_0}\ll 1 \,.
\ee
It is easy to show that the three inequalities \bref{condi}-\bref{kk-res}
can be satisfied simultaneously if we choose the three parameters $N, \Phi_0$ and
$\T$ to verify
\be \label{final-condi}
\frac{e^{-3\Phi_0}}{\T^2}\,\ll \, N \, \ll e^{-\Phi_0} \ll \T \,.
\ee
We shall see in section 3 that a further
restriction will have to be imposed in order to study the quark-antiquark potential.
Note that \bref{final-condi} is saying that the price we have
to pay to decouple the KK states is to set $\T$ as the largest length scale
of the problem, so we cannot use this to end up with a 'realistic'
field theory.

\subsubsection{Properties of the solution and UV/IR mixing} \label{ch6:sec:uv-ir}

As we mentioned already, the NC-MN solution \bref{nc-mn} reduces to the commutative one
when we send $\T$ to zero. This corresponds to the fact that, classically,
noncommutative theories reduce to commutative ones in this limit.
As we saw, this remark does not hold quantum-mechanically and constitutes
one of the most interesting facts of the NC field theories, which is
related to the so-called UV/IR mixing.

Let us devote our attention to review the metric and the field content
of the noncommutative case and carefully analyze if it reduces
to its commutative counterpart in the deep IR.
Thus we are interested
in the $\rho\rightarrow 0$ limit of \bref{nc-mn}. The key observation
is that the function $h(\rho)$ tends to the constant value $h(0)=1+\T^2e^{2\Phi_0}$.
Thus the coefficient multiplying the noncommutative coordinates
 $dx_2^2+dx_3^2$ becomes a \emph{constant}, which
could have been absorbed in a rescaling of the coordinates from the very beginning
\be
\hat{x}^i = {x^i \over h(0)}\,, \espai\espai\espai i=2,3\,.
\ee
We would like to clarify that although the NC metric seems to tend to the commutative
one, {\it its derivatives do not}. This can be easily inferred for example from the
value of the scalar of curvature at $\rho=0$ \bref{curva}, which does depend on $\T$.
In general, all objects constructed from derivatives of the metric may differ
from their commutative counterparts. The same observation applies to the following
analysis for the rest of the fields.

Consider now the $B$-field, which
tends to a constant in this limit, and so it becomes pure gauge. We could have started from
the beginning with a gauge-related $\hat{B}$-field
\be
\hat{B}_{[2]} = B_{[2]} + d\left(  {\T\over \T^2 +e^{-2\Phi_0}} \, x^2 \, dx^3\right)
\ee
that would vanish in the deep IR.
In any case, the gauge-invariant field strength $H_{[3]}=dB_{[2]}$  vanishes
for $\rho \rightarrow 0$. Let us now analyze the dilaton. In this limit, we
obtain
\be
e^{2\tPhi} \, \longrightarrow \,{e^{2\Phi_0} \over h(0)}
\ee
which just amounts to a redefinition of the value of the dilaton at the origin.
Furthermore, the field strength $F_{[3]}$ that couples magnetically to the D5
is unchanged everywhere.

Let us now analyze the remaining $C_{[4]}$ that couples to the D3 branes.
It is easy to see that in the deep IR limit it does not vanish. Since
this is not a gauge-invariant statement, we can look at its field strength\footnote{Indeed, one should make it
self-dual by defining $\tilde{F}_{[5]}= \undos ( F_{[5]}+ *F_{[5]} )$ but
the following discussion is not affected. Signs are chosen according to
the conventions of~\cite{Breckenridge:1997tt}.},
\be
F_{[5]}=dC_{[4]}-\undos\left( B_{[2]}\wedge F_{[3]} -C_{[2]}\wedge H_{[3]}\right)\,.
\ee
Upon substitution we obtain $F_{[5]}=-B_{[2]}\wedge F_{[3]}$,
which tends to
\be
F_{[5]} \longrightarrow -{\T \over e^{-2\Phi_0} +\T^2}\,\, dx^2\wedge dx^3 \wedge F_{[3]}\,.
\ee
when $\rho \rightarrow 0$ and, therefore, does not vanish.
Indeed, this statement is still
coordinate dependent. One way to make it more rigorous is to construct
scalar quantities out of $F_{[5]}$. One could for example compute
$F^2_{[5]}$ where all indices are contracted with the inverse metric.
Performing this calculation in the $\caln=4$  duals of
\cite{Maldacena:1999mh,Hashimoto:1999ut}, where the configuration corresponds
to D3-D1 bound states instead of D5-D3, one finds that the D1 field
strength vanishes quickly in the IR, while the D3 one remains finite.
Furthermore, one could compute it as well in the case of D5-D3 in flat space,
or more generally in the rest of Dp-D(p-2),
directly from~\cite{Breckenridge:1997tt}. The result is again that the lowest brane
field strength vanishes at the origin, while the one of the Dp remains.
Nevertheless, in our case, the square of $F_{[5]}$ remains constant
too, so that the D3 field strength does not vanish!
Therefore,
in the deep IR limit, all fields reduce to the commutative result
except for the metric and the $F_{[5]}$. For the latter, this difference
has its origin in the fact that the MN metric is completely regular at the origin.

Presumably,
this could be a signal of the UV/IR mixing that is expected to occur
in $\caln=1$  and $\caln=2$  theories. The observation
that in the large $N$ non-planar diagrams are sub-leading with respect
to the planar ones, so that noncommutative effects should not be
visible, might not apply here because from \bref{final-condi}
our solution does not necessarily require to send $N$ to infinity, and it is reasonable
to see a different IR behavior from the commutative case.

\subsection{Quark-antiquark potential} \label{ch6:sec:q-qbar-potential}

In this section we obtain the quark-antiquark potential in the $\caln=1$  $SU(N)$
field theory by examining the behavior of the Wilson loop. We follow
the standard prescription  originally given in~\cite{Maldacena:1998im}.

\begin{figure}[here] \begin{center}
    \includegraphics[width=8 cm,height=6 cm]{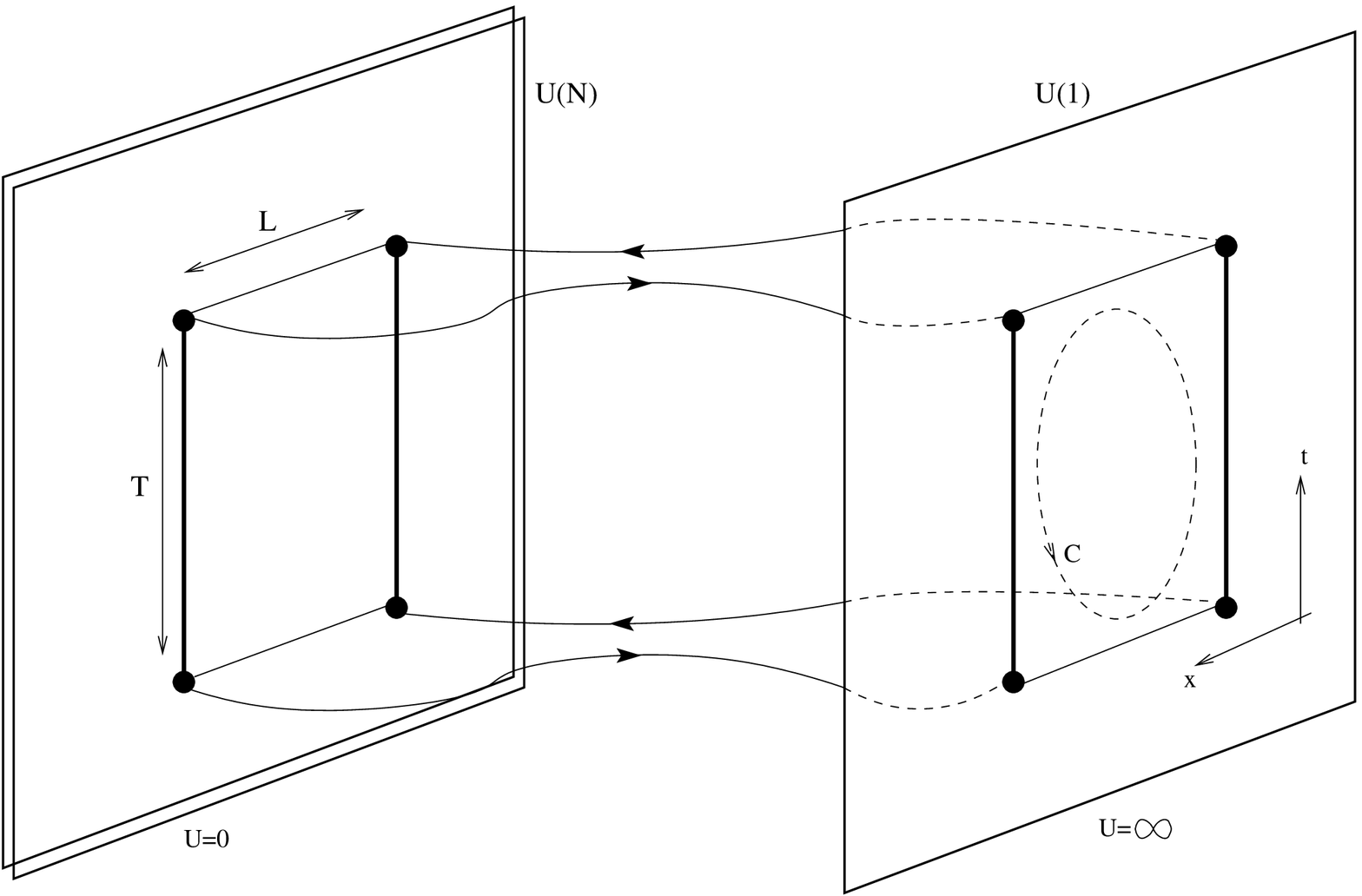}
    \caption{Above, two oppositely oriented strings with
    one end on the isolated brane and one in the stack. They provide
    a pair of massive $q\bar{q}$. Below, the stack is replaced
    by \5ads and forming a bound state is energetically favored. One is
    left with a minimal worldsheet with boundary $\calc$ at infinity.\label{wilsonloop} }
    \includegraphics[width=8 cm,height=6 cm]{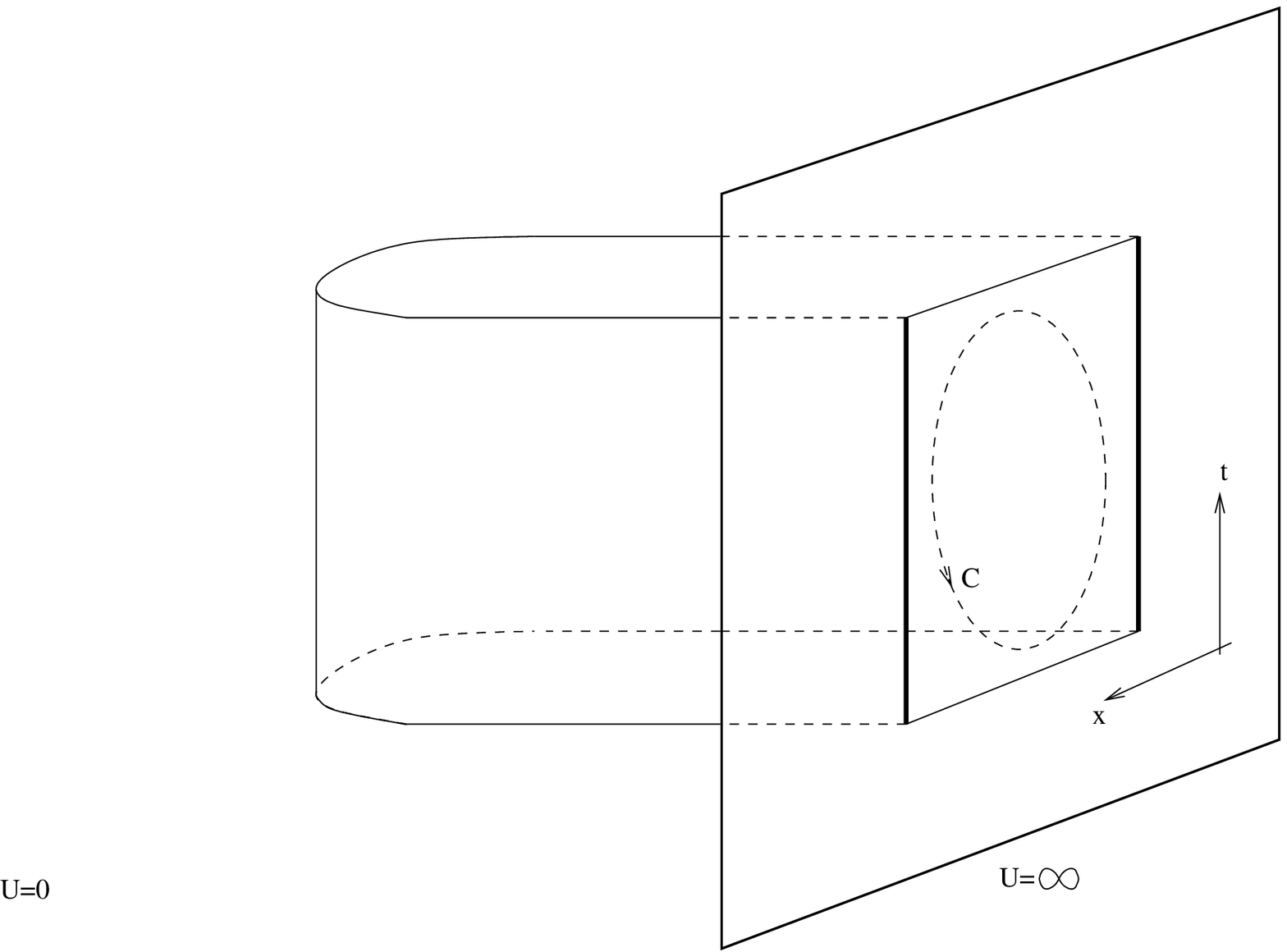}
    \end{center} \end{figure}

The standard way to check if a theory is confining is to introduce
an external (non-dynamical) quark-antiquark pair separated a distance $L$.
It is well-known that the potential $V(L)$
between them can be obtained from the expectation value of the Wilson loop
\be \label{wloop}
W(\calc) = \Tr \left[ P \, \exp \left(i \oint_{\calc} A\right) \right]\,,
\ee
by means of the formula
\be \label{vev-wilson}
\langle W(\calc) \rangle \sim e^{-T \, V(L)}\,.
\ee
In these formulae, $P$ denotes the path-ordered integral
of the gauge connection $A$ along the  contour $\calc$ shown
in fig.~\ref{wilsonloop}.

In order to compute the value of \bref{vev-wilson} in the string
theory side, we need to know to which sort of field it couples.
To this end, consider pulling away one brane from a stack of
$N$ D-branes. The gauge group is broken $U(N+1)\dreta U(N) \times U(1)$
and the open strings attached between the stack and the single brane
have excitations that correspond to $W$-bosons, with mass proportional
to the separation. The endpoints of these open strings transform
in the (anti-)fundamental of $U(N)$, so they look like massive
(anti)quarks from the point of view of an observer in the stack.
To make these quarks non-dynamical, all we need to do is to pull
the brane infinitely far away, so that its mass is higher than any
scale we are interested in.

When the stack is replaced by their \5ads background, the two
strings find it energetically favorable to form a bound state
(fig.~\ref{wilsonloop}).
These considerations led Maldacena~\cite{Maldacena:1998im} to propose
that the Wilson loop \bref{wloop}
acts as a source for open string worldsheet that ends at the boundary of AdS
on the contour $\calc$. Extending the AdS/CFT dictionary, he proposed
that the vev of $W(\calc)$ can be computed in the string theory
side by considering the string partition function in \5ads with
the condition that we have a string worldsheet ending on the loop
$\calc$. Such a partition function is given
in the supergravity approximation simply by the area of the
worldsheet with those boundary conditions $A(\calc)$,
so that from \bref{vev-wilson} we have
\be
\langle W(\calc) \rangle \sim e^{-T \, V(L)} \sim e^{-A(\petit{\calc)}}\,,
\ee
which allows us to straightforwardly compute the $q\bar{q}$-potential.

\newpage
\newpage

\subsubsection{Evaluation of the Wilson loop} \label{ch6:sec:wilson-loop}

In the case at hand the Wilson loop average is obtained
by minimizing the Nambu-Goto action
in the presence of the $B_{[2]}$ field background
\begin{equation}
\label{action}
S = \frac{1}{2\pi \alpha^\prime} \int d\tau d\sigma \left( \sqrt{-\rm{det}\, g}
+ B_{\mu \nu} \partial_\tau X^\mu \partial_\sigma X^\nu \right)\,,
\end{equation}
for an open string worldsheet with the mentioned boundary conditions. Explicitly, we want the
boundary to define a rectangular loop in the $(X^0,X^3)$-plane with lengths
$(T,L)$. Indeed, if we want to  account for the influence of the $B$-field
we need to take a non-static configuration in which the quarks acquire
a velocity $v$ in the NC plane. We therefore take the following configuration
\begin{equation} \label{confi}
X_0 = \tau,\quad X_2=v \tau\,, \quad X_3=\sigma,\quad \rho=\rho(\sigma)\,,
\ee
with
\be
-L/2<\sigma<L/2, \,\,\, 0<\tau <T \,.
\end{equation}
Plugging \bref{confi} and the NC background \bref{metrica} in the action,
we obtain
\begin{equation}
S = \frac{T}{2\pi } \int_{-{L\over 2}}^{L\over 2}
 d\sigma \left( H^{-1/2}(\rho) \left(1-\frac{v^2}{h(\rho)}\right)^{1/2}
\left( N{\rho^\prime}^2+\frac{1}{h(\rho)} \right)^{1/2} -\frac{\T}{H(\rho)+\T^2} v \right)\,,
\label{thelagr}
\end{equation}
where $\rho^\prime := \partial_\sigma \rho$ should be understood
hereafter and $H(\rho):= e^{-2 \Phi}$. In the large $T$ limit,
the unrenormalized potential for the $q\bar{q}$ system appears as $S=T \ V_{\rm unren}$.
Notice that in the above expression there are two controllable parameters:
the noncommutativity strength $\T$ and the velocity $v$ of the quarks.
From now on, we shall impose the non-supraluminical requirement $\vert v\vert < 1$,
which ensures
$\left( 1-\frac{v^2}{h(\rho)}\right) > 0$.

\begin{figure}[t]
\begin{center}
\includegraphics[width=8cm,height=8cm]{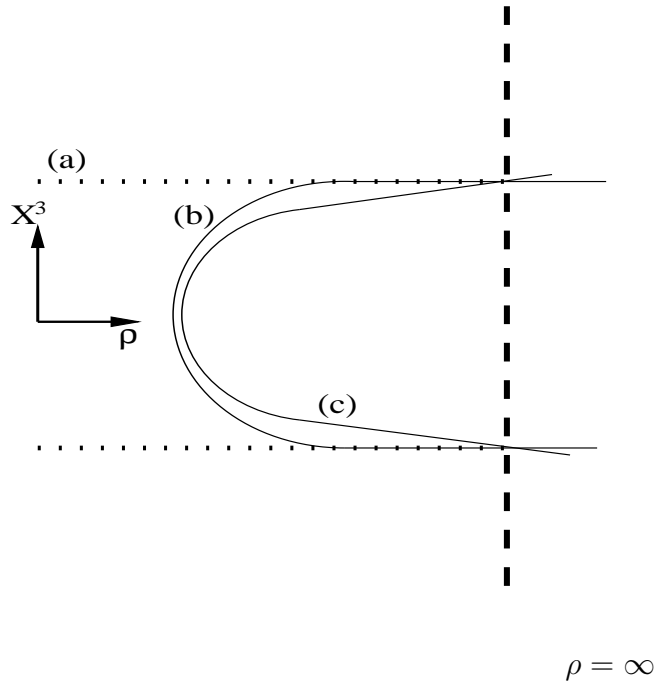}
\end{center}
\hspace{10cm} $\rho = \infty$
\caption{Different configurations for the open string worldsheet
in the evaluation of the Wilson loop.
(a) corresponds to the
subtraction of ``two bare quarks''.
(b) is the only
allowed configuration (fine tuned) that leads finite
 results, for both the potential and the quarks distance. (c)
is an example of a configuration that would not cancel
 the divergences at $\rho\rightarrow\infty$. The
difference between configurations (b) and (c) is
that the first one hits the brane at right angles and,
therefore, asymptotes to (a).
\label{fine}}
\end{figure}

We can think of the integrand for $V_{\rm unren}$ as a Lagrangian density in
classical mechanics with $\sigma$ as the
evolution parameter.
Since this Lagrangian density does not depend explicitly on $\sigma$,
its associated Hamiltonian is a conserved quantity on the extremal of
the action:
\begin{equation}
\label{hamiltonian}
-\frac{1}{h(\rho) H^{-1/2}(\rho)} \left(1-\frac{v^2}{h(\rho)}\right)^{1/2}
\left( N{\rho^\prime}^2+\frac{1}{h(\rho)} \right)^{-1/2} + \frac{\T}{H(\rho)+\T^2} v \equiv
{\rm ct}\,.
\end{equation}
To proceed we evaluate the constant at a special point $\rho_0$ defined as follows.
Locate the boundary of the worldsheet at some distance $\rho_{\rm max}$ from the origin, to
be sent to infinity at the end of the calculations. As we increase $\sigma$, the worldsheet
approaches the origin though the embedding $\rho(\sigma)$ until it reaches a minimum value
$\rho_0$. By symmetry of the background, this must happen at $\sigma=0$, so that
$\rho_0=\rho(0)$ and $\rho'(0)=0$. Evaluating \bref{hamiltonian} at $\rho_0$ and
solving for $\rho'$ we obtain
\begin{equation}
\label{distance}
\rho^\prime = \pm \left(\frac{H(\rho)\left[H(\rho_0)-H(\rho)\right]}{N}\right)^{1/2} \frac{1}{\alpha^2+\alpha v
\left[\alpha^2+H(\rho_0)\right]^{1/2}+H(\rho)}\,,
\end{equation}
where we have defined the \emph{effective} or boosted noncommutative parameter as
\[
\alpha^2:=\frac{\T^2}{1-v^2}\,.
\]
Equation \bref{distance} can be used to obtain an implicit relation between
the quark separation and $\rho_0$,
\begin{equation}
\label{L}
L(\rho_0)=2\sqrt{N} \int_{\rho_0}^{\rho_{\rm max}} d\rho\,
\frac{\alpha^2+\alpha v \left[\alpha^2+H(\rho_0)\right]^{1/2}+H(\rho)}{\left(H(\rho)\left[H(\rho_0)
-H(\rho)\right]\right)^{1/2}}\,.
\end{equation}
Similarly, we can plug equations \bref{hamiltonian} and \bref{distance} into \bref{thelagr}
to obtain a relation between the unrenormalized potential and $\rho_0$,
\be
\label{potential}
V_{\rm unren}(\rho_0) = {\sqrt{N} \over  \pi} \int_{\rho_0}^{\rho_{\rm max}}\, d\rho
\left(  \frac{\T^2 + (1-v^2) H(\rho_0)}{H(\rho)\left[H(\rho_0)-H(\rho)\right]} \right)^{1/2}.
\ee
Now, from fig. \bref{figdila}, we see that $H(\rho)$ decreases
very fast, so that $H(\rho_0) \gg H(\rho)$ for sufficiently large $\rho$.
As a consequence, (\ref{potential}) diverges as we let $\rho_{\rm max}\rightarrow \infty$,
which is interpreted as due to the presence of the two bare quark masses at
the endpoints of the string. To extract just the potential, we proceed to subtract
this contribution as usual~\cite{Maldacena:1998im,Drukker:1999zq}.
We therefore repeat the calculation for the following configuration
\begin{equation}
X_0=\tau\,,\quad X_2=v\tau\,,\quad X_3\equiv{\rm constant}\,,\quad \rho=\sigma\,,
\end{equation}
which corresponds to a straight worldsheet of a
string stretching from the initial stack of $N$ D-branes to the single one located at
infinity (see fig. (\ref{fine}.a)). Subtracting this contribution
we obtain the following regularized quark-antiquark potential
\bea
V_{{\rm ren}}&=& {\sqrt{N} \over  \pi} \left\{ \int_{\rho_0}^{\rho_{\rm max}}\, d\rho\,
\left(  \frac{\T^2 + (1-v^2) H(\rho_0)}{H(\rho)\left[H(\rho_0)-H(\rho)\right]} \right)^{1/2} \right.
\nn
\label{Ereg}
&& \left. ~~~~~~~~~~~~-\int_0^{\rho_{\rm
    max}} d\rho\,
\left(\frac{(1-v^2) H(\rho)+\T^2}{H(\rho)\left[H(\rho)
+\T^2\right]}\right)^{1/2}\right\}.
\eea
It is easy to check that in the commutative limit $\T=0$, both $V_{{\rm ren}}$
and $L$ remain finite as we let $\rho_{\rm max}$ grow to infinity.
Nevertheless, arbitrary values of $\T$ require a further
restriction for the potential to be well-defined. We discuss this issue and its
physical interpretation in the next subsection.

\subsubsection{The fine tuning} \label{ch6:sec:fine-tuning}

Consider, for a generic value of $\T$, the distance between the endpoints of the string in
the $X^3$ axis (\ref{L}). We want to keep $L$ finite as we move the boundary to
${\rho_{\rm max}} \rightarrow \infty$. Since in this limit
$H(\rho)\rightarrow 0$, we need
\be \label{v-con}
 \alpha^2+\alpha v \left[\alpha^2+H(\rho_0)\right]^{1/2} = 0\,.
\ee
The equation admits two solutions. The first one is $\alpha=0$, which corresponds
to the commutative case, and imposes no restrictions on $v$. This was to be expected,
since in the absence of $B$-field, Lorentz symmetry is restored in the whole flat part
of the brane, and two quarks moving at the same velocity are equivalent to
two static quarks. Nevertheless, in the presence of a $B_{23}$, the Lorentz symmetry
is broken, and equation \bref{v-con} selects
\begin{equation}
v= - \frac{\T}{\sqrt{H(\rho_0)}}\,.
\label{finet}
\end{equation}
Since, by equation \bref{L}, $L$ determines $\rho_0$, we see that
the velocity must be fine tuned with respect to the strength of
the $B$-field and the distance between quarks.
Remarkably, the same fine tuning reappears again when
we consider the
renormalized potential (\ref{Ereg}). To obtain a finite potential after the
subtraction we need both integrands in  (\ref{Ereg}) to cancel each other when
${\rho_{\rm max}} \rightarrow \infty$. This imposes the condition
\be
\frac{\T^2 + (1-v^2) H(\rho_0)}{H(\rho_0)} = 1\, \espai\espai \Rightarrow \espai
\espai v^2=  \frac{\T^2}{H(\rho_0)}\,\,,
\label{finet2}
\ee
which is consistent with (\ref{finet}). Therefore, the fine tuning solves
simultaneously the problem of fixing the distance between quarks at
the boundary at infinity,  and the problem of finiteness the potential.
Despite being an ad hoc requirement, the fine tuning is necessary to provide
a dual supergravity interpretation of the Wilson loop in the field theory.

The physical interpretation is somewhat analogous to the situation when a
charged particle enters a region with a constant magnetic field. In that case,
there is also a fixed relation -say, a fine tuning- between the three relevant
parameters: the radius of the circular orbit, the velocity, and the strength of
the magnetic field. As in our case, such a particle would not feel the
presence of the magnetic field if it did not have a non-zero velocity
transverse to it, which explains why we chose a non-static configuration.

Notice that implementing the fine tuning in \bref{distance} shows
that now the endpoints of the string hit the boundary at ${\rho_{\rm max}} \rightarrow \infty$
at right angles, as depicted in fig. (\ref{fine}.b). This is the only way of
keeping finite the quarks distance. For instance, the configuration
(c) in fig. \bref{fine} would not lead to a finite result.
In turn, this explains why the fine tuned configuration
allows for a finite renormalized potential, since it is the only
one that provides an asymptotic coincidence with the configuration that one
needs to subtract.

We conclude this subsection by studying the consequences of the requirement that $v<1$.
The fine tuning demands then that
$\T^2 < H(\rho_0)$. Since $H(\rho)$ is monotonically
decreasing and tends to zero at infinity, this inequality implies two things. The first one
is that $H(0)=e^{-2\Phi_0}$ must also satisfy the inequality, so that we need $\T^2<e^{-2\Phi_0}$.
This enters in contradiction with the requirements \bref{final-condi} of section~\ref{ch6:sec:validity}.
Therefore, to properly study the Wilson loop, we have to abandon one of the following
requirements: smallness of the dilaton, smallness of the curvature, or KK modes decoupling.
If, as in~\cite{Maldacena:2000yy}, we only disregard the KK condition, we then need to impose
\be
1 \, \ll \, \T \, < \, e^{-\Phi_0} \, \ll N \,.
\ee
The second one is that $\rho_0$ has an upper bound $\rho_w$, for which $H(\rho_w)=\T^2$.
Choosing $\rho_0>\rho_w$ would lead to supraluminical velocities
\footnote{
Having~\cite{Dhar:2000nj} in mind, we just mention
that $\rho_w$ has the property that the warp factor $e^{\Phi}h^{-1}$ in front
of the NC directions of the metric \bref{nc-mn} acquires its maximum value.}.
It is easy to see that an upper limit on $\rho_0$ implies a lower limit
on the quark separation $L$. Seeking for an understanding of this lower
limit for $L$, it is tempting to think that this could
be related to the fact that gauge invariant objects in NC theories involve
open Wilson lines (see section~\ref{ch6:sec:rho-energy-relation}), which exhibit a relation between
their lengths and their momentum through
\begin{equation}
\label{incerrel}
\Delta l^{\mu}=\T^{\mu\nu}k_{\nu}\,.
\end{equation}
In our case the length $L$ is along $X^3$ whereas the velocity is along $X^2$,
in agreement with our NC parameter $\T^{23}$. A complementary consideration~\cite{Matusis:2000jf}
is that relation (\ref{incerrel}) gives the size of the particle in the $X^3$ direction when
it has a given momentum along $X^2$. However all these are still vague
arguments that we do not claim as conclusive.

\subsubsection{The results} \label{ch6:sec:results}
\begin{figure}[t]
\begin{center}
\includegraphics[width=7cm,height=7cm]{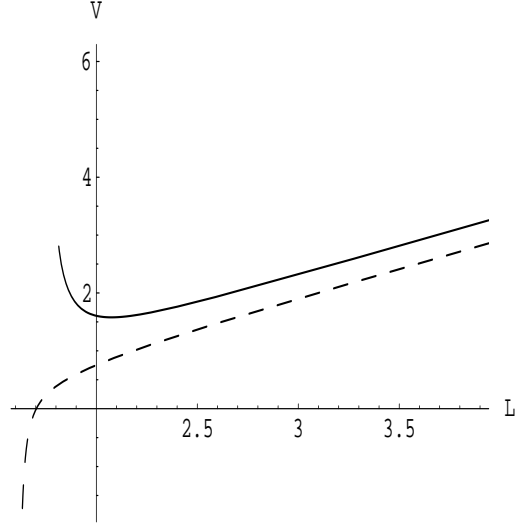}
\end{center}
\hspace{0.5cm}
\caption{Quark-antiquark
potential versus their separation. The dashed line
corresponds to the commutative case, while the full curve depicts the corresponding NC one.
At large distance, both theories confine, while as we move the quarks closer, the UV
physics give a completely different behavior.}
\label{fig2}
\end{figure}

Once the necessity for the fine tuning has been discussed, we proceed
to apply it to our formulas \bref{L} and \bref{Ereg} to obtain the
simplified expressions for the quarks distance and the renormalized
potential:

\bea
\label{L2}
L&=&2\sqrt{N} \int_{\rho_0}^{\infty} d\rho
\left(\frac{H(\rho)}{H(\rho_0)-H(\rho)}\right)^{1/2}\,,
\\ \nn \nn
V_{\rm ren} &=& \frac{\sqrt{N}}{\pi}
\left\{ \int_{\rho_0}^{\infty} d\rho\,
\sqrt{\frac{H(\rho_0)}{H(\rho)\left[H(\rho_0)-H(\rho)\right]}} \right. \nn
&& \left. \espai -\int_0^{\infty} d\rho\,
\sqrt{\frac{{\T^2\over H(\rho_0)}\left[H(\rho_0)-H(\rho)\right] +  H(\rho)}
{H(\rho)\left[\T^2 + H(\rho)\right]}}
\right\}\,.
\label{pot} \\ \nonumber
\eea
Both equations can be used to obtain $V_{\rm ren}$ as a function of $L$. Although the
relation cannot be given algebraically, one can make a numerical plot to
study the phases of the theory. In fig.\bref{fig2}, we present
the plot of the renormalized potential against the distance between quarks
in both the commutative (dashed line) and the NC (full line) backgrounds.

The first immediate observation is that both theories exhibit the same behavior
in the IR. At large separation, the potential is linear in both cases and, restoring
$\alpha'$ factors, we obtain
\be
V_{\rm ren}(L) \, \approx \, {e^{\Phi_0}\over 2\pi \alpha'}\,L \,,
\ee
independent of the value of $\T$. Indeed, this result
can be proven analytically, and does not rely only on the numerical analysis.

Nevertheless, as we move the quarks closer, the two theories exhibit
a very different behavior. In the NC case, the potential becomes extremely
repulsive, presumably due to the expected effects of the noncommutative
uncertainty relations at short distances. On the other hand, the commutative
potential starts deviating from the linear behavior in the opposite way,
although this happens in a region where the commutative dilaton \bref{com-dilato}
is not small anymore, and so the calculation should have been continued in the NS5 S-dual
picture.

\subsection{Gauge theory physics from noncommutative MN} \label{ch6:sec:physics-nc-mn}

In this section we try to extract more information of the noncommutative
gauge theory from the proposed supergravity dual. Our discussion will
be parallel to that in~\cite{DiVecchia:2002ks,Apreda:2001qb}, where they studied
the commutative Maldacena-\nun solution.
We will follow the conventions of~\cite{DiVecchia:2002ks}

\subsubsection{NC Yang-Mills coupling as a function of $\rho$} \label{ch6:sec:coupling}

Let us then begin with the discussion on the Yang-Mills coupling for the commutative
case.
The proposal in
\cite{DiVecchia:2002ks} is that one can obtain $g_{\rm YM}$ as a function of $\rho$ by the following procedure.
Consider the DBI action of a D5 in the background of MN. Take the $\a'\rightarrow 0$
limit and promote the abelian fields to transform in the adjoint of $SU(N)$.
That would give a $SU(N)$ Yang-Mills action in the curved space that the
D5 are wrapping, which in our case is $\CR^4 \times S^2$.
Since we are interested in the IR of the gauge theory, we take a limit in which
the volume of the $S^2$ is small, so that the action becomes, upon an $S^2$ reduction,
a four dimensional $\caln=1$  $SU(N)$ SYM with the following bosonic structure
\be \label{Action}
S[A_{\mu}]=-{1\over 4g_{\rm YM}^2}\int_{\CR^4} d^4x \, \, F_{\a\b}^A F_{A}^{\a\b}\,,
\espai \espai \espai \a,\b=0,1,2,3\,.
\ee
Indeed, one would get a series of corrections from the KK modes of the $S^2$
which, as discussed in section~\ref{sugra},
decouple under a certain choice of $N$, $\T$ and $\Phi_0$.
The YM coupling
appearing in \bref{Action} is essentially given by the inverse volume of the
$S^2$ measured with the ten-dimensional commutative metric, and it depends
on the radial coordinate $\rho$.\footnote{In the original paper of Maldacena-\nun
the YM coupling was calculated directly in the gauged supergravity.
At the end of the day, it differed from the one in~\cite{DiVecchia:2002ks} by the fact
that the volume of the $S^2$ was calculated with the seven-dimensional metric.
The remarkable matching of~\cite{DiVecchia:2002ks} with the field theory result seems
to select their method.}

Let us first adapt this method in order to obtain
$\hat{g}_{\rm YM}(\rho)$ for the NC-MN solution \bref{nc-mn}. We should
now expand the DBI including the background $B$-field. Actually, in the low-energy
limit, we have seen that the theory becomes noncommutative.
When the dilaton is independent of the gauge theory coordinates, \ie at zero momentum,
the DBI action with a constant magnetic $B$-field gives,
in the low-energy limit,
the same quadratic terms as its noncommutative version
\be \label{nc-DBI}
S_{\rm DBI}[\hA_{\mu}]={\tau_5 \over  G_s} \int_{\CR^4\times S^2} d^6x
\,\,\sqrt{\det \left(P[G]+2\pi\hF
\right)_{*}}
\ee
where $G_s$, $\hF$ and $G_{\mu\nu}$ are the effective coupling constant, field
strength and metric seen by the open strings in a $B$-field background,
and $\tau_5$ stands for $(2\pi)^{-5}$.
All products in \bref{nc-DBI} are understood as Moyal $*$-products with noncommutative
parameter $\T^{\mu\nu}$.
We recall that the relations between the open string quantities and the
closed string ones $e^{\hat{\Phi}}$, $F$ and $g_{\mu\nu}$ are
\begin{eqnarray} \label{sw-map}
G_{\mu\nu}&=& g_{\mu\nu}- (f^{*}B)_{\mu\rho} g^{\rho\lambda}(f^*B)_{\lambda\nu}\,, \nn
\hF&=& {1\over 1+ F\T}\,F\,, \nonumber\\
G_s&=&e^{\Phi}\left({ \det G \over \det g}\right)^{1/4}\,, \nn
\T^{\mu\nu}&=&-g^{\mu\rho} (f^{*}B)_{\rho\lambda} G^{\lambda\nu}\,.
\end{eqnarray}
In order to correctly identify the $\hg_{\rm YM}$ for the noncommutative theory,
we use the noncommutative action and variables. Expanding \bref{nc-DBI}
and plugging in our background \bref{nc-mn} we obtain
\be 
S[\hA_{\mu}]=-{1\over 4\hg_{\rm YM}^2}\int_{\CR^4} d^4x \, \, \hF_{\a\b}^A *
\hF_{A}^{\a\b}\,,
\espai \espai \espai \a,\b=0,1,2,3
\ee
with the following expression for the noncommutative YM coupling
\be
{1\over \hg^2_{\rm YM}(\rho)}= {2\pi^2 \tau_5 \over N^2 G_o} e^{2\Phi(\rho)}
e^{-4g(\rho)}  \int_{S^2} d\t d\phi \sqrt{P[G]}\,.
\ee
By explicit calculation, it turns out
that the
Yang-Mills coupling can be written in the following way
\be
{1\over \hg^2_{\rm YM}(\rho)}={N\over 32 \pi^2} Y(\rho) \int_0^\pi \, d\t \,\,
 \sin\t \left[ 1+{\cot^2 \t\over Y(\rho)}\right] ^{\undos}
\ee
where we defined
\be
Y(\rho)=4\rho \coth 2\rho -1\,.
\ee
By comparison with~\cite{DiVecchia:2002ks}, we see that the relation between $\hg_{\rm YM}$ and
the radial coordinate turns out to be identical to that of $g_{\rm YM}$!

\subsubsection{Relation between $\rho$ and the energy} \label{ch6:sec:rho-energy-relation}

To go further and obtain the $\b$-function, we still need to find the
relation between $\rho$ and the energy scale of the dual field theory.
In the commutative MN, there are two basic lines of argument that lead
to the same conclusions. We briefly review them in order to be applied
to the noncommutative case.

\avall

({\it i}) The authors of~\cite{DiVecchia:2002ks} observe that $\caln=1$  SYM theories
have a classical $U(1)_R$ symmetry which is broken at the quantum
level (and after considering non-perturbative effects) to $Z_2$.
An order parameter is the vacuum expectation value of
the gaugino condensate $<\l^2>$, i.e. if $<\l^2>\ne 0$,
the symmetry is broken. To relate this phenomenon to the supergravity
side, one is guided by the fact that we know how the $U(1)_R$ symmetry
acts, since it simply corresponds to rotations along the angle $\psi$.
It is easy to realize that such rotations are an isometry of the metric
if and only if the supergravity field $a(\rho)$ appearing in \bref{nc-mn}
is zero\footnote{Note that both $a(\rho)$ and $\psi$
appear in \bref{nc-mn} in an implicit way through the
definition of the gauge field $A$ and the left-invatiant
forms $\omega$, see \bref{definemya} and \bref{leftinvform}.}.
Therefore one is led to conjecture that $a(\rho)$ is the
supergravity field dual to the gaugino condensate. The argument finishes by
noticing that since $<\l^2>$ has protected dimension three, it must
happen that \footnote{The proportionality coefficient is 1 from
explicit calculations~\cite{Hollowood:1999qn}.}
\be
<\l^2>=\Lambda^3
\ee
where $\Lambda$ is the dynamically generated scale.
This leads to the following implicit relation between $\rho$ and
the field theory scale $\mu$
\be
\label{afield}
a(\rho) \propto {\Lambda^3\over \mu^3}\,.
\ee

\avall

({\it ii}) A slightly different argument is given in~\cite{Apreda:2001qb}.
The authors
first expand $g_{\rm YM}(\rho)$ for large $\rho$ (in the UV)
where it can be compared to perturbative results of the gauge theory,
\ie with $g_{\rm YM}(\mu/\Lambda)$. This immediately gives the searched relation
$\rho=\rho(\mu/\Lambda)$, valid in the UV region. Indeed, they also identify
$a(\rho)$ as dual to the gaugino condensate by
trying to guess what is the exact form of the mass term for the
gauginos in the four-dimensional $\caln=1$  SYM. Gauge invariance
of the Lagrangian must involve couplings to the gauge field through covariant
derivatives. This fact, together with the detailed knowledge of how the twisting
of the field theory is performed, allowed the authors to find that the Lagrangian
must involve a term like
\be
a(\rho) \, \bar{\l} \l \,.
\ee
Applying standard arguments of the original AdS/CFT correspondence one would conclude
again that $a(\rho)$ is the supergravity field dual to the gaugino condensate.

\avall

We now try to adapt these arguments to our NC-MN solution. The first important remark is that
noncommutative gauge theories do not have local gauge-invariant operators
\cite{Gross:2000ba, Das:2000md}.
Terms like $\Tr (\hF_{\mu\nu} * \hF^{\mu\nu})$ are only gauge invariant after integration
over all the space. This fact increases the difficulty to associate the dual supergravity fields,
since they should act as sources of gauge-invariant operators.
Nevertheless, since translations are still a symmetry of the
theory, there must exist gauge-invariant operators local in momentum space. Such operators
involve the so-called open Wilson lines, whose length must
be proportional to the transverse momentum. Explicitly, if we name $\Delta l^{\mu}$
the separation between the endpoints of an open Wilson line, and $k_{\mu}$ its momentum
in the noncommutative directions, gauge-invariance requires
\be \label{nc-relation}
\Delta l^{\mu}=\T^{\mu\nu}k_{\nu}\,.
\ee
Several scattering computations~\cite{Liu:2001ps,Okawa:2000sh,Das:2001uz} seem to
confirm that a general supergravity field $h$
couples to the noncommutative version of the ordinary operator to which it coupled
when $\T=0$  via
\be
\int d^dk \, h(-k) \hat{\cO}(k)\,.
\ee
The noncommutative operator $\hat{\cO}(k)$ is defined from its commutative local one $\cO(x)$
by inserting the mentioned Wilson line $W[x,\cC]$ and Fourier-transforming\footnote{
We refer to {\it e.g.}~\cite{Gross:2000ba, Das:2000md, Liu:2001ps} for further discussions on the
ambiguity of the insertion of $\cO(y)$ along the contour $\cC$, and for general aspects
of open Wilson lines. We shall only make use of a few of their properties.}
\be
\hat{\cO}(k)=\Tr P_{*} \int d^dx \, [W(x,\cC) \cO(y)]* e^{ikx}
\ee
where $\cC$ is a straight path connecting the endpoints separated according to \bref{nc-relation}
and $y$ is an arbitrary point along $\cC$.

The observation is that the relevant fields appearing in our background \bref{nc-mn} do not depend
on the noncommutative coordinates $(x^2,x^3)$, so that their Fourier-transforms would
involve a delta function in momentum space. In other words, we just need the zero-momentum
couplings, where the length of the Wilson lines vanishes, and $\hat{\cO}$ reduces to $\cO$.

We are now ready to apply the arguments ($i$) and ($ii$) to our case. As far as $U(1)_R$
symmetry breaking in the supergravity solution is concerned, nothing changes
with respect to the commutative case. Again, shifts of $\psi$ are an isometry of the NC metric
if and only if $a(\rho)=0$. This is due to the fact that the only change in the metric
is a factor of $h^{-1}(\rho)$ in front of $\, dx_{2,3}^2$.

For the same reason,
the whole structure of the twisting of the normal bundle to the $S^2$ inside
the Calabi-Yau threefold is also unchanged. At zero-momentum in the noncommutative directions,
gauge invariance in the field theory demands again that the fermionic couplings to
the gauge fields appear only via covariant derivatives. So it looks like the arguments
of ($i$) and ($ii$) lead again to conjecture that $a(\rho)$ is dual to the gaugino condensate.

Indeed, independently of this relation, one could proceed as in~\cite{Apreda:2001qb} and
expand $\hg_{\rm YM}(\rho)$ at very large $\rho$. In that region, the theory is in the UV
and perturbative calculations should be trustable. The field theory results
(see the review~\cite{Szabo:2001kg} and references therein) show
that the perturbative NC $U(N)$ $\b$-function is identical to the $SU(N)$ commutative
one.\footnote{Recall that the $U(1)$ degrees of freedom inside a noncommutative $U(N)$
gauge theory do not decouple and, unlike the commutative case, they run with
the same $\b$-function as the rest of the $SU(N)/Z_N$~\cite{Armoni:2000xr}.} So both the supergravity
behavior of $\hg_{\rm YM}(\rho)$ and the perturbative behavior of the NC $\b$-function
are identical to the commutative case. The conclusion is that the relation
between $\rho$ and $\Lambda/\mu$ is also unchanged.

Summarising, it seems like the NC $\b$-function calculated from \bref{nc-mn}
and the commutative one extracted from the commutative MN are identical.
Hence
the same results found in~\cite{DiVecchia:2002ks,Apreda:2001qb} hold in our case. We just recall that
properly choosing the proportionality function in (\ref{afield})
\cite{Olesen:2002nh, Marotta:2002ns}
one remarkably obtains the whole perturbative NSVZ $\b$-function
\be
\beta(g_{\rm YM}) = -\frac{3g^3_{\rm YM} N}
{16\pi^2}\left[1-\frac{N g^2_{\rm YM}}{8 \pi^2}\right]^{-1}\,,
\ee
and the authors of~\cite{DiVecchia:2002ks} even identified
the contribution of (presumably) non-perturbative
fractional instantons.

\subsubsection{Phase diagrams} \label{ch6:sec:phase-diagram}

Before finishing, it is worth to restore units and
analyze the relevant scales present in the problem.
The first comment is that our supergravity solution corresponds already to a
near horizon limit. This implies that we have implicitly taken $l_s \rightarrow 0$.
Restoring units in our background \bref{nc-mn} is equivalent to replacing
\be
\rho \rightarrow l_s \rho\,, \espai\espai \T \rightarrow {\T \over l_s^2}\,,
\ee
with $\rho$ and $\T$ acquiring now units of energy and energy$^{-2}$ respectively.
Notice that there are four dimensionful parameters in the problem, namely,
$l_s$, $\Lambda$, $\Theta$ and the mass of the KK modes $m_{KK}$. As mentioned,
in units of energy, $m_s=1/l_s$ is much larger than the rest of scales.

\begin{figure}[h]
\begin{center}
\vspace{-0.25cm}
\includegraphics[width=12cm,height=1.3cm]{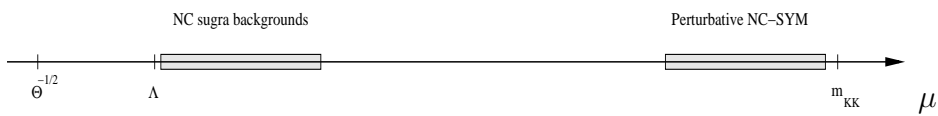} $\mu$
\end{center}
\caption{Flow diagram of the theory on the branes.\label{esca}}
\end{figure}

Consider the flow from high-energy to low-energy (see figure~\ref{esca}). The analysis
of section \ref{ch6:sec:validity} guarantees that, in the decreasing the energy, the first scale that
we find is $m_{KK}$, which is proportional to the inverse volume of the $S^2$. As
we cross this point, the little string theory on the large $S^2$ becomes an
effective four-dimensional $\caln=1$  NC-SYM. The gauge theory is in the perturbative
regime as long as we keep the energy much larger than $\Lambda$.
As we keep decreasing the energy, we approach $\mu \gtrsim \Lambda$ (which happens
about $\rho=0$ according to \bref{afield}), perturbation theory breaks down and
the theory is best described in terms of our NC supergravity background.
Finally, equation \bref{final-condi} tells us that the noncommutative scale
$1/ \sqrt{\T}$ is still at lower energies.

\newpage
\section{The supergravity dual of a NC $\caln=2$ SYM in 2+1} \label{ch6:sec:nc-gm}

\subsection{Introduction and a little bit of chronology} \label{ch6:sec:chronology}

Let us now switch to the $\caln=2$ SYM in 2+1 that we
studied in the previous chapter. Recall that the supergravity
solution was found in 8d gauged supergravity and then
uplifted to 11d. When going to IIA to study the moduli
space of the theory we mentioned that there were two immediate
choices of circles to compactify on: one preserved all supersymmetry
and the other one none. It is time to justify these statements
and to study the impossibility of obtaining the sugra duals
of NC theories in gauged supergravities.

To this end, we will construct the NC deformation of the
11d solution \bref{gib1}-\bref{probeta} using the first
method explained in section~\ref{ch6:sec:method-1}, \ie brute force.
The ansatz will be performed directly in 11d based on the
intuition of how the solution should be. We must say here
that the first ansatz we tried was in the 8d gauged supergravity,
a technique which had been used in {\it all} the previous
wrapped brane solutions found. After some time playing with
the Killing spinor equations we finally proved that there was
no supersymmetric solution in 8d supergravity with the
isometries that the configuration required!

The approach presented here is different for the sake of clarity.
The 11d ansatz will lead us to
a set of coupled first order equations by demanding supersymmetry.
These will tell us the precise form of the 11d Killing spinors
and, from them, we will acquire a better understanding of our
8d problems. At the same time, they will allow us to understand
the supersymmetry loss in going to type IIA.

The next section is rather technical. The reader who is not interested
in the details may just take a look at the equations \bref{ncd6-11}-\bref{bigspin} and then
jump to the section~\ref{ch6:sec:susy-without-susy}.

\subsection{11d solution of flat NC D6-branes} \label{ch6:sec:11d-solution-flat}

Before making the ansatz that will lead us to the sugra solution
of wrapped D6-branes in the presence of a magnetic $B$-field, we
need to understand how similar solutions look like when
the branes are flat in flat space. We will call the latter solutions
{\it NC flat D6-branes} and the former {\it NC wrapped D6-branes}.

So let us start with the 11d description of the ordinary flat D6
branes \bref{nonnearh}.
Noncommutativity will be put along the $(x^5,x^6)$ plane by
introducing a nonzero flux of $B_2$ along it; this will
explicitly break the $SO(1,6)$ isometry of the worldvolume to an
$SO(1,4)\times SO(2)$. Uplifting this IIA vocabulary to an 11d one,
the ansatz for the metric must be
\bea
ds^2_{(11)}&=&\tau^2(r)\left[
dx^2_{0,4}+ \sigma^2(r) dx^2_{5,6}+H\left(dr^2+ r^2
[d\t^2+\sin^2 \t d\phi^2] \right)\right]
\nn
\label{ans}
&&~~+ \tau^{-4}(r) R^2 H^{-1}
\left(d\psi+\cos\t d\phi \right)^2.
\eea
Note that the factor in front of the $U$(1) fiber is related
to the one in front of the other ten coordinates because of the
uplifting ansatz \bref{reduction}.
We also make an ansatz for the three-form that respects the $U$(1) monopole fibration
\be
A_{[3]}=\chi(r)\,\, dx^5 \we dx^6 \we (d\psi+\cos\t d\phi). \ee
We will determine the functions of our ansatz by demanding that the
supersymmetry transformations admit a non-trivial Killing spinor.
Since the background is bosonic, we just need to care about the
gravitino variation
\begin{equation} \label{gravitino6}
\delta \Psi _A  =  D_A\epsilon -\frac{1}{288}
\left( \Gamma _A{}^{BCDE}-8\delta _A{}^{[B}\Gamma ^{CDE]}\right)F_{BCDE}
 \epsilon,
\end{equation}
where $D_A= (\partial_A+ \frac 14 \omega_A{}^{CD}\Gamma_{CD})$
is the covariant derivative in flat coordinates and $F_{BCDE}$ is the four form
field strength.

In what follows it will be very important to make clear the
vielbein basis that we are using, since the explicit form of the
Killing spinors depends on it. We choose the following vielbein
for the diagonal part of \bref{ans}
\bea
e^a&=&\tau(r)\,dx^a \sac  ~~~~~a=0,..,4 \nn
e^{i}&=&\tau(r)\sigma(r)\, dx^{i} \sac i=5,6 \nn
e^7 &=& \tau(r)H^{\undos}(r) \, dr \,,
\eea
while for the squashed $S^3$ we take
\bea
 e^8&=&\tau(r) H^{\undos}(r) r \, \et^1 \,, \nn
e^9&=& \tau(r) H^{\undos}(r) r \, \et^2 \,, \nn
e^T&=& \tau^{-2}(r)H^{-\undos}(r)R \, \et^3 \,,
\label{vi}
\eea
with $\et^i$ the usual vielbeins of a round $S^3$
\be
\et^1=d \t \sac \et^2=\sin\t d\phi \sac
\et^3=d\psi+\cos\t d\phi.
\ee
Now we proceed to analyze the
supersymmetry variations. Due to the $SO(1,4)$ symmetry, the
equations for $A=0,1,2,3,4$ are equivalent. If we assume  that
the Killing spinors do not depend on the coordinates
$\{x^0,...,x^6\}$, these equations can be written as
\be\label{gamma1}
 \left( \cos\alpha \G_{D6}+\sin\alpha \G_{D4}\right) \e
=- \e,
\ee
with
\bea
\cos\alpha &= & {\chi'\over \chi} \tau^3
H R^{-1}r^2 \sac  \sin\alpha \,= \, -6 \tau^3 \chi^{-1}
\tau' \sigma^2 H^{\undos} r^2 \,, \nn
\Gamma_{D6} &\equiv& \Gamma_{0123456}
\sac ~~~~~~~~~ \Gamma_{D4}\equiv\Gamma_{01234T}.
\label{cos}
\eea
Since $\{\G_{D6},\G_{D4}\}=0$,
equation \bref{gamma1} is telling us that we are obtaining a
non-threshold bound state of D6-D4 from a IIA point of view, or
a bound state MKK-M5 from an M-theory one~\cite{Townsend:1997wg}.
To proceed, note that the equation \bref{gamma1} can be rewritten as
\be
 e^{-\alpha \G_{56T}} \e=-\G_{D6}\e, \ee
whose most general solution is
\be \label{analog}
 \e\,\,=\,e^{{\alpha \over 2}\G_{56T}}\, \tsp(r,\t,\phi,\psi)
\sac \mbox{with} ~~
\G_{D6}\tsp(r,\t,\phi,\psi)\,=\,-\tsp(r,\t,\phi,\psi).
\ee
Note that
the angle $\alpha$ is a function of $r$. At this point we need to make
an ansatz for $\tsp$. Experience suggests
\be
\label{spinorans}
\tsp(r,\t,\phi)=
f(r)e^{{\t \over 2}\G_{78}}
e^{{\phi \over 2}\G_{89}} \e_0,
\ee
where $\e_0$ is a constant spinor verifying $\G_{D6}\,\e_0\,=\,-\e_0$.
Plugging our ansatz in the remaining supersymmetry variations we obtain
the following set of first order, coupled, non-linear BPS equations
\bea
0&=& 3{\tau' \over \tau}+{\sigma'\over \sigma} \,,
\nn
0&=&\chi\chi'-6R^2H^{-1}\sigma^4 {\tau'\over \tau} \,,
\nn
0&=& {3\tau'\over \tau} +{\chi'\over 2\chi}+{H'\over 2H}  \label{tre} \,.
\eea
The general solution can be explicitly found and it depends
on three arbitrary constants. Two of them can be
fixed by demanding that the solution reduces to the commutative
one \bref{flatcase} when the $A_{[3]}$ is set to zero
(commutative limit).
The remaining arbitrary constant has a
physical meaning: it is the strength of the noncommutativity,
that we call $\T$. The solution is then
\bea
\tau(r)&=&h^{{1\over 6}} \,, \nn
\sigma(r)&=&h^{-\undos} \,, \nn
\chi(r)&=&-{\T R \over H h}\,, \nn
f(r)&=&h^{1\over 12}(r) \,, \nonumber
\eea
where $h(r)$ is the equivalent of equation \bref{last} for our case, \ie
\bea
h(r)=1+\T^2 H^{-1} \,.
\eea
Summarizing, the 11d metric, 3-form and the Killing spinors are given by
\zavall
\bea
 ds^2_{(11)}&=& h^{1\over 3}\left(-dx_{0,4}^2+
h^{-1} dx_{5,6}^2
+ H [dr^2+r^2 d\Omega_2^2] \right) \,,\nn
&& +
H^{-1}h^{-2/3}R \left( d\psi +\cos\t d\phi \right)^2
\label{nonnearh}
\\ \nn
A_{[3]}&=&-{\T R \over Hh} \;\; dx^5 \wedge dx^6\wedge \left(
d\psi+\cos\t d\phi\right)\,,
\\ \nn
\label{spinsol1}
\e(r,\t,\phi,\psi)&=&h^{1\over 12}(r) e^{{\alpha (r)\over 2}
\G_{56T}} e^{{\theta\over 2} \G_{78}} e^{{\phi\over 2}\G_{89}}
\e_0\,,
\eea
with the 1/2-preserving projection
\be
\G_{D6}\e_0=-{\e}_0 \,,
\ee
and the definitions
\be
\cos\alpha =
h^{-1/2} \sac \sin\alpha=\T (Hh)^{-\undos}\,.
\ee

\zavall

This solution describes the whole geometry of $N$ flat NC D6-branes
and the number of independent Killing spinors is 16. The
configuration  corresponds to a bound state of $N$ MKK monopoles
and $N$ M5 branes, or a bound state of $N$ D6-D4 branes in type IIA.
If we want to use this background {\it \`a la } AdS/CFT
to study the dual NC field theory, we must take
the near horizon limit, which consists of taking $\alpha'
\rightarrow 0$ keeping fixed
\be
\label{nhlimit}
u={r\over \alpha'} \sac \tilde{\T}=\alpha' \T \sac
g^2_{YM}= g (\alpha')^{3/2}.
\ee
After a change of radial variable
\be
u={y^2 \over 4 N \gym}\,, \nonumber
\ee
the metric and the three-form become
\bea
ds^2_{11}&=& h^{1/3} \left[dx_{0,4}^2 +
 h^{-{1}} dx_{5,6}^2 + dy^2+ {y^2\over 4} \left( d\Omega_{(2)}^2 +
h^{-1} [ d\psi+\cos\theta d\phi ]^2 \right)\right] \nn \label{nearh}
\\
A_{[3]}&=& -{\tilde{\T} \over 4N\gym^2} \, {y^2\over h} \,\, dx^5\wedge
dx^6 \wedge \left( d\psi +\cos\theta d\phi\right) \,,
\eea
with
\be
h(y)=1+\left({ \tilde{\T}\,\, y \over 2N\gym^2}\right)^2.
\ee
Recall that had we been in the commutative case, this near horizon
limit would have yielded the locally flat geometry with an ALE
singularity, so we would have found 32 locally preserved supersymmetries;
sixteen of them would be killed however by global identifications.
In our NC case we do not even find this enhancement at the local
analysis that we have just performed.

Let us consider in detail this commutative limit in both
the near horizon and the full geometry. Sending $\T \rightarrow 0$
implies $h \rightarrow 1$. In such limit,
the full geometry \bref{nonnearh}
collapses to eq.\bref{flatcase} and the 16 spinors become simply
\be \label{spinsol2}
\e(\t,\phi)= e^{{\theta\over 2} \G_{78}}
e^{{\phi\over 2}\G_{89}} \e_0, \espai\espai \mbox{with} \espai\espai
\G_{D6}\e_0=-{\e}_0.
\ee
On the other hand, in the commutative
limit, the near horizon region \bref{nearh} becomes the aforementioned $A_{N-1}$
singularity. Apart from the previous 16 spinors, it also admits
the following 16 ones
\be
\e(\psi)=e^{-{\psi\over 2}\G_{89}}\e_0, \espai\espai \mbox{with}
\espai\espai \G_{D6}\e_0=\e_0 \label{enhance}.
\ee
Note that they
have a different eigenvalue with respect to $\G_{D6}$. Modding
out by the $Z_N$ global identifications brings the number of
supersymmetries back to 16. Only for $N=1$, flat space, we
have a true enhancement of susy.

\subsection{11d solution of wrapped NC D6-branes} \label{ch6:sec:11d-solution-wrapped}

The analysis perform in order to obtain the NC deformation of the
flat D6 background will make it easier to find the corresponding
one for D6-branes wrapping a \kah four-cycle in a $CY_3$.
Again for the sake of simplicity we consider the case when the
4-cycle is an $S^2\times S^2$.

So we reconsider the background we obtained in \bref{gib1}-\bref{probeta}.
Let us turn on  a $B$-field along the $(x_1,x_2)$ plane.
As before, we explicitly break the worldvolume $SO(1,2)$ symmetry to
$R\times SO(2)$. As in the unwrapped case, we will also make use of
the fact that, in 11d, the factors in front of the 10d part of
the metric and in front of the $U$(1) fiber are related
through the lifting ansatz \bref{reduction}.
Therefore, our ansatz for the bosonic fields is
\footnote{We use the
definitions of \bref{petercond} for the functions $f(r,\t)$, $m(r,\t)$ and $B_{[1]}$.}
\bea
ds^2_{(11)}&=&\tau^2(r,\t)\left[-dx_0^2+\sigma^2(r,\t)
dx_{1,2}^2+{3\over 2}(r^2+l^2)ds^2_{cycle}+U^{-1}dr^2 \right.
\nn
&& \left. + {r^2
\over 4}\left( d\t^2 + m B_{[1]}^2\right)\right] +
+\tau^{-4}(r,\t) \,\tH^{-1} \,\left[d\phi-Uf^{-1} \cos\t
B_{[1]}\right]^2 \,,
\nn \nn
A_{[3]}&=&\chi(r,\t) \,\, dx^1\wedge dx^2\wedge \left[d\phi-Uf^{-1} \cos\t B_{[1]}\right].
\eea
Note that we allow the functions of the ansatz to depend on $\t$.
Now we proceed to make an ansatz for the spinor.
Just like in the NC flat case, we expect to obtain a projection
signaling a bound state of MKK-M5, so we impose
\be \label{const1}
\left(\cos\a \G_{D6} + \sin\a \tilde{\G}_{D4} \right) \e = \e,
\ee
for some angle $\a(r,\t)$ to be determined. Notice that since now the $B$-field
will be along $(x^1,x^2)$, we expect the D4 to span the directions
$\{x^0x^3x^4x^5x^6\}$, so that $\tilde{\G}_{D4}=\G_{03456T}$.
As in the unwrapped case, see \bref{gamma1} \bref{analog}, equation \bref{const1}
implies
\be
\e(r,\t,\phi,\psi)=
e^{{\a(r,\t)\over 2}\G_{12T}} \tsp(r,\t,\phi,\psi) \sac
\mbox{with} \espai\G_{D6}\,\tsp\,=\,\tsp.
\ee 
Now we are ready to obtain the BPS equations by imposing that the
supersymmetry variation of the gravitino \bref{gravitino6} vanishes.
The most immediate relations come from making them compatible for $A=0$ and $A=1,2$
and give
\footnote{We use primes for $\partial_r$ and
dots for $\partial_{\t}$. Also, the integration constant is set to one
in order to recover the commutative case when the three-form vanishes.}
\be  3\,{\tau' \over \tau}+{\sigma'\over \sigma}=0,
\espai \espai\espai  3\,{\dtau \over \tau}+{\dsigma\over \sigma}=0,
\ee
whose integration yields $\sigma=\tau^{-3}$.
The $A=5,6$ equations imply
that
\be \label{firsts}
\left(\partial_{\psi}+{\G_{36}\over 2}\right)\e=0 \sac  \mbox{and} \espai
\G_{36}\,\e=\G_{45}\,\e \,,
\ee
while the $A=7$ equation implies
\be
\tau^{-6}+\tH\tau^6\chi^2\,=\,1 \,.
\ee
Now taking a linear combination
of the $A=1,3,9$ equations, and assuming that the spinor
does not depend on the fiber coordinate $\phi$,
one reaches another constraint analogous to
\bref{gamma1}
\be \label{const2}
\left( \cos\b \G_{3689}+ \sin\b \G_{3679} \right)\e=-\e,
\ee
with
\be
\cos\b= U^{\undos}f^{-\undos}\cos\t, \sac
\sin\b=f^{-\undos} \sin\t.
\ee
Since the matrices $\G_{3689}$ and $\G_{3679}$ anticommute, we
can proceed as in \bref{analog}, and rewrite
this equation as
\be
e^{-\beta \G_{78}}\e=-\G_{3689}\e,
\ee
whose most general solution is
\be
\e(r,\t,\psi)=e^{{\a(r,\t)\over 2}\G_{12T}} e^{{\b(r,\t)\over 2}\G_{78}}
\tte(r,\t,\psi) \sac
\mbox{with} \espai \G_{D6}\,\tte\,= -\G_{3689}\,\tte\,=\,\tte.
\ee
Plugging this into \bref{firsts} allows us to write
down the final ansatz for the spinor:
\be
\tte(r,\t,\psi)=\g(r,\t)
e^{-{\psi\over 2}\G_{89}}\e_0 \sac \mbox{with}\espai
\G_{D6}\e_0= -\G_{3689}\,\e_0\,=\,\e_0.
\ee

The first order BPS equations are
\bea
0&=& 6 {\tau'\over \tau}+{\chi'\over \chi}+{\tH ' \over \tH}
\,,\nn 0&=&
\dot{\a}-\undos \tH^{\undos}\tau^6 \dot{\chi}\,,\nn 0&=&
\a'-\undos \tH^{\undos}\tau^6 \chi' \,,\nn 0&=&
{\g'\over \g} -{\tau' \over 2\tau} \,,\nn 0&=&
{\dot{\g}\over \g} -{\dot{\tau} \over 2\tau} \,.
\eea
Luckily, they can be solved analytically and, after fixing the integration
constants to reproduce the commutative case when $A_{[3]}$ vanishes, one
obtains
\bea
\tau&=&\th^{1\over 6} \,, \nn \chi&=&-{\T \over \tH \th} \,, \nn
\g&=&\th^{1\over 12} \,, \nn  \cos\a&=&-\th^{-\undos}\,, \nn
\sin\a&=&-\T (\tH \th)^{-\undos}\,,
\eea
with
\be
\th(r,\t)=1+\Theta^2 \tH^{-1}(r,\t).
\ee
So the whole solution for the metric, three-form and Killing spinor is
\bea
\label{ncd6-11}
ds^2_{(11)}&=&\th^{1\over 3} \left(-dx_0^2+ \th^{-1}
dx_{1,2}^2+{3\over 2}(r^2+l^2)ds^2_{cycle}+U^{-1}dr^2 \right.
\nn &&
\espai ~~~ \left. +
{r^2 \over 4} [ d\t^2 + m B_{[1]}^2]\right)
+\th^{-{2\over 3}} \tH ^{-1} \left(d\phi
 -{U f^{-1}\cos\t} B_{[1]} \right)^2 ,
\nn \\
A_{[3]}&=&-{\T\over \tH \th} \, dx^1\wedge dx^2 \wedge\left(
d\phi
 -{U f^{-1}\cos\t} B_{[1]} \right) \,,
\\ \nn
\e(r,\t,\psi)&=&\th^{1\over 12}(r,\t)e^{{\a(r,\t)\over 2}\G_{12T}} e^{{\b(r,\t)\over 2}\G_{78}}
e^{-{\psi\over 2}\G_{89}}\e_0  \,, \label{bigspin}
\eea
with the constant spinor $\e_0$ subject to the following 1/8-preserving constraints
\be
\G_{D6}\,\e_0\,= \,\e_0, \sac
\G_{36} \,\e_0 \,= \,\G_{45} \,\e_0 \, = \,\G_{89} \,\e_0.
\ee
Note that the introduction of the $B$-field has not
broken any extra supersymmetry as expected from the open string
picture analysis, so the configuration
still preserves 4 real supercharges.
This 11d background should be
dual in the IR to the 2+1 $\caln=2$ $U$(N) field theory with only a vector multiplet,
and with noncommutativity along
the $(x^1,x^2)$ plane.

This solution is an M-theory vacuum with fluxes. The topology is $\CR^3\times \mathbb{X}_8$,
with $\mathbb{X}_8$ the non Ricci-flat internal manifold.
$\mathbb{X}_8$ consists of a complicated four dimensional fibration over
the \kah base space $S^2\times S^2$.
Remarkably, we can smoothly send to zero the noncommutativity, so that
the $A_{[3]}$ flux goes to zero and $\mathbb{X}_8$ becomes an $SU$(4)-holonomy Calabi-Yau
four-fold. From a $IIA$ perspective it describes a non-threshold bound
state of D6-D4 branes with the $D4$  wrapped around
the four-cycle. We can describe the configuration by the
commonly used arrays as follows,

\begin{center}
\be
\begin{array}{c | c c c c c c c c c c}
\mbox{IIA} &x^0 & x^1 & x^2 & \t_1 & \t_2 & \phi_2 & \phi_1 & r &\t &
\psi \\  \hline &&&&&&&&&& \\
\mbox{D6}& -&-&-&-&-&-&-&&& \\
\mbox{D4}&-& & &-&-&-&-&&&\\
\end{array}\ee
\end{center}

\begin{center}
\be\begin{array}{c | c c c c c c c c c c c c}
\mbox{11d} &x^0 & x^1 & x^2 & \t_1 & \t_2 & \phi_2 & \phi_1 & r &\t & \psi & \phi
\\ \hline &&&&&&&&&& \\
\mbox{MKK}& -&-&-&-&-&-&-&&& &\\
\mbox{M5}&-& & &-&-&-&-& & & & -\\
\end{array}\ee
\end{center}

\avall

\subsection{Susy without susy when going to type IIA and to 8d gauged sugra} \label{ch6:sec:susy-without-susy}

Having obtained the Killing spinors in 11d directly, we are in the position
to discuss which kind of compactifications will preserve or destroy
a fraction of supersymmetry. The method was explained in section~\ref{ch5:sec:susy-without-susy}
and the following will provide a good set of examples of how the
'susy without susy' phenomenon works. The steps to follow are
\num{
\item Put all bosonic fields in a form that fits in the
ansatz to reduce. The most important point is that in these
type of ansatz {\it the elementary field is not the metric but
the vielbeins}. For example, in going to type IIA, the last
vielbein $e^T$ must be chosen such that it does not depend on
$x^T$. On the other hand, when going to 8d, the last three
vielbeins must be given in terms of  $SU(2)$ invariant
1-forms.
\item The chosen vielbeins provide a base of the tangent space.
We must express the Killing spinors in this base and then
look at how many of them are left invariant under $U(1)$
or $SU(2)$ transformations when going to IIA or 8d, respectively.\footnote{This
statement can be made more rigorous by computing
the more intrinsic Lie-Lorentz derivative~\cite{Ortin:2002qb} with
respect to the Killing vectors. For the cases considered in this paper,
such derivative collapses to the usual one.}
}
Let us algorithmically apply this procedure case by case.
\tem{
\item {\bf Flat D6-branes.}
   \tem{
   \item {\bf From 11d to IIA.} Note that the flat NC D6-branes
   background \bref{nonnearh} has at least two different
   $U$(1) isometries, generated by the Killing vectors
   $\partial_{\psi}$ and $\partial_{\phi}$. The amount of
   supersymmetry preserved is different along them. First of
   all, the vielbeins \bref{vi} we used in the computation
   are suitable for a reduction on both circles, as they do not
   depend on $\phi$ nor $\psi$. In this base, the 16 Killing spinors
   \bref{spinsol1} depend on $\phi$ but not $\psi$. Thus a reduction
   along $\phi$ kills all supersymmetries whereas along $\psi$ they
   are all preserved. The former leads to the IIA geometry
   found in {\it e.g.}~\cite{Larsson:2001wt}. The latter
   yields an interesting type IIA background as it is
   solution of the equations of motion which is not supersymmetric,
\bea
ds^2_{IIA}&=& \l^{\undos}
h^{1\over 3}\left(dx_{0,4}^2+h^{-1}dx_{5,6}^2+ H[dr^2+r^2 d\t^2]\right)
\,, \nn
&& ~~+~\l^{-\undos}h^{-{1\over 3}}R r^2 \sin^2\t \, d\psi^2 \label{pordios}
\,, \\
e^{4\Phi/3}&=& \l(r,\t) \,, \nn
B_{[2]}&=&-{\Theta R\over Hh}\cos\t \, dx^5\wedge dx^6 \,, \nn
C_{[1]}&=& \l^{-1}H^{-1}h^{-{2\over 3}}R\cos\t \, d\psi  \,, \nn
C_{[3]}&=&{\Theta R \over Hh}\, dx^5\wedge dx^6 \wedge d\psi\,,
\label{pordios2}
\eea
with
\be
\l(r,\t)\equiv Hh^{1\over 3}r^2 \sin^2\t+H^{-1}h^{-{2\over
3}}R \cos^2\t.
\ee

There is here a peculiarity. As we mentioned, the near-horizon of
the commutative limit of the 11d solution is locally flat space.
In this case there are 16 extra Killing spinors preserved (those
in \bref{enhance}). These are only $\psi$-dependent and survive
a $\phi$-reduction. Therefore, the $\T \dreta 0$ limit
of the IIA background \bref{pordios}-\bref{pordios2} does
preserve 16 supersymmetries.

\item {\bf From 11d to 8d.} Here we will encounter a relevant
novelty with respect to all other wrapped brane configurations
obtained in the literature. The compactification of
M-theory on an $SU$(2) manifold, as it was worked out
in ~\cite{Salam:1985ft}, requires the use of a
vielbein base for the $S^3$ which is not the one we used
in \bref{vi}. Instead, one has to use the $SU(2)$ invariant
one-forms $w^i$. Our conventions for this section are such that\footnote{Note that the signs
have been chosen so that both basis share the same orientation.}
\bea
w_1&=&-\cos\psi d\theta - \sin\theta \sin\psi d\phi \nn
w_2&=&-\sin\psi d\theta + \sin\theta \cos\psi d\phi \nn
w_3&=&-d\psi-\cos\theta d\phi. \label{vi3}
\eea
So instead of \bref{vi}, one should use
\bea
\hat{e}^8&=&\tau(r) H^{\undos}(r) r \, w^1 \nn
\hat{e}^9&=&\tau(r) H^{\undos}(r) r \, w^2 \nn
\hat{e}^T&=&\tau^{-2}(r)H^{-\undos}(r)R\, w^3 \,.
\label{vi2}
\eea
We will call \bref{vi2} the
$w$-base and \bref{vi} the $e$-base. It is easy to work out the
form of the spinor in this new base, since we have just performed
a local Lorentz transformation which can be shown to consist
of a rotation of $\pi$ along $x^9$, followed by a rotation of angle
$-\psi$ along $x^T$. The Killing spinors transform with
the (inverse) spin $\undos$ representation of such rotations
\be
\label{rotation} \e'\,=\,e^{-\psi {\G_{89}\over
2}}e^{\pi{\G_{T8}\over 2}}\,\e\,=\, \G_{T8}e^{\psi {\G_{89}\over 2}}\,\e.
\ee
Applying this to the Killing spinors \bref{spinsol2} we see that
they all become $\{\t,\phi,\psi\}$-dependent in the $\w$-base,
which means that a compactification to 8d supergravity will
not preserve a single supersymmetry. Note that all the bosonic
fields do fit in the reduction ansatz, so that we still obtain
a solution of the 8d gauged sugra equations,
\bea \label{oneone}
ds^2_{(8)}&=&{g\over 4}y \,h^{1/3} \left( dx_{0,4}^2+h^{-1} dx_{5,6}^2
+ dy^2\right)
\nn
e^{2\phi\over 3}&=&{g\over4} y
\nn
e^{\lambda}&=&h^{1/6}
\nn
G_{[2]}&=&-{\T g^2\over 16 N \gym^2}\,{y^2 \over h} \,\, dx^5\wedge dx^6
\\
G_{[3]}&=&-{\T g \over 4 N\gym^2}{y\over h^2} \,\,
dx^5\wedge dx^6\wedge dy,
 \label{two}
\eea
where $\lambda$ is a scalar field  on the coset space $\frac{SL(3,R)}{SO(3)}$
and $G_{[2]}$ and $G_{[3]}$ are field strength forms of 8d Sugra.

Its is now understandable what happened when we tried to find the solution
in 8d. Despite the fact that the whole solution \bref{oneone}-\bref{two}
fits in the ansatz for the bosonic fields that we were considering, there
was no hope to solve it by imposing supersymmetry. Had we worked at the
level of the equations of motion, we could have succeeded though.

}
\item {\bf Wrapped D6-branes}
\tem{ \item {\bf From 11d to IIA.}
            The vielbeins we used in the computations are suitable
            for reducing along both $\phi$ and $\psi$, but
            the 11d Killing spinors \bref{bigspin} depend only
            on  $(\t,\psi)$. Thus a reduction along $\psi$
            destroys all supersymmetry,\footnote{The resulting background
            is, in the commutative limit,
            the one we used in section~\ref{ch5:sec:zero-d} and yielded a zero-dimensional
            moduli space of the commutative $\caln=2$ SYM.}
            whereas one along $\phi$ will produce a type IIA solution
            preserving the four supercharges. Explicitly,\footnote{The
            commutative limit of this background
            was used in section~\ref{ch5:sec:two-d} and yielded a two-dimensional
            \kah moduli space of the commutative $\caln=2$ SYM.}
\bea \label{nc-metrica}
ds^2_{IIA}&=&e^{2 \Phi/3}\th^{1\over 3} \left(-dx_0^2+ \th^{-1}
dx_{1,2}^2+{3\over 2}(r^2+l^2)ds^2_{cycle} \right.
\nn
&& \espai\espai \left. +U^{-1}dr^2+
{r^2 \over 4} [ d\t^2 + m B_{[1]}^2]\right)
\,, \nn
e^{4\Phi/3}&=&\th^{-{2\over 3}} \tH ^{-1}
\,, \nn
B_{[2]}&=&-{\T\over \tH \th}\, dx^1\wedge dx^2
\,, \nn
C_{[1]}&=&-Uf^{-1}\cos\t\,B_{[1]}
\,, \nn
C_{[3]}&=&-{\T\over \tH \th}Uf^{-1}\cos\t\,\,\, dx^1\wedge dx^2 \wedge B_{[1]}.
\eea

\item {\bf From 11d to 8d.} When reducing to 8d gauged sugra, we find
a big difference between the commutative
and the NC cases, so we analyze them separately.
As discussed above, to see amount of  supersymmetry preserved in the $SU(2)$
compactification, we have to transform the spinors to the $SU$(2) left-invariant
$w$-base \bref{vi3}. To do so, we need to apply
the rotation \bref{rotation} to the Killing spinors.

If we are in the commutative case, it is easy to see that the corresponding
spinors become constant, independent of all the $S^3$ angles.
Therefore, the compactification can be performed preserving all four supersymmetries.
This is what allowed us to find such solution
using 8d supergravity.

On the other hand, in the NC case, it can be checked that not even the
metric can be put in a form that satisfies the reduction ansatz,
so the compactification is simply not possible.
As a consequence, the NC wrapped D6 solution \bref{ncd6-11} could have
never been found with the usual gauged supergravity method.
}}

\chapter{Hamiltonian Formalism for nonlocal theories}\label{ch:hamiltonian}

In this section we will develop a Hamiltonian formalism for theories
that are non-local in time. Our main concern will be to establish
a solid formalism and to apply it to NC theories with
electric noncommutativity. Throughout this section, the term
{\it non-local} will always implicitly mean {\it non-local in
time}; spatial non-locality is well understood and its
Hamiltonian formalism is straightforward. The whole chapter
is based on the papers containing the original construction
of the formalism~\cite{Llosa:1993sj,Gomis:2000gy} and the results
reported in~\cite{Gomis:2000sp}.

\section{Definition and examples of non-local theories} \label{ch4:sec:definition}

To simplify the discussion, let us start considering classical mechanics.
Standard Lagrangians are functions of $\{q(t),\dq(t),...,q^{(n)}\}$ with
$n$ finite. In other words, they depend on the value of a set of functions
{\it at a given point}, and hence the name of local Lagrangians.
The ones we want to deal with here depend on a whole piece of
trajectory about a given point $t$, so that we can write
\be \label{lnon}
L^{non}(t)=L([q(t+\sigma)]) \,,
\ee
with $\sigma$ being extendable as far as differentiability of $q(t)$ holds.
The best one can do, if Taylor's theorem applies, is to write
$L^{non}$ as a function of all time derivatives of $q$ at a given $t$.

The Euler-Lagrange (EL) equation is obtained by functional variation of
\bref{lnon}
\be\label{el-eq}
\int dt \cale (t,t';[q]) = 0 \sac \cale (t,t';[q]) \equiv {\d L^{non}(t) \over \d q(t')} \,.
\ee
The main qualitative difference with respect to the standard
cases is that the familiar existence and unicity theorems do not apply here.
This is because \bref{el-eq} is not a differentiable system. The
physical consequence of this is rather deep. One is (probably) used
to giving a set of initial conditions at some initial time and then
to interpreting the EL equations as univocally dictating the future
of the system. In our case, however, the 'initial conditions' are actually
the whole trajectory! Furthermore, not any trajectory is a good 'initial
condition' as it may not verify \bref{el-eq}. The point of view should
then be modified and the dynamics are summarized by saying that \bref{el-eq}
defines the hypersurface of allowed trajectories
in the space of all possible ones.

Needless to say, equation \bref{el-eq} reduces to the standard EL equations
if $L^{non}$ is actually local. In such case we can write
\bea
& &L(t)= L(q(t),\dq(t),...,\overset{(n)}{q}(t)) \hspace{.3cm} \Rightarrow
\hspace{.3cm} \cale (t,t';[q]) = \sum_{m=0}^n {\pa L \over \pa \overset{(m)}{q}(t)}
{d^m\over dt^m} \d (t-t') \,, \nonumber
\eea
so that \bref{el-eq} yields the familiar equations of motion
\be
\sum_{m=0}^n\left(-{d^m\over dt^m} \right) {\pa L \over \pa \overset{(m)}{q}(t)} =0 \,.
\ee

Examples of truly non-local theories are
\begin{enumerate}
\item Fokker-Wheeler-Feynman electrodynamics,
\item Regularized local field theories,
\item Some models of meson-nucleon interaction,
\item Semiclassical gravity,
\item String field theory,
\item The p-adic string,
\item Electric NC theories.
\end{enumerate}
Although we will mainly be concerned with the application of our
formalism to the last case, it has been recently applied~\cite{Gomis:2003xv}
to the study of tachyon condensation in the framework of cases 5 and 6.
{\vskip 1cm}
\caixa{\cent{{\bf Why is it important to have a Hamiltonian formalism?}}

First of all, the standard method to quantize a theory needs to go
through a Hamiltonian treatment; the Hamiltonian functional is promoted to
an operator on a Hilbert space and Poisson (or Dirac) parenthesis
are promoted to commutators. For nonlocal Lagrangians, there is nothing
analog to a Legendre transformation and not even the phase space is well
defined. As a bypass, one could take the Lagrangian Path Integral representation
as the definition of the quantum field theory
\be \label{pathi}
Z=\int [d\phi] \e^{-i \int L} \,,
\ee
but even this expression is normally derived from an Path Integral in
phase space and the existence of an hermitian Hamiltonian.
One of the properties that will be shown is that,
in our formalism, \bref{pathi} is obtained after integrating out
the momenta in a well-defined phase-space Hamiltonian path integral.}

\section{An equivalent first order Lagrangian} \label{ch4:sec:1sr-order}

The idea of~\cite{Llosa:1993sj} comes from trying to view the equation \bref{el-eq} still
as an 'evolution equation'. Let us imagine that we propose a piece of trajectory
$[q(\sigma)]$ as initial condition.
If we rewrite \bref{el-eq}  as (sometimes
this will only be possible implicitly)
\be
\ddot{q}_{\sigma}(t)=\calf(t,[q(\sigma)]) \,,
\ee
where $\calf$ is an integro-differential operator, then we can think
of it as determining the $t$-evolution of each point in $q(\sigma)$, which
we write as $q_{\sigma}(t)$. The problem of not having given a proper
initial condition $[q(\sigma)]$ is then that it can happen that
\be
q_{\sigma}(t) \neq q(\sigma+t) \,,
\ee
as we illustrate in figure~\ref{couldhappen}.
\clfigu{9}{7}{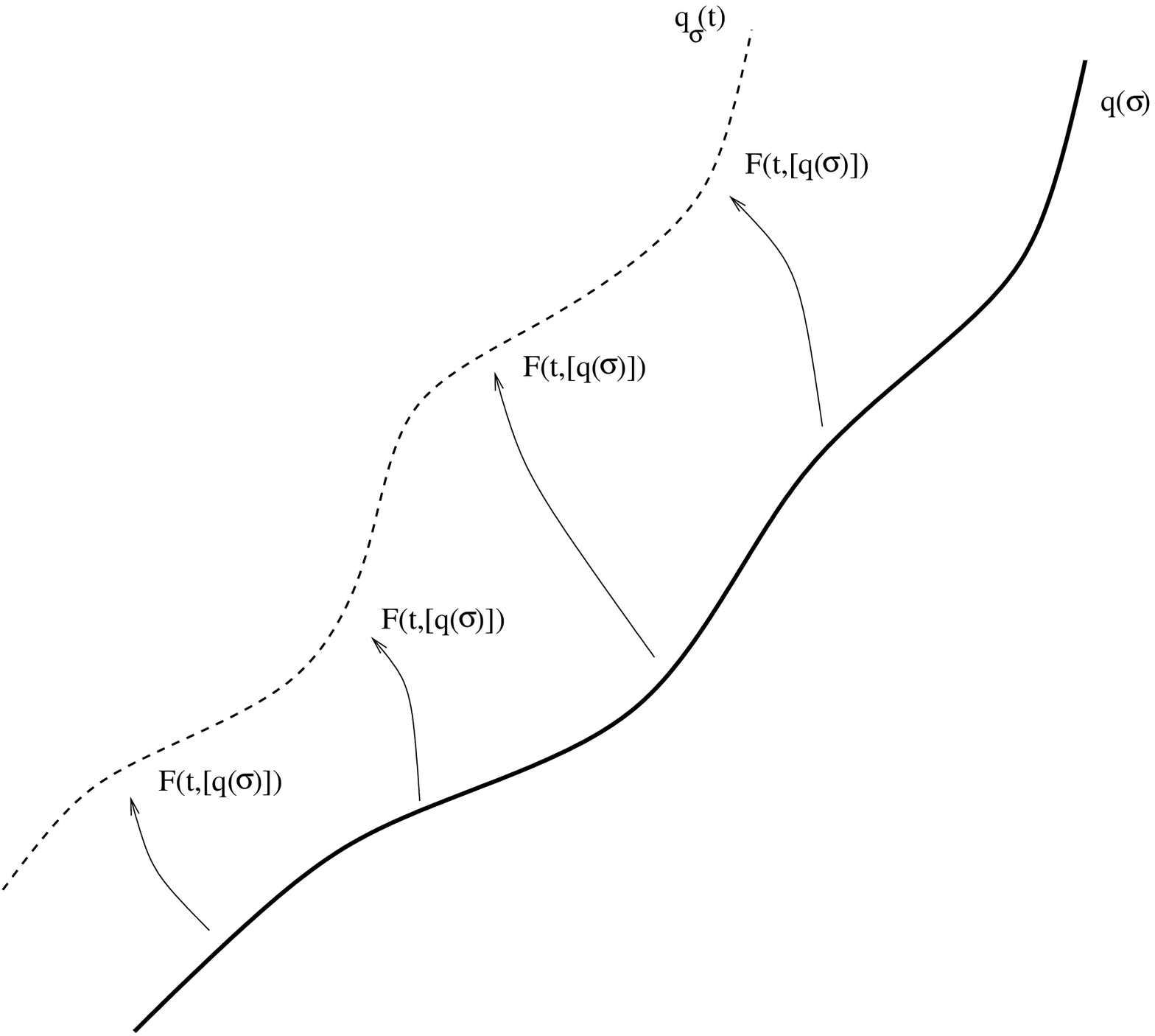}{The condition \bref{cons-one} forces
these two trajectories to coincide.}{couldhappen}

So, if we think of the EL equation as an equation for a function
of {\it two variables} $\calq(t,\sigma)$, with the boundary
condition that at $t=0$ we recover our proposed
solution \be \calq(0,\sigma)=q(\sigma)\,, \label{ini-con}\ee then all we need to
impose is the constraint that the EL equation must
be compatible with $\calq(t,\sigma)$ being a function only of $t+\sigma$:
\be \label{cons-one}
{\pa \over \pa(t-\sigma)} \, \calq(t,\sigma)=0 \,.
\ee
Note that although we replace $q(t)$ by a function of two
variables, the constraint \bref{cons-one} immediately
makes it into a function of a single one.
Of course this would be just a way of rephrasing the problem
if it was not for the great simplifications that this new
point of view will bring.

The equivalent Lagrangian $\tL[\calq(t,\sigma)]$ proposed in~\cite{Llosa:1993sj} is then
\be \label{new-laga}
\tL[\calq(t,\sigma)]=L^{non}[\calq(t,\sigma)] +\int d\sigma \,\mu(t,\sigma) [ \dot{\calq}(t,\sigma)-
\calq'(t,\sigma) ] \,,
\ee
where $\dot{}$ and $'$ stand for $\pa/\pa t$ and $\pa/\pa \sigma$ respectively.
The field $\mu$ is non-dynamical and it just introduces \bref{cons-one}
as a primary Lagrangian constraint. This enables us to pass all
the original nonlocality in time of $q(t)$ to nonlocality in $\sigma$
of $\calq(t,\sigma)$ by replacing everywhere in $L^{non}[\calq(t,\sigma)]$
\be
q(t+\rho) \rightarrow \calq(t,\sigma+\rho) \,.
\ee
But then, \bref{new-laga} is actually a first order Lagrangian!
This allows for a straight-forward development of a Hamiltonian
formalism.

Before going on, we would like to motivate from a more direct point
of view the form of the final 1+1 Lagrangian \bref{new-laga}.
Consider starting from a non-local Lagrangian, thought of as
a function of all the derivatives of $q$ at some point, \ie
$L(t)= L(q(t),\dq(t),...,\overset{(\infty)}{q}(t))$.
If we try to apply the usual procedure that makes it into
a first order one, we should introduce an infinite set of
new variables $\{q_1,q_2,...\}$ together with an infinite
set of constraints forcing
\be
q_{n+1} = {d\over dt}~ q_n \sac n=1,...,\infty\,.
\ee
These can be implemented at the Lagrangian level by introducing
an infinite set of Lagrange multipliers $\mu_n$, so that we would
end up with a Lagrangian
\be \label{intuitive}
L_{non} = L(q_0,q_1,...,q_{\infty}) + \sum_{n=0}^{\infty}
\left( {dq_n\over dt}-q_{n+1} \right) \mu_n (t) \,.
\ee
This is not quite the same as \bref{new-laga} yet, but if
we assume that $Q(t,\s)$ and $\mu(t,\s)$ can be expanded as
\be \label{expandable}
Q(t,\s)=\sum_{n=0}^{\infty} e_n(\s) q_n(t) \sac
\mu(t,\s)=\sum_{n=0}^{\infty} e_n(\s) \mu_n(t)\,,
\ee
with $e_n(\s) = {\s^n\over n!}$, then it is immediate to
check that the 1+1 Lagrangian \bref{new-laga} reduces to
\bref{intuitive}.
We remark that, however, the 1+1 Lagrangian admits richer
dynamics than \bref{intuitive}, as it may admit solutions
which are not expandable as in \bref{expandable}. It such
cases, one must remain in 1+1 to study the system and
proceed to its Hamiltonian formalism. We will return
to the issue of reducing back to 1 dimension in section~\ref{ch4:sec:reducing}.

Let us go on now with the 1+1 Lagrangian \bref{new-laga}.
Being first order, it allows for a straight-forward development of a Hamiltonian
formalism. In this case one has to take into account the various
set of constraints and proceed with the well-known Dirac formalism.
We refer the reader to~\cite{Llosa:1993sj} for details and we simply
quote here the results, which can be summarized in two steps:
\enub
\item The Hamiltonian for a classical mechanics problem of $N$ particles is
\bea \label{h}
H(t)&=&\int d\s ~[~\calp{^i}(t,\s)\calq{_i}'(t,\s)~-~\d(\s)\call(t,\s)~] \,,
\eea
with $i= 1,...,N$ and where the Lagrangian density $\call(t,\s)$ is constructed from the
original non-local one $L^{non}$ by performing the following replacements
\bea
q_i(t)&\to&\calq_i(t,\s) \,, \nn
\frac{d^n}{dt^n}q_i(t) &\to &\frac{\pa^n}{\pa\s^n}\calq_i(t,\s) \,, \nn
q_i(t+\rho) &\to &\calq_i(t,\s+\rho).
\label{rep}
\eea
\item There is one Hamiltonian constraint per particle given by\footnote{The
symbol $\approx$ is used for equations that must hold only on the
phase space hypersurface defined by the constraints.}
\be
\vp{^i}(t,\s)=\calp{^i}(t,\s)~-~\int d\s'~\chi(\s,-\s')~\cale{^i}(t;\s',\s)~
\approx~0,
\label{vp}
\ee
where
\be
\cale{^i}(t;\s',\s)=\frac{\d \call(t,\s')}{\d \calq{_i}(t,\s)}\sac
\chi(\s,-\s')=\frac{\e(\s)-\e(\s')}{2} \,,
\ee
and $\e(\s)$ is just the sign distribution.
\enue

In principle this is all we need to know to define
a Hamiltonian formalism. Everything now follows
from
\enub
\item The Hamilton equations of motion
\bea
\dot \calq{_i}(t,\s)&=& \calq_i'(t,\s),
\label{Qdot}\\
\dot \calp{^i}(t,\s)&=& {\calp^i}'(t,\s)+\frac{\d \call(t,0)}{\d \calq{_i}(t,\s)}=
{\calp^i}'(t,\s)+\cale{^i}(t;0,\s), \nn
\label{Pdot}
\eea
\item The compatibility of these equations of motion with the
constraint \bref{vp}; in other words, from demanding that
the evolution dictated by the equations of motion does not
move the system away from the hypersurface in phase space
determined by the constraints. This leads us to
\bea
\dot\vp{^i}(t,\s)~\approx~\d(\s)~[\int d\s'~\cale{^i}(t;\s',0)]\approx 0.
\label{EOM}
\eea
We should require  further consistency conditions of this constraint.
Repeating this we get an infinite set of Hamiltonian constraints
which can be expressed collectively as
\bea
\tilde\vp{^i}(t,\s) \equiv \int d\s'\cale{^i}(t;\s',\s)&\approx&0
,~~(-\infty<\s<\infty).
\label{EOMf}
\eea
If we use \bref{Qdot} and \bref{rep} it reduces to the EL equation
\bref{el-eq} of $q_i(t)$ obtained from $L^{non}(t)$.
\enue
{\vskip 1cm}
{\bf Summary:}

Equation \bref{EOMf} is precisely what we were looking for at the beginning,
since now we see that the new 1+1 Hamiltonian system incorporates the EL equation
as a constraint on the phase space. It can actually be taken as a proof that
our Hamiltonian formalism is equivalent to the original nonlocal system
in one dimension less. The advantage is that we are now dealing with a
system which is {\it local in time}. Issues like the construction of the
conserved charges, the BRST quantization and the field-antifield formalism
follow naturally in the 1+1 formalism.

\section{Reducing back the fake non-local theories} \label{ch4:sec:reducing}

One of the self-consistency tests that one must ask to
the $d+1$ formalism\footnote{We indistinctively use the name $1+1$
or $d+1$ for the formalism with one extra dimension, as it equally
applies to mechanics and field theory.} is that all its phase space quantities
reduce back to the ones we would compute in $d$ dimensions
when the theory we are dealing with is actually local.
We stress that, at the moment, the only consistent formalism
in the literature for nonlocal theories is in $d+1$, and
that one must remain in $d+1$ to perform any phase space
analysis. The reduction to $d$ dimensions is not possible
in general, one exception being obviously theories that
are actually local. 
In this section we describe how this reduction works.

Let us then consider a regular higher derivative theory described by the
Lagrangian $L(q,\dot q,\ddot q,...,q^{(n)})$ and proceed to the $1+1$ formalism
as if we were dealing with a non-local Lagrangian. When we embed
the higher order theory in the non-local setting, our phase space
$T^*J(t)=\left\{ Q(t,{\s}), P(t,{\s})\right\}$
becomes infinite dimensional.\footnote{In this subsection we
consider only the one particle case for simplicity, so that
we remove all subscripts $i$ from the formulas in the
previous section.}
We expand the phase space quantities in the Taylor basis~\cite{Marnelius:1974yt}
\be
Q(t,{\s}) = \sum_{m=0}^{\infty}~e_m({\s})~q^m(t) \sac
P(t,{\s}) = \sum_{m=0}^{\infty}~e^m({\s})~p_m(t),
\label{momentahigh}
\ee
where $e^\l({\s})$ and $e_\l({\s})$ are orthonormal bases
\bea
e^\l({\s})&=&(-\pa_{\s})^\l{\d}({\s}),~~~~e_\l({\s})~=~\frac{{\s}^\l}{\l!}.
\eea
Note that the coefficients in \bref{momentahigh} become new canonical variables
\bea
\{q^m(t),~p_n(t)\}&=&{{\d}^m}_n \,.
\label{symplectich}
\eea
We can now rewrite the Hamiltonian \bref{h} and the two momentum constraints
\bref{vp} and \bref{EOMf} in this new basis:
\bea
H(t) &=&\sum_{m=0}^{\infty}~p_m(t)~q^{m+1}(t)~-~L(q^0,q^1,...,q^n)\,,
\\
\vp_m(t)&=&p_m(t)-\sum_{\l=0}^{n-m-1}
(-D_t)^{\s}~\frac{\pa L(t)}{\pa q^{\l+m+1}(t)}\approx 0\,,
\label{vpmm} \\
\psi^m(t)&=&(D_t)^m\;[\sum_{\l=0}^n~ (-D_t)^\l
 \frac{\pa L(t)}{\pa q^\l(t)}]\approx 0 \,,
\label{EOMm}
\eea
where
\bea
D_t &=& \sum_{r=0} q^{r+1}\frac{\pa}{\pa q^r}.
\label{defD}
\eea
These constraints \bref{vpmm} and \bref{EOMm} are
second class and thus they can be  used to reduce
the dimension of the phase space. It will happen
that the reduction is so large that it will turn it into
a finite dimensional one, leading to the ordinary Ostrogradski \ham formalism.
The operator $D_t$ defined in \bref{defD} will become
a time evolution operator for the $q$'s using the first set of the Hamilton equations
\bea
\dot q^r&=&q^{r+1}.
\label{qmdot}
\eea
Using this in \bref{vpmm} the constraints
 $\vp_{m},\;(0\leq m \leq n-1) $ coincide with the definition of
the Ostrogradski momenta
\bea
p_m&\sim& ~\sum_{\l=0}^{n-m-1}
(-\pa_t)^{\s}~\frac{\pa L(t)}{\pa(\pa_t^{\l+m+1} q(t))}\sac
0\leq m \leq n-1.
\label{vpmm2}
\eea
These $n-1$ equations allow to solve
for $q^\l,\;(n\leq \l \leq 2n-1)$ as functions of
canonical pairs $\{q^j,p_j\},\;(0\leq j\leq n-1)$,
\be
q^\l\approx q^\l(q^0,q^1,...,q^{n-1},p_0,p_1,...,p_{n-1}) \sac
n\leq \l \leq 2n-1.
\label{qmm}
\ee
They are combined with the constraints $\vp_\l,\;(n\leq \l \leq 2n-1)$
\bea
\vp_{\s}=p_\l\approx 0\sac n\leq \l \leq 2n-1,
\eea
to form a second class set and can be used to eliminate the canonical pairs
 $\{q^\l,p_\l\}\;(n\leq \l\leq 2n-1)$.

If we take into account \bref{qmdot}
the constraint \bref{EOMm} for $m=0$ becomes the Euler-Lagrange
equation for the original higher derivative Lagrangian,
\bea
\psi^0&\sim&\;\sum_{\l=0}^n~ (-\pa_t)^\l
 \frac{\pa L(t)}{\pa(\pa_t^\l q(t))}=0.
\label{EOM0}
\eea
The  constraints \bref{EOMm} for $m>0$ are the time derivatives of the
Euler-Lagrange equation \bref{EOM0} expressed in terms of $q$'s.
For a non-singular theory, all such constraints can be rewritten as
\bea
q^\l&-&q^\l(q^0,q^1,...,q^{n-1},p_0,p_1,...,p_{n-1})\approx 0 \sac
\l \geq 2n
\label{qmmm}
\eea
and can be paired with the constraints $\vp_{\s},\;(\l\geq 2n)$
\bea
\vp_{\s}=p_\l\approx 0 \sac \l\geq 2n \,,
\eea
forming another set of second class constraints.
They can then be used to eliminate the canonical pairs $\{q^\l,p_\l\}\;(\l\geq 2n)$.

In this way the infinite dimensional phase space is reduced to a finite
 dimensional one.
The reduced phase space is coordinated  by
$T^*J^n=\{ q^l,p_l\}$ with $l=0,1,...,n-1$.
All the  constraints are second class and
the Dirac bracket for these variables has the standard form,
\bea
\{q^m,p_n\}^*&=&{{\d}^m}_n \sac \{q^m,q^n\}^*=\{p_m,p_n\}^*=0.
\label{symplectichh}
\eea
Finally, The Hamiltonian \bref{h} in the reduced space becomes
\bea
H(t)&=&\sum_{m=0}^{n-1}~p_m(t)~q^{m+1}(t)~-~L(q^0,q^1,...,q^n) \,,
\label{hhh}
\eea
where the $q^n$ are expressed using \bref{qmm} as functions of the reduced
variables in $T^*J^n$.

\caixa{
Note that if we consider the limit $n$ going to infinity,
the constraints \bref{vpmm} and \bref{EOMm} do not allow, in general,
to reduce the dimensionality of the infinite dimensional phase space of
the non-local system via Dirac brackets. This shows that
the Ostrogradski \ham formalism does not generalize properly for
 truly  non-local theories.}

\section{A proper Path Integral quantization} \label{ch4:sec:pathintegral}

Let us consider the Hamiltonian path integral quantization of the  $1+1$
dimensional field theory associated with the \ham (\ref{h}) for
$L^{\rm non}(t)$. The path integral in the presence of the two
constraints is given by
\bea \calz &=&
\int [d P(t,{\l})][dQ(t,{\l})] \,\, \mu \nn
&\times&
\exp \left\{i\int dt d{\l} \left(
P(t,{\l})[\dot Q(t,{\l})-Q'(t,{\l})]+\tilde L(t){\delta({\l})}
\right) \right\}. \nn
\label{pathh}
\eea
The integration is performed over the constrained phase space
 thanks to the measure~\cite{Faddeev:1969su,Senjanovic:1976br}
\bea
\mu&=& \det
\left( \begin{array}{cc} \{\vp,\vp\}&\{\vp,\psi\} \\ \{\psi,\vp\}&\{\psi,\psi\} \end{array} \right)
 ~{\d}(\vp){\d}(\psi) \,.
\eea

Using the expansions of the previous section,
it is immediate to show that this path integral reduces
to the Ostrogradski one in the cases that we deal with
a local theory. However, the opposite is not true, in the
sense that if we start with the Ostrogradski Path Integral
and we just let $n\rightarrow \infty$, we recover
\bref{pathh} {\it without the measure} $\mu$. The system
does not have therefore the necessary constraints to
be equivalent to the Lagrangian formalism.

We conclude by noting that the Hamiltonian Path Integral
\bref{pathh} reproduces the correct Lagrangian path integral
after integrating out the momenta.

\section{Hamiltonian symmetry generators} \label{ch4:sec:symmetry-generators}

For local theories, symmetry properties of the system are typically examined
using the N\"oether theorem.
In \ham formalism the relation between
symmetries and conservation laws has been discussed extensively for
singular\footnote{We recall that a Lagrangian is said to be singular
when $\det \left({\pa L \over \pa \dq \pa \dq}\right)=0$ which implies that, in trying
to build a phase space, not all the momenta can be solved in terms
of the velocities.}
Lagrangian systems, for example
\cite{Konopleva:1981ew}\cite{Pons:1999az}.
In this section, we develop a formalism to treat the case of non-local theories.

Suppose we have a non-local \lag like \bref{lnon} which  is invariant under
some transformation $\d q(t)$ up to a total derivative,
\bea
\d L^{non}(t)&=&\int dt'~\frac{\d L^{non}(t)}{\d q{_i}(t')}~
{\d q{_i}(t')}~=~\frac{d}{dt}k(t).
\label{DL}
\eea
Now we move to our $1+1$ dimensional theory and take profit
of the fact that it was local in the evolution time $t$.
Therefore, we can construct the corresponding symmetry
generator in the \ham formalism in the usual way
\bea
G(t)&=&\int d\s ~[~\calp{^i}(t,\s)\d \calq{_i}(t,\s)~-~\d(\s)\cf(t,\s)~],
\label{G}
\eea
where $\d \calq_i(t,\s)$ and  $\cf(t,\s)$ are constructed from
 $\d q(t)$ and  $k(t)$ respectively by the same replacements \bref{rep}.
The quasi-invariance of the non-local \lag \bref{DL}, translated to the $1+1$ language,  means
that
\bea
 \int d\s'~\frac{\d\call(t,\s)}{\d \calq{_i}(t,\s')}{\d \calq{_i}(t,\s')}&=&
\pa_\s \cf(t,\s).
\label{paF}
\eea

When the original non-local \lag has a gauge symmetry the $\d q_i(t)$
and  $k(t)$ contain an arbitrary function of time $\l(t)$ and its
$t$ derivatives.
In $\d \calq_i(t,\s)$ and  $\cf(t,\s)$ the $\l(t)$ is replaced  by
$\Lam(t,\s)$ in the same manner  as  $q_i(t)$ is replaced  by
$\calq_i(t,\s)$ in \bref{rep}.
However in order for the transformation generated by \bref{G} to be
a symmetry of the Hamilton equations,
$\Lam(t,\s)$ can not be an arbitrary function of $t$ but it should satisfy
\bea
\dot\Lam(t,\s)&=&\Lam'(t,\s)
\label{Lamdd}
\eea
as will be shown shortly.
This restriction on the parameter function $\Lam$ means that the
transformations generated by $G(t)$ in the $\dpo$ dimensional
\ham formalism are {\it rigid} transformations in contrast to the
original ones for the non-local theory which are {\it gauge} transformations.
In the appendix~\ref{ch4:sec:u1-c} we will see how this {rigid} transformations in
the $\dpo$  dimensional \ham formalism are reduced to the usual
{gauge} transformations
in $d$ dimension for the $U(1)$ Maxwell theory.
\medskip

The generator $G(t)$ generates the transformation of $\calq_i(t,\s)$,
\bea
\d \calq{_i}(t,\s)&=&\{\calq{_i}(t,\s),~G(t)\},
\eea
corresponding to the transformation $\d q_i(t)$ in the non-local Lagrangian.
It also generates the transformation of the
momentum $\calp^i(t,\s)$ and so, of any functional of the phase space variables.
In particular, we will see that, as consistency demands, the Hamiltonian
\bref{h}
and the constraints \bref{vp} and \bref{EOMf} are 
invariant, in the sense
that their symmetry 
transformation vanishes on the hypersurface of
phase space determined by the constraints.
Let us state a series of results and properties of our gauge generator.

\medskip

{\it a) G(t) is a conserved quantity}

\small
\bea
\frac{d}{dt}G(t)&=&\{G(t),H(t)\}~+~\frac{\pa}{\pa t}G(t)
\label{dtG1}\\
&=& \int d\s d\s'\left[\calp{^j}(t,\s) \left( \frac{\d(\d \calq{_j}(t,\s))}
{\d \calq{_i}(t,\s')}\calq{_j}'(t,\s')-
\pa_\s\d(\s-\s'){\d \calq{_j}(t,\s')}
\right.\right.\nn &&\left.\left.~~~~~~~~~+~
\frac{\d(\d \calq{_j}(t,\s))}{\d \Lam(t,\s')}\dot\Lam(t,\s')\right)~-~
\d(t,\s)~\left(\frac{\d \cf(t,\s)}{\d \calq{_i}(t,\s')}\calq{_i}'(t,\s')
\right.\right.\nn &&\left.\left.~~~~~~~~~-~
\frac{\d(\call(t,\s))}{\d \calq{_i}(t,\s')}{\d \calq{_i}(t,\s')}+
\frac{\d \cf(t,\s)}{\d \Lam(t,\s')}\dot\Lam(t,\s')\right)\right]~=~0.
\label{dtG}\eea
\normalsize
The last term of \bref{dtG1} is an explicit $t$ derivative
through $\Lambda(t,\s)$. In order to show \bref{dtG}
we need to use the symmetry condition \bref{paF} and
the condition on $\Lambda(t,\s)$ in \bref{Lamdd}.

\medskip
\medskip

{\it b) All the constraints are invariant under the symmetry transformations.}
{\vskip 0.2 cm}

Let us show first the invariance of \bref{EOMf}, which is
nothing but the invariance of the equations of motion, as was to expected
for $G(t)$ generating a symmetry,
\small
\bea \lefteqn{
\{\7\vp{^i}(t,\s),G(t)\}
=
\int d\s' [\calp{^j}(t,\s')\d \calq{_j}(t,\s')-\d(\s')\cf(t,\s')]\}}
\nn
&=&
\int d\s' d\s'' \frac{\d^2\call(t,\s'')}{\d \calq{_j}(t,\s')\d \calq{_i}(t,\s)}
\d \calq{_j}(t,\s')=
\int d\s' \frac{\d \7\vp{^j}(t,\s')}{\d \calq{_i}(t,\s)}\d \calq{_j}(t,\s')
\nn
&=&
-\int d\s'  \7\vp{^j}(t,\s')\frac{\d (\d \calq{_j}(t,\s'))}
{\d \calq{_i}(t,\s)}
\approx0,
\label{vptG}
\eea
\normalsize
where we have used an identity obtained from \bref{paF},
\be
\int d\s d\s'\cale{^j}(t,\s,\s')\d \calq{_j}(t,\s')=
\int d\s'\7\vp{^j}(t,\s')\d \calq{_j}(t,\s')~=~0.
\label{IpaF}
\ee
We now prove the invariance of the remaining constraint
\bref{vp}. Using \bref{paF} and \bref{IpaF},
\small
\bea
&&
\{\vp{^i}(t,\s),G(t)\}
=
\nn&&
=-\int d\s'\vp{^j}(t,\s')\frac{\d (\d \calq{_j}(t,\s'))}{\d \calq{_i}(t,\s)}-
\int d\s'[\int d\s''\chi(\s',-\s'')\cale{^j}(t;\s'',\s')
\frac{\d (\d \calq{_j}(t,\s'))}{\d \calq{_i}(t,\s)}
\nn&&~\hskip 22mm~
-\d(\s')
\frac{\d(\cf(t,\s'))}{\d \calq{_i}(t,\s)}+
\int d\s''\chi(\s,-\s'') \frac{\d\cale{^i}(t;\s'',\s)}{\d \calq{_j}(t,\s')}
\d \calq{_j}(t,\s')]
\nn
&&
=-\int d\s'\vp{^j}(t,\s')\frac{\d (\d \calq{_j}(t,\s'))}{\d \calq{_i}(t,\s)}+
\int d\s'\chi(\s,-\s')\7\vp{^j}(t,\s')
\frac{\d (\d \calq_{j}(t,\s'))}{\d \calq{_i}(t,\s)}~\approx~0. \nonumber
\label{vpG}
\eea
\normalsize
Thus we have shown that the constraint surface defined by
$\vp\approx\7\vp\approx 0$
is invariant under the transformations generated by $G(t)$.

\medskip
\medskip

{\it c) Our Hamiltonian \bref{h} is the generator of time translations.}

\medskip

Consider a non-local Lagrangian in \bref{lnon} that does not depend on
$t$ explicitly, so that time translation is a symmetry of the Lagrangian.
To show that the generator of such a symmetry is our Hamiltonian $H$ in
\bref{h} and that it is conserved,
we should simply show that we recover its expression \bref{h} from the general
form of the generator \bref{G}. Indeed,
the \lag changes as $\d L^{non}=\vep \dot L^{non}$ under a time translation
$\d q_i(t)=\vep \dot q_i(t)$. The corresponding generator in the present
formalism is then, using \bref{G}
\bea
G_H(t)&=&\int d\s ~[~\calp{^i}(t,\s)(\vep  \calq{_i}'(t,\s))~-~\d(\s)(\vep\call
(t,\s))~],
\label{GH}
\eea
which is $\vep$ times the Hamiltonian \bref{h}, as we wanted to show.
Then, property $b)$ applied to this case brings us back
something that we imposed in the construction of the formalism:
the constraints hypersurface is stable under time evolution.

As our Hamiltonian in the $1+1$ theory is the generator of time
translations, it should considered as giving
the energy of the system.
If we had started with a local theory, but still we had insisted
on using the $1+1$ formalism, the system of constraints would
the allow us to reduce to one dimension less and to recover
the standard Hamiltonian of such local theory. We will
explicitly show in the appendix~\ref{ch4:sec:u1-c} how this works
for a $U(1)$ {\it commutative case}.
Nevertheless, for a truly non local theory, there is no such a
simplification and the phase space typically remains infinite dimensional.
Our discussion then shows that
it is our Hamiltonian \bref{h} the one that we should
use for computing the energy of the system.

\medskip

{\bf Summary:}

We have constructed the \ham symmetry generators of
a general non-local theory working in a $\dpo$ dimensional space.  In
this formulation original {gauge} symmetries in $d$ dimensions
are {rigid} symmetries in the $\dpo$ dimensional space.
The next section is an illustration of how our formalism
can be applied to a NC gauge theory, like the ones we discussed
in the previous chapter.

\section{ $U(1)$ non-commutative gauge theory} \label{ch4:sec:u1-nc}

Let us apply all the machinery of the new \ham formalism
to one of the nonlocal theories that we studied in chapter~\ref{ch:nc-theories}:
a $U(1)$ NC gauge theory in an electric background.
The action is
\be \label{Maxwell}
S=-\frac14~ \int d^dx~\8 F_{\mu\nu}\8 F^{\mu\nu},
\ee
where $\8 F_{\mu\nu}$ is the field strength of the $U(1)$ NC
gauge potential $\8 A_\mu$ defined by{\footnote
{Once again, we recall that we put "hats" on NC fields, see
section~\ref{ch3:sec:nc-gauge}.}}
\bea
\8 F_{\mu\nu}&=&\pa_\mu \8 A_\nu-\pa_\nu \8 A_\mu-
i[\8 A_\mu,\8 A_\nu].
\label{defF}
\eea
In this section, all commutators are defined using the NC *-product
that we defined in \bref{Moyal}, so that $[f,g] \equiv f*g-g*f$.

Here we are mainly interested in the most
general case of {\it space-time} non-commutativity with
$\theta^{0i}\neq 0$.{\footnote {A \ham formalism for the magnetic
theory ($\theta^{0i}=0$) was analyzed in~\cite{Dayi:2000ht}.}} This means
that the action \bref{Maxwell} contains an infinite number of time
derivatives of the field $\8 A$, and it is therefore nonlocal in time.
Let us now obtain the main equations that we will need later
to translate to the $d+1$ formalism.
\enub
\item
The EL equation of motion is
\bea
\8 D_\mu\8 F^{\mu\nu}&=&0,
\label{ELeq}
\eea
where the covariant derivative is defined by
$\8 D=\pa-i[\8 A,~~]$.
\item
The gauge transformation is
\bea
\d \8 A_\mu&=&\8 D_\mu \8\l \label{u1trans}
\eea
and it satisfies a non-Abelian gauge algebra,
\bea
(\d_{\8\l} \d_{\8\l'} -\d_{\8\l'}\d_{\8\l}) \8 A_\mu&=&
-i\8 D_\mu [\8\l,\8\l'].
\eea
Since the field strength transforms covariantly as
$\d \8 F_{\mu\nu}=-i[ \8 F_{\mu\nu},\8\l]$,
the \lag density of \bref{Maxwell} transforms as
\bea
\d (-\frac14~\8 F_{\mu\nu}\8 F^{\mu\nu})&=&
\frac{i}{2}~[\8 F_{\mu\nu},\8\l]~\8 F^{\mu\nu}.
\label{Dlag}
\eea
Using $\int d x(f*g)=\int d x(fg)$ and the associativity of the
$*$-product, \bref{Dlag}  becomes a total divergence, as
was to be expected for \bref{u1trans} being a symmetry. We have just
proven that the action \bref{Maxwell} is invariant under
the $U(1)$ NC transformations.
\enue
\subsection{Going to the $d+1$ formalism} \label{ch4:sec:u1-dp1}

Not only the \lag \bref{Maxwell} is  non-local, but also
the NC gauge transformation \bref{u1trans} is since,
for electric backgrounds ($\theta^{0i}\neq 0$),
it contains time derivatives of infinite order in $\8 \l$.
Let us now proceed to construct the Hamiltonian and the generator
for the $U(1)$ NC theory using the
formalism introduced in the last section.
\bitem
\item
We associate a $\dpo$ gauge potential\footnote{From
now on we will use calligraphic letters for fields in the $d+1$ formalism.}
 $\8{\cala}_\mu(t,\s,\bx)$
to the $d$ dimensional one $\8A_\mu(t,\bx)$.
\item
We regard $t$ as the evolution ``time''. For convenience
of notation, we relabel $\s \rightarrow x^0$. This is the same $\s$
appearing in $q_i(t,\s)$ in the last section.
\item
The other $(d-1)$ spatial coordinates ${\bf x}$ correspond to
the indices $i$ of $q_i(t,\s)$.
The signature of $\dpo$ space is $(-,-,+,+,...,+)$.
\eitem
\caixa{As explained above, the aim will be to convert the original motion in time
into motion in $x^0$ of $\8{\cala}_\mu(t,x^0,\bx)$, but not into
motion in $t$. Despite its signature, we will often refer to $x^0$ as
another spatial coordinate, and we will reserve the name of {\it time}
for $t$.}

\zavall

The canonical system equivalent to the non-local action \bref{Maxwell}
is defined by the Hamiltonian \bref{h} and the two constraints,
 \bref{vp} and \bref{EOMf}. For our present theory,
the Hamiltonian reads
\bea
H(t)&=&\int d^dx\;[{\8{\it \Pi}}^\nu(t,x)\pa_{x^0}\8 {\cala}_\nu(t,x)-
\d(x^0)\call(t,x)],
\label{Ham}
\eea
where ${\8{\it \Pi}}^\nu$ is a momentum for $\8 {\cala}_\nu$ and
\bea
\call(t,x)&=&-\frac14~\8 \calf_{\mu\nu}(t,x)\8 \calf^{\mu\nu}(t,x),
\\
\8 \calf_{\mu\nu}(t,x)&=&\pa_\mu \8 {\cala}_\nu(t,x)-
\pa_\nu \8 {\cala}_\mu(t,x)-i
[\8 {\cala}_\mu(t,x),\8 {\cala}_\nu(t,x)].
\eea
Note that by using \bref{Qdot}, {\it the
$*$-product involves now only $x^\mu=(x^0,\bx)$
but it does not involve $t$}. Thus it
contains spatial derivatives of infinite order but no
{time} derivative. The same applies for the Hamiltonian, it contains
no derivative with respect to $t$, and so it is a good phase-space quantity,
a function of the canonical pairs
$\{\8 {\cala}_\mu(t,x),\8 {\it\Pi}^\mu(t,x)\}$ with Poisson bracket
\bea
\{\8 {\cala}_\mu(t,x),\8 {\it\Pi}^\nu(t,x')\}&=&
{\d_\mu}^\nu~ \d^{(d)}(x-{x}').
\eea
The momentum constraint  \bref{vp} is
\small
\bea
\vp^\nu(t,x)&=&\8 {\it\Pi}^\nu(t,x)+\int dy~\chi(x^0,-y^0)\;
\8 \calf^{\mu\nu}(t,y)\;\8 \cald^y_\mu\d(x-y)
\nn
 &=&
\8 {\it\Pi}^\nu(t,x)+\d(x^0)\8 \calf^{0\nu}(t,x)-\frac{i}{2}
\left(\e(x^0)[\8 \calf^{\mu\nu},\8 {\cala}_\mu]
-[\e(x^0)\8 \calf^{\mu\nu},\8 {\cala}_\mu]\right)\approx 0.
\nonumber
\label{vpu1}
\eea
\normalsize
while the constraint \bref{EOM}, which followed from the consistency of
the one just obtained, turns out to be
\bea
\7\vp^\nu(t,x)&=&\8 \cald_\mu \8 \calf^{\mu\nu}(t,x)\approx 0.
\label{EOMu1}
\eea
Note that, as expected for a theory with gauge invariance,
these constraints are reducible since $\8\cald_\mu\7\vp^\mu\equiv 0$.
The Hamilton equation \bref{Qdot}, which now reads
\bea
\pa_{t}{\8{\cala}}_\mu(t,x)&=&\{\8 {\cala}_\mu(t,x), H(t)\}~=~
\pa_{x^0}\8 {\cala}_\mu(t,x) \,,
\eea
together with the identification \bref{rep}, $\8 {\cala}_\mu(t,x^\nu)=\8
A_\mu(t+x^0,\bx)$, can be used in \bref{EOMu1} to recover the
original equations of motion.
Finally, since the \lag of \bref{Maxwell} translational invariant,
the Hamiltonian \bref{Ham}, as well as the
constraints \bref{vpu1} and \bref{EOMu1}, are conserved.

Let us now show how to compute the generator of the NC $U(1)$
transformation. We just apply \bref{G} to our case,
\bea
G[\8\Lam]&=&\int d^dx[\;\8 {\it\Pi}^\mu \d \8 {\cala}_\mu-\d(x^0)\;\cf^0\;],
\label{defG}
\eea
where the last term must be evaluated from the surface term appearing
in the variation of the Lagrangian, \ie we must read it from the RHS
of \bref{Dlag}. Instead of rewriting \bref{Dlag} as $\pa_{\mu}$ of
some $k^\mu$, it is easier to integrate by parts the second term
in \bref{defG} as follows
\be
\int d^dx[-\d(x^0)\;\cf^0\;]=
\int d^dx[\frac{\e(x^0)}{2}\;\pa_{\mu}\cf^\mu\;]=
\int d^dx[\frac{\e(x^0)}{2}\;\d \call\;].
\nonumber
\ee
We can then plug immediately the RHS of \bref{Dlag} (after the usual
replacements needed to go to the $d+1$ formalism).
The final expression for the $U(1)$ generator is then
\bea
G[\8\Lam]&=&\int d^dx\left[\8 {\it\Pi}^\mu \8 \cald_\mu\8\Lam~+~
\frac{i}{4}{\e(x^0)} \8 \calf_{\mu\nu}~[\8 \calf^{\mu\nu},\8\Lam]\right],
\label{defG2}
\eea
where, as discussed in \bref{Lamdd}, $\8\Lam(t,x^\mu)$ is  an arbitrary
function satisfying
\bea
\dot{\8\Lam}(t,x^\mu)&=&\pa_{x^0}\8\Lam(t,x^\mu) \,.
\label{Lamdd1}
\eea
The generator can be expressed as a linear
 combination of the constraints,
\bea
G[\8\Lam]&=&\int d^dx~\8\Lam\left[-(\8 \cald_\mu\vp^\mu)-\d(x^0)\7\vp^0+
\frac{i}{2}\left(\e(x^0)[\7\vp^\nu,\8 {\cala}_\nu]-[\e(x^0)\7\vp^\nu,
\8 {\cala}_\nu]
\right)\right].
\nn
\label{GLam}
\eea
The fact that the generator \bref{GLam} is a sum of constraints
shows explicitly
the conservation of the generator on the constraint surface.
It is not hard to check that $G[\8\Lam]$ is conserved
\bea
\frac{d}{dt}G[\8\Lam]&=&\{G[\8\Lam],H\}~+~\frac{\pa}{\pa t}G[\8\Lam]~=~0
\label{dtGLam} \,,
\eea
in agreement with \bref{dtG}.

\vskip 4mm

We will conclude this section by computing the energy of a
given field configuration. For this, we need to isolate the
part of the Hamiltonian which does not vanish in the
constraint surface. It turns out that the Hamiltonian
can be rewritten as
\bea
H&=&G[\8{\cala}_0]\;+\;\int d^dx\;\vp^i\;\8\calf_{0i}\;+\;E_L,
\eea
where the first term is the $U(1)$ generator \bref{GLam}
replacing the parameter $\8\Lam$ by $\8{\cala}_0$;
it therefore vanishes in the constraint surface.
The last term $E_L$ is then the only relevant piece. Its explicit form is
\bea
E_L&=&
\int d^dx\; \d(x^0)\;\{\frac12 {\8\calf_{0i}}^2+\frac14 {\8\calf_{ij}}^2 \}
\nn
&& +
\frac{i}{2}\int d x\;\8{\cala}_0\big\{
\frac12 [\8\calf^{ij}, \e(x^0)\8\calf_{ij}]-[ \8\calf^{0i},\e(x^0)
\8\calf_{0i}]\big\}
\nn&&+
\frac{i}{2}\int d x\;\8{\cala}_j\big\{
[\8\calf_{0i}, \e(x^0)\8\calf^{ij}]-[\e(x^0) \8\calf_{0i},\8\calf^{ij}]\big\}.
\label{jeje}
\eea
This expression is useful, for example,
to evaluate the energy of classical configurations of the theory.
The two terms in the first line have the same form as the "energy" of
the commutative $U(1)$ theory. The novelty are the last two lines,
which are non-local contributions. They need to be taken into account
except for two cases in which they identically vanish
\enub
\item in magnetic backgrounds $\t^{0i}=0$,
\item in $t$ independent solutions of ${\cala}_\mu$.
\enue

 \vskip 6mm

\section{Seiberg-Witten map, gauge generators and Hamiltonians} \label{ch4:sec:sw-map}
\indent

One of the advantages of having a well defined Hamiltonian
and phase space formalism is that it allows us to apply
the whole machinery that was developed in classical mechanics
(canonical transformation, Hamilton-Jacobi equation, action-angle
variables...). In this section we give a new interpretation to
the Seiberg and Witten map that we discussed in section~\ref{ch3:sec:sw-map}.
We will show that, in the $\dpo$ formalism,  it can be seen as a simple
phase space canonical transformation.  This makes
explicit the physical equivalence of describing the
action in terms ordinary and NC fields.
By finding the corresponding
generating functional, we will able to map quantities between  both theories.
In particular, we will show how the gauge generator and the \ham obtained
in the previous section for the $NC$ case are mapped to those of the
commutative theory.

\zavall
\zavall

\caixa{In the following discussions we keep terms only up to the first order in
$\t$. All equations implicitly omit any higher powers.
The problem of finding exact results is just technical and it is
probably not too difficult since it just boils down to
finding the exact generating functional.}

\zavall
\zavall

We recall from section~\ref{ch3:sec:sw-map} that the SW map among quantities
in both pictures looks like
\bea
\8 A_\mu & =& A_\mu+\frac12\t^{\r\s} A_\s(2\pa_\r A_\mu-\pa_\mu A_\r)
\,,
\label{Ahat} \\
\8 F_{\mu\nu}&=&F_{\mu\nu}+\t^{\r\s} F_{\r\mu}F_{\s\nu}-
\t^{\r\s}A_{\r}\pa_\s F_{\mu\nu} \,,\\
\8\l(\l,A) &= &\l+\frac12\t^{\r\s} A_\s\pa_\r\l \,,
\label{lamhat}
\eea
so that under a commutative $U(1)$ transformation of $\d A_\mu=\pa_\mu\l$, the
mapped $\8 A_\mu$ transforms as
\be
\d\8 A_\mu=
\pa_\mu\{\l
+\frac12\t^{\r\s} A_\s\pa_\r\l
\}+\t^{\r\s} \pa_\s\l\pa_\r A_\mu~
=~\8 D_\mu\8\l.
\label{DAhat}
\ee

Let us proceed to the $\dpo$ formalism.
As in the previous section, we denote the $\dpo$ dimensional potentials
$\8{\cala}_\mu(t,x)$ and ${\cala}_\mu(t,x)$ corresponding to $d$ dimensional
ones $\8A_\mu(t,\bx)$ and $A_\mu(t,\bx)$ respectively.\footnote{Remember,
hats for fields in the non-commutative theory,
and calligraphic letters for fields in the $d+1$ formalism}
The way to realize the SW map as a canonical transformation
is by means of the following generating function
\be
W({\cala},\8{\it\Pi})=\int dx\;\8{\it\Pi}^\mu\left[
{\cala}_\mu+\frac12\t^{\r\s} {\cala}_\s(2\pa_\r {\cala}_\mu-\pa_\mu {\cala}_\r)
\right]\;+\;W^0({\cala}),
\label{GF}
\ee
where $W^0({\cala})$ is an arbitrary function of ${\cala}_\mu$ of order $\t$.
Any choice of $W^0({\cala})$ leads to the correct transformation
of ${\cala}_\mu$ as in \bref{Ahat}
\bea
\8{\cala}_\mu&=&
{\cala}_\mu+\frac12\t^{\r\s} {\cala}_\s(2\pa_\r {\cala}_\mu-\pa_\mu {\cala}_\r) \,.
\label{hAhat}
\eea
At the same time, the canonical transformation
determines the relation between ${\it\Pi}^\mu$ and $\8{\it\Pi}^\mu$,
conjugate momenta of ${\cala}_\mu$ and $\8 {\cala}_\mu$ respectively,
\bea
{\it\Pi}^\mu& = &\8{\it\Pi}^\mu   + \frac12\8{\it\Pi}^\s\t^{\r\mu}
(2\pa_\r {\cala}_\s-\pa_\s {\cala}_\r) \nn &-&\pa_\r(\t^{\r\s} {\cala}_\s\8{\it\Pi}^\mu)
+\frac12 \pa_\r (\8{\it\Pi}^\r\t^{\mu\s}{\cala}_\s)+
\frac{\d W^0({\cala})}{\d {\cala}_\mu},
\nonumber
\eea
which can be inverted, to first order in $\t$,
\bea
\8{\it\Pi}^\mu &= &{\it\Pi}^\mu + \t^{\mu\r}{\it\Pi}^\s \calf_{\r\s}+
{\it\Pi}^\mu\frac12\t^{\r\s}\calf_{\r\s} \nn &+&\t^{\r\s} {\cala}_\s\pa_\r {\it\Pi}^\mu
-\frac12 (\pa_\r {\it\Pi}^\r)\t^{\mu\s}{\cala}_\s-
\frac{\d W^0({\cala})}{\d {\cala}_\mu}.
\label{Pihat}
\eea
Note that the canonical transformation, \bref{hAhat} and \bref{Pihat},
is independent of the concrete theories we are considering as we have not
even needed to specify the action yet.
Now that the canonical transformation is defined, we can use it
to translate any phase space function from one picture to another.

\para{Mapping the gauge generator.}

We obtained the gauge generator
for the NC $U(1)$ theory in \bref{defG2}. Let us apply the canonical
transformation to map it to the commutative picture.
It is straightforward to show that, if we choose $W^0({\cala})=0$,
the first part of the gauge generator is mapped to
\bea
\int dx[~\8 {\it\Pi}^\mu \8 \cald_\mu\8\Lam(\Lam,{\cala})~]
&=&
\int dx~[\; {\it\Pi}^\mu \pa_\mu\Lam\;],
\label{defGc}
\eea
where
\bea
\8\Lam(\Lam,{\cala})=\Lam+\frac12\t^{\r\s} {\cala}_\s\pa_\r\Lam,~~~~~~~~~~~
\dot \Lam=\pa_{x^0}\Lam.
\label{lamhat2}
\eea
This result is again independent of the specific form of Lagrangian for $U(1)$
NC and commutative gauge theories. However,
the second part of the NC gauge generator, whose
origin was a surface term in the action,
does depend on the specific theory we are considering.
For the $U(1)$ NC  theory,
this term is  nothing but the Lagrangian
dependent term in \bref{defG2}, which expanded to first order in $\t$ reads
\bea
\frac{i}{4}\int dx\;{\e(x^0)} \8 \calf_{\mu\nu}[\8 \calf^{\mu\nu},\8\Lam]\;
=~
\frac{1}{4}\int dx\;{\d(x^0)}\t^{0i}
  \calf_{\mu\nu} \calf^{\mu\nu}\pa_i\Lam.
\eea
In this case the generator of $U(1)$ NC transformations
can be mapped to the commutative one
\petit{\bea
G[\8\Lam(\Lam,{\cala})]&=&
\int dx\; \{{\it\Pi}^0 \pa_0\Lam\;+({\it\Pi}^i+
\frac{1}{4}{\d(x^0)}\t^{0i}
  \calf_{\mu\nu} \calf^{\mu\nu})\pa_i
\Lam\} \nn && -\int dx \frac{\d W^0({\cala})}{\d {\cala}_\mu}\pa_\mu\Lam
\,=\,
\int dx\; [~{\it\Pi}^\mu \pa_\mu\Lam~]
\label{Gcom}
\eea}
if we choose the arbitrary function in the canonical transformation to be
\bea
W^0({\cala})&=&\frac14\int d x~\d(x^0)~\t^{0\mu}{\cala}_\mu \calf_{\r\s}\calf^{\r\s}.
\label{W00}
\eea
The right hand side of \bref{Gcom} is the well-known generator of the
$U(1)$ commutative theory (see appendix~\ref{ch4:sec:u1-c}).

\para{Mapping the Hamiltonian.}

Now we would like to see what is the form of the commutative
$U(1)$ Hamiltonian obtained from the NC one \bref{Ham} under the SW map,
\bref{hAhat} and \bref{Pihat}. A short calculation yields
\be
H_{com}
=\int dx\;[{{\it \Pi}}^\nu(t,x) {\cala}'_\nu(t,x)~-~
\d(x^0) \call_{com}(t,x)] \,,
\label{abelianham}
\ee
where
\be
\call_{com}(t,x)= -\frac14 \calf^{\nu\mu}\calf_{\nu\mu}-\frac12\calf^{\mu\nu}
\t^{\r\s} \calf_{\r\mu}\calf_{\s\nu}+
\frac18 \t^{\nu\mu}\calf_{\nu\mu}\calf_{\r\s}\calf^{\r\s}.
\label{U1lag}
\ee

Let us pause for a second. What should we have expected to obtain by this
mapping? As we discussed in section~\ref{ch3:sec:sw-map}, commutative fields
are ruled by the usual commutative product of functions and the
usual gauge transformations. However, in their effective actions
one has the annoying presence everywhere of the $B$-field, whose
relation to $\t$ in the decoupling limit is $B=\t^{-1}$. Can
we recover this point of view from the commutative U(1) Hamiltonian
that we have obtain?

The answer is affirmative.
The expression \bref{abelianham} is nothing but the $ \dpo$ dimensional Hamiltonian
that we would have obtained from an abelian $U(1)$ gauge theory with \lag
\be
L_{com}(t,\bx)=
-\frac14 F^{\nu\mu}F_{\nu\mu}-
\frac12 F^{\mu\nu}\t^{\r\s} F_{\r\mu}F_{\s\nu}+
\frac18 \t^{\nu\mu}F_{\nu\mu}F_{\r\s}F^{\r\s} ,
\label{U1BIa}
\ee
in $d$ dimensions. And, as expected,
one can check that this Lagrangian is, up to a total derivative, the
expansion of the usual Born-Infield action to order $F^3$
\bea
L_{com}&\sim& -\sqrt{-\det(\h_{\mu\nu}-\t_{\mu\nu}+F_{\mu\nu})} |_{
\, O (F^3)} \,.
\label{relBI}
\eea

\vskip 6mm

\section{BRST symmetry} \label{ch4:sec:BRST}

In what follows, we will complete our exploitation  of
the $\dpo$ formalism and setup future studies of nonlocal theories
by means of some useful tools that were developed for local ones,
concentrating on their BRST and field-antifield properties.
The already familiar NC $U(1)$ theory will serve as an example
all the way through.
So, in what follows, one will find
\bitem
\item A study of the BRST symmetry~\cite{Becchi:1974xu}\cite{Tyutin:1975qk} at
classical and quantum levels. We will construct
the BRST charge and the BRST invariant Hamiltonian
working with the $\dpo$ dimensional formulation, and we will
check the nilpotency of the BRST generator.
\item
In order to map the BRST charges and Hamiltonians
of the commutative and NC $U(1)$ gauge theories,
we will generalize the SW map to a canonical transformation
in the superphase space.
\item
In the last section we will study the BRST symmetry at \lag level using the
field-antifield formalism~\cite{z}\cite{Batalin:1981jr}\footnote{See
\cite{Henneaux:1992ig}\cite{Gomis:1995he}\cite{Weinberg:1996kr}
for reviews on this subject.}.
We will construct the solution of the
classical master equation in the classical and gauge fixed basis.
As this is a study at the Lagrangian level, the $d+1$ formalism
will not be required. Still, we will be able to
realize the SW map as an antibracket canonical transformation.
\eitem

\subsection{Hamiltonian BRST charge} \label{ch4:sec:ham-BRST}

The BRST symmetry at classical level encodes the classical
gauge structure through the nilpotency of the BRST transformations of
the classical fields and ghosts~\cite{Kugo:1982hm}\cite{Batalin:1985qj}\cite{Fisch:1990rp}.
BRST transformations of the classical fields are constructed from the
original gauge transformation simply by changing the gauge parameters
by ghost fields.

Let us consider again the $U(1)$ NC theory still in $d$ dimensions.
Its BRST transformations are then
\bea
\d_B\8  A_\mu & = &\8 D_\mu \8 C \sac \d_B \8 C~=~-i \8 C*\8 C,
\\
\d_B \8 {\overline C} & = &\8 B \sac \d_B \8 B~=~0,
\eea
where $\8 C, \8{\overline C}, \8B $ are the ghost, antighost and auxiliary field
respectively.
These are again a symmetry of the \lag associated
to \bref{Maxwell}, since its change under the BRST
transformations is
\be
\d_B L~=~
\frac{i}{2}~[\8 F_{\mu\nu},\8 C]~\8 F^{\mu\nu} \,,
\label{Dlagc}
\ee
which, as in \bref{Dlag}, can be shown to be a total divergence.
To fix the gauge symmetry, we add a 'gauge fixing' term to
the \lag $\8L_{gf+FP}$ with the requirement that it must
be of the form $\d_B\8\Psi$.
If we choose the gauge fixing fermion to be
\be
\8 \Psi= \8{\overline C}~(\partial^\mu \8 A_\mu +\alpha \8 B) \,,
\label{GFF}
\ee
then the $\8 L_{gf+FP}$ is, up to a total derivative,
\bea
\8 L_{gf+FP}&=&-~
\partial^\mu\8{\overline C}~ \8 D_\mu \8 C~+~
\8 B~(\partial^\mu \8 A_\mu +\alpha \8 B).
\eea
By construction, this term does not spoil the symmetry. Indeed
\bea
\d_B \8 L_{gf+FP}&=&
\pa^\mu(\8 B \8 D_{\mu}\8 C).
\label{Dlaggfc}
\eea

\vskip 6mm

{\bf Going to $d+1$.}

In order to construct the generator of the BRST transformations and the
BRST invariant Hamiltonian we should use the
$\dpo$ dimensional formulation.
We use calligraphic characters for the $\dpo$ dimensional fields
$\8{\calc}, \8{\overline {\calc}}, \8{\calb} $,
corresponding to the  $d$ dimensional ones $\8 C, \8{\overline C}, \8B $,
respectively. The resulting BRST invariant Hamiltonian is given by
\bea
H(t)&=&H^{(0)}~+~H^{(1)} \,,
\label{Hamc01}\nn
H^{(0)}&=&\int dx\;[{\8{\it \Pi}}^\nu(t,x)\8 {\cala}'_\nu(t,x)~+~
{\8{ \calp_c}}(t,x)\8 {\calc}'(t,x)~-~\d(x^0)\8 \call^0(t,x)],
\nn
H^{(1)}&=&\int dx\;[{\8{\calp}}_{\8{\calb}} \8{\calb}'(t,x)~+~
{\8{ \calp_{\overline {\calc}}}}(t,x)\8{\overline {\calc}}'(t,x)~-~
\d(x^0)\8 \call_{gf+FP}(t,x)]\,, \nonumber
\label{Hamc2}
\eea
while the  BRST charge is
\bea
Q_B&=&Q_B^{(0)}~+~Q_B^{(1)} \,,
\label{BRSTQ}\\
Q_B^{(0)}&=&\int dx\left[\8 {\it\Pi}^\mu \8 \cald_\mu\8{\calc}-i\8{ \calp_{{\calc}}}*
\8{\calc}*\8{\calc}+\frac{1}{2}{\e(x^0)} \d_B \8 \call^0(t,x)\right]\,,
\\
Q_B^{(1)}&=&\int dx\left[~\8{ \calp_{\overline {\calc}}} ~\8 {\calb}~+~
\frac{1}{2}{\e(x^0)} \d_B \8 \call_{gf+FP}(t,x)\right]\,.
\label{defG2c}
\eea
$Q_B$ is  an analogue of the BFV charge~\cite{Fradkin:1975cq}\cite{Batalin:1977pb}
 for our NC $U(1)$ theory.
$H^{(0)}$, $Q_B^{(0)}$ are the "gauge unfixed" and
$H$, $Q_B$ are "gauge fixed" Hamiltonians and BRST charges respectively.
Finally,
using the graded symplectic structure of the superphase space~\cite{Casalbuoni:1976tz}
\petit{\bea
\{\8{\cala}_\mu(t,x),\8{\it\Pi}^\nu(t,x')\}&=&{\d_\mu}^\nu~ \d^{(d)}(x-{x}')
,~~~
\{\8{\cal {C}}(t,x),\8{\cal P}_{\8{\cal {C}}}(t,x')\}~=~
\d^{(d)}(x-{x}'),
\nn
\{\8{{\overline {\calc}}}(t,x),\8{\cal P}_{\8{\overline{{\calc}}}}(t,x')\}&=&
\d^{(d)}(x-{x}'),~~~~~~~~
\{\8{{\calb}}(t,x),\8{\cal P}_{\8{{\calb}}}(t,x')\}~=~
\d^{(d)}(x-{x}'),
\nonumber
 \eea} we have
\be \nonumber
\{H^{(0)},Q_B^{(0)}\}~=~\{Q_B^{(0)},Q_B^{(0)}\}~=~0,
\ee
and
\be \nonumber
\{H,Q_B\}~=~\{Q_B,Q_B\}~=~0.
\ee
Thus the BRST charges are nilpotent and the Hamiltonians are
BRST invariant both at the gauge unfixed and the gauge fixed
levels.


\subsection{Seiberg-Witten map in superphase space} \label{ch4:sec:SW-superphase}

Now we would like to see how the BRST charges and the
BRST invariant Hamiltonians
of the NC and commutative gauge theories are related.
In order to do that we will extend the SW map to a
canonical transformation in the superphase space
$\{{\cala},{\calc},{\overline{\cal {C}}},{\calb},\iPi,{\cal P}_{\cal {C}},
\calp_{\overline{\cal {C}}},\calp_{{\calb}}\}$.
We introduce the generating function
\petit{\bea
W({\cala},{\calc},{\overline{\cal {C}}},{\calb},\8\iPi,\8{\cal P}_{\cal {C}},
\8\calp_{\overline{\cal {C}}},\8\calp_{\calb})
&=&\int dx\;\left[\8\iPi^\mu\left(
{\cala}_\mu+\frac12\t^{\r\s} {\cala}_\s(2\pa_\r {\cala}_\mu-\pa_\mu {\cala}_\r)
\right)\;\right.
 \nn
&+&\left. \;\8{\cal P}_{\cal {C}}
\left({\calc}+\frac12\t^{\r\s} {\cala}_\s\pa_\r {\calc} \right)~+~
\8{\cal P}_{\overline{\cal {C}}}{\overline{\cal {C}}}~+~
\8{\cal P}_{{\calb}}{{\calb}}\right]
\nn &+&
W^0({\cala},{\calc})~+~W^1({\cala},{\calc},{\overline{\cal {C}}},{\calb}).
\nonumber
\label{GF3}
\eea}
As before, $W^0({\cala},{\calc})$ depends on the
specific form of the Lagrangian and the novelty is the
appearance of $W^1({\cala},{\calc},{\overline{\cal {C}}},B)$,
whose form depends also on the gauge fixing. For the $U(1)$
NC theory and for the gauge fixing \bref{GFF}, we must choose
\bea
W^0({\cala},{\calc})&=&\frac14\int d x~\d(x^0)~\t^{0\mu}{\cala}_\mu \calf_{\r\s}
\calf^{\r\s}
\label{W0}
\eea
as in \bref{W00} and
\bea
W^1({\cala},{\calc},{\overline{\cal {C}}},{\calb})&=&\int dx ~\frac{1}{2}{\e(x^0)}
\left[\pa^\mu \{\frac12\t^{\r\s}{\cala}_\s(2\pa_\r{\cala}_\mu-\pa_\mu{\cala}_\r)\}
{\calb}\right.
\nn &+&\left.
\{\frac12\t^{\r\s}{\cala}_\s(2\pa_\r{\cala}_\mu-\pa_\mu{\cala}_\r)\pa_\s{\calc}+
\frac12\t^{\r\s}{\cala}_\s\pa_\mu\pa_\r{\cal {C}}~\}~\pa^\mu\overline{\calc}~\right].
\nonumber
\label{solw2}
\eea
The transformations are obtained from the generating function by
\bea
\8\Phi^A=\frac{\partial_l W}{\partial \8 P_A},~~~~~\quad
P_A=\frac{\partial_r W}{\partial \Phi^A},
\eea
where $\Phi^A$ represent any fields, $P_A$ their conjugate momenta,
and $\pa_r$ and $\pa_l$ are right and left derivatives respectively.
Explicitly we have
\bea
\8 {\cala}_\mu&=&{\cala}_\mu+\frac12\t^{\r\s} {\cala}_\s(2\pa_\r {\cala}_\mu-
\pa_\mu {\cala}_\r),
\\
\8{\cal {C}}&=&{\cal {C}}+\frac12\t^{\r\s}{\cala}_\s\pa_\r{\cal {C}},
\label{chata}
\\
\8{\overline{\calc}}&=&{\overline{\calc}},
\\
\8{\calb}&=&{\calb},~~~~~~~~~~~~~
\eea
and
\bea
\8\iPi^\mu&=&\iPi^\mu+\t^{\mu\r}\iPi^\s \calf_{\r\s}+
\iPi^\mu\frac12\t^{\r\s}\calf_{\r\s}
+\t^{\r\s} {\cala}_\s\pa_\r \iPi^\mu
-\frac12 (\pa_\r \iPi^\r)\t^{\mu\s}{\cala}_\s
\nn
&&~~+\frac12{\cal P}_{{\calc}}\t^{\mu\s}\pa_{\s}{\calc}~-~
\frac{\d (W^0+W^1)}{\d {\cala}_\mu},
\\
\8{\cal P}_{{\calc}}&=&{\cal P}_{{\calc}}+\frac12\t^{\r\s} \partial_\rho
({\cal P}_{{\calc}}{\cala}_\s)~-~
\frac{\d_r (W^0+W^1)}{\d {\calc}},
\\
\8{\cal P}_{\overline{\calc}}&=&{\cal P}_{\overline{\calc}}~-~
\frac{\d_r W^1}{\d {\overline{\calc}}},
\\
\8{\cal P}_{{\calb}}&=&{\cal P}_{{\calb}}~-~
\frac{\d_r W^1}{\d {\calb}}.
\eea
These transformations allow us to map the NC quantities to
the commutative ones.
\medskip

\noindent
{\it Mapping the BRST charge.}

\noindent
The NC BRST charge \bref{BRSTQ} becomes
\bea
Q_B&=&Q_B^{(0)}~+~Q_B^{(1)}~=~
\int dx [\iPi^\mu  \pa_\mu{\calc}~+~{ \calp_{\overline {\calc}}}  {\calb}~-~
\d(x^0) {\calb}\pa^0 {\calc}~]
\nn\\&=&
\int dx [\iPi^\mu  \pa_\mu{\calc}~+~{ \calp_{\overline {\calc}}}  {\calb}~+~
\frac{1}{2}{\e(x^0)} \d_B \call_{gf+FP}(t,x)~],
\label{totQBgf}
\eea
where $\call_{gf+FP}(t,x)$ is the abelian gauge fixing
Lagrangian
\bea
\call_{gf+FP}&=&-\partial^\mu\overline {{\calc}}~ \pa_\mu  {\calc}~+~
{\calb}~(\partial^\mu  {\cala}_\mu +\alpha  {\calb}).
\label{laggf0}
\eea

\medskip

\noindent
{\it Mapping the Hamiltonian.}

\noindent
The total NC $U(1)$ \ham \bref{Hamc01} becomes
\be
H=\int dx\;[{{\it \Pi}}^\nu  {\cala}'_\nu +{{ \calp_{\calc}}} {\calc}'
+{{ \calp_{\overline{\calc}}}} {\overline{\calc}}'+{{\it P}}_{\calb} {\calb}'
-\d(x^0) (\call_{com}+\call_{gf+FP}) ].
\label{totHam}
\ee
Remember that $\call_{com}$ is the $U(1)$ \lag in the commutative given in \bref{U1lag}.

Summarizing, we have been successful in mapping the NC and commutative
charges in the $d+1$ formalism by generalizing the SW map to a
canonical transformation in the superphase space.

\vskip 6mm
\subsection{Field-antifield formalism for $U(1)$  non-commutative theory} \label{ch4:sec:field-antifield}

The field-antifield formalism allows us to study the BRST symmetry
of a general gauge theory by introducing a canonical
structure at a Lagrangian level~\cite{z}\cite{Batalin:1981jr}\cite{Henneaux:1992ig}\cite{Gomis:1995he}.
The classical master equation in the classical basis encodes the
gauge structure of the generic gauge theory~\cite{Batalin:1985qj}\cite{Fisch:1990rp}.
The solution of the classical master equation in the gauge fixed basis
gives the ``quantum action'' to be used in the path integral quantization.
Any two solutions of the classical master equations
are related by a canonical transformation in the antibracket sense~\cite{Batalin:1981jr}.

Here we will apply these ideas to the $U(1)$  NC theory.
Since we work at a \lag level there is no need
to go to the $d+1$ formalism, and we stay in $d$ dimensions.
In the classical basis the set of fields and antifields are
\be
\Phi^A=\{ \8 A_\mu, \8 C\},~~~~~~~ \Phi^*_A=\{ \8 A^*_\mu,\8 C^* \}.
\ee
The solution of the classical master equation
\be
(S,S)=0,
\ee
is given by{\footnote{
As usual in the antifield formalism, $d$ dimensional
integration is understood in summations.}}
\be
S[\Phi,\Phi^*]=I [\8 A]+\8 A^*_\mu \8 D^\mu \8 C-i\8 C^* (\8 C * \8 C),
\ee
where $I [\8 A]$ is the classical action and the antibracket
$ (~~,~~)$ is  defined by
\be
(X,Y)=\frac {\partial_r X}{\partial \Phi^A}
\frac {\partial_l Y}{\partial \Phi^*_A}-
\frac {\partial_r X}{\partial \Phi^*_A}
\frac {\partial_l Y}{\partial \Phi^A}.
\ee

The gauge fixed basis can be analyzed by introducing  the antighost
and auxiliary fields and the corresponding antifields. It can be
obtained from the classical basis by considering a canonical transformation in
the antibracket sense
\bea
\Phi^A&\longrightarrow& \Phi^A
\nn
\Phi^*_A&\longrightarrow& \Phi^*_A +\frac {\partial_r \Psi}
{\partial \Phi^A} \,,
\eea
which is generated by
\be
\8\Psi=  \8{\overline C}~(\partial^\mu \8 A_\mu +\alpha \8 B),
\ee
where $\8{\overline C}$ is the antighost and $\8 B$ is the auxiliary field.
In the new gauge fixed basis, we then have
\bea
S[\Phi,\Phi^*]=\8I_\Psi +\8 A^{*\mu} \8 D_\mu \8 C- i\8 C^* (\8 C * \8 C)
+{\8 {\overline C}}^* \8 B,
\label{SPP}
\eea
where $\8I_\Psi$ is the ``quantum action'', given by
\bea
\8I_\Psi= I[\8 A]+(-\partial_\mu \8{\overline C}~\8D^\mu\8 C+
\8 B~\partial_\mu\8 A^\mu+\alpha\8 B^2).
\label{IPSI}\eea
The action $\8I_\Psi$ has well defined propagators and is the starting
point of the Feynman perturbative calculations.
\vspace{3mm}

Now we would like to study what is the SW map in the space of
fields and antifields. We first consider it in the classical basis.
 In order to do that we construct a canonical
transformation in the antibracket sense
\be
\8\Phi^A=\frac{\partial_l F_{cl}
[\Phi,\8 \Phi^*]}{\partial \8 \Phi^*_A},\quad
\Phi^*_A=\frac{\partial_r F_{cl}[\Phi,\8 \Phi^*]}{\partial \Phi^A},
\ee
where
\be
F_{cl}
=\8 A^{*\mu} \left(A_\mu+\frac12\t^{\r\s} A_\s(2\pa_\r A_\mu-\pa_\mu A_\r)
\right)+\8 C^*(C+\frac12\t^{\r\s} A_\s\pa_\r C).
\label{FCL}\ee
The NC and commutative gauge structures are then are mapped to each other
\be
 \8 A^*_\mu \8 D^\mu \8 C- i\8 C^* (\8 C * \8 C)=A^*_\mu\partial^\mu C.
\ee

We can generalize the previous results to the gauge fixed
basis. In this case the transformations of the antighost
and the auxiliary field sectors should be taken into account.
The generator of the
canonical transformation is modified from \bref{FCL} to
\be
F_{gf}
=F_{cl}+
\left(\8{\overline C}^*+\frac12\t^{\r\s}\pa^\mu \left(
A_\s(2\pa_\r A_\mu-\pa_\mu A_\r)\right)
\right)\overline {C}+\8B^* B.
\label{SWgf}
\ee
Note that the additional term gives rise to new terms in $A^{*\mu}$ and
$\overline {C}^*$
while the others remain the same as in the classical basis.
In particular
\bea
\8{\overline {C}}&=&\overline {C},~~~~~~\8B~=~B.
\eea
Using the canonical transformation we can express  \bref{SPP} and \bref{IPSI} as
\bea
S[\Phi,\Phi^*]=I_\Psi + A^{*\mu} \pa_\mu C+{\overline{C}}^*  B \,,
\label{SPPgf}
\eea
where
\bea
I_\Psi= I[\8 A(A)]+ (-\partial_\mu \overline{ C}~\pa^\mu C+
B~\partial_\mu A^\mu+\alpha B^2) \,,
\label{IPSIgf}
\eea
and $I[\8 A(A)]$ is the classical action in terms of $A_\mu$.
This is indeed the quantum action for the commutative $U(1)$ BRST
invariant action in the gauge fixed basis.
In this way the canonical transformation \bref{SWgf}
maps the  $U(1)$ NC structure of the $S[\Phi,\Phi^*]$ into the
commutative one in the gauge fixed basis.

\vskip 6mm

\section{Discussion} \label{ch4:sec:discussion}

In this chapter the \ham formalism of the non-local theories has been
discussed by using a $\dpo$ dimensional formulation~\cite{Llosa:1993sj}\cite{Gomis:2000gy}.
For any given non-local \lag in $d$ dimensions the corresponding Hamiltonian
in $d+1$ is defined in \bref{h}.
The equivalence with the original non-local theory is
ensured by imposing two constraints \bref{vp} and \bref{EOMf} consistent
with the time evolution.
The degrees of freedom of the extra dimension (denoted by the coordinate $\s$)
have their origin in the infinite degrees of freedom associated with the
non-locality.
The fact that we have been led to a theory with ``two times''
should be intimately related to their acausality~\cite{Seiberg:2000gc}\cite{Alvarez-Gaume:2000bv} and
non-unitarity~\cite{Gomis:2000zz}\cite{Aharony:2000gz}. It remains however as an interesting
exception the case of light-like NC theories.

The $d+1$ formalism is also applicable to {\it local} and
higher derivative theories.
In these cases the set of
constraints can be used to reduce the redundant degrees of freedom
of the infinite dimensional phase space,
reproducing the standard $d$ dimensional formulations~\cite{Gomis:2000gy}.

We have analyzed the symmetry generators of non-local
theories in the \ham formalism. In particular, we have
shown that the Hamiltonian is the conserved charge under
time translations, justifying its interpretation as
the energy of a configuration. We exemplified the
formalism by applying it to the electric NC $U(1)$ gauge theory.
We remark that gauge transformations in $d$ dimensions are
described as a rigid symmetry in $\dpo$ dimensions.
The generators of
{\it rigid} transformations in $\dpo$ dimensions turn out to be the
generators of {\it gauge}
transformations when the reduction to $d$ dimensions can be performed
as is shown for the $U(1)$ commutative gauge theory in the appendix~\ref{ch4:sec:u1-c}.

Within this formalism, we
reinterpreted the Seiberg-Witten map  as a canonical transformation.
This allowed us to map the Hamiltonians and the
gauge generators of NC and
commutative theories. We exemplified this by explicitly mapping the
$U(1)$ NC action to the commutative DBI one to order $F^3$.

The BRST symmetry has been analyzed at Hamiltonian and Lagrangian levels,
and functionals in the commutative and NC pictures have been mapped
by extending the SW map to a canonical transformation of the ghosts
in the super phase space.
Purely at a Lagrangian level, using the field-antifield formalism,
we have seen how the solution of the
classical master equation for NC and
commutative theories are related by a canonical transformation in the
antibracket sense. This result shows that the antibracket cohomology
classes of both theories coincide in the space of non-local functionals.
The explicit forms of the antibracket canonical transformations could be
useful to study the observables, anomalies, etc. in the $U(1)$ NC theory.

\chapter{Conclusions}\label{ch:conclusions}

In this thesis we have performed a little tour about
two of the main open branches of String Theory that followed
the discovery of D-branes: the gauge/string correspondence and
noncommutative theories. At the end we managed to somehow
close the loop and link them two, trying to shed new light
on controversial issues like the UV/IR mixing of NC theories
from the point of view of its closed string dual.

During the tour we stopped at some points that we found interesting
to be studied on their own. One of the issues that we needed to
face repeatedly was how to stabilize a D-brane when the target space
or the embedding are not flat. During the thesis we have seen
at least four qualitatively different ways to achieve it:
\tem{
\item Either in flat space or in a target space of the style
$\CR^{1,1} \times \CM_8$, with $\CM_{8}$ any special holonomy
manifold, a D2 brane can be stabilized in a tubular shape
with an arbitrary cross section. In the case that $\CM_8$ is
not completely flat, the only requirement is that the
longitudinal direction lives in the $\CR$ factor.
The result confirms the understanding that the Poynting
vector generated by the electromagnetic fields (or, equivalently,
the local density of F1 and D0-branes)
can be chosen to locally compensate not only the D2 tension, but
also the extra gravitational effect produced by the
curvature of $\CM_8$. Furthermore,
the D-brane picture allowed us to find new supergravity
solutions of type IIA which describe the backreaction of these
generalized supertubes, providing backgrounds whose preserved
fraction of supersymmetry ranges from 1/32 to 1/8.

\item In a completely counterintuitive manner, the D-brane
worldvolume can be an $AdS_{p'}\times \Sigma^{q'}$ manifold of
an $AdS_p\times S^q$ background, with $\Sigma^{q'}$ a minimal
submanifold of the $S^q$. In all the cases we studied, $\S^{q'}$
had actually the maximal volume within its homology class, and there
was no topological obstruction at all for $\S^{q'}$ to collapse
into a point. The stability of all the examples considered could
be understood by reducing the
worldvolume theory on $\S^{q'}$ and checking in the effective Lagrangian in $AdS_{p'}$
if there was any tachyonic fluctuation violating the corresponding
Breitenlohner-Freedman bound. In most cases, stability
 follows from supersymmetry, as most of the stable embeddings can be understood
as arising from supersymmetric brane configurations in flat space in which
on type of branes is replaced by its background. We exploited this mechanism
to try to embed branes in a stable but non-supersymmetric way, with the aim
to provide holographic duals of superconformal field theories in which
non-supersymmetric matter is added. We mentioned however that this is
work in progress and that we can not be conclusive about the success
of this possibility yet.

\item Angular momentum alone can stabilize a relativistic surface
of any dimension,
as long as the dimension of target space
is large enough. Although in section~\ref{ch2:sec:rotating-strings}
we have just exploited this for strings in type IIB, by S-duality
we would obtain a D1, and by T-dualities we would obtain a general
Dp-brane with one compact rotating direction. Applying the results of~\cite{Hoppe:1987vv},
more directions could become involved in the rotation as long as
they form a minimal surface within the corresponding sphere.
It is remarkable that this phenomenon already happens in flat space
where it is perturbatively stable~\cite{Demkin:1995td}
and completely independent of supersymmetry.

\item The last possibility is maybe the more intuitive:
D-branes can wrap calibrated cycles of special holonomy manifolds
and still preserve some supersymmetry. The cycles prevent the collapse
of the brane in a fine-tuned manner, as they
realize geometrically the twisting of the gauge theory that lives
on the worldvolume.
First we reviewed that
the holonomy must be 'special' in order for the target space to preserve susy
 and the cycle must be 'calibrated' in order for some supersymmetry
to be linearly realized in its worldvolume.
}
Let us expand on the latter possibility. The understanding of the
brane/tar\-get-space setup provided the necessary intuition,
together with the technical improvement provided by gauged supergravities, to
construct the string dual of an $SU(N)$ $\caln=2$ SYM theory in 2+1.
One of the nice points about this construction is that it involved
D6-branes. In this particular duality, they wrap a four-dimensional
\kah cycle $\S_4$ in a Calabi-Yau threefold. The following two facts
\tem{
\item D6-brane solutions without extra fluxes lift to a purely gravitational 11d backgrounds,
\item our particular D6 configuration has a 2+1 flat noncompact part,
so that in 11d the solution is $\CR^{1,2}\times X_8$,
}
imply that $X_8$ has to be a special holonomy manifold. As
the number of  preserved supersymmetries is 4, the table in
page \pageref{special-hol-table} implies that $X_8$ has to
be a $CY_4$. In other words, the cycle, its transverse space and
the M-theory circle have to conspire in order to produce a noncompact
8d manifold with $SU(4)$ holonomy. Our particular $CY_4$ \bref{gib1}-\bref{probeta}
is of a conical type with constant radial sections being a $U(1)$
fibration over an $S^2 \times \S_4$ base. The complete analytic
solution of the BPS equations involve an integration constant
that essentially measures the radius of the $S^2$ at the origin.
Unless this constant is set to zero, the metric does not suffer
from a conical singularity at the origin and it turns out
to be locally regular everywhere.

Having obtained the desired 11d metric we compactified along
a circle to analyze the problem in type IIA. There we performed
a probe analysis in order to obtain the moduli space in the Coulomb
branch of the
dual field theory and we found that it was a 2d \kah space
as naively expected for an $\caln=2$ gauge theory in 2+1. We recall that
although it has long been known that instanton contributions
completely destroy the moduli space, their effects are exponentially
suppressed with $N$. They are then expected to be invisible in the supergravity
approximation in agreement with our result.
Indeed, the shape of the moduli space that we get from supergravity resembles
very much the classical plus 1-loop moduli space (see fig. \ref{radi}
in page \pageref{radi}) that is obtained from the
field theory. The latter is known to have a singularity as the vev
of the scalar approaches the bare coupling, a sign that more loops
should be taken into account. The probe result seems to be able to
smooth this singularity and push it to the origin, where the vev of
the scalar is zero. At this point supergravity is not valid as
the curvature of the background diverges. However, we can approach
as much as required by taking $g_s N$ large enough. It could be
that supergravity is giving us the all-loops correction to the moduli
space, although this is just a suggestive possibility.

\bigskip

Purely within the AdS/CFT correspondence we considered in detail the
possibility of testing the duality in sectors where supersymmetry
could be absent. The GKP ideas have allowed to produce a number shortcuts
to bypass the quantization of the $\s$-model in various RR-backgrounds.
Essentially one just needs to identify the appropriate $\s$-model soliton
that carries a set of charges and compute their classical value. When
these numbers are large, the quantum $\s$-model corrections can happen
to be small, yielding a non-perturbative prediction for the dual field
theory operators.

A set of qualitatively new solutions are strings that rotate in the $S^5$
with three angular momenta. The results that the GKP method provides here
are surprising as they show an exact agreement with perturbative computations in the CFT side.
When all momenta are large, the string states (or its dual operators)
are very far from the BMN vacuum (or the BMN operators), so they were
originally thought be providing tests of the AdS/CFT 'far from supersymmetry'.

We have shown however that precisely at large momenta the strings
approach another set of supersymmetric states, preserving 1/8 or
1/4 supersymmetry for 3 or 2 non-vanishing momenta respectively.
We have proven this result both from a worldsheet $\k$-symmetry
computation and from a standard BPS argument that follows from
the background $PSU(2,2|4)$ superalgebra.
Asymptotic restoration of supersymmetry suggests, among other things,
that all these configurations must become stable in the limit,
and this can be proven perturbatively
by noting that the mass of all the tachyonic modes (in the unstable
cases) go to zero in the limit.

We noted that, exactly as in the BMN case, the exact supersymmetric string is one
for which all its energy is due to rotation; this means that the string becomes
effectively tensionless. That the large angular momentum limit is also a tensionless
limit is a non-trivial feature at all. In flat space, an increase of the angular
momentum is always accompanied by an increase of the length of the soliton in
such a way that the kinetic energy remains comparable to the energy due to the tension.
One might suspect that all we need to do is to put the string to rotate in a compact
space, just like the $S^5$, as the length of the string soon reaches a maximum value.
But even this intuition can fail. There is no reason why the soliton should
be able to absorb the extra energy by speeding up without growing. Giant gravitons
provide an example: the giant D3-branes have an upper bound on the angular momentum
that corresponds to the point in which they have maximal volume in the $S^5$; beyond
that angular momentum the solution simple does not exist.

The spinning strings are very special this sense. Let us consider a string with
only two equal angular momenta $J_1=J_2=J$ for simplicity, and consider an
almost collapsed string with very low $J$. It is easy to see that, just as in
flat space, $J$ depends on an inverse power of the angular velocity $\w$, so that we
need to decrease $\w$ in order to increase $J$. As we do so the string starts expanding.
This phase is qualitatively identical to the string rotating in flat space.
However, there is a value of $J$ for which the string reaches its maximal size
within the $S^5$. Beyond this critical point, the system suffers a kind of
'phase transition' in which $J$ suddenly starts growing with $\w$. This is the
second phase which is absent for giant gravitons and the responsible for
allowing the string to absorb more and more kinetic energy until it effectively
behaves as a tensionless string.

Somewhat as an aside, let us mention that there seems to be a paradox here.
On the one hand, the action for a tensionless string is invariant under conformal rescalings
of the background metric. On the other hand, we know that \5ads is conformal
to flat 10d space. How do we explain that the string becomes supersymmetric
in the former and not in the latter? The solution is to recall that
\5ads is conformal but not {\it superconformal}
to flat 10d space, as explained in~\cite{Bandos:2002nn}.

The reason why the tensionless property is so relevant is that the perturbative
field theory calculations always come in a power series of $\l=\gym^2 N$. At best,
they may acquire important combinatoric factors which soften the effective
relevant coupling; this is the case of BMN and of the operators
dual to the 3-angular momenta strings, where $\l_{eff} \sim \l/J^2$. Therefore,
the field theory expansion is in terms of the ratio between the string tension
and its kinetic energy. Any chance to test the duality by means of perturbative
CFT computations must then involve
an almost tensionless soliton in the string theory side.

As an open question, it remains to understand better if it is
the supersymmetry or the tensionless property what really makes
the comparison possible. As we discussed in detail in section
\ref{ch2:sec:discussion}, one of the main points to explain
is the fact that the $\s$-model quantum corrections do not spoil
the classical result at any order. The pulsating string of~\cite{Engquist:2003rn}
seems to provide an example of a successful matching for configurations
arbitrarily far from being BPS which, nevertheless, become
asymptotically tensionless. But as we discussed, they did not check
the $\s$-model corrections and, at this stage, their matching could be
a coincidence. The more recent results of~\cite{Bigazzi:2004yt}, in which similar
rotating strings solutions are considered in the Maldacena-\nun background
seem to indicate that only in those cases where supersymmetry is
asymptotically restored can one really neglect the $\s$-model corrections.

One of the most interesting developments along these lines consists on trying
to extract the string $\s$-model action directly from the gauge theory.
In the latter one needs to consider a limit of the dilatation
operator (interpreted as a certain spin-chain Hamiltonian)
in which the angular momenta are large keeping $\l/J^2$ fixed.
The resulting Hamiltonian has been matched
to the one arising from taking the same limit in
the string $\s$-model~\cite{Kruczenski:2003gt,Kruczenski:2004kw}
in the case of two angular momenta and to second order in $\l/J^2$.
If the dilatation operator could be computed at all loops,
it would then be possible to reconstruct the whole $\s$-model
from the gauge theory. The correspondence would then go beyond
the matching of particular string states to particular SYM
states.

\bigskip

In the branch of noncommutative theories we first investigated
the toy model of a non-relativistic NC $\phi^4$ theory
at the quantum level. Apart from the UV/IR mixing, we
showed that, despite their completely different treatment
of time and space, the same rules of their relativistic counterparts
apply: they are unitary except in electric backgrounds.
However, they do not share the property that unitarity
can be restored at the one-loop level by the addition of
new asymptotic states (the $\chi$-particles). This is not
a big problem anyway as it would simply
imply that there is no possibility to detect the undecoupled
string modes by looking only at the field theory. In this case
we would be talking about undecoupled excitations of a non-relativistic
string theory~\cite{Gomis:2000bd}.

Motivated by a proper quantization of the electric NC theories
and, in general, of any field theory non-local in time, we have
elaborated on the Hamiltonian formalism for such theories that
was introduced in \cite{Llosa:1993sj} and further developed in
\cite{Gomis:2000gy}. The formalism requires the introduction
of an extra timelike coordinate with respect to which the
Lagrangian is finally local. Its Hamiltonian formalism
is essentially based on, apart from a certain Hamilton functional,
a set of constraints that guarantee the equivalence with
the original non-local theory in one dimension less.
Within this framework we first provided the correct construction
of the symmetry generators. It is remarkable that gauge transformations
of the original Lagrangian end up as global transformation in
the d+1 formalism. We have also shown that our Hamiltonian
is indeed the generator of time translations, justifying its
property of giving the energy of a given field configuration.
In general, there are some contributions to the energy
which give what one would have naively found by
putting $*$-products in the expression for the energy of
a commutative theory. However, there are some other contributions
which are purely $(d+1)$-dimensional and they do not vanish
in a generic configuration. We checked this explicitly
for an electric NC $U(1)$ gauge theory (its energy functional
is given in equation \bref{jeje}). We completed the our study
of non-local theories with a BRST and field-antifield
analysis within this formalism.

\bigskip

Finally, we have been able to study the string duals of NC field theories
by means of D-branes wrapped in calibrated cycles in backgrounds
with non-zero $B_2$-fields. The first lesson we learnt was a technical
one: gauged supergravities do not help anymore in finding the
supergravity solutions. The reason why is that the (in this cases)
eleven-dimensional solution preserves a set of Killing spinors
that do not survive the (in this case $SU(2)$) compactification.
Remarkably, the resulting background does solve the gauged-supergravity
equations of motion but it does not preserve supersymmetry.
Thus one must choose between
\tem{
\item solving coupled {\it second order} partial differential equations
in 8d supergravity (which is simpler than 11d supergravity),
\item or solving coupled {\it first order} BPS equations directly
in 11d supergravity.
}
We managed to use this second option and produce the supergravity
dual of the NC deformation of the $\caln=2$ in 2+1 that we had
previously studied. By reducing along the appropriate circle
(avoiding the phenomenon of susy without susy) we obtained
a rather complicated IIA solution in which the metric depends
on two coordinates: one is the transverse direction to the D-branes
inside the $CY_3$ and one outside it. A simple probe analysis
showed that the moduli space exactly coincides with the commutative
one, a signal that the UV/IR mixing  leaves it unaffected.

By using an improved method based on a chain of T-dualities and the
addition of a constant $B_2$-field in one of the steps, we were able to
analyze the NC deformation of the Maldacena-\nun background, which
is dual to an $\caln=1$ $SU(N)$ gauge theory in 3+1 dimensions.
Because this theory (the commutative one) shares so many features of QCD,
the possibility of studying it non-perturbatively has received a lot of attention
in the last years. We have analyzed a series of features of its NC
version with the following results,
\tem{
\item there seems to be stronger effects of the UV/IR mixing. This is
the only case we are aware of in which the background does not reduce
to its commutative one in the deep IR, as can be seen by looking at
coordinate invariants such as the curvature scalar. Neither the RR-forms
completely reduce to the commutative ones.

\item the computation of the Wilson line shows confinement at large distances
with exactly the same string tension as in the commutative version of the theory.
The UV behavior does vary a lot, as expected in a region were the noncommutativity
scale is perfectly visible.

\item the computation of the $U(N)$ NC $\beta$-function shows that
it coincides exactly with the $SU(N)$ commutative one. This is already
to be expected from general perturbation theory
considerations~\cite{Armoni:2000xr,Matusis:2000jf}. Essentially,
the $U(1)$ and the $SU(N)/Z_N$ factors inside the NC $U(N)$ run
with the same coupling constant, and the divergent contributions
to it come from planar diagrams (as $\T$ acts as a cut off for non-planar ones).

\item the introduction of the NC scale $\Theta$ makes it possible to
decouple the unwanted KK modes. This condition, together with
negligibility of string loop corrections and the condition that the
curvatures be small can be written, in units where $l_s=1$,
\be
\frac{e^{-3\Phi_0}}{\T^2}\,\ll \, N \, \ll e^{-\Phi_0} \ll \T \,.
\ee
Thus we see that the only way to decouple the KK modes is to let
$\T$ be the largest scale of the problem. This shows that we cannot
use NC deformations of backgrounds with decoupling problems
if our aim is to end up with a realistic theory. However, this
procedure could help to clarify the role of the KK modes in the future.
}
Note that the fact that the background does not tend to its commutative
version in the deep IR does not seem to affect the deep IR
behavior of the confining string nor the $\b$-function. It would
be interesting to think of a way of probing this difference in the
supergravity side and then translating it into some field theory effect.

\medskip

The future of the research lines studied here seems to strongly depend
on our capability to understand string theory in backgrounds with RR fluxes,
as these appear inevitably in the closed string description of D-branes.
Progress along this direction would extend the validity of the stringy
calculations in a fashion similar to the BMN correspondence. Indeed, not
only the AdS/CFT duality would be much better understood, but
also the more general gauge/string correspondence, and the latter would immediately
improve our understanding of non-perturbative phenomena in QCD-like
theories. It may well happen that when quantization in such backgrounds
will be possible, all the closed string duals obtained so far
will receive a renewed interest from field theorists.
Finally, let us remark that all these considerations do not depend
on whether string/M-theory is the final theory describing nature
(if it describes it at all!), a fact that makes all results obtained
in this context more solid and that allows the gauge/string correspondence
to constitute an almost independent field on its own.

\appendix

\newcounter{zahler}
\setcounter{zahler}{0}

\renewcommand{\theequation}{A-\arabic{equation}}
\renewcommand{\thesection}{\Alph{zahler}}

\addtocounter{zahler}{1}
\setcounter{equation}{1}

\section{Superconformal algebra, representations and BPS operators} \label{ch2:sec:algebra}

It is of crucial importance to understand a few properties of
the superconformal algebra and its representations. Let us start
with recalling the commutation relations of the algebra. We will
write them in a schematic form,
\be
[M,P]  =  P \sac [M,K]=K  \sac [M,M]=M \sac [M,D] = 0 \nonumber \ee\be
[ D ,K ] = K \sac [D,P]=-P \sac [P,K]=M+D \,. \setcounter{equation}{1}
\ee
The first line just defines the tensorial behavior of the generators and
the second one assigns conformal weights $[K]=-1$ and $[P]=+1$.
Unitarity of the representation requires that all its states
have conformal weight (eigenvalue under $D$) $\Delta \ge 0$, so
by successive application of $K$ to any definite scaling state
we obtain one which is annihilated by all $K$'s; such a
state is called a  {\it conformal primary state}.

The supersymmetric extension of the conformal algebra introduces
Poincar\'e $Q$ and conformal $S$ fermionic generators whose
(anti)commutation relations read
\be
[D,S]=\undos S \sac [D,Q]=- \undos Q \sac \{Q,Q\}=P \sac \{S,S\}=K \nonumber
\ee\be
[K,Q]=S \sac [P,S]=Q \sac \{Q,S\}=M+D+R \,.
\ee
Again, the first two relations imply that $[S]=-1/2$ and $[Q]=1/2$. Repeating
the argument above, we can apply $S$ to any state until we reach one
annihilated by all the $S$'s. Such a state is called {\it superconformal
primary} if, in addition to this, it is not the case that this
state can be written as $Q$ acting on another state. This subtlety
is important because the construction of a superconformal multiplet
typically begins with the identification of such a superconformal
primary, and by successive application of the rest of the generators,
including the $Q$'s. If our candidate to start this procedure was
annihilated by all the $S$'s but it was the $Q$ of some other state,
then we would not construct the true whole multiplet but just a
part of it, since $Q^2=0$. To summarize:
\be
\mbox{Superconformal primary state } |O\rangle \,\,\,\,
\Longleftrightarrow  \,\,\,\, S_\a |O\rangle =0
, \,\, |O\rangle \neq Q |O'\rangle \nonumber
\ee

If the superconformal primary state  we start with is not annihilated
by any of the $Q$'s then the multiplet constructed upon it is
called a {\it long multiplet}. Some special cases occur when
the superconformal primary state is itself annihilated by some
of the $Q$'s, in which case its multiplet contains much less states
than a long one. It can be proven that the only possibilities are
that $1/n$ of the $Q$'s annihilate it with $n=2,4,8$. Such
superconformal primary states are called $1/n$ {\it chiral
primary states} and their corresponding supermultiplets
are called $1/n$-BPS multiplets. The name BPS is justified by the fact that the
superconformal primary states in these multiplets have the
smallest possible conformal dimension among all states with
the same remaining quantum numbers. It can be further proven
that the only nonzero charges of chiral primary state can
be $R$-symmetry charges. If we name their $SO(2)^3 \subset SO(6)$
Cartan charges by $(J_1,J_2,J_3)$, then the chiral primaries
saturate the BPS bound
\be
\Delta \ge |J_1| + |J_2| + |J_3| \,.
\ee
Having three, two or one nonzero charges $J_i$ turns the
corresponding multiplet into a $1/8$-, $1/4$- or $1/2$- BPS
one.
It is easy to intuitively understand how such bounds arise, for if
a state is annihilated by all of the $S$'s and at least one of the $Q$'s
then the $\{Q,S\}=M+D+R$ anticommutator automatically provides a
relation among the conformal dimension and the $R$-symmetry charges.
Being a purely algebraic relation, the conformal dimension
of such states does not depend on the YM coupling, and it is
therefore an exact non-renormalized statement. We remark
that this property is automatically inherited by the
descendants of the primary.

\medskip
\medskip

\noindent
{\bf From states to operators.} \label{ch2:sec:states-ops}

\noindent
We have avoided mentioning about operators in the discussion above.
However, conformal theories are known to admit a 1-to-1 map between
states in the radial quantization $\ket{O}$ and local operators $O(x)$,
the map being given by $\ket{O} = \lim_{x\rightarrow 0} O(x)$.

It is then immediate to ask what are the operators that correspond
to the {\it superconformal primary states}. A first intuitive argument
to find them is that such operators must be build only out of
scalar fields. The reason follows from observation of the supersymmetry
transformation of the \4n fields
\be
[Q, \phi] \sim \l \sac \{Q,\l\} \sim F_2 + [\phi,\phi] \sac
\{Q , \bar{\l} \} \sim D_{\mu} \phi \sac [Q,A_1] \sim \bar{\l}. \nonumber
\ee
As we are looking
for fields that are not the $Q$ of anything then they cannot be
fermions nor gauge fields, for they appear in the RHS of the $Q$-transformation
for scalars and fermions respectively. The same argument forces
the scalars to be symmetrized, as $[\phi^i,\phi^j]$ appears
in the $Q$-transformation of the fermions.
The simplest superconformal primaries are then writable as
\be \label{singletrace}
O_{1,n} = \str \left(\phi^{i_1}\phi^{i_n} \,...\,\phi^{i_n}\right) \,,
\ee
where $\str$ stands for the symmetrized trace and the subscripts in $O_{p,n}$ indicate
that the operator contains $p$ traces and $n$ scalars, so that the operator
\bref{singletrace} is a single-trace operator. Multiple trace operators
can be build by products of single-trace ones and they allow for the appearance
of partially anti-symmetrized representations of SO(6). The operators $O_{p,n}$
exhaust the list of known superconformal primary operators.

Let us now look for {\it chiral primary} operators, \ie we further require
that the superconformal primary $O_{p,n}$
is annihilated by at least one of the $Q$'s. The answer is less intuitive
so we just give the results:\footnote{In what follows we will often
use Dynkin labels instead of $SO(2)^3$ charges; the former are written
in squared brackets $[p_1,p_2,p_3]$ and the latter in parenthesis
$(J_1,J_2,J_3)$. The relation between the two is that $(J_1,J_2,J_3)$
corresponds to $[J_2-J_3,J_1-J_2,J_2+J_3]$ for $J_1\ge J_2 \ge J_3$.}
\tem{\item {\bf 1/2-BPS Operators.} All we need to do is to select the
traceless combination of an operator $O_{p,n}$. This means that
they all transform in the $[0,n,0]$ Dynkin irrep of SO(6). In terms
of complexified scalar fields
\be
X=\phi^1+i\phi^2 \sac Y=\phi^3+i\phi^4 \sac Z=\phi^5+i\phi^6\,,
\ee
the single and double trace $1/2$-BPS operators are simply
\bea
\calo_{1,J}&=&\Tr (\overset{J}{\overbrace{ X \, ...\, X}}) 
\nn
\calo_{2,J}&=& \sum_{p=0}^{J} a_{p} \,
\Tr (\overset{J-p}{\overbrace{ X \, ...\, X}})
\Tr (\overset{p}{\overbrace{ X \, ...\, X}}) \,,
\eea
where in the double trace operator one needs to choose the coefficients
$a_p$ in order to make the operator transform in the $[0,J,0]$
irrep.
\item {\bf 1/4-BPS Operators.} In this case the $1/4$ requirement
implies that one must select the $[l,k,l]$ Dynkin irrep
of SO(6) with $k+2l$ being equal to the number of scalars $n$ in
the operator. For $l=0$ we recover the $1/2$ operators, and
for $l\neq 0$ we clearly see that we need at least 2 traces.
So $1/4$-BPS operators are those $O_{p,n}$ with $p\ge 2$
that transform in $[l,k,l]$ Dynkin irreps with $k+2l=n$.
\item {\bf 1/8-BPS Operators.} In this case the $1/8$ requirement
implies that one must select the $[l,k,l+2m]$ Dynkin irrep
of SO(6) with $k+2l+3m=n$. We now see that we need at least 3 traces.
}
We summarize this discussion in the following table, which we adapt from \cite{D'Hoker:2002aw}.
The table refers to the chiral primary representative of the irrep and
it shows the number of $Q$'s that annihilate it, and its $R$-quantum
numbers in two formats, according to its charges under $SO(2)^3 \subset SO(6)$
or according to its Dynkin labels.
\begin{table}[h]
\begin{center}
\begin{tabular}{|c|c| c|c|c|c|c|} \hline
 Operator type & $\# Q$  & $SO(2)^3$ charges & Dynkin labels & dimension
$\Delta$ \\ \hline \hline
identity & 16  & $(0,0,0)$  & $[0,0,0]$             & 0        \\ \hline
1/2 BPS  & 8   & $(J_1,0,0)$  & $[0,k,0]$, \ $k\geq 2$  & $k $     \\ \hline
1/4 BPS  & 4   & $(J_1,J_2,J_2)$  & $[\ell,k,\ell]$, \ $\ell \geq 1$  & $k+2\ell$
\\ \hline
1/8 BPS  & 2   & $(J_1,J_2,J_3)$ & $[\ell,k,\ell+2m]$ & $k+2\ell+3m$, \ $m\geq 1$
\\ \hline
non-BPS  & 0       & any      & any          & unprotected \\ \hline
\end{tabular}
\end{center}
\caption{Characteristics of BPS and Non-BPS multiplets}
\label{table:3}
\end{table}

\addtocounter{zahler}{1}
\renewcommand{\theequation}{B-\arabic{equation}}
\newpage
\section{Generalization to strings with 3 independent angular momenta} \label{ch2:sec:generalization}

In this appendix we describe more general string solutions
which carry 3 independent angular momenta. The trick consists
on winding the string differently in the 3 planes contained in $\CR^6$,
as proposed in~\cite{Arutyunov:2003uj}.
We also redo the $\kappa$-symmetry analysis.
As we know that they all become tensionless in the limit,
we will start directly with the Lagrangian and $\kappa$-symmetry
transformations suitable for tensionless strings. This will
simplify the problem considerably.

Let us rewrite them metric \bref{metric} as
\be
\label{relevant}
ds^2= -dt^2 + d\t^2+\sin^2 \t d\phi^2 +
\sum_{i=1}^3 a_i^2 (d\a_i)^2 \,,
\ee
where $a_1 = \cos \t$,  $a_2=\sin\t \cos\phi$, $a_3=\sin\t \sin \phi$,
and $\a_i$ are polar angles in three orthogonal planes.
In order to avoid coordinate singularities, we will assume here that
$a_1 a_2 a_3$ is non-zero, which corresponds to the assumption that
the angular momentum two-form has maximal rank.

Let $(\tau,\s)$ be the worldsheet coordinates. In the gauge
$t=\tau$ the phase-space form of the Lagrangian density for a
tensionless string in the above background is
\be
L = p_\t \dot \t + p_\phi \dot\phi + \sum_i p_i \dot \alpha_i - H
-s\left(p_\t \t'+p_\phi \phi'+ \sum_i p_i \alpha_i'\right) \,,
\ee
where $s$ is the Lagrange multiplier for the string reparametrization
constraint, and $H$ is the Hamiltonian density
\be
H= \sqrt{ p_\t^2 + \frac{p_\phi^2}{\sin\t^2} +
\sum_i \frac{p_i^2}{a_i^2}} \,.
\ee
Note that the $\alpha_i$ equation of motion is
$\dot p_i = \left(sp_i\right)'$,
from which it follows that the angular momentum,
$J_i = \oint d\sigma \, p_i$, is a constant of motion.

We seek solutions for which $\t$ and $\phi$ are constant and
$p_\t=p_\phi=0$. This requires, in order to solve the
corresponding equations of motion, that we set
\be
\label{one}
a_i^2 = |p_i| \Big/ \sum_j |p_j| \,,
\ee
from which we see that
\be \label{ham}
H= \sum_i |p_i| \,.
\ee

Given \bref{one}, the $p_i$ equations of motion reduce to
$\dot\alpha_i -s \alpha_i' = 1$, while the constraint imposed by
$s$ is $\sum_i p_i \alpha'_i =0$.
We may choose a gauge for which  $s=0$, in which case the equations
above have the solution\footnote{This is not the unique solution, but
it is the one of relevance when one considers the tensionless string
as a limit of the tensionful string.}
\be \label{solution-three}
\alpha_i = t + m_i \sigma \sac p_i = J_i /2\pi \,,
\ee
for integers $m_i$  satisfying $\sum_i m_i J_i  = 0$.
Because each $p_i$ has a definite sign, integration of \bref{ham} yields
the total energy
\be
E = |J_1|+ |J_2|+ |J_3| \,.
\ee
An argument analogous to that of section \ref{ch2:sec:algebra} shows that this energy
saturates a  BPS bound implied by the $PSU(2,2|4)$
supersymmetry algebra of \adss{5}{5}. Because of this,
the rotating strings are supersymmetric. Since we have assumed that
all $J_i$ are non-zero, the fraction of supersymmetry preserved is
1/8, as we will confirm below.
If only two $J$'s are non-zero, then 1/4 supersymmetry is
preserved. And if only one $J$ is non-zero, then 1/2 supersymmetry is
preserved.

Note that the strings with two equal angular momenta considered
in section~\ref{ch2:sec:ads-beyond} are
a subclass of these more general ones; as soon
as we the winding numbers $m_i$ are different,
one can have three different angular momenta. The novelty
is that now it is not true anymore that the string lies in
a two-plane at each moment; it describes a closed curve that
spans (in general) the whole $\CR^6$.

\medskip

Let us now redo the $\kappa$-symmetry analysis of section \ref{ch2:sec:k-symmetry} for
these new cases.
The condition for a IIB superstring to be supersymmetric in
the ultra-relativistic limit in which it becomes null is
\be
p_M \, e^M_{\,\,\,\,\,\,A} \, \G^A \, \epsilon =0 \,,
\label{susy}
\ee
where $p_M$ is the ten-momentum and $e^M_{\,\,\,\,\,\,A}$ is the obvious
orthonormal frame associated to the metric \bref{relevant}.
The spinor \bref{ads5-spinor} in our renamed coordinates is
\be
\e=e^{i \frac{t}{2} \tilde{\G}} e^{i \frac{\t}{2} \G_{\phi 123}}
e^{\frac{\phi}{2}\G_{\t\phi}} e^{\fc{i}{2} \left(
\a_1 \G_{\t\phi 23} + \a_2 \, i \G_{2 \t} + \a_3 \, i \G_{3\phi}
\right)} \, \e_0 \,,
\label{epsilon}
\ee
where recall that $\tilde{\G}$ commutes with all other matrices in the problem
and $\G_i \equiv \G_{\a_i}$. It is clear from this expression that
the matrices occurring inside the brackets in the last exponential
generate, when acting on $\e_0$, rotations in each of the three
orthogonal two-planes parametrized by $\a_i$. Note that they all
square to unity (so they have eigenvalues $\pm 1$), they are
mutually-commuting, and the product of any two of them yields the
third one, up to a sign.

For the rotating string solutions of the previous section, the
supersymmetry condition \bref{susy} reduces to
\be
a_i \, \G_{ti} \, \epsilon = \epsilon \,,
\label{aaa}
\ee
where we have assumed, for definiteness, that all $J_i$ are positive.
If the angular momentum two-form has maximum rank then all $a_i$
are non-zero, and equation \bref{aaa} is equivalent to
\be
i\G_{2\t} \, \e_0 = \e_0 \sac
i\G_{3\phi} \, \e_0 = \e_0 \sac
\G_{t1} \, \e_0 = \e_0 \,.
\label{susybis}
\ee
Let us first show that \bref{susybis} implies \bref{aaa}.
We see from the first two conditions in \bref{susybis}
that $\e_0$ is an eigen-spinor of the last exponential in
equation \bref{epsilon}, so this exponential cancels on both sides
of \bref{aaa}. The first exponential also cancels because $\tilde{\G}$
commutes with all other matrices in equation \bref{aaa}, so the latter
may be rewritten as
\be
a_i \, \G_{ti} \, \e_0 = e^{-\frac{\phi}{2}\G_{\t\phi}}
e^{i \t \G_{\phi 123}} e^{\frac{\phi}{2}\G_{\t\phi}} \, \e_0 =
\left( a_1 + a_2 \, i \G_{\phi 123} +
a_3 \, i \G_{123\t} \right) \, \e_0 \,,
\ee
which is identically satisfied by virtue of \bref{susybis}.

With some further algebra, it can also be shown that \bref{aaa} implies
\bref{susybis}, but we will not do this here. Instead, we note that,
being mutually-commuting projections, the conditions \bref{susybis}
imply that the fraction of preserved supersymmetry is 1/8,
as expected for a string carrying three independent angular
momenta. We also recall that, as observed above, they imply that
$\e_0$ is invariant (up to a phase) under rotations in the
$\a_i$-planes associated to the non-zero components of the angular
momentum two-form.

\addtocounter{zahler}{1}
\renewcommand{\theequation}{C-\arabic{equation}}
\setcounter{equation}{1}

\newpage
\section{Conventions for the Maldacena-\nun background} \label{ch6:sec:conventions}
In this appendix we collect the conventions and definitions
necessary to read the Maldacena-\nun background in \bref{metrica}-\bref{com-dilato},
referring the reader to the original references for a careful derivation.
There are two functions of the radial variable $\rho$, which are given by
\bea
e^{2g(\rho)}&=&\rho \coth 2\rho - {\rho^2 \over \sinh^2 2\rho} -{1\over 4}\,, \nn
\label{A}
a(\rho)&=&{2\rho \over \sinh 2\rho}\,.
\eea
The $SU(2)$ gauge-field $A$ is parametrized by
\be \label{definemya}
A=\undos\left[\s^1 a(\rho) d\theta+\s^2 a(\rho) \sin\t d\phi + \s^3 \cos\t d\phi\right]\,,
\ee
where
\be
A=A^a{\s^a\over 2}\sac F=F^a {\s^a\over 2} \,,
\ee
and $F^a_{\mu\nu}=\partial_{\mu}A^a_{\nu}-\partial_{\nu}A^a_{\mu}
+\epsilon^{abc}A^b_{\mu}A^c_{\nu}$ should be understood.
Finally, the $SU(2)$ left-invariant one forms parametrizing the transverse $S^3$ are
\bea
w^1+i w^2 &=& e^{-i\psi}\left( d\t_1 +i \sin d\phi_1 \right) \,, \nn
\label{leftinvform}
w^3&=&d\psi+cos\theta_1 d\phi_1.
\eea

\addtocounter{zahler}{1}
\renewcommand{\theequation}{D-\arabic{equation}}
\setcounter{equation}{1}

\newpage
\section{$U(1)$ commutative Maxwell theory in $\dpo$ dimensions} \label{ch4:sec:u1-c}

Our $d+1$ formalism can also be used for describing ordinary
local theories. As an example of this, we will show
how the $U(1)$ commutative (and therefore local)
Maxwell theory is formulated using
the $\dpo$ dimensional canonical formalism developed for non-local
theories and see
how it is reduced to the standard canonical formalism
in $d$ dimensions.

The canonical $d+1$ system
is defined by the Hamiltonian \bref{h} and two constraints,
 \bref{vp} and \bref{EOM}.
The Hamiltonian is
\bea
H&=&\int d^dx\;[{\it\Pi}^\nu(t,x)\pa_{x^0}{\cala}_\nu(t,x)-\d(x^0)\call(t,x)~],
\label{Hama}
\eea
where
\bea
\call(t,x)&=&-\frac14~\calf_{\mu\nu}(t,x)\calf^{\mu\nu}(t,x),
\\
\calf_{\mu\nu}(t,x)&=&\pa_\mu {\cala}_\nu(t,x)-\pa_\nu {\cala}_\mu(t,x).
\eea
The momentum constraint  \bref{vp} is
\bea
\vp^\nu(t,x)&=&{\it\Pi}^\nu(t,x)+\int dy~\chi(x^0,-y^0)\;
\calf^{\mu\nu}(t,y)\;\pa^y_\mu\d(x-y)
\nn\\
 &=&
{\it\Pi}^\nu(t,x)+\d(x^0)\calf^{0\nu}(t,x)\approx 0
\label{vpu1a} \,,
\eea
and the constraint \bref{EOM} is
\bea
\7\vp^\nu(t,x)&=&\pa_\mu \calf^{\mu\nu}(t,x)\approx 0.
\label{EOMu1a}
\eea
The generator of the {$U(1)$} transformation
is given, using \bref{G}, by
\bea
G[\Lam]&=&\int dx[\;{\it\Pi}^\mu \pa_\mu \Lam\;].
\label{defGa}
\eea
It can be expressed as a linear combination of the constraints,
\bea
G[\Lam]&=&\int dx~\Lam\left[-(\pa_\mu\vp^\mu)-\d(x^0)\7\vp^0\right].
\label{GLama}
\eea
The Hamiltonian is expressed using the constraints and the $U(1)$ generator
as
\be
H=G[{\cala}_0]\;+\;\int dx\;\vp^i\;\calf_{0i}\;+\;
\int dx\; \d(x^0)\;\{\frac12 {\calf_{0i}}^2+\frac14 {\calf_{ij}}^2 \}.
\label{Hamb}
\ee
\medskip

The  Hamiltonian \bref{Hamb} as well as the
constraints \bref{vpu1a} and \bref{EOMu1a} contain
no time ($t$) derivative and they are functions of the canonical pairs
$({\cala}_\mu(t,x),{\it\Pi}^\mu(t,x))$.
They are conserved since the Maxwell \lag in $d$ dimensions
has time translation invariance. The $U(1)$ generator is also conserved,
 even without using the constraints,
\bea
\frac{d}{dt}G[\Lam]&=&\{G[\Lam],H\}~+~\frac{\pa}{\pa t}G[\Lam]~=~0
\label{dtGLam2} \,,
\eea
for
$\Lam(t,x)$ satisfying \bref{Lamdd},
\be \dot \Lam=\pa_{x^0}\Lam \,, \ee
in agreement with \bref{dtG}.
The condition on $\Lam$ implies that the $U(1)$ transformations
in the $\dpo$ dimensional canonical formulation
are not {gauge} but {rigid} ones. We will see below how the
gauge transformations are recovered when reducing back
to the original $d$ dimensional formalism.
\vskip 4mm

In cases where our  Lagrangians are local
or higher derivative ones we can expand our fields
as in \bref{momentahigh} to reduce back to $d$ dimensions,
\be
{\cala}_\mu(t,x) \equiv \sum_{m=0}^{\infty}~e_m(x^0)~A_\mu^{(m)}(t,\bx) \sac
\iPi^\mu(t,x) \equiv\sum_{m=0}^{\infty}~e^m(x^0)~\iPi^\mu_{(m)}(t,\bx),
\label{Taylor}\ee
The fields $(A_\mu^{(m)}(t,\bx),\iPi^\mu_{(m)}(t,\bx))$ are
the new symplectic coordinates in $d$ dimensions.
In terms of them, the constraint \bref{vpu1a} can be expressed as
\bea
\vp^\mu(t,x)&=&\sum_{m=0}^{\infty}e^m(x^0)\vp^\mu_{(m)}(t,\bx),
\\
\vp^0_{(m)}(t,\bx)&=&\iPi^0_{(m)}(t,\bx)~=~0
\sac (m\geq 0),
\label{vp0m}\\
\vp^i_{(0)}(t,\bx)&=&\iPi^i_{(0)}(t,\bx)-
({\cala}_i^{(1)}(t,\bx)-\pa_i{\cala}_0^{(0)}(t,\bx))~=~0,
\label{vpi0}\\
\vp^i_{(m)}(t,\bx)&=&\iPi^i_{(m)}(t,\bx)~=~0
\sac (m\geq 1).
\label{vpim}
\label{vpu1acomp}
\eea
and the constraint \bref{EOMu1a} as
\bea
\7\vp^\mu(t,x)&=&\sum_{m=0}^{\infty}e_m(x^0)\7\vp^{\mu{(m)}}(t,\bx),
\\
\7\vp^{i(m)}(t,\bx)&=&\pa_j\left(\pa_j{\cala}_i^{(m)}(t,\bx)-\pa_i{\cala}_j^{(m)}
(t,\bx)\right)\nn
& -&\left({\cala}_i^{(m+2)}(t,\bx)-\pa_i{\cala}_0^{(m+1)}(t,\bx)\right)=0
\label{tvpim} , (m\geq 0)
\\
\7\vp^{0(m)}(t,\bx)&=&\pa_i\left({\cala}_i^{(m+1)}(t,\bx)-\pa_i{\cala}_0^{(m)}(t,\bx)
\right)=0 , \,(m\geq 0).
\label{tvp0m}\label{EOM1acomp}
\eea
Due to the identities
\bea
\7\vp^{0(m+1)}(t,\bx)&=&\pa_i\7\vp^{i(m)}(t,\bx) \sac (m\geq 0)\,,
\eea
the only independent constraint of \bref{tvp0m} is the $m=0$
case. It can be expressed, using \bref{vpi0}, as the gauss law constraint,
\bea
\7\vp^{0(0)}(t,\bx)&=&\pa_i
\iPi^i_{(0)}(t,\bx)~=~0.~
\label{gauss}\eea

Following the Dirac's standard procedure for dealing with constraints \cite{Dirac:1950pj},
we classify them
and eliminate the second class ones.
The constraints \bref{vpim} with $(m\geq 2)$ are paired with
the constraints \bref{tvpim} with $(m\geq 0)$ to form second class sets.
They are used to eliminate the canonical pairs
$({\cala}_i^{(m)}(t,\bx),\iPi^i_{(m)}(t,\bx)),(m\geq 2)$,
\bea
{\cala}_i^{(m)}(t,\bx)&=&
\pa_j(\pa_j{\cala}_i^{(m-2)}(t,\bx)-\pa_i{\cala}_j^{(m-2)}
(t,\bx))+\pa_i{\cala}_0^{(m-1)}(t,\bx),
\nn
\iPi^i_{(m)}(t,\bx))&=&0,~~~~~~~~~~~(m\geq 2).
\eea
The constraints \bref{vpim} with $(m=1)$ and
\bref{vpi0} are paired to a second class set and
can be used to eliminate
$({\cala}_i^{(1)}(t,\bx),\iPi^i_{(1)}(t,\bx))$,
\bea
{\cala}_i^{(1)}(t,\bx)&=&\iPi^i_{(0)}(t,\bx)+\pa_i{\cala}_0^{(0)}(t,\bx),~
\\
\iPi^i_{(1)}(t,\bx)&=&0.
\eea

After eliminating the canonical pairs
$({\cala}_i^{(m)}(t,\bx),\iPi^i_{(m)}(t,\bx)),(m\geq 1)$
using the second class constraints, the system is described
in terms of the canonical pairs
$({\cala}_i^{(0)}(t,\bx),\iPi^i_{(0)}(t,\bx))$ and
$({\cala}_0^{(m)}(t,\bx),\iPi^0_{(m)}(t,\bx)),(m\geq 0)$.
The Dirac brackets among them remain the same as the Poisson brackets.
Remember that the $d$ dimensional fields can be read from \bref{ini-con} as
\be
A_\mu(t,\bx)={\cala}_\mu(t,0,\bx)~=~{\cala}_\mu^{(0)}(t,\bx) \sac
\Pi^\mu(t,\bx)=\iPi^\mu_{(0)}(t,\bx).
\ee
The remaining constraints are \bref{gauss} and \bref{vp0m},
\bea
\pa_i\iPi^i_{(0)}(t,\bx)~=~0,~~~~~~~~~
\iPi^0_{(m)}(t,\bx)~=~0\sac (m\geq 0)\,.
\label{firstcc}
\eea
They are first class constraints.
The Hamiltonian \bref{Hamb} in the reduced variables is
\bea
H(t)&=&\int d\bx\left[~
\sum_{m=0}^{\infty}~{\cala}_0^{(m+1)}(t,\bx)\iPi^0_{(m)}(t,\bx)~-~
{\cala}_0^{(0)}(t,\bx)(\pa_i\iPi^i_{(0)}(t,\bx))\right.
\nn\\
&&\left.~+~
\frac12 (\iPi^i_{(0)}(t,\bx))^2~+~
\frac14 (\pa_j{\cala}_i^{(0)}(t,\bx)-\pa_i{\cala}_j^{(0)}(t,\bx))^2\right].
\label{hh}
\eea
The $U(1)$ generator \bref{GLama} is
\be
G[\Lam]=\int d\bx~
\left[~
\sum_{m=0}^{\infty}~\Lam^{(m+1)}(t,\bx)\iPi^0_{(m)}(t,\bx)~-~
\Lam^{(0)}(t,\bx)(\pa_i\iPi^i_{(0)}(t,\bx))\right],
\label{Lamlam}
\ee
where
\be
\Lam(t,\l)= \sum_{m=0}^{\infty}~\Lam^{(m)}(t,\bx)e_m(x^0)\sac
{\rm and}~~~~\dot\Lam^{(m)}(t,\bx)=\Lam^{(m+1)}(t,\bx).
\ee

The first class constraints $\iPi^0_{(m)}(t,\bx)=0,(m\geq 0)$ in
\bref{firstcc} mean that ${\cala}_0^{(m)}(t,\bx),$ $(m\geq 0)$ are the
gauge degrees of freedom and we can  assign to them
{ any function of $\bx$
for all values of $m$ at given time}
$t=t_0$. It is equivalent to saying that we can assign
{any function of time
to ${\cala}_0^{(0)}(t,\bx)$ for all value of $t$},
due to the equation of motion
$\dot{\cala}_0^{(m)}(t,\bx)={\cala}_0^{(m+1)}(t,\bx)$.
In this way we can understand that the Hamiltonian \bref{hh}
is equivalent to the standard form of the canonical Hamiltonian of
the Maxwell theory,
\bea
H(t)&=&\int d\bx\left[~
\dot A_0(t,\bx)\Pi^0(t,\bx)~-~
A_0(t,\bx)(\pa_i\Pi^i(t,\bx))
\right.\nn\\&&\left.~+~\frac12 (\Pi^i(t,\bx))^2~+~
\frac14 (\pa_jA_i(t,\bx)-\pa_iA_j(t,\bx))^2\right]
\label{hhm} \,,
\eea
in which $A_0(t,\bx)$ is an arbitrary function of time.
In the same manner the $U(1)$ generator \bref{Lamlam} is
\bea
G[\Lam]&=&\int d\bx~
\left[~
\dot\l(t,\bx)\Pi^0(t,\bx)~-~
\l(t,\bx)(\pa_i\Pi^i(t,\bx))\right],
\label{Lamlamm}
\eea
in which the gauge parameter function $\l(t,\bx)\equiv\Lam^{(0)}(t,\bx)$
is regarded as any function of time.
\vskip 6mm


\backmatter

\bibliographystyle{tonim}
\bibliography{Finalbib}

\end{document}